\documentclass[11pt,a4paper]{scrartcl}

\usepackage{CLICdp}
\usepackage{amssymb}
\usepackage{CLICdp_definitions}
\usepackage{float}
\usepackage{adjustbox}
\usepackage{sansmath}
\usepackage{pgf-pie}
\usepackage{pgfplots, pgfplotstable}
\usetikzlibrary{positioning}
\usepackage{threeparttable}
\usepackage{multirow}
\usepackage{makecell}
\usepackage{etoolbox}
\usepackage{afterpage}
\robustify\bfseries

\newcommand{\imgc}[1]{(image credit: #1)}
\newcommand{\imcl}{\imgc{CLIC}}
\newcommand{\imdp}{\imgc{CLICdp}}

\title{The Compact Linear $\Pep\Pem$ Collider (CLIC)\newline 2018 Summary Report}

\collName{The CLIC and CLICdp collaborations}

\addauthor{
T.K.~Charles, 
P.J.~Giansiracusa,
T.G.~Lucas,
R.P.~Rassool,
M.~Volpi$^{1}$\\
\emph{University of Melbourne, Melbourne, Australia}
}

\addauthor{
C.~Balazs\\ 
\emph{Monash University, Melbourne, Australia}
}

\addauthor{
K.~Afanaciev, 
V.~Makarenko\\
\emph{Belarusian State University, Minsk, Belarus}
}

\addauthor{
A.~Patapenka, 
I.~Zhuk\\
\emph{Joint Institute for Power and Nuclear Research - Sosny, Minsk, Belarus}
% ...
}

\addauthor{
\begin{flushleft}
{$^{1}$}Also at CERN, Geneva, Switzerland\\
% ...
\end{flushleft}
}

\abstract{The Compact Linear Collider (CLIC) is a \si{\TeV}-scale high-luminosity linear \epem collider under development at CERN.
Following the CLIC conceptual design published in \num{2012}, this report provides an overview of the CLIC project, its current status, and future developments.
It presents the CLIC physics potential and reports on design, technology, and implementation aspects of the accelerator and the detector.
For an optimal exploitation of its physics potential, CLIC is foreseen to be built and operated in stages, at centre-of-mass energies of \SI{380}{\GeV}, \SI{1.5}{\TeV} and \SI{3}{\TeV}, respectively, for a site length ranging from \SI{11}{\km} to \SI{50}{\km}. CLIC uses a two-beam acceleration scheme, in which normal-conducting high-gradient \SI{12}{\GHz} accelerating structures are powered via a high-current drive beam. For the first stage, an alternative with X-band klystron powering is also considered.
CLIC accelerator optimisation,
technical developments and system tests have resulted in significant progress in recent years. 
Moreover, this has led to an increased energy efficiency (power around \SI{170}{\mega\watt})
for the \SI{380}{\GeV} stage, together with a reduced cost estimate at the level of \mbox{\num{6} billion \si{CHF}}.
The detector concept, which matches the physics performance requirements and the CLIC experimental conditions, has been refined using improved software tools for simulation and reconstruction. Significant progress has been made on detector  technology developments for the tracking and calorimetry systems.
A wide range of CLIC physics studies has been conducted, both through full detector simulations with overlay of beam-induced backgrounds, and through parametric studies, together providing a broad overview of the CLIC physics potential. Each of the three energy stages adds cornerstones of the full CLIC physics programme, such as Higgs width and couplings, top-quark properties, Higgs self-coupling, direct searches, and many precision electroweak measurements. The interpretation of the combined results gives crucial and accurate insight into new physics, largely complementary to LHC and HL-LHC. 
The construction of the first CLIC energy stage could start by \num{2026}. First beams would be available by \num{2035}, marking the beginning of a broad CLIC physics programme spanning \SIrange{25}{30}{\years}.
}

\graphicspath{ {./logos/}{./figures/} }

\newlength{\abc}
\AtBeginDocument{%
\renewcommand{\ref}[1]{\mbox{\Cref{#1}}}

}

\begin{document}

\pagenumbering{roman}
\begin{titlepage}
\noindent
\setlength{\unitlength}{1mm}
\begin{center}
\begin{tikzpicture}[overlay]
\node[anchor=west] at (5,0) {CERN-2018-005-M};
\node[anchor=west] at (5,-0.5) {14 December 2018};
\node at (0,-4) {\includegraphics[width=15cm]{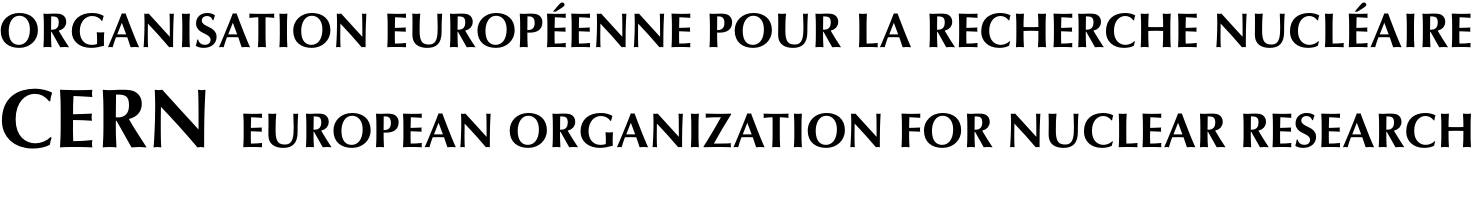}};
\node at (0,-11) {\includegraphics[width=11.25cm]{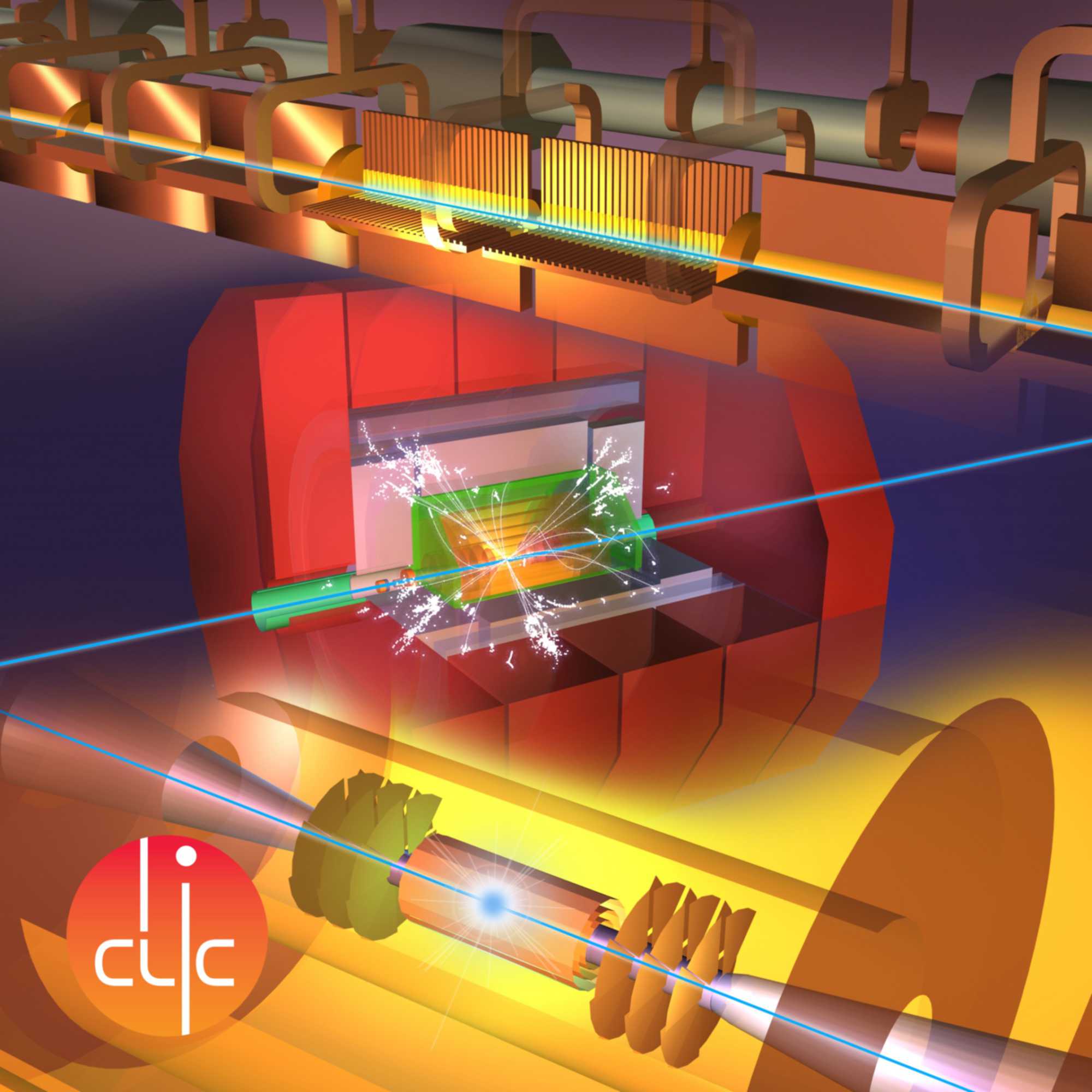}};
\node at (0,-19) {\huge\bfseries {\scshape The Compact Linear $\Pep\Pem$ Collider (CLIC)}};
\node at (0,-20) {\huge\bfseries {\scshape 2018 Summary Report}};
\node at (0,-23) {GENEVA};
\node at (0,-23.5) {2018};
\end{tikzpicture}
\end{center}
\newpage

\thispagestyle{empty}
\mbox{}
\vfill

\begin{flushleft}
CERN Yellow Reports: Monographs\\
Published by CERN, CH-1211 Geneva 23, Switzerland\\[3mm]

\begin{tabular}{@{}l@{~}l}
  ISBN & 978--92--9083--506--6 (paperback) \\
  ISBN & 978--92--9083--507--3 (PDF) \\
  ISSN & 2519-8068 (Print)\\ 
  ISSN & 2519-8076 (Online)\\
  DOI & \url{https://doi.org/10.23731/CYRM-2018-002}\\
\end{tabular}\\[3mm]
Accepted for publication by the CERN Report Editorial Board (CREB) on 10 December 2018\\
Available online at \url{http://publishing.cern.ch/} and \url{http://cds.cern.ch/}\\[3mm]

Copyright \copyright{} CERN, 2018\\[1mm]
\raisebox{-1mm}{\includegraphics[height=12pt]{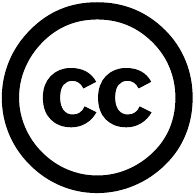}}
 Creative Commons Attribution 4.0\\[1mm]
Knowledge transfer is an integral part of CERN's mission.\\[1mm]
CERN publishes this volume Open Access under the Creative Commons Attribution 4.0 license\\
(\url{http://creativecommons.org/licenses/by/4.0/}) in order to permit its wide dissemination and use.\\
The submission of a contribution to a CERN Yellow Report series shall be deemed to constitute the contributor's agreement to this copyright and license statement. Contributors are requested to obtain any clearances that may be necessary for this purpose.\\[5mm]

This volume is indexed in: CERN Document Server (CDS), INSPIRE.\\[5mm]

This volume should be cited as:\\[1mm]
The Compact Linear Collider (CLIC) -- 2018 Summary Report, edited by P.N.~Burrows, N.~Catalan~Lasheras, L.~Linssen, M.~Petri\v{c}, A.~Robson, D.~Schulte, E.~Sicking, S.~Stapnes, CERN Yellow Reports: Monographs, Vol. 2/2018, CERN-2018-005-M (CERN, Geneva, 2018). https//doi.org/10.23731/CYRM-2018-002
\end{flushleft}

\cleardoublepage

\thispagestyle{empty}
\vspace*{3cm}
\begin{center}
  \large{\bfseries\sffamily Abstract}
\end{center}
\begin{quotation}
\noindent\MyAbstract
\end{quotation}
\vfill

\begin{center}
{\large {\bfseries\sffamily Corresponding editors}} 
\vspace*{0.25cm}

{
Philip N.\ Burrows (University of Oxford), Nuria Catalan Lasheras (CERN), \\ Lucie Linssen (CERN), Marko Petri\v{c} (CERN), \\ Aidan Robson (University of Glasgow), Daniel Schulte (CERN), \\Eva Sicking (CERN), Steinar Stapnes (CERN)
}
\end{center}
\end{titlepage}

\cleardoublepage
\textcolor{white}{ }
\thispagestyle{empty}
\newpage

\begin{center}
  \Large\bfseries\sffamily\MyCollName
\end{center}

{
\setcounter{footnote}{0}\def\@currentlabel{}
\begingroup\def\thefootnote{\arabic{footnote}}
\def\@makefnmark{\hbox{$^{\@thefnmark)}$}}
\large
{
\MyAuthors
\par}
\endgroup

\vspace{-5mm}
\clearpage
\normalsize
\pagebreak

\cleardoublepage
\textcolor{white}{ }
\thispagestyle{empty}

\tableofcontents
\clearpage

\pagenumbering{arabic}

\section{Introduction}
\label{sec:introduction}

The Compact Linear Collider (CLIC) is a multi-\si{\TeV} high-luminosity linear \epem collider under development by the CLIC accelerator collaboration~\cite{clic-study}.
CLIC uses a novel two-beam acceleration technique, with normal-conducting accelerating structures operating in the range of \SIrange{70}{100}{\mega\volt/\meter}.
Detailed studies of the CLIC physics potential, design of a detector for CLIC, and R\&D on detector technologies are performed by the CLIC detector and physics (CLICdp) collaboration~\cite{clic-study}.
The CLIC Conceptual Design Report (CDR) was published in \num{2012}~\cite{cdrvol1,cdrvol2,cdrvol3}. 
The main focus of the CDR was to demonstrate the feasibility of the CLIC accelerator at high energy (\SI{3}{\TeV}) and to confirm that high-precision physics measurements can be performed, 
despite the luminosity spectrum and the presence of particles from beam-induced background. 

Following the completion of the CDR, 
detailed studies on Higgs and top-quark physics, with particular focus on the first energy stage of CLIC, concluded that the optimal centre-of-mass energy for the first stage 
is \SI{380}{\GeV}.
As a result, a comprehensive optimisation study of the CLIC accelerator complex was performed, 
by scanning the full parameter space for the accelerating structures, and by using the CLIC performance, 
cost and energy consumption as a gauge for operation at \SI{380}{\GeV} and \SI{3}{\TeV}. 
The results led to optimised accelerator design parameters for the proposed CLIC staging scenario, 
with operation at \SI{380}{\GeV}, \SI{1.5}{\TeV} and \SI{3}{\TeV}, as reported in~\cite{StagingBaseline}.

This report summarises progress and results of the CLIC studies at the time of submitting input to the update of the European Strategy for Particle Physics, in December \num{2018}. 
The report describes recent achievements in accelerator design, technology development, system tests and beam tests.
Large-scale CLIC-specific beam tests have taken place, for example, at the CLIC Test Facility CTF3 at CERN~\cite{Geschonke2002}, 
at the Accelerator Test Facility ATF2 at KEK~\cite{Kuroda2016,Okugi2016}, at the FACET facility at SLAC~\cite{FACET} 
and at the FERMI facility in Trieste~\cite{FERMI}. 
Crucial experience also emanates from the expanding field of Free Electron Laser (FEL) linacs and recent-generation light sources. 
Together they provide the demonstration that all implications of the CLIC design parameters are well understood and reproduced in beam tests. Therefore the CLIC performance goals are realistic.
An alternative CLIC scenario for the first stage, where the accelerating structures are powered by X-band klystrons, is also under study.
The implementation of CLIC near CERN has been investigated. Principally focusing on the \SI{380}{\GeV} stage, this includes civil engineering aspects, electrical networks, cooling and ventilation,
installation scheduling, transport, and safety aspects. 
All CLIC studies have put emphasis on optimising cost and energy efficiency, and the resulting power and cost estimates are reported.  

Since the completion of the CDR, the CLIC detector was further optimised through a broad range of simulation studies, resulting in the CLICdet detector design~\cite{CLICdet_note_2017, CLICdet_performance}.
In order to enlarge the angular acceptance of the detector, the final focusing quadrupoles are now placed outside the detector in the accelerator tunnel. 
The software suite for simulation and event reconstruction was modernised and tuned for use with CLICdet. Detector technology developments have focused on the most challenging aspects of the experiment, 
namely the light-weight silicon vertex and tracker system and the highly-granular calorimeters. 
The detector R\&D activities have resulted in technology demonstrators, showing that the required performance is already achievable or will be achieved in the next phase, compatible with the CLIC timescale~\cite{ESU18RnD}.

The physics potential at the three CLIC energy stages has been explored in much detail.
The first stage of CLIC provides collisions at $\roots=\SI{380}{\GeV}$.
This gives access to Higgs boson measurements through Higgsstrahlung and WW-fusion~\cite{ClicHiggsPaper},
thereby providing accurate model-independent measurements of the Higgs couplings and the Higgs width. 
Precision top-quark physics~\cite{ClicTopPaper} is also addressed at this energy, 
and a fraction of the running time will be devoted to a threshold scan of top-quark pair production around \SI{350}{\GeV}.

The second stage, with collisions foreseen at \SI{1.5}{\TeV}, opens the energy frontier,
allowing for the discovery of new physics phenomena~\cite{cdrvol3, ESU18BSM}.
The double Higgsstrahlung process $\epem\to\PZ\PH\PH$ can be observed 
and additional Higgs and top-quark properties are within reach, 
such as the Higgs self-coupling and rare Higgs branching ratios. 
The third stage at \SI{3}{\TeV} further enlarges the CLIC physics potential,
e.g.\ allowing discovery of new electroweak particles or dark matter candidates, 
which may be more easily observed at CLIC than at the HL-LHC. 
The \SI{3}{\TeV} stage also provides the best sensitivity to new physics processes at much higher energy scales, via indirect searches.

\ref{sec:physics} of this report gives an overview of physics measurements at CLIC, demonstrating how they improve our knowledge of the Standard Model and 
how CLIC results have an impact on understanding the nature and scale of new physics Beyond the Standard Model (BSM).
\ref{sec:accelerator} provides an overview of the CLIC accelerator design and performance 
at \SI{380}{\GeV} for both the two-beam baseline design and the klystron-based option.
This section also describes the path to the higher energies, \SI{1.5}{\TeV} and \SI{3}{\TeV}, and gives an overview of the key CLIC technology developments. 
Referring to beam experiments and hardware tests,~\ref{sec:accelerator} also describes key achievements providing evidence that the CLIC performance goals can be met.

In~\ref{sec:detector} the CLIC detector and its performance results through simulation and event reconstruction are described. 
The progress made and the status of the detector technology developments are summarised. 
\ref{sec:project} describes the present plans for the implementation of CLIC, with emphasis on the 380\,\GeV stage. 
It reports on civil engineering and schedule aspects, and provides estimates of the energy consumption and of the cost for construction and operation. 
Physics motivation and options to expand the energy reach of CLIC using future technologies are discussed in~\ref{sec:opportunities}. 
The CLIC objectives for the period \numrange{2020}{2025} are outlined in~\ref{sec:objectives}. 
\ref{sec:summary} provides a short summary of this report.

\clearpage
\section{CLIC physics overview}
\label{sec:physics}

\subsection{CLIC physics exploration at three energy stages}

CLIC's physics programme will substantially improve our knowledge and probe
the open questions arising from the LHC and other particle physics facilities, and from
related astronomical observations. 
As CLIC gives access to a very wide range of energies, it can reach unprecedented precision
in the properties and interactions of the Higgs boson, top quark, and electroweak gauge bosons.
It has the potential to make discoveries of new states that are inaccessible at
other facilities, including the possible discovery of dark matter, and could potentially 
give some experimental insight to cosmological questions such as the stability
or instability of the vacuum and the origin of the baryon asymmetry.
The science program ranges from the `guaranteed physics' of precision studies of the
Standard Model (SM), which through effective field theory interpretations give access
beyond the capacity of other facilities to new physics at high scales,  
to `prospect physics' of directly producing new states or observing new interactions. 

\paragraph{Detector and experimental environment}
A single optimised CLIC detector, CLICdet~\cite{CLICdet_note_2017}, has been refined from 
the two CLIC detector concepts \clicsid and \clicild that were adapted from the 
International Linear Collider (ILC) concepts as described in the CLIC CDR~\cite{cdrvol2}. 
All three detector designs are compatible with the experimental conditions at CLIC
and satisfy the performance requirements driven by the physics objectives,
which are described in~\ref{sec:expcond,sec:CLICdet}.
Common software tools for \clicsid and \clicild perform full simulation~\cite{Mokka, Graf:2006ei},
digitisation and reconstruction~\cite{MarlinLCCD, Graf:2011zzc},
particle flow reconstruction~\cite{thomson:pandora, Marshall:2013bda, Marshall2013153, Marshall:2015rfaPandoraSDK},
and vertexing and heavy flavour tagging~\cite{Suehara:2015ura}.
The corresponding, and partially overlapping, software tools for CLICdet are described
in~\ref{sec:CLICdetPerformance}. Full simulations of the \clicild and \clicsid detector
concepts have been used for the physics projections in the areas of Higgs and top-quark physics
reported in this section. Physics background processes, as well as the CLIC experimental
conditions and luminosity spectrum have systematically been taken into account.
The dark matter search using the mono-photon signature (see~\ref{sec:physics_bsm}) is the first full simulation study
using CLICdet. Recent phenomenological studies to explore the BSM potential of CLIC often
use parameterised detector performance derived from full simulations,
for example through the fast simulation package Delphes~\cite{deFavereau:2013fsa, Schnoor2018}.

\paragraph{Staging and polarisation}
\label{sec:polarisation}
The total integrated luminosities for each energy stage are summarised in~\ref{tab:clicstaging}.
Each stage takes seven or eight years and stages are separated by around two years, resulting in
a total programme of \SIrange{25}{30}{\years}.

\begin{table}[htp]\centering
  \caption{Baseline CLIC energy stages and integrated luminosities for each stage in the updated scenario~\cite{Roloff:2645352}. \label{tab:clicstaging}}
  \begin{tabular}{SSS}\toprule
   {Stage} & {$\sqrt{s}$ [\si{\TeV}]} & {$\mathcal{L}_{\textrm{int}}$ [\si{\per\ab}]} \\
    \hline
   1 &  0.38{ (and 0.35)} &  1.0 \\
   2 &  1.5             &  2.5 \\
   3 &  3.0             &  5.0 \\
    \bottomrule
  \end{tabular}
\end{table}

The staged approach allows optimal exploitation of the CLIC physics capabilities.
For the initial stage, a centre-of-mass energy of $\roots=\SI{380}{\GeV}$
gives access to SM Higgs physics and top-quark physics, and
provides direct and indirect sensitivity to BSM effects.
A top-quark pair-production threshold scan around \SI{350}{\GeV} is also foreseen.
The second stage at \SI{1.5}{\TeV} opens more Higgs production channels including $\ttbar\PH$,
double-Higgs production, and rare decays, and allows further direct sensitivity to many 
BSM models.  The ultimate stage at \SI{3}{\TeV} gives the best sensitivity to many new
physics scenarios and to the Higgs self-coupling.
The energies of the second and third stages are benchmarks, and can be optimised
in light of new physics information.
For many of the studies reported here, an earlier energy staging
baseline of $\sqrt{s}=\SI{350}{\GeV}$, \SI{1.4}{\TeV}, and \SI{3}{\TeV} was assumed;
it was as a consequence of these and other studies that the present initial stage energy
of $\roots=380\,\GeV$ was adopted in order to optimise the physics reach of CLIC. 

The CLIC baseline specifies \SI{\pm80}{\percent} electron polarisation, and no positron polarisation.
At the initial energy stage, equal amounts of \SI{-80}{\percent} and \SI{+80}{\percent} polarisation running
are foreseen. 
For the two higher-energy stages, a compromise is required between 
the strong enhancement in single and double-Higgs production
through $\PW\PW$-fusion that comes by running with \SI{-80}{\percent} electron polarisation,
and the full reach to BSM effects, which requires some running with \SI{+80}{\percent} electron polarisation.
The baseline is to share the running
time for \SI{-80}{\percent} and \SI{+80}{\percent} electron polarisation in the ratio 80:20.

The following sections discuss the Higgs physics, top-quark physics, and BSM physics reach of CLIC
across the three energy stages. 

\subsection{Higgs physics potential}
\label{sec:physics_higgs}

A detailed understanding of the Higgs sector is one of the highest priorities in
particle physics.  While the discovery of a Higgs boson confirms the
electroweak symmetry breaking mechanism, the nature of the particle is still
to be determined: whether it is the fundamental scalar of the SM, or a more
complex object, or part of an extended Higgs sector. 
The Higgs interactions could provide access to BSM physics that couples
to the SM only through the Higgs boson. 

Around \num{160000} Higgs bosons will be produced at the initial CLIC
energy stage, and millions will be produced at $\roots = \SI{3}{\TeV}$.  All collider
events will be read out, without online event filters, and high acceptance and event 
selection efficiencies will result in very large datasets available
for analysis.  The large datasets and the large energy range will allow a comprehensive
Higgs programme with unique reach.
The staging of the collider is crucial to the Higgs physics programme, with some
measurements accessible only at the initial energy, while other measurements are
enabled or improved by the high-energy running.  

Different BSM scenarios result in different patterns of modifications to the
Higgs couplings from their SM values; for new physics at the \si{\TeV} scale this
is typically at the percent level and so measurement of the couplings with
similar or better precision is necessary.
The CLIC Higgs reach has been comprehensively investigated using full simulation studies~\cite{ClicHiggsPaper}.
Some updated studies and the
sensitivities resulting from the updated luminosity staging scenario are summarised
in this section.
BSM Higgs scenarios are further explored in~\ref{sec:physics_bsm}.

\paragraph{Higgs production}
Cross sections for the main Higgs production processes are shown in~\ref{fig:higgs_xs}. 
At the initial CLIC energy stage the dominant Higgs production mechanism is 
the Higgsstrahlung process, $\epem\to\PZ\PH$.
These events can be selected based only on the decay products of the $\PZ$ boson, enabling
measurements of the total production cross section and hence the Higgs branching ratios and width. 
This allows a model-independent determination of the Higgs couplings, without any assumptions 
about BSM invisible decays of the Higgs boson; a feature that is unique to lepton colliders.
Higgs production at the initial energy stage also has a significant contribution from
the $\PW\PW$-fusion process, $\epem\to\PH\PGne\PAGne$, and the combined study of
the Higgsstrahlung and $\PW\PW$-fusion processes improves the precision of the Higgs
couplings and width measurements. 
Around $\roots=\SI{1.5}{\TeV}$ and at $\roots=\SI{3}{\TeV}$ Higgs production is dominated by
$\PW\PW$-fusion; at these energies also the $\PZ\PZ$-fusion process, $\epem\to\PH\epem$,
becomes significant, and $\PQt\PAQt\PH$ and direct double-Higgs production become accessible.

\begin{figure}[hbt]
\begin{center}
\includegraphics[width=0.55\linewidth]{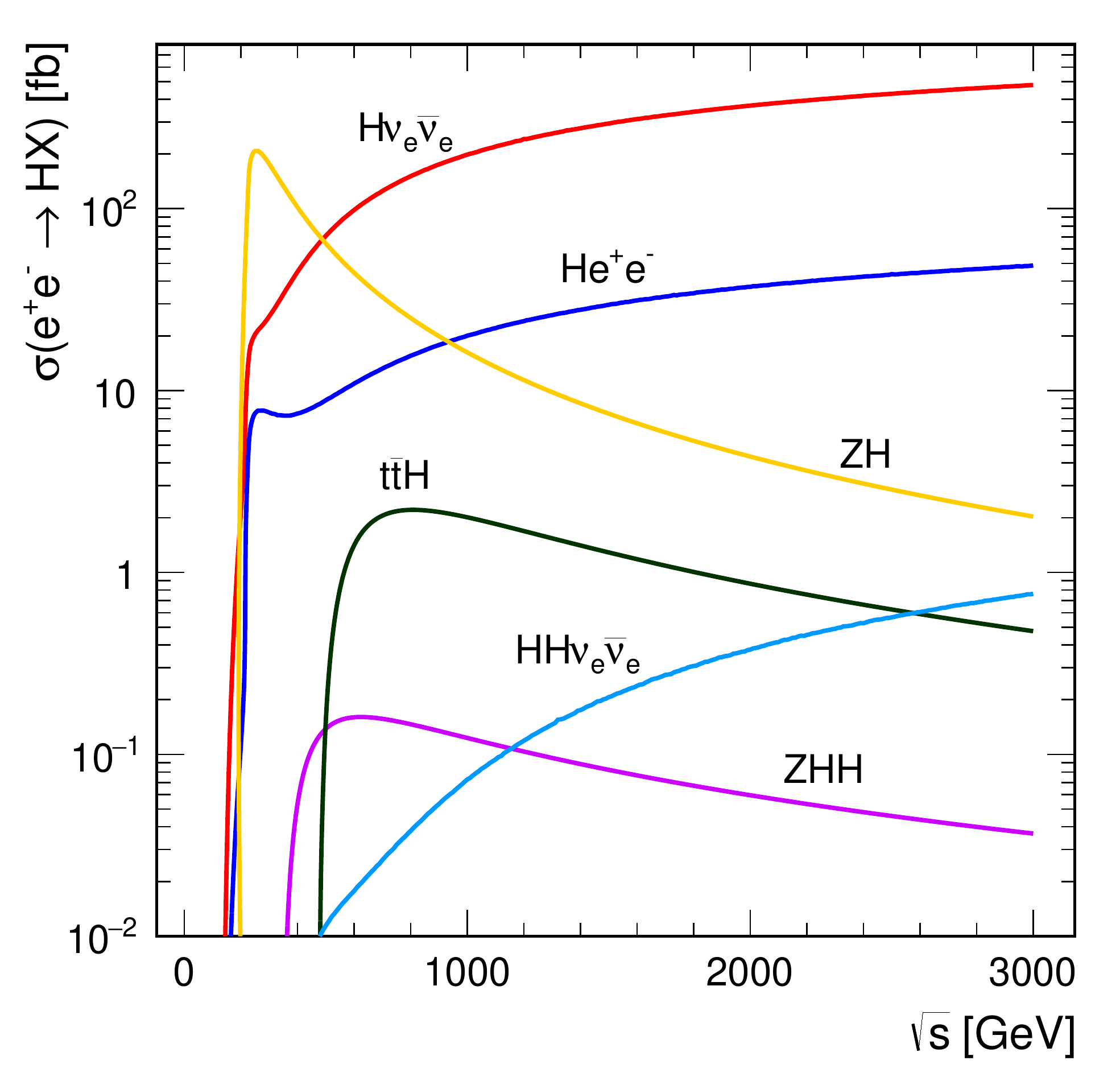}
\caption{Production cross section as a function of centre-of-mass energy for the main Higgs production processes at an \epem collider.  The values shown correspond to unpolarised beams with initial-state radiation and do not include the effect of beamstrahlung. \imdp}
\label{fig:higgs_xs}
\end{center}
\end{figure}

\paragraph{Invisible Higgs decays}
The recoil mass distribution from $\Pep\Pem\to\PZ\PH$
events can be used to search for BSM decay modes of the Higgs boson into 
`invisible' final states~\cite{Thomson:2015jda}.
For \SI{1}{\per\ab} at $\roots=\SI{350}{\GeV}$ the upper limit
on the invisible Higgs branching ratio, obtained from Higgsstrahlung events with hadronic $\PZ$ boson decays, is 
$\br{\PH\to \mathrm{invis.}}< \SI{0.69}{\percent}$ at \SI{90}{\percent} C.L.

\paragraph{Higgs couplings}
Measurements of Higgs production cross sections times branching fractions
to many final states can be combined to extract the Higgs couplings and widths.  
Precisions extracted from a model-independent
global fit, described in~\cite{ClicHiggsPaper}, are given
in~\ref{fig:MIResultsPolarised8020,tab:MIResultsPolarised8020}.
The fit assumes the current baseline scenario of operation with \SI{-80}{\percent} (\SI{+80}{\percent}) electron beam polarisation for
  \SI{80}{\percent} (\SI{20}{\percent}) of the collected luminosity at the second
  and third energy stages. 
Each energy stage contributes significantly to the Higgs programme;
the initial stage provides $\sigma_{\PH\PZ}$ and couplings to most fermions and
bosons, while the higher-energy stages improve them and add the top-quark 
and muon couplings.  The initial stage is required, to allow
the model-independent coupling fits to be performed at all energy stages. 
Precisions extracted from a model-dependent
global fit, also described in~\cite{ClicHiggsPaper}, where it is assumed that there are no non-Standard-Model Higgs decays,
are given in~\ref{fig:MDResultsPolarised8020,tab:MDResultsPolarised8020}.
This fit also assumes the current beam polarisation scenario. 
Already after the initial energy stage, in many cases the CLIC precision is
significantly better than for the HL-LHC~\cite{CMS_HLLHC_Higgs}, and improves further with the higher-energy running.

\begin{minipage}{\linewidth}
  \begin{minipage}{0.495\textwidth}
   \centering
    \begin{table}[H]
      \begin{tabular}{lS[table-format=1.1]S[table-format=2.1,table-space-text-post = \si{\percent}]S[table-format=1.1]}
        \toprule
        Parameter & \multicolumn{3}{c}{Relative precision}\\
        \midrule
        & {\SI{350}{\GeV}} & {+ \SI{1.4}{\TeV}} & {+ \SI{3}{\TeV}}\\
        &{\SI{1}{\per\ab}}& {+ \SI{2.5}{\per\ab}}& {+ \SI{5}{\per\ab}}\\
        \midrule
        $\gHZZ$ & 0.6\si{\percent} & 0.6\si{\percent} & 0.6\si{\percent} \\
        $\gHWW$ & 1.0\si{\percent} & 0.6\si{\percent} & 0.6\si{\percent} \\
        $\gHbb$ & 2.1\si{\percent} & 0.7\si{\percent} & 0.7\si{\percent} \\
        $\gHcc$ & 4.4\si{\percent} & 1.9\si{\percent} & 1.4\si{\percent} \\
        $\gHTauTau$ & 3.1\si{\percent} & 1.4\si{\percent} & 1.0\si{\percent} \\
        $\gHMuMu$ & {$-$} & 12.1\si{\percent} & 5.7\si{\percent} \\
        $\gHtt$ & {$-$} & 3.0\si{\percent} & 3.0\si{\percent} \\
        \midrule
        $g^\dagger_{\PH\Pg\Pg}$ & 2.6\si{\percent} & 1.4\si{\percent} & 1.0\si{\percent} \\
        $g^\dagger_{\PH\PGg\PGg}$ & {$-$} & 4.8\si{\percent} & 2.3\si{\percent} \\
        $g^\dagger_{\PH\PZ\PGg}$ & {$-$} & 13.3\si{\percent} & 6.7\si{\percent} \\
        \midrule
        $\Gamma_{\PH}$ & 4.7\si{\percent} & 2.6\si{\percent} & 2.5\si{\percent} \\
        \bottomrule
      \end{tabular}
      \caption{ }\label{tab:MIResultsPolarised8020}
    \end{table}
  \end{minipage}
  \begin{minipage}{0.495\textwidth}
    \centering
    \begin{figure}[H]
      \includegraphics[width=\linewidth]{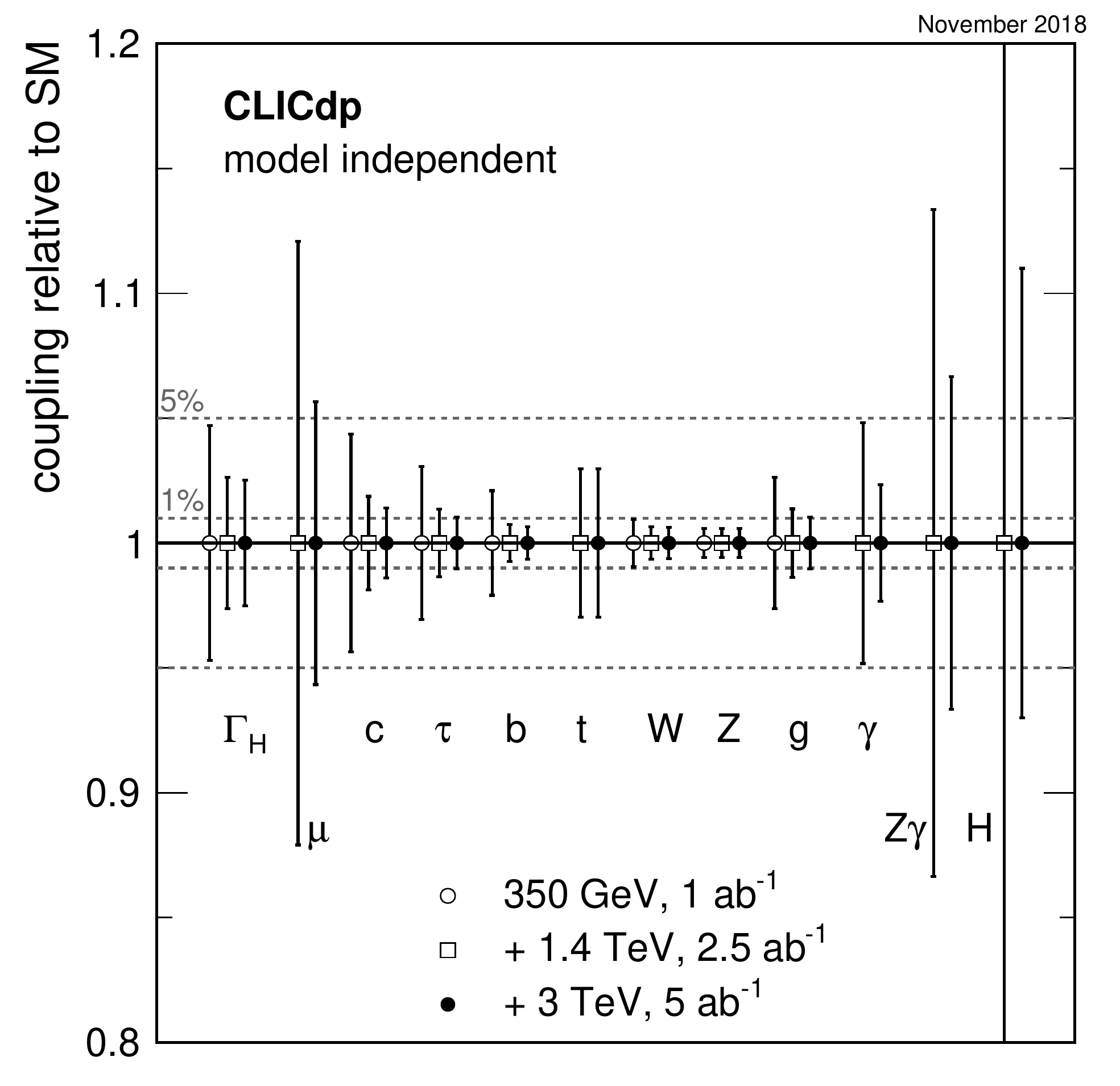}
      \caption{ }\label{fig:MIResultsPolarised8020}
    \end{figure}
  \end{minipage}
  \vspace{0.2cm}

        CLIC results of the model-independent fit to the Higgs couplings.
        For $\gHtt$, the $\SI{3}{\TeV}$ case has not yet been studied. The three
        effective couplings $g^\dagger_{\PH\Pg\Pg}$, 
        $g^\dagger_{\PH\PGg\PGg}$ and $g^\dagger_{\PH\PZ\PGg}$ are also included in the fit. 
        Operation with \SI{-80}{\percent} (\SI{+80}{\percent}) electron beam polarisation is assumed for
        \SI{80}{\percent} (\SI{20}{\percent}) of the collected luminosity above \SI{1}{\TeV}, corresponding
        to the baseline scenario. \imdp \\

  \begin{minipage}{0.495\textwidth}
   \centering
    \begin{table}[H]
      \begin{tabular}{lS[table-format=1.1]S[table-format=2.1,table-space-text-post = \si{\percent}]S[table-format=1.1]}
        \toprule
        Parameter & \multicolumn{3}{c}{Relative precision}\\
        \midrule
        & {\SI{350}{\GeV}} & {+ \SI{1.4}{\TeV}} & {+ \SI{3}{\TeV}}\\
        &{\SI{1}{\per\ab}}& {+ \SI{2.5}{\per\ab}}& {+ \SI{5}{\per\ab}}\\
        \midrule
        $\kappa_{\PH\PZ\PZ}$ & 0.4\si{\percent}& 0.3\si{\percent}& 0.2\si{\percent}\\
        $\kappa_{\PH\PW\PW}$ & 0.8\si{\percent}& 0.2\si{\percent}& 0.1\si{\percent}\\
        $\kappa_{\PH\PQb\PQb}$ & 1.3\si{\percent}& 0.3\si{\percent}& 0.2\si{\percent}\\
        $\kappa_{\PH\PQc\PQc}$ & 4.1\si{\percent}& 1.8\si{\percent}& 1.3\si{\percent}\\
        $\kappa_{\PH\PGt\PGt}$ & 2.7\si{\percent}& 1.2\si{\percent}& 0.9\si{\percent}\\
        $\kappa_{\PH\PGm\PGm}$ & {$-$} & 12.1\si{\percent}& 5.6\si{\percent}\\
        $\kappa_{\PH\PQt\PQt}$ & {$-$} & 2.9\si{\percent}& 2.9\si{\percent}\\
        $\kappa_{\PH\Pg\Pg}$ & 2.1\si{\percent}& 1.2\si{\percent}& 0.9\si{\percent}\\
        $\kappa_{\PH\PGg\PGg}$ & {$-$} & 4.8\si{\percent}& 2.3\si{\percent}\\
        $\kappa_{\PH\PZ\PGg}$ & {$-$} & 13.3\si{\percent}& 6.6\si{\percent}\\
        \bottomrule
      \end{tabular}
      \caption{ }\label{tab:MDResultsPolarised8020}
    \end{table}
  \end{minipage}
  \begin{minipage}{0.495\textwidth}
    \centering
    \begin{figure}[H]
      \includegraphics[width=\linewidth]{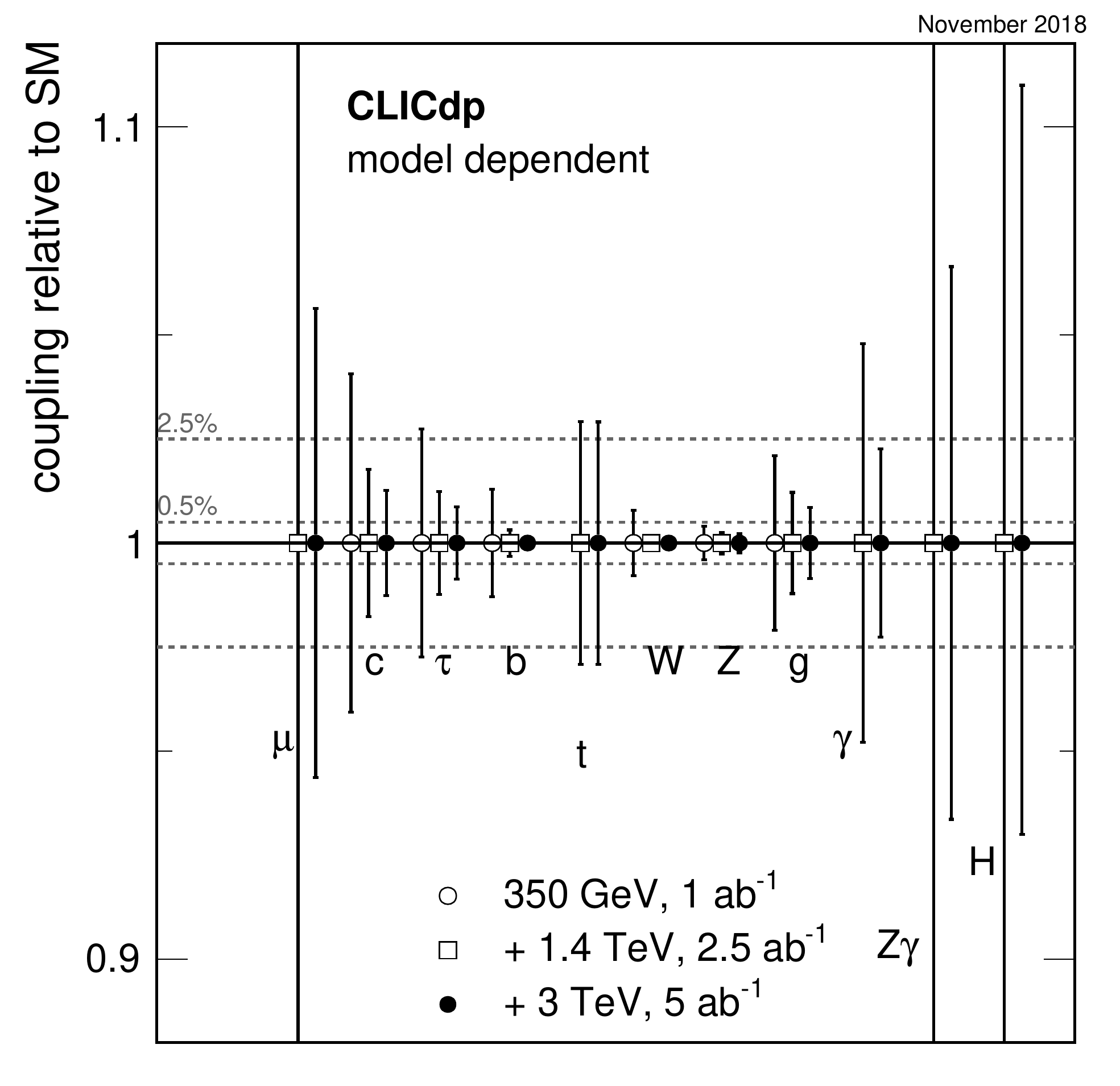}
      \caption{ }\label{fig:MDResultsPolarised8020}
    \end{figure}
  \end{minipage}
  \vspace{0.2cm}

  CLIC results of the model-dependent fit to the Higgs coupling, without theoretical uncertainties. 
  For $\kappa_{\PH\PQt\PQt}$, the $\SI{3}{\TeV}$ case has not yet been
  studied. Operation with $\SI{-80}{\percent}$ (\SI{+80}{\percent}) electron beam polarisation is assumed for
  \SI{80}{\percent} (\SI{20}{\percent}) of the collected luminosity above \SI{1}{\TeV}, corresponding to the baseline scenario. \imdp
\end{minipage}

\paragraph{Higgs self-coupling}
Centre-of-mass energies of \SI{1.4}{\TeV} or \SI{1.5}{\TeV} and \SI{3}{\TeV} give access to
double-Higgs production processes, which are sensitive to the trilinear Higgs
coupling $\lambda$ at tree level.
The second energy stage allows a 5$\sigma$-observation
of the double Higgsstrahlung process $\epem\to\PZ\PH\PH$ and provides evidence for the $\PW$ boson fusion process 
$\epem\to\PH\PH\PGne\PAGne$ with a significance of $3.6\sigma$ assuming the SM value of $\lambda$. 
At $\roots=\SI{3}{\TeV}$ the vector-boson fusion process $\epem\to\PH\PH\PGne\PAGne$ is the leading double-Higgs production process 
and its cross section can be measured with a
precision $\Delta\sigma/\sigma=\SI{7.4}{\percent}$ at $\roots=\SI{3}{\TeV}$.
The cross section dependence on $\lambda$ is not monotonic, so that using only the
cross section information to extract $\lambda$, or equivalently the variation
$\Delta\kappa_{\lambda}$ from the Standard Model value $\kappa_{\lambda}=1$, leads to an ambiguity as 
shown in the~\ref{fig:chi2ofHHa}.
This can be resolved
by adding the differential information including the invariant mass $M(\PH\PH)$ in $\epem\to\PH\PH\PGne\PAGne$ at $\roots=\SI{3}{\TeV}$, and the cross section measurement of the double
Higgsstrahlung process $\PZ\PH\PH$ at $\roots=\SI{1.4}{\TeV}$.  This removes the ambiguity and increases the sensitivity as shown in~\ref{fig:chi2ofHHb}.
The $\epem\to\PH\PH\PGne\PAGne$ process also contains the quartic vertex $\PH\PH\PW\PW$.
\begin{center}
The final sensitivity that can be reached on the trilinear self-coupling $\lambda$ 
is $[\SI{-7}{\percent},\SI{+11}{\percent}]$~\cite{DoubleHiggsNote}.
\end{center}
The analysis can also be used to set two-dimensional constraints on $\lambda$ and $\kappa_{\PH\PH\PW\PW}$. 

  \begin{figure}[hbt]
    \centering
    \begin{subfigure}{.45\textwidth}
        \includegraphics[width=\linewidth]{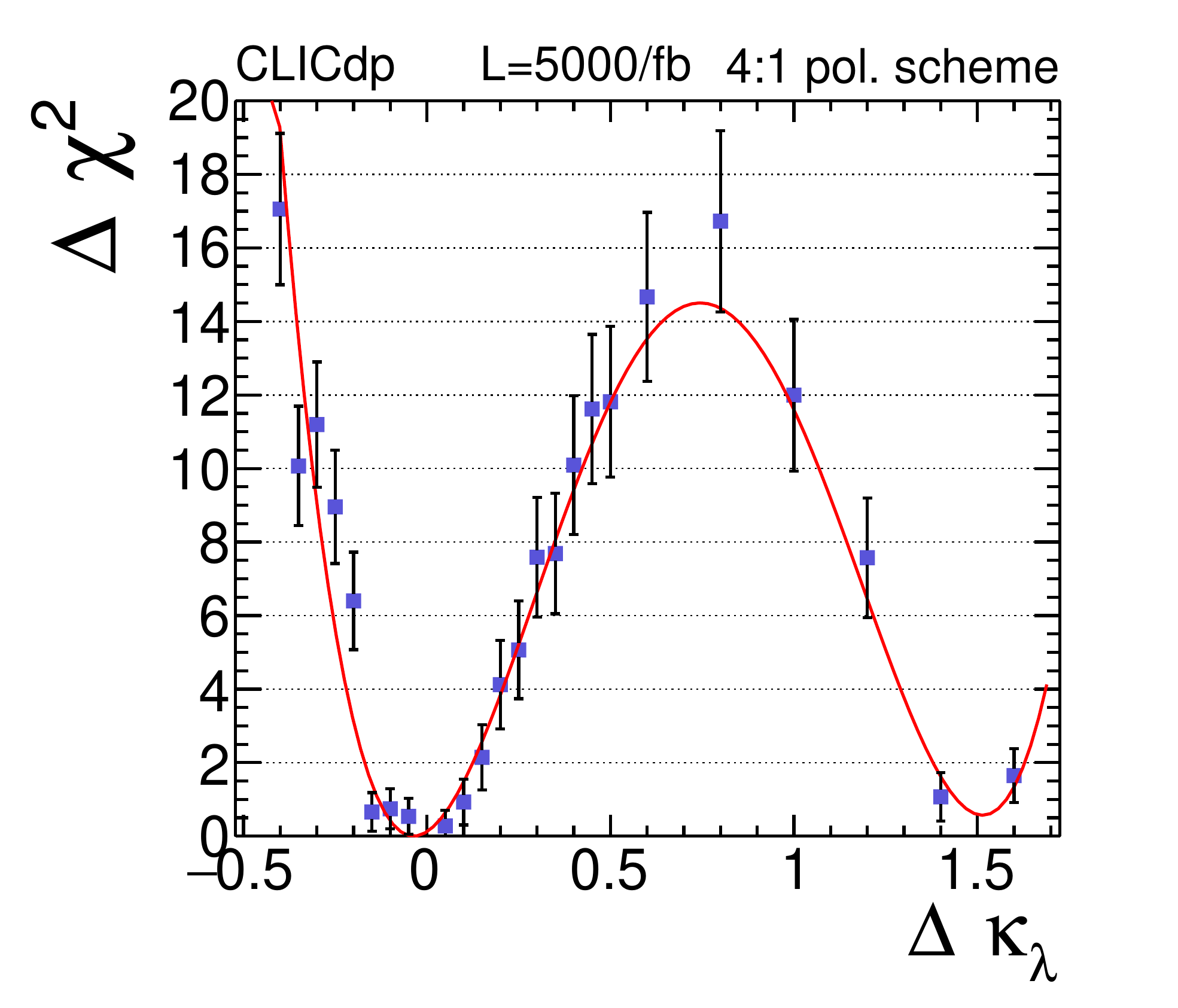}
        \caption{}\label{fig:chi2ofHHa}
    \end{subfigure}
    \hspace{.01\linewidth}
    \hspace{.01\linewidth}
    \begin{subfigure}{.45\textwidth}
        \includegraphics[width=\linewidth]{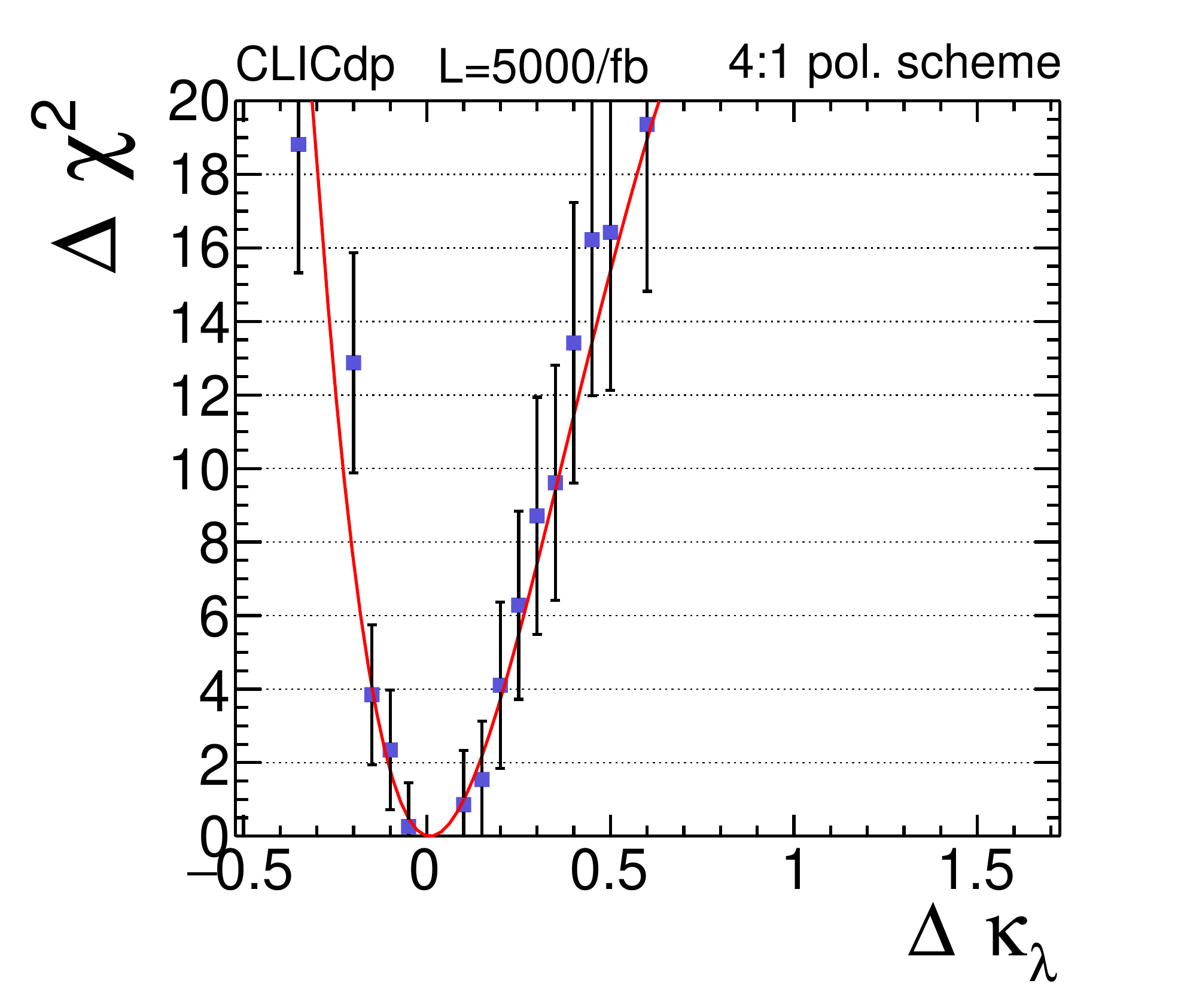}
        \caption{}\label{fig:chi2ofHHb}
    \end{subfigure}
    \caption{Nominal $\Delta \chi^{2}$ distributions from template fitting with different values of the Higgs self-coupling $\lambda$, shown for the variation $\Delta\kappa_{\lambda}$ from the SM value of $\kappa_{\lambda}=1$ and using (a) only cross section information for the $\PH\PH\PGne\PAGne$ process at \SI{3}{\TeV}; (b) additionally using the differential distribution $M(\PH\PH)$ in $\PH\PH\PGne\PAGne$ at \SI{3}{\TeV} and the cross section measurement of $\PZ\PH\PH$ at \SI{1.4}{\TeV}. In this case the ambiguity is removed and the sensitivity is increased. \imdp }
    \label{fig:chi2ofHH}
  \end{figure}

\paragraph{Composite Higgs} 
Constraints from the CLIC Higgs measurements, along with measurements of 
Drell-Yan production ($\epem\to f \bar{f}$) and $\PW\PW$ production, can be interpreted 
in concrete classes of BSM
scenarios; for example, scenarios where the known particles are in fact composite bound states.
Composite Higgs frameworks could address the electroweak naturalness problem.
In these frameworks the Higgs boson is a pseudo-Nambu-Goldstone
boson of an underlying strongly-interacting composite sector.
The 5$\sigma$ discovery reach in the plane of the mass scale $m_*$ and coupling strength
parameter $g_*$ that characterise the Higgs composite sector is shown in
\ref{fig:compositeness_higgs}~\cite{ESU18BSM}. 
CLIC will discover Higgs compositeness if the compositeness scale is below \SI{8}{\TeV}.
Scales up to \SI{40}{\TeV} can be discovered, in particularly favourable conditions,
for large composite sector couplings $g_*\simeq 8$.  For comparison, the model-independent
projected HL-LHC {\em{exclusion}} reach is only around \SI{3}{\TeV}, and for
$g_*\simeq 8$ the maximum HL-LHC reach is around \SI{7}{\TeV}~\cite{ESU18BSM}. 

\begin{figure}[hbt]
   \centering
      \includegraphics[width=.45\linewidth]{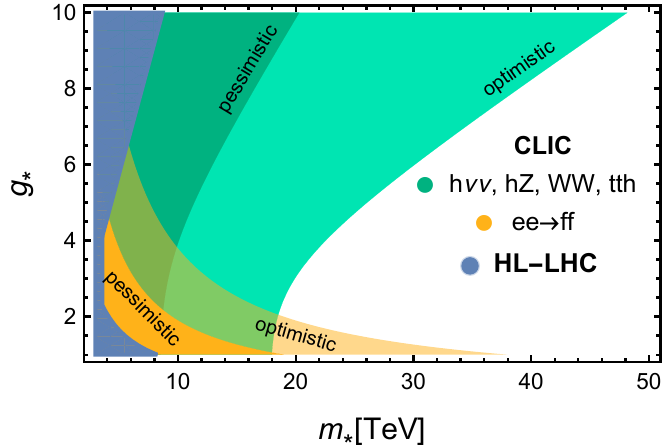}
\caption{Higgs compositeness: `Optimistic' (light colour) and `pessimistic' (dark colour) $5\sigma$ discovery regions for Higgs compositeness derived from a combined fit to Higgs and $\PW\PW$ production (green) and Drell-Yan processes (orange); and the HL-LHC 95\% C.L. exclusion reach. From~\cite{ESU18BSM}. }
\label{fig:compositeness_higgs}
\end{figure}

\subsection{Top-quark physics potential}
\label{sec:physics_top}

The top quark is the heaviest known fundamental particle and occupies an important role in
many BSM theories; it therefore provides unique opportunities to test the SM and probe signatures of BSM effects.
So far, top quarks have been produced only in hadron collisions, whereas studying their 
properties in electron-positron collisions
will provide a new set of complementary and improved-precision measurements.
Cross sections for the main top-quark pair production processes are shown in~\ref{fig:top_xs}. 
Each stage of CLIC provides sensitivity to different aspects of top-quark production and
properties: at the initial stage this includes top-quark mass
measurements and cross section and asymmetry measurements, while
higher-energy stages additionally allow studies including direct access to the top Yukawa coupling and CP properties in the
$\PQt\PQt\PH$ coupling.
The \SI{3}{\TeV} stage is most favourable for study of top-quark pair production
via vector boson fusion.
A broad set of measurements has been investigated for CLIC~\cite{ClicTopPaper}; they are summarised in the following sections.

\begin{figure}[hbt]
\begin{center}
\includegraphics[width=0.5\linewidth]{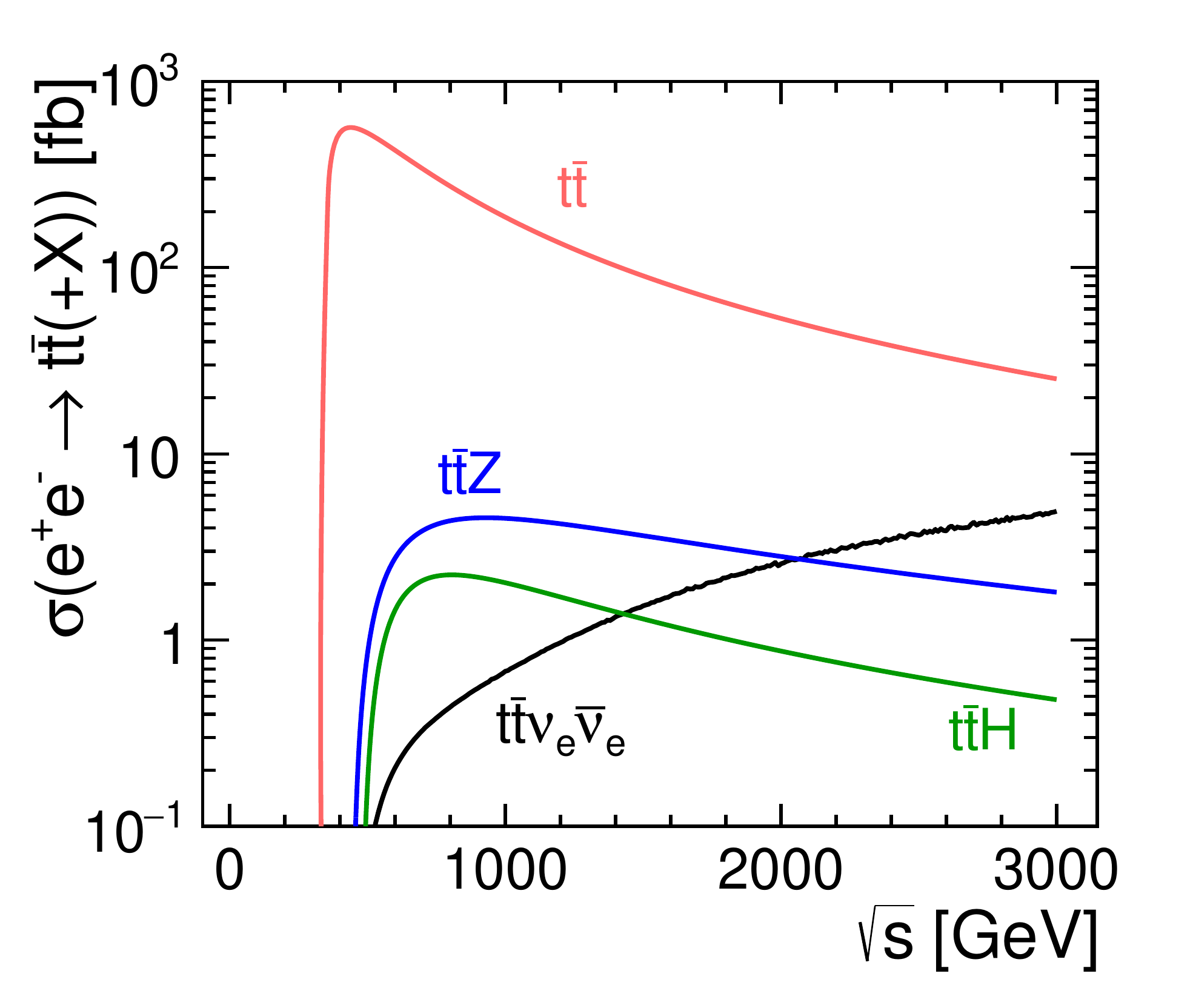}
\caption{Production cross section as a function of centre-of-mass energy for the main top-quark pair production processes at an \epem collider.  Leading-order values for unpolarised beams with initial-state radiation are shown, not including the effect of beamstrahlung. \imdp}
\label{fig:top_xs}
\end{center}
\end{figure}

\paragraph{Top-quark mass} 
Several complementary methods allow precise determinations of the top-quark mass at CLIC:
a threshold scan; measurements of the cross section for radiative events; and direct
reconstruction.
Direct reconstruction of the top-quark decay products yields a statistical precision
of around \SI{30}{\MeV}; a systematic accuracy that matches the statistical accuracy would require
control of the jet energy scale at the level of \SI{0.02}{\percent}, and the theoretical
interpretation of the measurement introduces a large systematic uncertainty. 
Fitting templates to cross sections as a function of centre-of-mass energy
has the advantage that the extracted mass is well defined theoretically,
which significantly reduces the overall uncertainty.
This can be done for radiative events above threshold, as a function of the effective \ttbar centre-of-mass energy \rootsprime
after radiation of an energetic initial state photon from the incoming electron or positron beam, $\epem\to\PQt\PAQt\PGg$.
This method leads to a total uncertainty on the top-quark mass of around \SI{140}{\MeV} with an integrated luminosity of \SI{1}{\per\ab}.
However, the most precise top-quark mass determination comes from an energy
scan around the top-quark pair-production threshold. 
The study assumes a scan collecting \SI{10}{\per\fb} at each of ten points in \roots, separated by \SI{1}{\GeV}.
A highly pure sample of top-quark events can be selected and used to measure the cross section at each point. 
The top-quark mass is extracted using a template fit to the measured cross sections as a function
of \roots\ as shown in~\ref{fig:TopThresholdScan}. 
The flexibility of the CLIC accelerator design is illustrated by potential optimisation of the
luminosity spectrum for the threshold scan as described in~\ref{sec:energyflexibility},
and shown by the two bunch charge options in~\ref{fig:TopThresholdScan}.
The cross section and the position and shape of the turn-on curve are strongly dependent
on the precise value of the top-quark mass and width, Yukawa coupling, and strong coupling $\alpha_s$.
The template line shapes can be computed at NNNLO QCD, taking into account NLO Higgs and electroweak
effects.  
The statistical uncertainty is \SI{20}{\MeV} for the reduced bunch charge option
and \SI{22}{\MeV} for the nominal luminosity spectrum.
The small theoretical uncertainty of around \SI{10}{\MeV} associated with converting the
measured quantity to the $\overline{\rm MS}$ (modified minimal subtraction) mass scheme means that 
a total uncertainty of around \SI{50}{\MeV} is thus feasible.

\begin{figure}[t!]
  \centering
     \begin{subfigure}{.48\textwidth}
        \includegraphics[width=\linewidth]{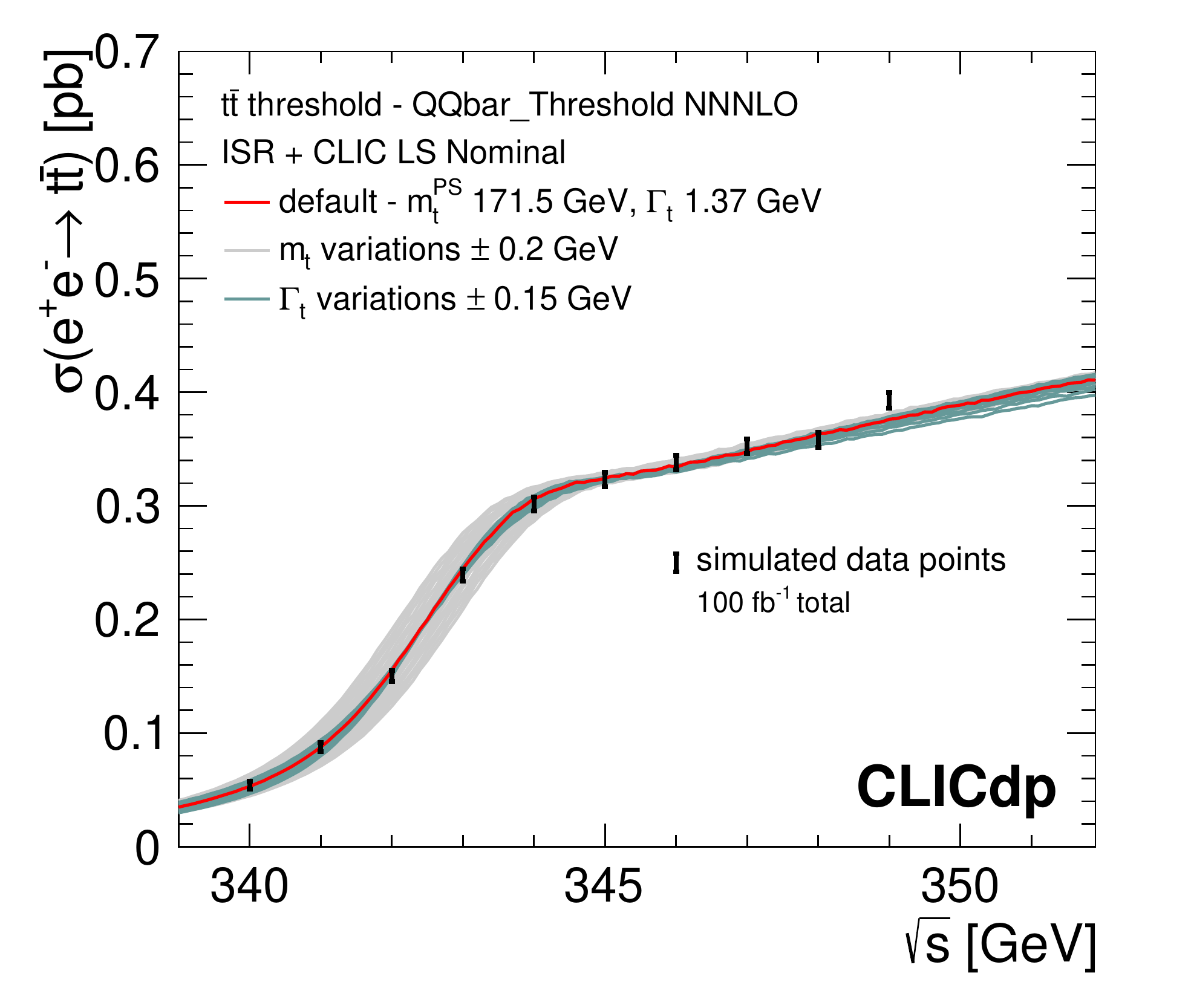}
        \caption{}
     \end{subfigure}
     \hspace{.01\linewidth}
     \begin{subfigure}{.48\textwidth}
        \includegraphics[width=\linewidth]{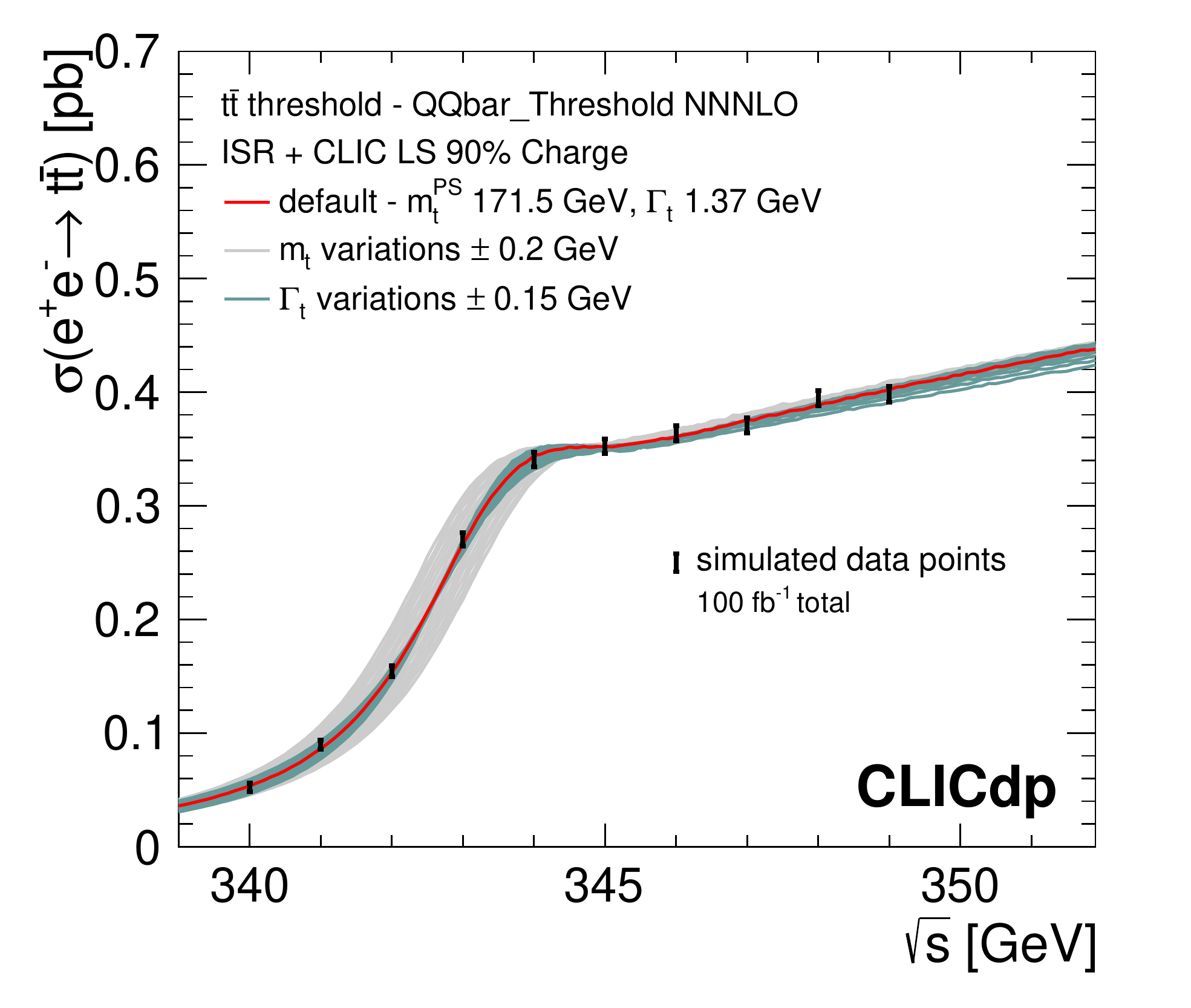}
        \caption{}
     \end{subfigure}
  \caption{Illustration of a top-quark threshold scan at CLIC with a total integrated luminosity of \SI{100}{\per\fb}, for two scenarios for the luminosity spectrum (LS): (a) nominal, and (b) `reduced charge'. The bands around the central cross section curve show the dependence of the cross section on the top-quark mass and width, illustrating the sensitivity of the threshold scan. The error bars on the simulated data points show the statistical uncertainties of the cross section measurement, taking into account signal efficiencies and background levels. From~\cite{ClicTopPaper}.}
\label{fig:TopThresholdScan}
\end{figure}

\paragraph{Top-quark pair-production properties} 
Top-quark pair production at all three energy stages contributes to the determination of 
electroweak couplings.

At the initial stage, semi-leptonic events, where the $\PW$ boson from one top-quark decay decays
hadronically and the other leptonically, are reconstructed by clustering into four jets, 
identifying a reconstructed lepton and jets that originate from $\Pb$-quarks, and performing a kinematic fit.
The efficient $\Pb$-tagging of two jets allows suppression of most of the backgrounds.  
Measurements of the cross section can be made with a statistical accuracy below \SI{1}{\percent} for
each electron beam polarisation, and the measured top-quark angular
distribution in semi-leptonic events gives the forward-backward asymmetry with a statistical
uncertainty of around \SIrange{3}{4}{\percent}.

At the higher energy stages the top quarks are produced with
significant boosts.  Jet substructure techniques~\cite{Kaplan:2008ie} have been successfully applied
to reconstructing boosted semi-leptonic events and measuring the forward-backward asymmetry.  
The total cross section and forward-backward asymmetries are extracted with precisions of 
$\Delta\sigma/\sigma = \SI{1.1}{\percent} \; (\SI{2.3}{\percent})$ and $\Delta A_{FB}/ A_{FB} = \SI{1.5}{\percent} \; (\SI{2.0}{\percent})$
at \SI{1.4}{\TeV}\ (\SI{3}{\TeV}) for an electron beam polarisation of \SI{-80}{\percent}.
All of these measurements, along with a set of ten statistically optimised observables defined
for each centre-of-mass energy and electron beam polarisation
on the $\epem\to \PQt\PAQt\to \PQb\PWp\PAQb\PWm$ differential distribution~\cite{Atwood:1991ka, Grzadkowski:2000nx, Janot:2015yza, Khiem:2015ofa}, 
are used in global fits to constrain possible BSM effects induced by heavy
new physics, as described in~\ref{sec:physics_bsm}.

\paragraph{Top-quark Yukawa coupling and CP properties}
Associated $\PQt\PAQt\PH$ production is accessible at the second CLIC stage.
Studies of $\PH\to\PQb\PAQb$ with fully-hadronic and semi-leptonic
$\PQt\PAQt$ decays
give an expected precision on the cross section at \SI{1.4}{\TeV}\
with \SI{2.5}{\per\ab} of \SI{5.7}{\percent}.
The cross section is sensitive to the strength of the Yukawa coupling and 
its measurement can be translated 
into a determination of the Yukawa coupling with a precision of \SI{2.9}{\percent}.
As an extra probe of the CP structure of the Standard Model, to search for
sources of CP violation that could give insight into the matter-antimatter asymmetry in the universe, 
the CP structure of the $\PQt\PQt\PH$ coupling can be investigated.  
A CP-odd admixture to the coupling can be parameterised with a mixing angle $\phi$ as
$-ig_{\PQt\PAQt\PH}(\cos\phi + i\sin\phi\gamma_5$); 
this alters the cross section, which can therefore be used to set limits
on the admixture.  Further sensitivity can be obtained from differential distributions,
for example the up-down asymmetry of the anti-top quark with respect to the
plane defined by the incoming electron and the top quark~\cite{Godbole:2011hw}.
The sensitivity is $\Delta\sin^2\phi < 0.07 \; (0.04)$ for the 
range $0<\sin^2\phi < 1$ ($0.3<\sin^2\phi < 1$). 

\paragraph{Compositeness interpretation}
Constraints from the global fit of optimised observables from $\PQt\PAQt$ production 
and from $\PQt\PAQt\PH$ production can also be interpreted in compositeness scenarios. 
Top-quark compositeness emerges naturally in the composite Higgs frameworks discussed
at the end of section~\ref{sec:physics_higgs}.
The $\PQt\PAQt$ and $\PQt\PAQt\PH$ sensitivities have been interpreted for the
discovery reach of top-quark compositeness scenarios at $\roots=\SI{3}{\TeV}$.
The 5$\sigma$ discovery reach in the plane of the mass scale $m_*$ and coupling strength
parameter $g_*$ that characterise the composite sector is shown in~\ref{fig:compositeness_limits_2}~\cite{Durieux:2018ekg,ClicTopPaper,ESU18BSM}, 
for optimistic and pessimistic values of coefficients of the top-philic operators.
In this framework the top-quark compositeness is characterised by couplings $y_L$ and $y_R$ that
control the strength of the mixing of the $\PQq_L$ doublet and the $\PQt_R$ singlet
with the composite sector.  The two benchmark scenarios considered are 
partial compositeness, where $y_L=y_R=\sqrt{y_Lg_*}$; and total $\PQt_R$ compositeness,
where $y_L=y_{\PQt}$ and $y_R=g_*$.
The strong sensitivity of the observables to effects that grow with centre-of-mass energy results
in a discovery reach for top compositeness beyond \SI{7}{\TeV}, and more than \SI{20}{\TeV} in
favourable conditions, at CLIC.  These values are higher than could be {\em{excluded}} at HL-LHC. 

\begin{figure}[hbt]
   \centering
   \begin{subfigure}{.48\textwidth}
      \includegraphics[width=\linewidth]{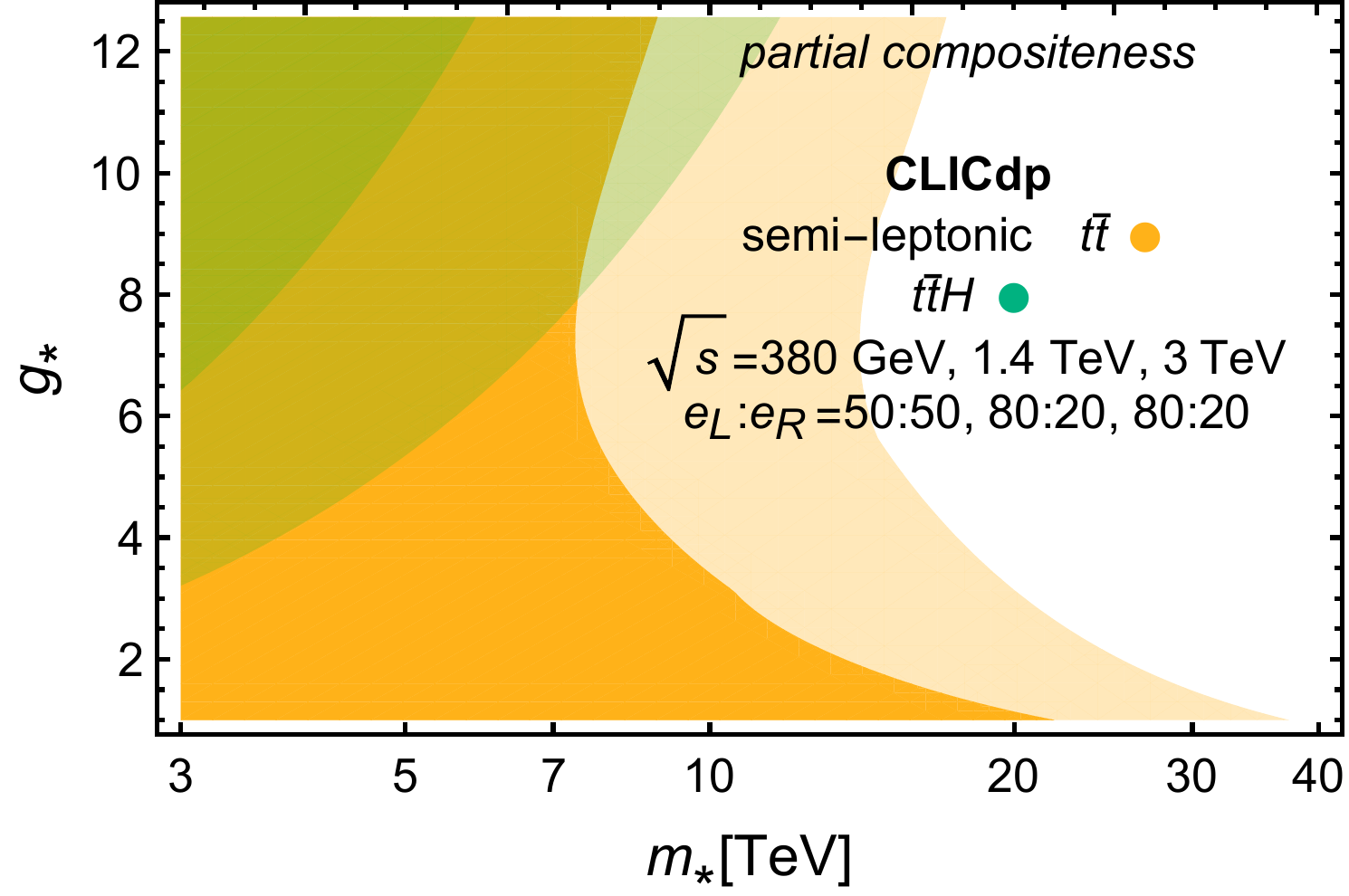}
      \caption{}
   \end{subfigure}
   \hspace{.01\linewidth}
   \begin{subfigure}{.48\textwidth}
      \includegraphics[width=\linewidth]{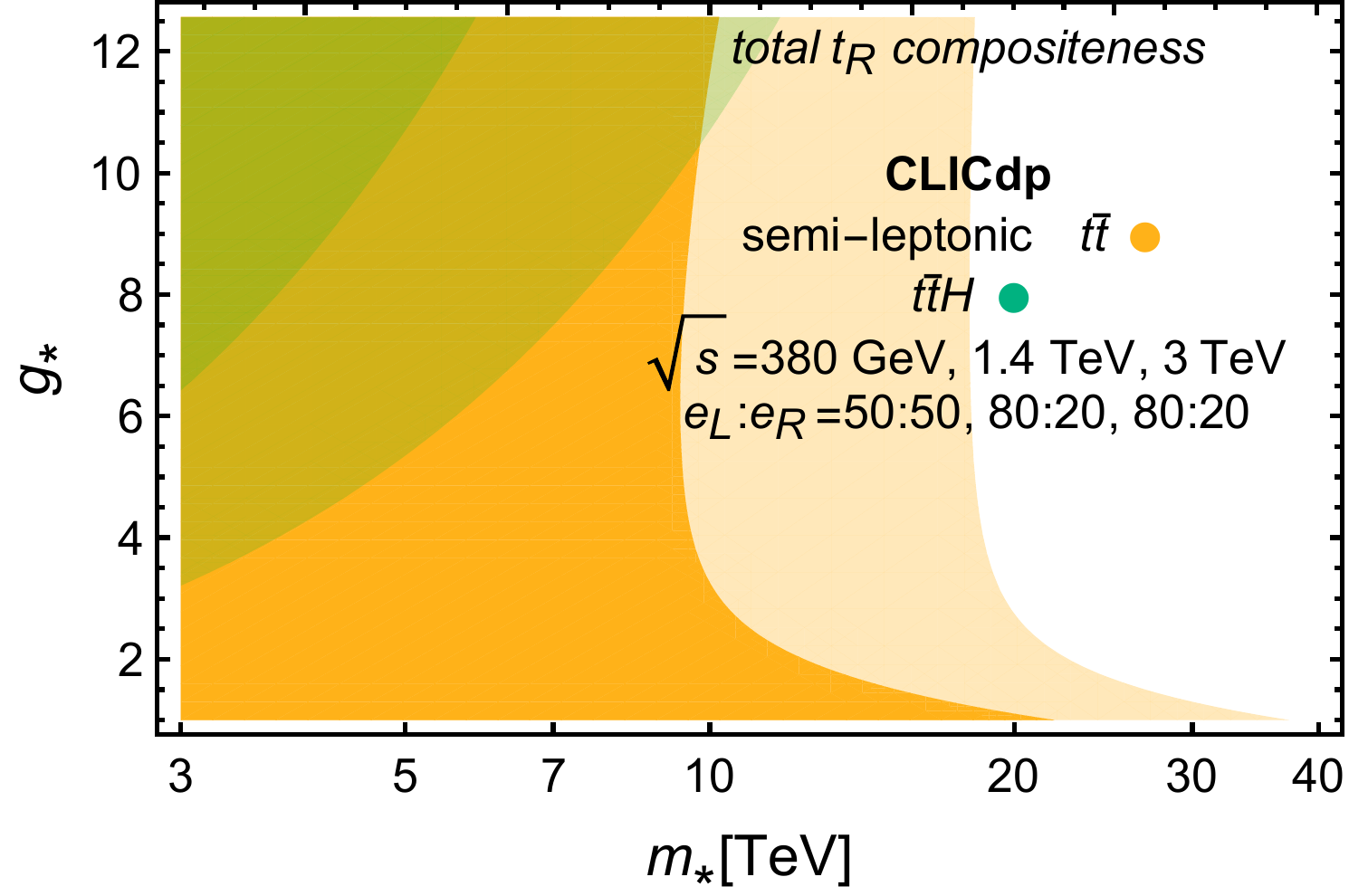}
      \caption{}
   \end{subfigure}
\caption{Top compositeness: `Optimistic' (light colour) and `pessimistic' (dark colour) $5\sigma$ discovery regions for (a) the partial compositeness, and (b) the total $t_R$ compositeness scenarios. The orange contours are derived from the $\ttbar$ global fit, while the green contours are derived from the top Yukawa analysis. From~\cite{ClicTopPaper}.}
\label{fig:compositeness_limits_2}
\end{figure}

\paragraph{Vector-boson fusion production of top-quark pairs}
The high-energy stages of CLIC allow the production of top-quark pairs through
vector boson fusion, which could reveal large BSM effects.
An important role is played by longitudinally-polarised vector bosons,
which at high energy are equivalent to the Higgs field.
In several new physics scenarios that address the naturalness problem,
Higgs boson and top-quark interactions receive large modifications, and 
a process like $\PWp\PWm\to\PQt\PAQt$ is directly sensitive to this,
for example through the modification of the $\PQt\PAQt$ scattering angle
in the centre-of-mass frame.
The process $\PWp\PWm\to\PQt\PAQt$ and its backgrounds have been studied 
at parton level at $\roots=\SI{3}{\TeV}$.
In \epem collisions at CLIC, the backgrounds to this process (including $\ttbar$ production)
can be suppressed to a negligible level. 
The vector-boson fusion process has been used to constrain
possible BSM effects as described in~\ref{sec:physics_bsm},
and in particular the high $M(\ttbar)$ region is sensitive to
BSM effects that grow with energy.

\subsection{Direct and indirect searches for BSM physics}
\label{sec:physics_bsm}

The exploration of physics beyond the Standard Model is well motivated by
problems that the Standard Model cannot address, such as the origin of
the weak scale, the nature of dark matter, and the
origin of the asymmetry between baryons and anti-baryons in the universe.
Typically, at \epem colliders new particles can be observed almost up to the kinematic
limit, e.g. $m\lesssim \roots/2$, and the flexibility to adjust beam energies
and polarisations at a linear collider potentially allows accurate determination of their quantum numbers. 
Earlier CLIC studies emphasised the capabilities for characterising BSM
particles expected to be discovered at the LHC or CLIC, and unveiling the underlying
BSM theory~\cite{Battaglia:2004mw, cdrvol2, cdrvol3}.
While this is still important, recent work has focused on the
prospects for discoveries at CLIC, in the event of no new physics being
observed at the LHC.  The goal is to go beyond attempting to quantify
the performance in terms of a few benchmark models or scenarios, and
to explore the landscape of fundamental physics as broadly as possible,
showing the reach for broad classes of theories.
With this in mind, many new physics models and scenarios have been investigated
and are gathered together in a dedicated report~\cite{ESU18BSM}. 

Potential new physics can be probed directly by searching for new states,
and indirectly through interpretation of precision measurements. 
The latter approach can be investigated through
the Standard Model Effective Field Theory (SM-EFT), which allows the
systematic parameterisation of BSM effects and their modification of SM processes.
Direct searches are explored in the context of general scenarios: for
example, extended Higgs sectors, which can be connected with models of
electroweak baryogenesis; and general dark matter models, such as Higgsinos
and minimal dark matter.  CLIC can discover the Higgsino for a mass of \SI{1}{\TeV},
which is the mass that it would need to have to be responsible for the observed
relic dark matter density. CLIC can also be conclusive on other relevant
and less standard dark matter scenarios. 
In the remainder of this section, these and other highlights from the
recent studies are discussed further in order to illustrate the range and reach of CLIC.

\paragraph{Global EW EFT analysis}
CLIC precision measurements can be sensitive to different
BSM contributions that arise from heavy BSM dynamics, associated with a mass
scale beyond CLIC's direct energy reach.
CLIC sensitivities to Higgs couplings, top-quark observables, $\PW\PW$ production,
and two-fermion scattering processes $\epem\to f \overline{f}$ where $f=\PQc, \PQb, \Pe, \PGm$, have all been put together
in a global fit using the SM-EFT.
The purpose of the global fit is to probe many classes of BSM theory at once,
in a model-independent fashion.  The SM-EFT extends
the dimension-four SM Lagrangian to include interaction operators of higher dimension $d>4$.
Here, BSM particles are implicit and must be light enough and strongly-enough
coupled to the SM to generate large enough operators to give visible effects,
while being heavy enough not to be produced directly.
The leading effects can be captured by dimension-6 operators:
\begin{equation*}\label{EFTLAG}
\mathcal{L}_{\rm eff} = \mathcal{L}_{\rm SM} + \frac{1}{\Lambda^2} \sum_{i} c_i \mathcal{O}_i+\cdots
\end{equation*}
for dimensionless coefficients $c_i$ and a common suppression scale $\Lambda$.
Through the global fit, limits are placed on the coefficients for operators or
combinations of operators, which can be translated into constraints on particular BSM models.

The CLIC sensitivity to the operator coefficients $c_i/\Lambda^2$ for one operator basis set
is shown in~\ref{fig:eft_limits_summary}.  Sensitivity to smaller values corresponds to
probing higher mass scales.  Sensitivities for the three CLIC energy stages are shown,
along with preliminary HL-LHC sensitivities for comparison.  
The HL-LHC sensitivities are shown for two scenarios of systematic uncertainties:
`S1', where LHC Run 2 systematics are kept constant with integrated luminosity;
and `S2', where Run 2 theoretical uncertainties are halved and experimental uncertainties
are scaled down with the square root of integrated luminosity until they reach
a defined minimum~\cite{CMS_HLLHC_Higgs, Gori:2650162}.
CLIC's measurements in the Higgs, top, and EW sectors at all three energy
stages are found to be highly synergistic.
\ref{fig:eft_limits_summary} shows that the initial stage of CLIC is already very
complementary to the HL-LHC for many of the operators. 
The high-energy stages, which
are unique to CLIC among all proposed \epem colliders, are found to be
crucial for the precision programme.
For example, the operators $c_{3\PW}, c_{2\PW}$, and $c_{2\PB}$ have effects that
grow with energy.  Hence, they can be probed better by observable measurements at
high energy with even moderate accuracy, than they can by very accurate Higgs measurements.
The operator $c_6$ relates to the Higgs self-coupling, and benefits from the
direct double-Higgs production available at high energies.
The overall result is to probe the SM-EFT much more precisely than is possible at the HL-LHC,
to mass scales well beyond the centre-of-mass energy of the collider.

In the following sections, the direct and indirect sensitivities
to some general models are discussed.

\begin{figure}[t]
\begin{center}
\includegraphics[width=\linewidth]{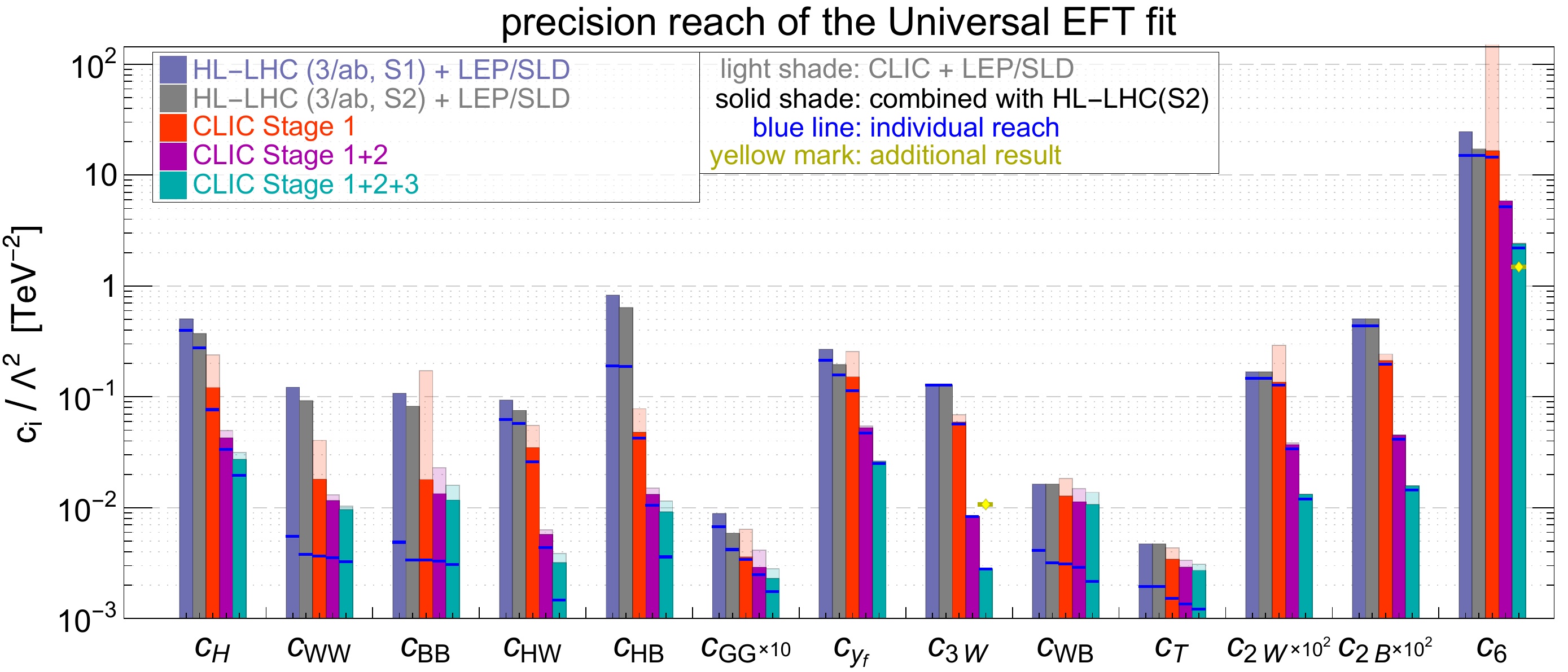}
\caption{Summary of the sensitivity to SM-EFT operators $c_i/\Lambda^2$ from a global analysis of CLIC's sensitivities to Higgs couplings, top-quark observables, $\PW\PW$ production, and two-fermion scattering processes $\epem\to f \overline{f}$, for three energy stages.  Smaller values correspond to a higher scale probed. Preliminary projections for HL-LHC are shown for comparison, under two systematic uncertainty scenarios (described in text).  Blue markers correspond to single-operator sensitivies, and yellow markers correspond to results from dedicated individual analyses (for example, the Higgs self-coupling analysis). From~\cite{ESU18BSM}. }
\label{fig:eft_limits_summary}
\end{center}
\end{figure}

\paragraph{Extended Higgs sectors}
In many extensions of the Standard Model the scalar sector is extended by
new states that are not charged under the Standard Model gauge group.
These states, referred to as `singlet' states, do not interact with
Standard Model gauge bosons and fermions at tree level, but may acquire such interactions
at loop level or through mixing with the Higgs boson radial mode. 
Through the couplings that the singlets `inherit', their 
production and decay can be like Higgs bosons. 
Experimentally, therefore, they may be searched for directly as resonances,
or indirectly through their modifications to the Higgs couplings.

In this example a singlet-like state $\PGf$ that is heavier than twice the
\SI{125}{\GeV} SM-like Higgs boson, decays to two SM-like Higgs bosons.
Owing to the large branching fraction $\PH\to\PQb\PAQb$, the most promising
final state comes from the decay to four $\PQb$-quarks,
which can be well identified at CLIC:
$\PGf\to\PH\PH\to\PQb\PAQb\PQb\PAQb$.
This has been analysed using a fast simulation of the CLIC detector, 
including background processes.  

The phenomenology of the model is determined by $\sin\gamma$, where 
$\gamma$ describes the mixing between the interaction eigenstates
$\PSh_0$ (a pure doublet of the weak interaction) and $S$ (a real scalar singlet) 
to give the mass basis $\PH(\SI{125}{\GeV})$ and $\PGf$:
\begin{equation*}
\label{eq:eigenstates-singlets}
\PH = \PSh_0 \cos\gamma  + \text{S} \sin\gamma, \;\;\;\;\;\; \PGf = \text{S} \cos\gamma - \PSh_0 \sin\gamma.
\end{equation*}

The parameter space of the mass of the heavy state $\PGf$, $m_{\PGf}$, and the mixing
$\sin^2\gamma$ can be explored, and 
the projected cross section bounds for the resonant final states
translated into \SI{95}{\percent} C.L. limits as shown in~\ref{fig:higgsSinglets}~\cite{Buttazzo:2018qqp,ESU18BSM}.

The CLIC Higgs coupling sensitivities also give an indirect constraint on $\sin^2\gamma$,
because the Higgs signal strengths are reduced universally by a factor $(1-\sin^2\gamma)$
from their SM values.  This constraint is complementary to the direct search constraint, and 
is also shown as a horizontal line in~\ref{fig:higgsSinglets}.

It is seen that the CLIC limits extend significantly beyond those projected for the HL-LHC,
to much lower values of the mixing:  for $m_{\PGf}<\SI{1}{\TeV}$, the mixing $\sin^2\gamma$
must be lower than around 0.0013.

\begin{figure}[h]
\begin{center}
\includegraphics[width=0.7\textwidth]{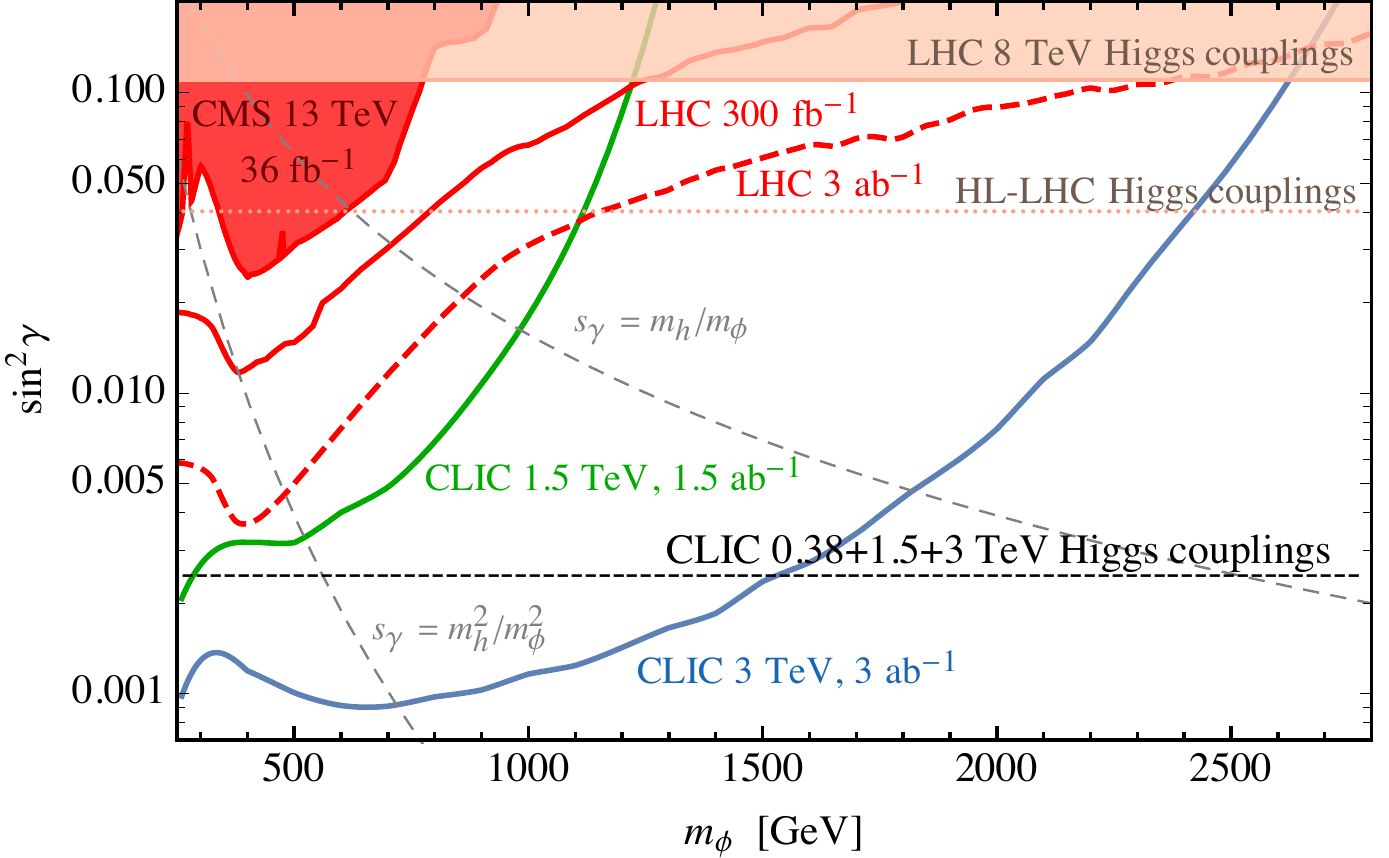}
\caption{Constraints on singlet-like states $\PGf$ at \SI{95}{\percent} C.L. in the plane $(m_{\PGf},\sin^2\gamma)$. The reach from the direct search for $\PGf\to\PH\PH\to\PQb\PAQb\PQb\PAQb$, for the second (green) and third (blue) CLIC stages are compared with the projections for LHC with a luminosity of \SI{300}{\per\fb} (solid red) and \SI{3}{\per\ab} (dashed red). Branching ratios of the $\PGf$ have been fixed as $\br{\PGf\to \PSh\PSh} = \br{\PGf\to \PZ\PZ} = \SI{25}{\percent}$. The horizontal line shows the indirect constraint from the CLIC Higgs couplings measurements (stat. only). The shaded regions are the present constraints from LHC direct searches for $\PGf\to \PZ\PZ$ (red) and Higgs couplings measurements (pink).  From~\cite{Buttazzo:2018qqp,ESU18BSM}. }
\label{fig:higgsSinglets} 
\end{center}
\end{figure}

{\bf \em Model interpretations}
Many BSM models contain extra singlets of the kind generically described above,
and so the same experimental measurements can be interpreted in different model frameworks.  
The next-to-minimal supersymmetric Standard Model, NMSSM, is the MSSM augmented by a single chiral super-field, and contains a scalar singlet.
The above results have been translated into the NMSSM parameter space of $m_{\PGf}$ and $\tan\beta$;
in this framework the $\roots=\SI{1.5}{\TeV}$ stage of CLIC is already more sensitive
than the HL-LHC over the whole parameter space, and the $\roots=\SI{3}{\TeV}$ stage
significantly extends the reach.  Similarly for Twin Higgs models, the $\roots=\SI{3}{\TeV}$
CLIC stage significantly surpasses the HL-LHC reach~\cite{Buttazzo:2018qqp,ESU18BSM}.

{\bf \em Relaxions}
The examples above discuss heavy singlets.  The case of light singlets can also
be examined.  One example comes from the relaxion mechanism, which stabilises
the Higgs mass dynamically and is interesting for addressing the hierarchy problem. 
Similarly to the heavy singlets, the relaxion $\PGf$ mixes with the Higgs and
inherits its couplings to SM fields.
The possible relaxion mass range spans from sub-\si{\eV} to tens of \si{\GeV}.
For masses in the \si{\GeV} range the relaxion is short-lived and decays inside the detector.
At CLIC, three approaches have been considered:
searching for the exotic decay $\PH\to\PGf\PGf$ directly via the decay $\PGf\to\PQb\PAQb$
(extrapolated from ILC studies), constraining the decay $\PH\to\PGf\PGf$ indirectly
via Higgs `untagged' decays that, while not invisible, consist of soft particles that are not
reconstructed as Higgs decays and appear as an extra invisible contribution to the Higgs width;
and directly searching via the recoil mass in the process
$\epem\to\PGf\PZ$.  
CLIC will have the potential to exclude masses down to \SI{20}{\GeV} with indirect
searches and further down to \SI{12}{\GeV} with direct searches~\cite{Frugiuele:2018coc,ESU18BSM},
significantly reducing the available model parameter space beyond the reach of the
HL-LHC in regions that are challenging for hadron colliders.

\paragraph{Dark matter}
The Higgsino is a compelling target of searches for supersymmetric extensions
of the Standard Model: it is strongly connected to the naturalness of the weak scale,
is important for gauge coupling unification, and is an ideal WIMP dark matter candidate.
Naturalness considerations motivate Higgsinos in the range $m_{\PGc} \lesssim \si{\TeV}$,
while the observed dark matter relic abundance singles out $m_{\PGc} \simeq \SI{1.1}{\TeV}$
for thermal Higgsino dark matter. High centre-of-mass lepton colliders such as CLIC
provide one of the best avenues for probing Higgsinos across this mass range.
A Higgsino that is not the lightest supersymmetric particle can decay into
a SM boson and missing energy, and these cases have been examined in previous studies~\cite{cdrvol2}.
Here, two complementary approaches to searches for Higgsinos in challenging scenarios are discussed:
searches for mono-photon signatures, and searches for disappearing stub track signatures.

{\bf \em Mono-photon signature} 
In the search for dark matter candidates using ISR photons, $\epem\to\PGc\PGc\PGg$, 
the experimental signature is a single photon reconstructed in the detector, while
the $\PGc$ particles escape undetected.  
A full simulation study at $\roots=\SI{380}{\GeV}$ has been carried out,
including the main SM background processes, for different values of $m_{\PGc}$.
The \SI{95}{\percent} upper cross section limit for the process $\epem\to\PGc\PGc\PGg$ is shown
as a function of $m_{\PGc}$ in~\ref{fig:monophotonHiggsino_a}~\cite{ESU18BSM}.
It can be seen that CLIC at $\roots=\SI{380}{\GeV}$ would be sensitive to the
$\epem\to\PGc\PGc\PGg$ process down to cross sections of \SIrange{5}{10}{\fb} in the
mass range from the LEP limit of around \SI{100}{\GeV} up to almost \SI{180}{\GeV},
depending on the systematic error assumption, 
and Higgsino pair production with an ISR photon would be excluded across
the entire mass range.
The mono-photon-based search at an \epem collider is complementary to mono-jet
searches at hadron colliders as the different production mechanism allows the coupling
to leptons to be probed.  Also, once a signal is established, the photon energy
distribution could be used to measure the mass of the dark matter candidate
in a way that would be very difficult at a hadron collider. 
For a \SI{120}{\GeV} Higgsino particle, a precision on its mass of about \SI{2}{\GeV}
is expected using the endpoint of the photon energy distribution.

\begin{figure}[h]
    \centering
    \begin{subfigure}{.43\textwidth}
       \includegraphics[width=\linewidth,trim=0 -8mm 10mm 0,clip]{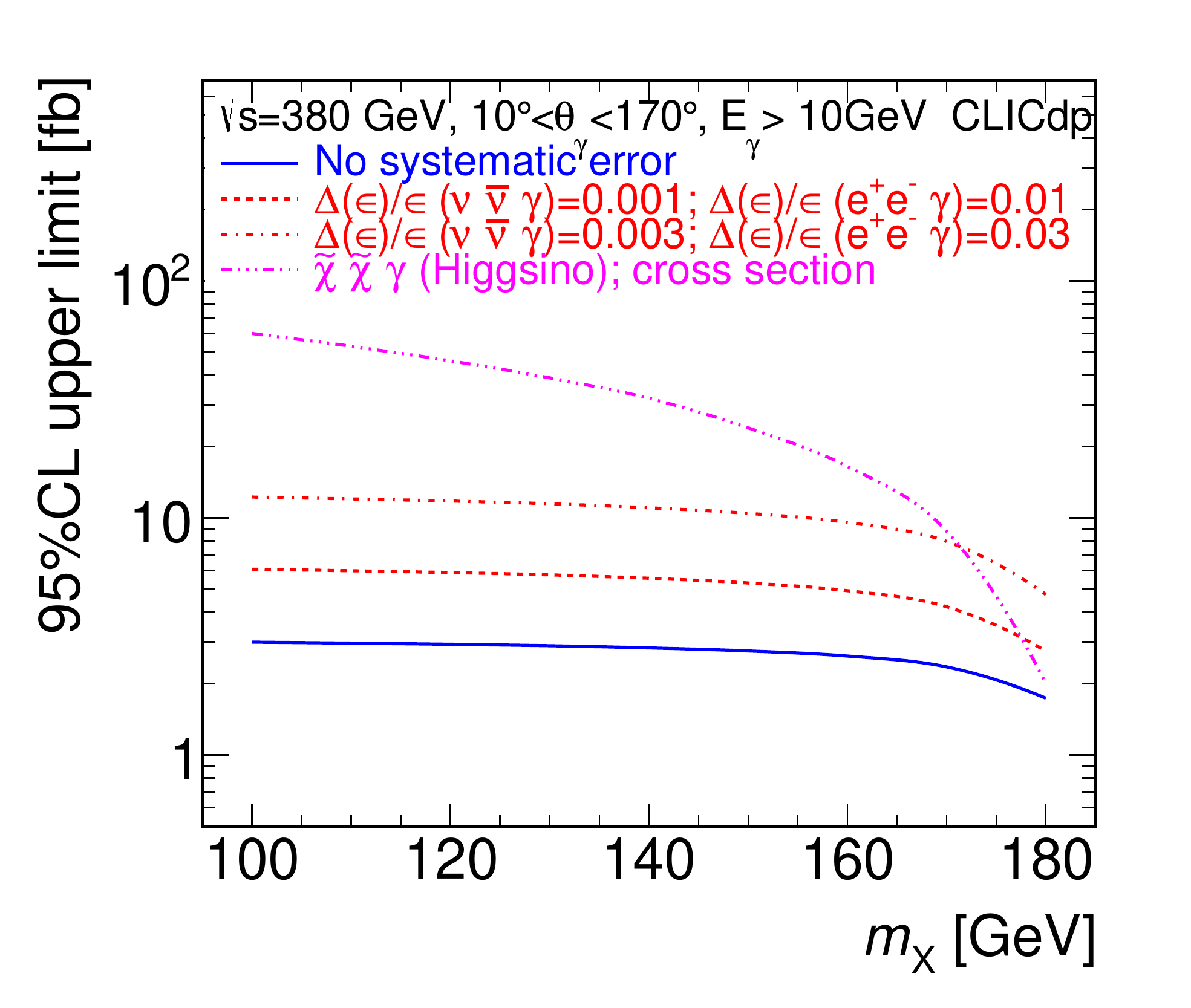}
       \caption{\label{fig:monophotonHiggsino_a}}
      \end{subfigure}
    \hspace{.01\linewidth}
    \begin{subfigure}{.54\textwidth}
       \includegraphics[width=\linewidth]{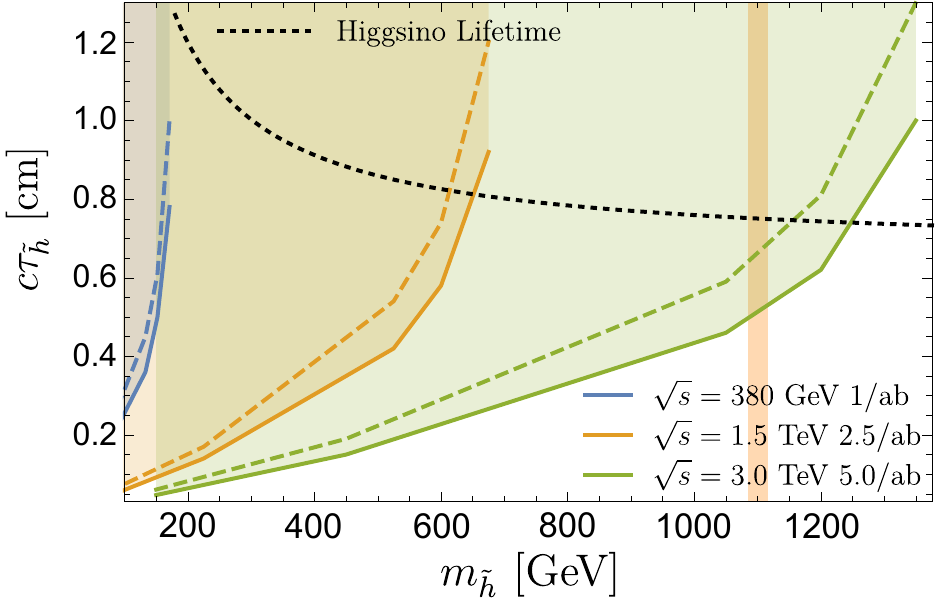}
       \caption{\label{fig:monophotonHiggsino_b}}
    \end{subfigure}
    \caption{(a): \SI{95}{\percent} C.L. upper limit on the $\epem \to \PGc\PGc\PGg$ cross section at $\roots = \SI{380}{\GeV}$ as a function of $m_{\PGc}$. The limit without including systematic uncertainties (blue) is compared to two different assumptions on the systematic uncertainty for the two main background processes (red). In addition, the cross section for Higgsino pair production with an ISR photon is shown (magenta).  (b): \SI{95}{\percent} C.L. exclusion contours in lifetime--mass for $N=3$ (solid) and $N=30$ (dashed) Higgsino events in the detector acceptance at the three stages of CLIC.  The black dashed line indicates the lifetime of the pure Higgsino state of a given mass, and the pure Higgsino thermal relic dark matter mass of \SI{1.1}{\TeV} is indicated by a vertical line. From~\cite{ESU18BSM}.}
   \label{fig:monophotonHiggsino}
\end{figure}

{\bf \em Disappearing track signature}
A very different experimental signature for dark matter production could be
`disappearing tracks', for example caused by a heavy charged particle passing
through the detector before decaying to a neutral particle that escapes detection.
Generically, `long-lived particles' could produce this signature.  
A particular example is the charged particle in a Higgsino multiplet, $\PGc^\pm$,
which can be only slightly heavier than the neutral components $\PGc^0_{1,2}$.
In this case the $\PGc^\pm$ travels a macroscopic distance, of order 1\,cm,
before decaying into an invisible $\PGc^0$ and SM states that are too soft to
reconstruct: $\PGc^\pm \to \PGppm \PGc^0$. 
The charged Higgsino lifetime makes it particularly challenging for LHC searches,
as the `charged stub' left by the Higgsino is short and is challenging to reconstruct 
among the large number of hits from pile-up.

CLIC prospects for probing the pure Higgsino via the production of charged
Higgsino pairs and the ensuing disappearing track signature have been assessed. 
The study does not use full simulation, but takes into account the CLIC detector
geometry by requiring that the track stub traverses enough of the CLIC tracker
to leave at least four hits.
The most sensitive analysis comes from requiring a signature of only one charged stub;
for which the \SI{95}{\percent} C.L. exclusion limits are shown in~\ref{fig:monophotonHiggsino_b}~\cite{ESU18BSM}.
Possible variations in the Higgsino lifetime, which may arise if the Higgsino
is not exactly a pure state, are also given. 
It is seen that for $\roots=\SI{3}{\TeV}$, the 1-stub strategy yields around 30
events in the acceptance up to the thermal dark matter target of $m_{\PGc}\simeq \SI{1.1}{\TeV}$.
A handful of events are produced in acceptance even if an ISR photon with $p_T > \SI{100}{\GeV}$ is required. 
CLIC should therefore be able to probe the thermal relic Higgsino dark
matter even with some level of background.
The large tracker volume, compact vertex detector geometry and low background 
conditions compared to those at hadron colliders give CLIC better sensitivity
for this type of signature than the HL-LHC.

\paragraph{Electroweak precision tests}
The process $\epem\to f \bar{f}$, where $f$ is a SM fermion, can be studied in depth
at CLIC, for example through the differential cross section with respect to polar angle,
$d\sigma/d\cos\theta$, and related asymmetries.  
Precision measurements of differential distributions and asymmetries are sensitive
to corrections induced by any new state $\PGc$ that has SM charges, which can modify the
EW gauge boson propagators. 
This has been investigated by adding form factors to the effective Lagrangian
and establishing \SI{95}{\percent} C.L. limits that could be set
through a combination of the $\Pe, \PGm, \PQb$, and $\PQc$ channels 
on the masses of different states $\PGc$.  
These limits are shown in~\ref{Plot:Results_CLUC_comb}~\cite{DiLuzio:2018jwd,ESU18BSM}, 
for the second and third stages of CLIC. 
The exclusions assume polarisation fractions $P(\Pem) = \SI{-80}{\percent}$ and $P(\Pep) = 0$.
Limits are given for different states $\PGc \sim (1,n,Y)$,
where the entries denote the $\rm{SU}(3)_c\times\rm{SU}(2)_L\times\rm{U}(1)_Y$
representation; and for different Lorentz representations: 
complex scalar (CS), Majorana fermion (MF), and Dirac fermion (DF). 
For example, the $n=2$ Dirac fermion corresponds to the Higgsino and the 
$n=3$ Majorana fermion to the wino.
The limits show that CLIC at $\roots=\SI{3}{\TeV}$ excludes the $n=3$ Dirac fermion
for $m_{\PGc} < \SI{0.8}{\TeV}$ and $\SI{1}{\TeV} < m_{\PGc} < \SI{2}{\TeV}$; and in addition
the intermediate region  $\SI{0.8}{\TeV} < m_{\PGc} < \SI{1}{\TeV}$ is covered by
CLIC at $\roots=\SI{1.5}{\TeV}$.
For the $n=3$ Dirac fermion, a mass of $m_{\PGc}=\SI{2}{\TeV}$ is needed to saturate
the dark matter relic abundance~\cite{ESU18BSM, DiLuzio:2018jwd}, and so CLIC covers the whole relevant
parameter space and fully tests that dark matter hypothesis.

\begin{figure}[ht]
   \centering
      \includegraphics[width=0.65\linewidth]{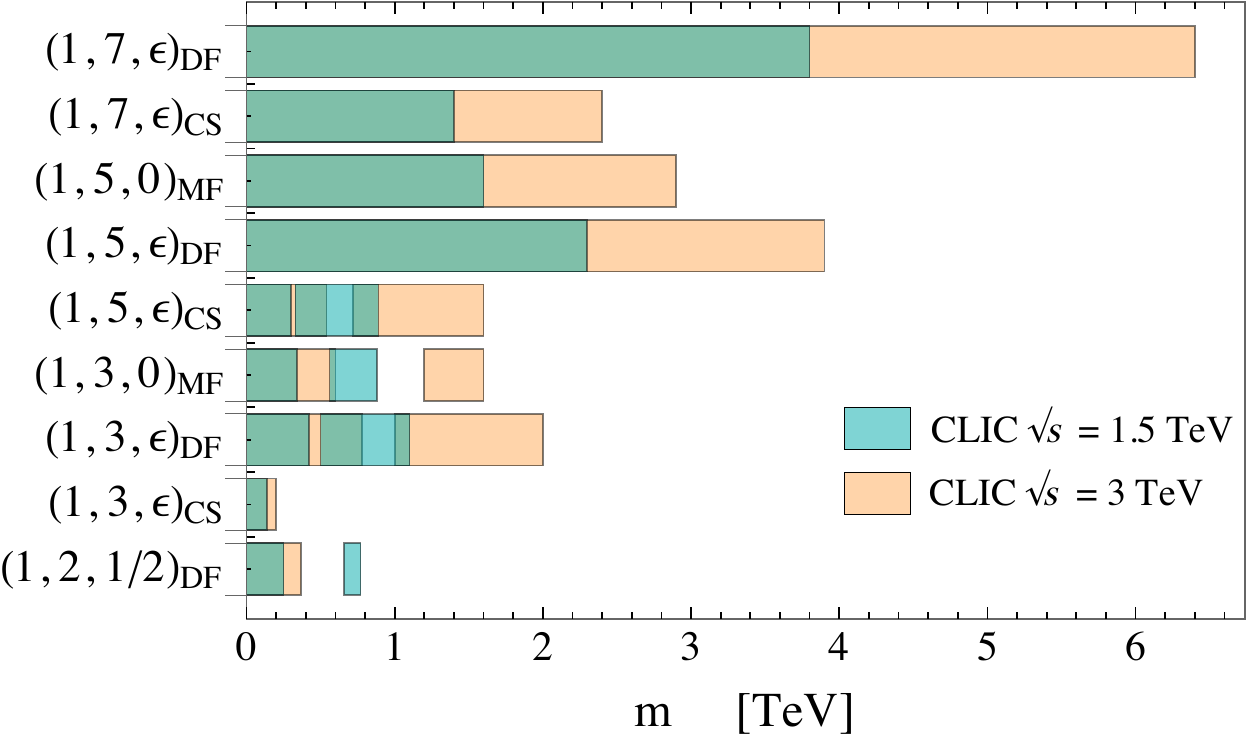}
   \caption{$\SI{95}{\percent} $ C.L. exclusion ranges on the mass of new SM-charged states from $\epem\to f \bar{f}$
   precision measurements for $P_{\Pem}=\SI{-80}{\percent}$ data at $\roots=\SI{1.5}{\TeV}$ (blue) and $\SI{3}{\TeV}$ (beige).
   Green regions would be excluded by data from both centre-of-mass energies. Exclusion regions are obtained by combining the $\Pe, \PGm, \PQb$, and $\PQc$ channels with \SI{0.3}{\percent} systematic error. $(1,n,Y)$ denotes the $\rm{SU}(3)_c\times\rm{SU}(2)_L\times\rm{U}(1)_Y$ representation of the state, and the mass exclusions are given for different Lorentz representations: complex scalar (CS), Majorana fermion (MF), and Dirac fermion (DF). $\epsilon$ denotes a milli-charge which has no bearing on collider physics, but ensures stability.  From~\cite{DiLuzio:2018jwd,ESU18BSM}.}
\label{Plot:Results_CLUC_comb}
\end{figure}

\paragraph{Electroweak phase transition}

The SM plus singlet scenario discussed previously can be used to probe 
the nature of the electroweak phase transition in the early universe.
As the universe expanded it cooled, and electroweak symmetry breaking occurred
when it became energetically favourable for the Higgs field to acquire a non-zero
vacuum expectation value.  
If there is a potential barrier separating the symmetric vacuum from the broken one,
the electroweak phase transition could be first order. 
The addition of a singlet can result in such a first-order electroweak phase transition. 
A strong first-order phase transition is a necessary condition for
electroweak baryogenesis, so 
the shape of the Higgs potential and the nature of the electroweak phase transition
is a critical open question that could shed light on the stability or instability of
the vacuum, and potentially on the origin of the baryon asymmetry.

CLIC can search for resonant double-Higgs production arising from a singlet,
as well as looking for deviations from SM predictions for the Higgsstrahlung
cross section and the Higgs self-coupling.  These measurements can set limits
on the parameter space for general scenarios that lead to a strong first-order
phase transition.
Examples are shown in~\ref{EWPT_Plot_500}~\cite{No:2018fev,ESU18BSM}.
The parameter space has been scanned for various fixed values of the singlet mass
and singlet mixing; shown here is the singlet mass $m_2=\SI{500}{\GeV}$ and
two values of the mixing: $\sin\,\theta=0.05$, and $0.1$. 
Values of $\sin\,\theta$ higher than around 0.2 will be excluded by HL-LHC.  
The parameter space is that of coefficients $a_2$, $b_3$, and $b_4$
of additional $T^2$ terms in a temperature-dependent effective potential
that is added to the Higgs potential and modifies the behaviour as the
early universe expands and cools.  
Points in the parameter space compatible with unitary, perturbativity, and
absolute stability of the EW vacuum have been identified (red circles),
and among them the points yielding a strong first-order phase transition
have been identified (green circles).
Constraints are provided by a fit to an overall rescaling of the Higgs couplings $\Delta\kappa$  
(grey regions), the Higgs self-coupling (black lines), and by searches for
resonant double-Higgs production in the 4$\PQb$ final state at $\roots=\SI{1.4}{\TeV}$ and $\SI{3}{\TeV}$
(orange and blue lines).  The excluded regions are the regions outside the pairs of lines,
for higher absolute values of $a_2$.

The three complementary methods are all competitive in the interesting region and 
in the examples in~\ref{EWPT_Plot_500} are able to exclude all of the 
interesting parameter space.
CLIC therefore provides a direct avenue to probe the nature of the EW phase
transition for non-minimal scalar sectors, and the possible origin of
the cosmic matter-antimatter asymmetry via electroweak baryogenesis.

\begin{figure}[h]
    \centering
    \begin{subfigure}{.48\textwidth}
        \includegraphics[width=\linewidth,trim=0 0 27cm 0,clip]{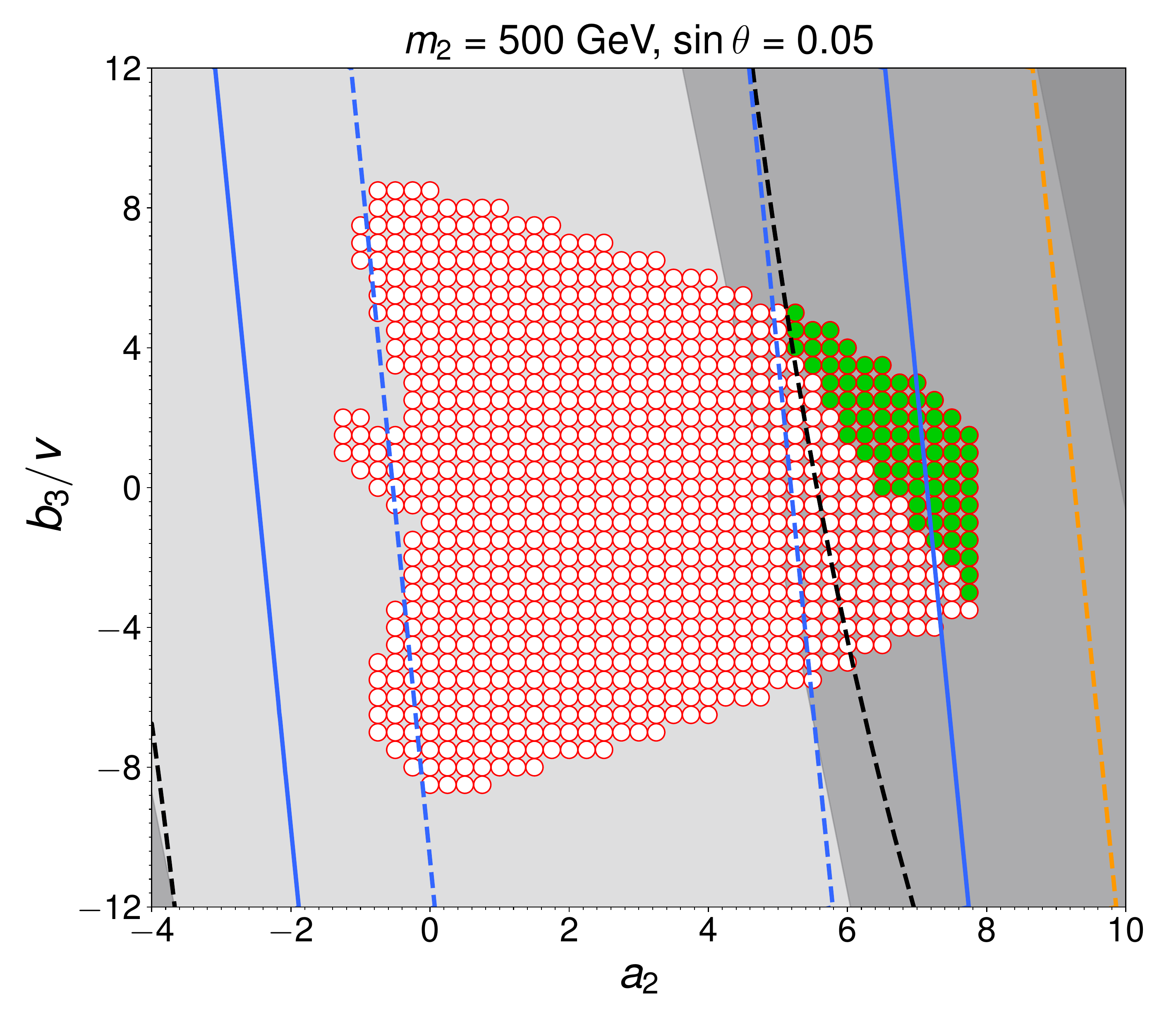}
        \caption{}
    \end{subfigure}
    \hspace{.035\linewidth}
    \begin{subfigure}{.45\textwidth}
        \includegraphics[width=\linewidth,trim=29cm 0 0 0,clip]{figures/physics/PLOT_EWPT_500_CLIC_Report_Stage123_NEW_181109.pdf}
        \caption{}
    \end{subfigure}

\caption{\small  
Region of parameter space in ($a_2$, $b_3/v$) for singlet mass $m_2=\SI{500}{\GeV}$
and singlet mixing (a) ${\rm sin}\,\theta = 0.05$ and (b) ${\rm sin}\,\theta = 0.1$ 
within the \SI{95}{\percent} C.L. sensitivity reach of resonant double-Higgs production
searches at CLIC.  Limits are given for $\roots=\SI{1.4}{\TeV}$ (orange) and $\roots=\SI{3}{\TeV}$ (blue)
for a $b$-tagging efficiency of \SI{70}{\percent} (solid) and \SI{90}{\percent} (dashed): the CLIC sensitivity
is the region not contained within each pair of sensitivity lines.
The red circles indicate the region compatible with the requirements of unitary,
perturbativity and absolute stability of the EW vacuum. 
The parameter $b_4$ has been scanned over. 
Overlaid is the region yielding a strongly first order EW phase transition (green points).
The dashed black lines are \SI{95}{\percent} C.L. lines corresponding to the \SI{68}{\percent} C.L.
CLIC sensitivity $[\SI{-7}{\percent},\SI{+11}{\percent}]$ on the Higgs self-coupling. 
The yellow band (only for $\sin \theta = 0.1$) corresponds to the
projected sensitivity of $\Pp\Pp\to \Ph_2 \to\PZ\PZ$ searches at HL-LHC.
The region within reach of a measurement of an overall scaling of Higgs couplings $\Delta\kappa$ 
from the first, second, and third CLIC stages is shown in dark, middle, and light grey, respectively.
From~\cite{No:2018fev,ESU18BSM}.}
\label{EWPT_Plot_500}
\end{figure}

\subsection{Overall CLIC physics reach}
This section has presented a selection of recent studies to illustrate the overall
CLIC physics reach.

In Higgs physics, the initial CLIC energy stage already provides coupling measurements
that are in many cases significantly more precise than for the HL-LHC.
The higher CLIC energy stages improve upon these, and give access to further couplings. 
Measured in \epem collisions, the couplings can be determined in a model-independent way;  
at CLIC they can reach precisions at the percent level,
while the model-dependent precisions reach the per-mille level.
The Higgs self-coupling can be determined at the level of $[\SI{-7}{\percent},\SI{+11}{\percent}]$.
CLIC can measure the top-quark mass at the level of \SI{50}{\MeV}, including
current theory systematics.  Measurements
of top-quark production and decay properties at all three energy stages can be used
to probe the top-quark couplings, and give sensitivity to potential new physics
scenarios such as top compositeness or CP violation in the $\PQt\PQt\PH$ coupling.  

All of these SM probes, together with measurements of diboson and Drell-Yan production, 
can be combined and interpreted in the SM-EFT framework.  In this way the 
precision studies at CLIC allow sensitivity to physics beyond the SM that originates 
from scales at tens of TeV, well above the centre-of-mass energy of the collider.
The high energy stages of CLIC, and the electron polarisation afforded by
the linear collider, gives sensitivity beyond that achievable at other colliders. 

General examples have been given to show that through direct searches,
CLIC is sensitive to new particles produced
in a wide range of new physics scenarios, including those that are challenging experimentally.
These scenarios include extended Higgs sectors and a variety of dark matter candidates.
Combining the direct and indirect searches provides a direct avenue to probe the
nature of the EW phase transition for non-minimal scalar sectors, and the possible
origin of the cosmic matter-antimatter asymmetry via electroweak baryogenesis.

Previous studies have demonstrated that if particles are discovered at CLIC or the LHC,
CLIC has the ability to measure their masses and couplings at the percent level or
better: typically much more precisely than is possible at a hadron collider~\cite{cdrvol2}.

A summary of these and further studies showing the overall reach for many aspects of potential
new physics is given in~\ref{tab:newPhysicsSummary}.
For example, new particles can be discovered directly over the whole CLIC kinematic reach, under many
new physics scenarios, and the sensitivity to high energy scales from EFT fitting is complemented by limits on new physics
scales in particular sectors arising from dedicated searches, such as those for lepton flavour violation. 

Overall, the stand-alone discovery and precision capacity of CLIC, complementary to that of
the HL-LHC, makes it an ideal facility for extending the search for physics beyond the SM.

\begin{table}[ht]\centering
  \caption{CLIC reach for new physics. Sensitivities are given for the full CLIC programme covering the three centre-of-mass energy stages at $\roots=\SI{380}{\GeV}$, \SI{1.5}{\TeV} and \SI{3}{\TeV} with integrated luminosities of \SI{1}{\per\ab}, \SI{2.5}{\per\ab}  and \SI{5}{\per\ab}, respectively. At \SI{380}{\GeV}, equal amounts of \SI{-80}{\percent} and \SI{+80}{\percent} polarisation running are assumed. Above \SI{1}{\TeV} a sharing in the ratio 80:20 is assumed between \SI{-80}{\percent} and \SI{+80}{\percent} electron polarisation.  All limits are at 95\% C.L. unless stated otherwise.\label{tab:newPhysicsSummary}}
\begin{adjustbox}{width=\linewidth}
\begin{tabular}{lll}\toprule
Process & HL-LHC & CLIC \\
   \hline
Heavy Higgs scalar mixing angle $\sin^2\gamma$  & $<\SI{4}{\percent}$ & $<\SI{0.24}{\percent}$ \\
Higgs self-coupling $\Delta\lambda$      & $\sim 50\%$ at \SI{68}{\percent} C.L. & $[\SI{-7}{\percent},\SI{+11}{\percent}]$ at \SI{68}{\percent} C.L.\\ 
\br{\PH\to \mathrm{invisible}}                                   &  & $<\SI{0.69}{\percent}$ at \SI{90}{\percent} C.L. \\
\hline
Higgs compositeness scale $m_*$  &  $m_*>\SI{3}{\TeV}$ & Discovery up to $m_*$ = $\SI{10}{\TeV}$ \\
                                 &  \;\;\;\;\; ($>\SI{7}{\TeV}$ for $g_*\simeq 8$) &  \;\;\;\;\; ($\SI{40}{\TeV}$ for $g_*\simeq 8$) \\
\hline
Top compositeness scale $m_*$    &  & Discovery up to $m_*$ = $\SI{8}{\TeV}$ \\
                                 &  &  \;\;\;\;\; ($\SI{20}{\TeV}$ for small coupling $g_*$) \\
\hline
Higgsino mass {\footnotesize{(disappearing track search)}}  & $>\SI{250}{\GeV}$ & $>\SI{1.2}{\TeV}$ \\ 
Slepton mass                                                     &  & Discovery up to $\sim 1.5$\,\TeV \\ 
RPV wino mass                                                    &  & $>\SI{1.5}{\TeV}$ ($\SI{0.03}{\meter}<c\tau<\SI{30}{\meter}$) \\ 
\hline
$\PZ'$ ({\footnotesize{SM couplings}}) mass            & Discovery up to 7\,TeV & Discovery up to 20\,TeV \\ 
\hline
NMSSM scalar singlet mass     &  $>\SI{650}{\GeV}$ ($\tan\beta=4$)  & $>\SI{1.5}{\TeV}$ ($\tan\beta=4$) \\ 
Twin Higgs scalar singlet mass                                    & $m_{\sigma} = f > 1$\,TeV & $m_{\sigma} = f > 4.5$\,TeV \\ 
\hline
Relaxion mass                             & $<\SI{24}{\GeV}$ & $<\SI{12}{\GeV}$ (all for vanishing $\sin\theta$) \\ 
Relaxion mixing angle $\sin^2\theta$                             &  & $\leq \SI{2.3}{\percent}$ \\
\hline
Neutrino Type-2 see-saw triplet                                  &  & $>\SI{1.5}{\TeV}$ ({\footnotesize{for any triplet VEV}}) \\
                                                                 &  & $>\SI{10}{\TeV}$ ({\footnotesize{for triplet Yukawa coupling $\simeq 0.1$}}) \\
\hline
Inverse see-saw RH neutrino                                     &  & $>\SI{10}{\TeV}$ ({\footnotesize{for Yukawa coupling $\simeq 1$}}) \\
\hline
Scale $V_{LL}^{-1/2}$ for LFV $(\bar{\Pe}\Pe)(\bar{\Pe}\PGt)$     &  & $>\SI{42}{\TeV}$ \\ 
\bottomrule
\end{tabular}
\end{adjustbox}
\end{table}

\cleardoublepage
\textcolor{white}{ }
\newpage
\section{CLIC accelerator design, technologies and performance}
\label{sec:accelerator}
\subsection{Introduction}
The aim of the studies for CLIC has been to develop the designs and the technologies to enable the building of a multi-\si{\TeV} electron--positron collider.
The feasibility of the concept and the technologies has been documented in the CLIC CDR~\cite{cdrvol1}.
As reported in~\ref{sec:physics}, the new staged approach to the project optimally addresses the physics needs.
The accelerator design, technologies and implementation have recently been optimised for the first energy stage at \SI{380}{\GeV}, while fully taking upgrades to \SI{1.5}{\TeV} and \SI{3}{\TeV} into account.
The following will summarise the CLIC design and parameters for the first energy stage, present a klystron-based alternative, and show the staging to higher energies.

\subsection{CLIC design and performance at \SI{380}{\GeV}}

\subsubsection{Design overview}
The schematic layout of the baseline CLIC complex for \SI{380}{\GeV} operation is shown in~\ref{scd:clic_layout} and the key parameters for all three energy stages are listed in~\ref{t:scdup1}. 
\begin{center}
\begin{figure}[h]
\includegraphics[width=\textwidth]{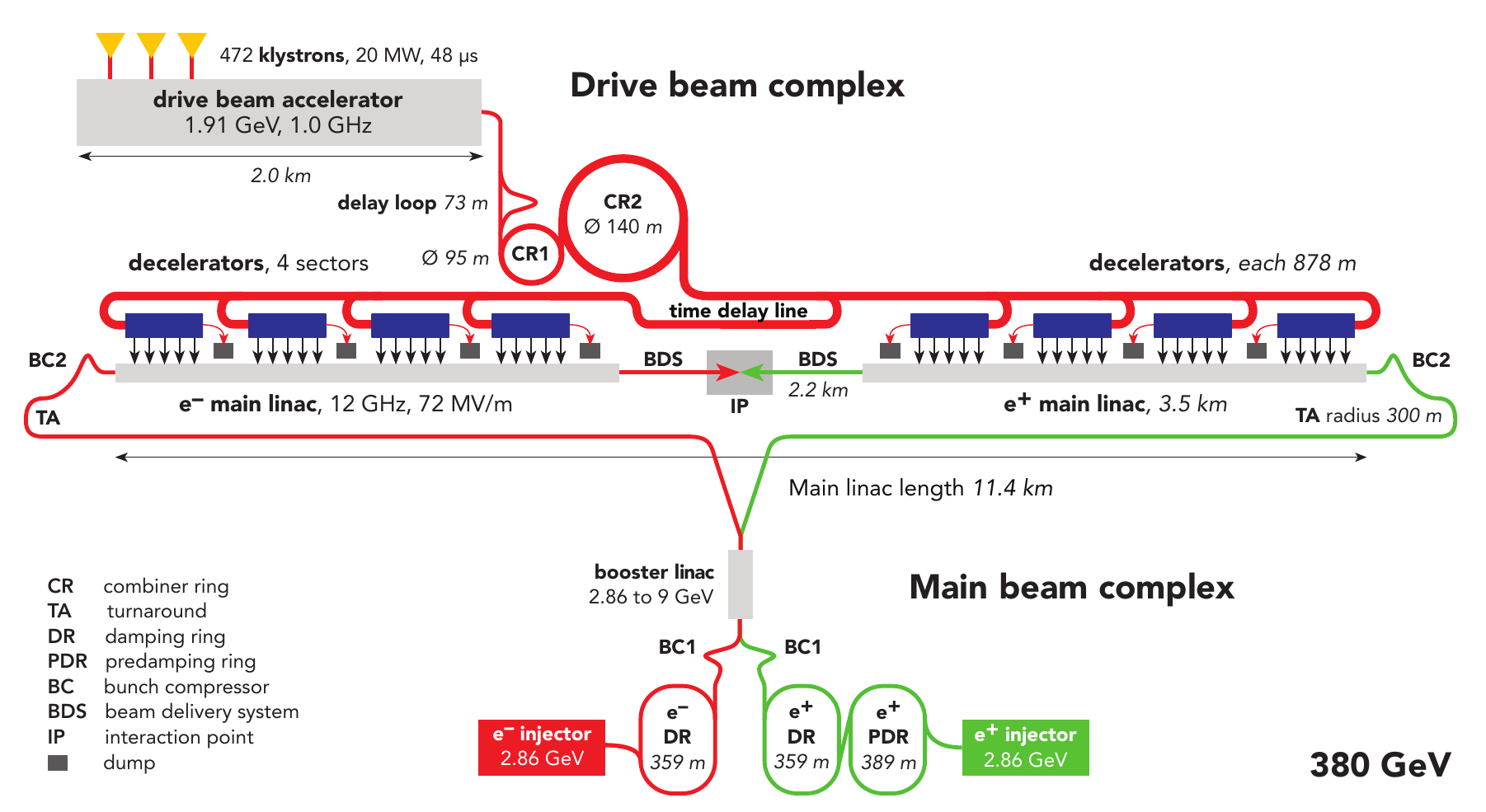}
\caption{Schematic layout of the CLIC complex at \SI{380}{\GeV}. \imcl}
\label{scd:clic_layout}
\end{figure}
\end{center}

The main electron beam is produced in a conventional radio-frequency (RF) source and accelerated to \SI{2.86}{\GeV}. The beam emittance is then reduced in a damping ring.
To produce the positron beam, an electron beam is accelerated to \SI{5}{\GeV} and sent into a crystal to produce energetic photons, which hit a second target and produce electron--positron pairs.
The positrons are captured and accelerated to \SI{2.86}{\GeV}. Their beam emittance is reduced first in a pre-damping ring and then in a damping ring. 
The ring to main linac system (RTML) accelerates the beams to \SI{9}{\GeV} and compresses their bunch length. The main linacs accelerate the beams to the beam energy at collision of \SI{190}{\GeV}.
The beam delivery system removes transverse tails and off-energy particles with collimators and compresses the beam to the small sizes required at the collision point.
After the collision the beams are transported by the post collision lines to the respective beam dumps.

The RF power for each main linac is provided by a high current, low-energy drive beam that runs parallel to the colliding beam through a sequence of power extraction and transfer structures
(PETS). The drive beam generates RF power in the PETS that is then transferred to the accelerating structures by waveguides.

The drive beam is generated in a central complex with a fundamental frequency of \SI{1}{\GHz}. A \SI{48}{\us} long beam pulse is produced in the injector and fills every other bucket, i.e.\ with a bunch spacing of \SI{0.6}{\meter}. Every \SI{244}{\ns}, the injector switches from filling even buckets to filling odd buckets and vice versa, creating \SI{244}{\ns} long sub-pulses.
The beam is accelerated in the drive-beam linac to \SI{1.91}{\GeV}. A \SI{0.5}{\GHz} resonant RF deflector sends half of the sub-pulses through a delay loop such that its bunches can be interleaved with those
of the following sub-pulse that is not delayed. This generates a sequence of \SI{244}{\ns} trains in which every bucket is filled, followed by gaps of the same \SI{244}{\ns} length.
In a similar fashion three of the new sub-pulses are merged in the first combiner ring. Groups of four of the new sub-pulses, now with \SI{0.1}{\meter} bunch distance, are then merged in the second combiner
ring. The final pulses are thus \SI{244}{\ns} long and have a bunch spacing of \SI{2.5}{\cm}, i.e. providing \num{24} times the initial beam current. The distance between the pulses has increased to \SI[product-units = single]{24 x 244}{\ns}, which corresponds to twice the length of a \SI{878}{\meter} decelerator. The first four sub-pulses are transported through a delay line before they are used
to power one of the linacs while the next four sub-pulses are used to power the other linac directly.
The first sub-pulse feeds the first drive-beam decelerator, which runs in parallel to the colliding beam. When the sub-pulse reaches the decelerator end,
the second sub-pulse has reached the beginning of the second drive-beam decelerator and will feed it, while the colliding beam has meanwhile reached the same location along the linac.

\begin{table}
\caption{Key parameters of the CLIC energy stages.}
\label{t:scdup1}
\centering
\begin{tabular}{l l l l l l}
\toprule
Parameter                  &   Symbol         &   Unit &    Stage 1 &   Stage 2 &   Stage 3 \\
\midrule
Centre-of-mass energy               & $\sqrt{s}$              &\si{\GeV}                                     & 380     & 1500          & 3000\\
Repetition frequency                & $f_{\text{rep}}$        &\si{\Hz}                                     & 50      & 50            & 50\\
Number of bunches per train         & $n_{b}$                 &                                              & 352     & 312           & 312\\
Bunch separation                    & $\Delta\,t$             &\si{\ns}                                      & 0.5     & 0.5           & 0.5\\
Pulse length                        & $\tau_{\text{RF}}$      &\si{\ns}                                      & 244     &244            & 244\\
\midrule
Accelerating gradient               & $G$                     &\si{\mega\volt/\meter}                        & 72      & 72/100        & 72/100\\
\midrule
Total luminosity                    & $\mathcal{L}$           &\SI{e34}{\per\centi\meter\squared\per\second} & 1.5     & 3.7           & 5.9 \\
Luminosity above \SI{99}{\percent} of $\sqrt{s}$ & $\mathcal{L}_{0.01}$    &\SI{e34}{\per\centi\meter\squared\per\second} & 0.9     & 1.4           & 2\\
Total integrated luminosity per year& $\mathcal{L}_{\rm int}$ &\si{\per\fb}                                  & 180     & 444           & 708 \\ 
\midrule
Main linac tunnel length                  &                         &\si{\km}                                      & 11.4    & 29.0          & 50.1\\
Number of particles per bunch       & $N$                     &\num{e9}                                      & 5.2     & 3.7           & 3.7\\
Bunch length                        & $\sigma_z$              &\si{\um}                                      & 70      & 44            & 44\\
IP beam size                        & $\sigma_x/\sigma_y$     &\si{\nm}                                      & 149/2.9 & $\sim$ 60/1.5 & $\sim$ 40/1\\
Normalised emittance (end of linac) & $\epsilon_x/\epsilon_y$ &\si{\nm}                                      & 900/20  & 660/20        & 660/20\\
Final RMS energy spread & & \si{\percent} & 0.35 & 0.35& 0.35 \\
\midrule
Crossing angle (at IP)              &                         &\si{\mrad}                                    & 16.5    & 20            & 20 \\
\bottomrule
\end{tabular}
\end{table}

\subsubsection{Main-beam design considerations and choices}
The CLIC target luminosity at \SI{380}{\GeV} is $\mathcal{L}=\SI{1.5e34}{\per\centi\meter\squared\per\second}$. The nominal beam parameters at the interaction point
are given in~\ref{t:scdup1}.
Reaching the energy goal requires achieving the target gradient in the accelerating structures. This in turn requires that the structures
can sustain the gradient and that the drive beam provides enough power. In addition, to reach the luminosity goal, the colliding beam needs to have a high current and
an excellent quality.
Thorough studies established a feasible concept for the \SI{380}{\GeV} stage~\cite{cdrvol1}. Based on this the first stage has been designed.
The key considerations are:
\begin{itemize}
\item The choice of bunch charge and length ensures stable transport of the beam. The main limitation arises from short-range wakefields in the Main Linac.
\item The spacing between subsequent bunches ensures that the long-range wakefields in the Main Linac can be sufficiently damped to avoid beam break-up instabilities.
\item The horizontal beam size at the collision point ensures that the beamstrahlung caused by the high beam brightness is kept to an
acceptable level for the given bunch charge. This ensures a luminosity spectrum consistent with the requirements of the physics experiments.
\item The horizontal emittance is dominated by single particle and collective effects in the Damping Rings and includes some additional contributions from the Ring To Main Linac.
\item The vertical emittance is given mainly by the Damping Ring and additional contributions from imperfections of the machine implementation.
The target parameters take into account budgets for detrimental effects from static and dynamic imperfections such as component misalignments and jitter.
\item The vertical beta-function is the optimum choice in terms of luminosity. The horizontal beta-function is determined by the combination of required beam size and horizontal emittance.
\end{itemize}
In summary, the parameters are largely determined by fundamental beam physics and machine design with the exception of the vertical emittance that is determined by imperfections. Without them a
luminosity of $\mathcal{L}=\SI{4.3e34}{\per\centi\meter\squared\per\second}$ would be achieved.

\subsubsection{Performance of the drive-beam concept}
\label{s:energy}
The successful technology demonstration of the CLIC accelerating gradient is discussed in~\ref{sec:acc-technologies}.
The main performance limitation arises from vacuum discharge, i.e. breakdowns; a rate of less than $\SI{3e-7}{\per\meter}$ is required for the target gradient of \SI{72}{\mega\volt/\meter}.
The key parameters for the accelerating structure and the beam have been optimised together.
In particular, structures with smaller iris apertures achieve higher gradients for the same breakdown rate, but they reduce
the maximum bunch charge for stable beam transport because they produce stronger wakefields.

To test the drive-beam concept, the third CLIC Test Facility (CTF3)~\cite{Geschonke2002} was constructed and operated by an international collaboration.
It has addressed the key points of the concept:
\begin{itemize}
\item The stable acceleration of the initial high-current drive beam in the accelerator.
\item The high transfer efficiency from the RF to the drive beam.
\item The generation of the final drive-beam structure using the delay loop and a combiner ring.
\item The quality of the final drive beam. In particular, feedback has been used to stabilise the drive-beam current and phase to ensure
correct main-beam acceleration. CTF3 achieved the drive-beam phase stability that is required for CLIC~\cite{c:phasetol,c:chetan_gm,c:phasetest}.
\item The use of the drive beam to accelerate the main beam and the performance of the associated hardware.
The main beam has been accelerated with a maximum gradient of \SI{145}{\mega\volt/\meter}.
\end{itemize}
CTF3 established the feasibility of the drive-beam concept and the ability to use this scheme to accelerate the main beam. \ref{fig:CTF3photo} shows the corresponding two-beam acceleration test stand in the CTF3 facility.

\begin{figure}[t]
\begin{center}
\includegraphics[width=0.6\textwidth]{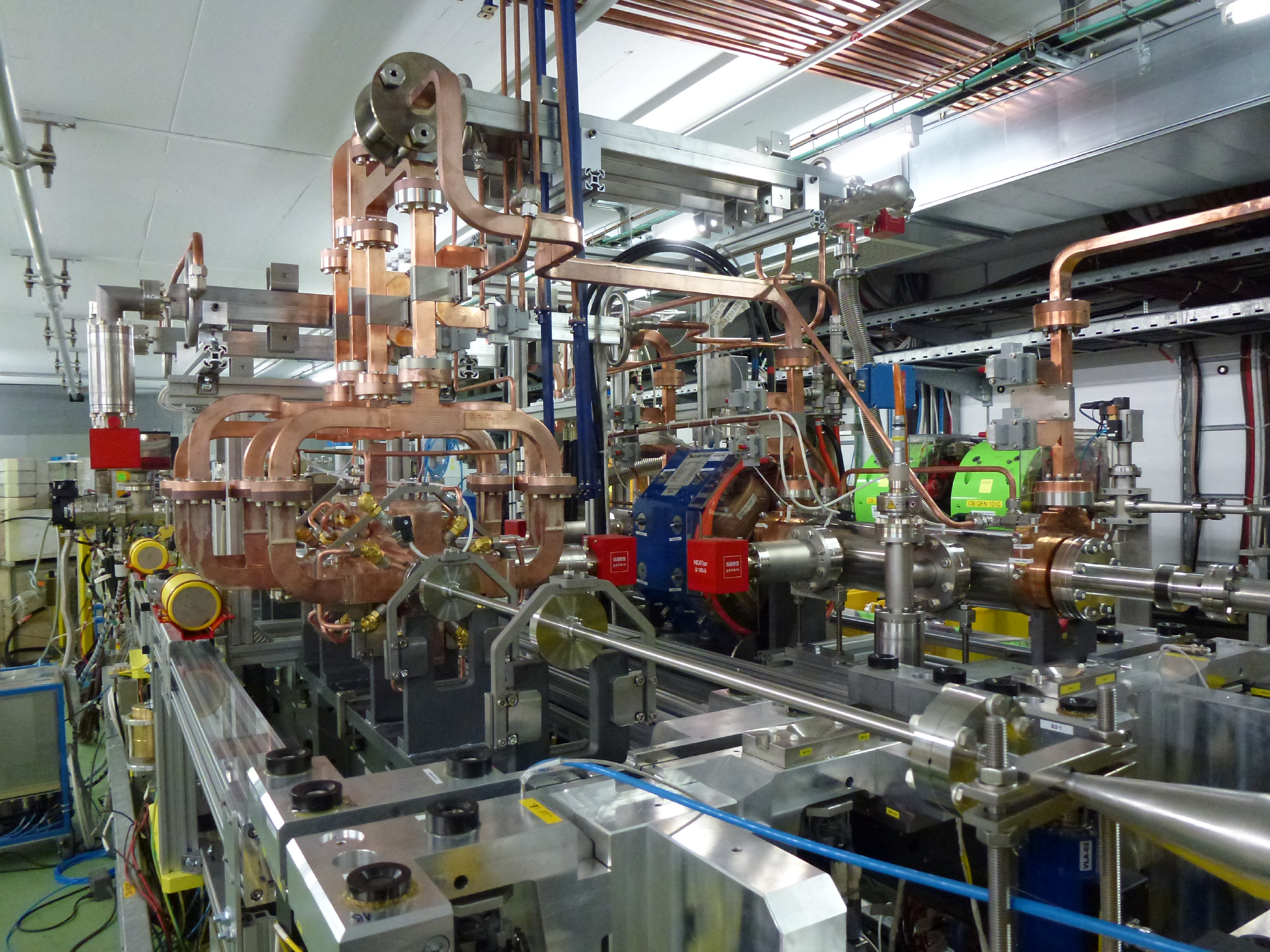}
\caption{The two-beam acceleration test stand in the CTF3 facility. The drive beam enters from the middle-right, while the probe (main) beam enters from the bottom-right. \imcl}
\label{fig:CTF3photo}
\end{center}
\end{figure}
It has also been instrumental in developing all the different hardware components that are essential for the scheme -- among them
the drive-beam gun, the bunch compressor, the drive-beam accelerating structures, RF deflectors, the PETS including a mechanism
to switch them off individually, the power distribution waveguide system, fast-feedback systems,
drive-beam current and phase monitors, as well as other instrumentation.
CTF3 stopped operation after successfully completing its experimental programme in December \num{2016} and a new facility, CLEAR,
has started to operate. It is re-uses the CTF3 main-beam installations and additional hardware to address further beam dynamics with
the focus on the main beam.

\subsubsection{Luminosity performance}
For static imperfections, the vertical emittance growth budgets are the same at \SI{380}{\GeV} and \SI{3}{\TeV} and they correspond to the values described in the CDR~\cite{cdrvol1}.
It is required that each system, i.e. RTML, Main Linac and BDS, remains within its emittance budget with a likelihood of more than \SI{90}{\percent} without further intervention.
The key static imperfection is the misalignment of the beamline components with respect to the design.
A sophisticated system has been developed and tested that provides a spacial reference frame with unprecedented accuracy, see~\ref{sec:acc-technologies}.
The Main Linac and BDS components are mounted on movable supports and can be remotely aligned with respect to the reference system.
In addition, the main linac accelerating structures are equipped with wakefield monitors that allow the measurement and correction of their offset with respect to the beam.
Dispersion-free steering, which has been successfully tested at the SLAC Facility for Advanced Accelerator Experimental Tests (FACET), will further reduce the emittance growth using high-resolution Beam Position Monitors (BPM).
In the BDS, additional tuning is required using optical knobs that move several multi-pole magnets simultaneously to correct the optics properties.

The performance specifications for the alignment systems and instrumentation are kept the same at \SI{380}{\GeV} and \SI{3}{\TeV} and correspond to the CDR description.
They are sufficient to achieve the required performance at \SI{3}{\TeV} and most of them could be relaxed for the first energy stage, typically by about a factor of two,
to meet the same emittance budget.
However, the original, better performances are required for the upgrade to the higher energy stages.
No substantial cost saving has been identified by relaxing the specifications for the first stage.
Therefore, it has been decided to ensure that the system is consistent with the final energy from the very beginning,
thus avoiding the need for upgrades of the already existing hardware. This also provides additional margin for achieving the required luminosity.

The tuning procedures for the RTML and the BDS have been improved compared to the CDR.
Studies of the static imperfections in the RTML~\cite{c:RTML_perf}, the Main Linac~\cite{c:neven} and the BDS~\cite{c:jim} show
that the target budgets can be met in each system with a margin; in the BDS, the tuning is now also much faster.
Combining these effects, one can expect an average luminosity of $\mathcal{L}=\SI{3e34}{\per\centi\meter\squared\per\second}$, about twice the
luminosity target at \SI{380}{\GeV}~\cite{c:jim}.

Also, for the dynamic imperfections the vertical emittance budgets are the same for \SI{380}{\GeV} and \SI{3}{\TeV}.
Key imperfections are the movement of components due to ground motion or technical noise, phase and amplitude jitter of the
drive beam, and potentially dynamic magnetic fields.

The level of ground motion is site dependent; measurements in the LEP tunnel showed
very small motion~\cite{c:gm:lep} while measurements in the CMS detector hall showed much larger motion~\cite{c:gm:cms}. 
With the new design of the final focus system, all relevant accelerator components are mounted in the tunnel of the collider, so one can expect
ground motion levels similar to the LEP tunnel. However, for the ground motion studies the level of the CMS detector hall has been used in order
to evaluate the robustness of the solutions. The ground motion is mitigated by the design of the magnets, a mechanical feedback that decouples them from the ground,
and by beam-based feedback on trajectories. Prototypes of the mechanical feedback have been tested successfully.
In the CDR, detailed studies of the \SI{3}{\TeV} stage showed that the performance goal can be met with margin. Studies of the \SI{380}{\GeV} case~\cite{c:chetan_gm}
confirm that ground motion will only use about \SI{10}{\percent} of the budget allocated to dynamic imperfections.

Dynamic magnetic stray fields deflect the colliding beams, leading to trajectory jitter and emittance growth, thus reducing luminosity.
Their impact is particularly large in the RTML and the BDS. In the latter they are more important at \SI{380}{\GeV} than at \SI{3}{\TeV} due to the lower beam energy.
A study in collaboration with experts from the Hungarian Geophysics Institute has commenced to investigate these fields and define the mitigation technologies.
The magnetic fields can originate from different sources: natural sources, such as geomagnetic storms; environmental sources, such as railway trains and
power lines and technical sources, i.e. from the collider itself.
A survey of natural sources showed that they should not affect the luminosity~\cite{c:balazs} and a measurement station has been established in the Jura mountains near CERN to collect long-term
regional data.
The study of the environmental and technical sources has started but is not yet complete. Preliminary estimates have been performed
using the magnetic field variations that were measured in the LHC tunnel. They concluded that a thin mu-metal shield of the drifts in the RTML and BDS
can bring the fields down to a level that does not impact luminosity~\cite{c:chetan_gm}.

Further development of the foreseen technical and beam-based imperfection mitigation systems should allow for a reduction in the emittance budgets
and an increase in the luminosity target. Also, new systems could be devised to this end. As an example, the addition of a few klystron-powered,
higher-frequency accelerating structures
could allow to reduce the energy spread of the colliding beams, which can improve the luminosity and also the luminosity spectrum for specific
measurements such as the top-quark threshold scan.

\subsubsection{Operation and availability}
The machine protection and operational considerations and strategies at \SI{380}{\GeV} are similar to those at \SI{3}{\TeV} and are described in the CDR~\cite{cdrvol1}.
Machine protection relies on passive protection and the processing of the diagnostics data between two beam pulses to generate a beam permit signal.

The tentative plan for the operation of CLIC includes a yearly shutdown of \SI{120}{\days}. In addition \SI{30}{\days} are foreseen for the machine commissioning, \SI{20}{\days} for
machine development and \SI{10}{\days} for planned technical stops. This leaves \SI{185}{\days} of operation for the experiments. The target availability for the experiments
during this period is \SI{75}{\percent}. Hence the integrated luminosity per year corresponds to operation at full luminosity for \SI{1.2e7}{\second}~\cite{Bordry:2018gri}.
An optimisation of the schedule has started which will also refine the trade-off between planned short stops and availability to reach the integrated luminosity goal. 

Different events can impact both operation and availability and can be roughly categorised as:
\begin{itemize}
\item
Events that do not require an intervention in the machine and are handled by the control system.
These include RF breakdowns in the accelerating structures, which will lead to a small energy error
and potentially slight transverse deflection of the beam. Typically this will happen only every \num{100} beam pulses and will be corrected by the feedback systems.
\item
Events that require a short stop on the machine but no intervention, such as a false trigger of the machine protection system -- e.g. caused by a single event upset.
In this case, the machine can be brought back to full intensity in a few seconds.
\item
Failures of machine components that might compromise the performance but do not require stopping the beam. This includes failures of klystrons or instrumentation.
These are mitigated by providing sufficient reserve.
\item
Failures that require to stop the beam and repair the machine. This is the case for failures of power converters.
\end{itemize}

Based on an assessment of the complexity of the different systems, an availability goal has been defined for each of them.
This allows investigation of individual systems and focus on the key issues.

A number of key failures has been studied in detail, in particular of magnet power converters and RF power systems.
In the drive-beam accelerator, a reserve of \SI{5}{\percent} RF units are installed and klystrons operate below their maximum power.
If one fails, the power of the others is increased accordingly.
Similarly, BPMs and orbit corrector failures in the main linac compromise the correction of ground motion. However, if \SI{10}{\percent} of them fail, the effect of ground motion is only increased by \SI{14}{\percent}.
During the technical stops failed klystrons and instrumentation can be replaced.
The CLIC lattice design has been optimised to minimise the impact of power converter failures. In particular in the drive beam, the many quadrupoles are powered in groups to minimise the number of
power converters and small trims adjust their strength as needed. Compared to individual powering, this strongly increases the mean time between failures,
since failures of trims can be mitigated to a large extent. A similar strategy is used for the main-beam quadrupoles.

Detailed studies will be required during the technical design phase covering all components to ensure that the availability goal can be met. Currently, considering key failures, no obstacle has been identified
to reaching the target availability.

\subsubsection{Energy flexibility}
\label{sec:energyflexibility}
The beam parameters can be adjusted to different physics requirements. In particular, the collision energy can be adjusted to the requirements by lowering the gradient in the main linacs accordingly.
For a significantly reduced gradient, the bunch charge will have to be reduced in proportion the energy to ensure beam stability. However, at this moment
the only operation energy different from \SI{380}{\GeV} that is required is around \SI{350}{\GeV} to scan the top-quark pair-production threshold. In this case, the bunch charge can remain constant.
The RF phases of the accelerating structures are slightly modified compared to the \SI{380}{\GeV} case in order to achieve an RMS beam energy spread of only \SI{0.3}{\percent}. This allows reaching 
a luminosity of $\mathcal{L}=\SI{1.5e34}{\per\centi\meter\squared\per\second}$, similar to the \SI{380}{\GeV} goal.

Also the beam energy can be reduced at the cost of some reduction in luminosity. For example, at the top-quark threshold,
one can reduce the bunch charge by \SI{10}{\percent} and increase its length by \SI{10}{\percent}. This would keep the wakefield effects in the main linac constant. This configuration slightly
reduces the luminosity to $\mathcal{L}=\SI{1.18e34}{\per\centi\meter\squared\per\second}$, but reduces the beam energy spread to \SI{0.2}{\percent}.
Similarly, it is possible to reduce the beamstrahlung by increasing the horizontal beam size, if the reduced luminosity is out-weighted by the improved luminosity spectrum.

Operation at much lower energies can also be considered.
At the Z-pole, between \SI{2.5}{\per\fb} and \SI{45}{\per\fb} can be achieved per year for an unmodified and a modified collider, respectively.

\subsubsection{Beam experiments}
Beam experiments and hardware tests provide the evidence that the CLIC performance goals can be met. Some key cases are discussed in the following:
\begin{itemize}
\item
The novel drive-beam scheme has been demonstrated in CTF3, as discussed in~\ref{s:energy}.
\item The Stanford Linear Collider (SLC)~\cite{c:slc}, the only linear collider so far, is a proof of principle for the linear collider concept and
contributed important physics data at the Z-pole. The SLC achieved collision beam sizes smaller than nominal, but did not reach the nominal bunch charge~\cite{c:nan}.
Two collective effects led to the charge limitations. They have been fully understood and are not present in the CLIC design.
\item
The electron polarisation that has been achieved at collision in SLC is similar to the CLIC goal.
\item
The strong beam--beam effect increases the luminosity in CLIC. This effect has been observed at the SLC, in agreement with the theoretical predictions~\cite{c:slc_bb}.
\item
Modern light sources achieve CLIC-level vertical emittances, in particular the Swiss Light Source and the Australian Light Source~\cite{c:ls1,c:ls2,c:ls3}.
\item
CLIC parameters require strong focusing at the IP. This focusing has been demonstrated at two test facilities, FFTB~\cite{Balakin1995} at SLAC and the Accelerator Test Facility 
ATF2~\cite{Kuroda2016,Okugi2016} at KEK.
The achieved vertical beam sizes were \SI{40}{\percent} and \SI{10}{\percent} above the respective design values for these test facilities. In the super B-factory at KEK
the beams will perform many turns through the final focus system, still one aims at beta-functions that are only a factor three larger
than in CLIC, and even smaller beta-functions similar to the CLIC values are being discussed~\cite{Thrane2017}.
\item
The use of beam-based alignment, i.e. dispersion free steering~\cite{Latina2014,Latina2014a}
to maintain small emittances in a linac has successfully been tested in FACET~\cite{FACET} and FERMI~\cite{FERMI}.
\item
The effective suppression of harmful long-range wakefields has been tested with beam in the CLIC accelerating structures~\cite{PhysRevAccelBeams.19.011001}.
\item
The novel precision pre-alignment system of CLIC and sophisticated beam-based alignment and tuning ensure the preservation of the beam quality during transport.
The alignment system is based on a concept developed for the LHC interaction regions, but with improved performance.
Prototypes have been built and successfully tested, see~\ref{sec:acc-technologies-stab}.
\item Quadrupole jitter has been an important source of beam jitter in the SLC. For CLIC this has been addressed by designing the
magnet supports to avoid resonances at low frequencies and by developing
an active stabilisation system for the magnets, which demonstrated a reduction of the jitter to the sub-nanometre regime, see~\ref{sec:acc-technologies-stab}.
\item
CLIC requires excellent relative timing at the \SI{50}{\fs} level over the collider complex.
CTF3 has demonstrated the phase monitor and correction with fast feed-forward. Modern Free Electron Lasers (FEL) have developed the technology to provide the timing reference over large distances.
\item
High availability is key to achieve the luminosity goal. The very reliable routine operation of light sources, FELs, the B-factories
and the LHC provide concepts to address this issue.
\end{itemize}

In conclusion, the CLIC parameters are ambitious but are supported by simulation studies, measured hardware performances and beam tests.
This gives confidence that the goals can be met. More details on the performance benchmarks can be found in the Project Implementation Plan~\cite{ESU18PiP}.

\subsection{A klystron-based CLIC at \SI{380}{\GeV}}
An alternative design for the \SI{380}{\GeV} stage of CLIC is based on the use of X-band klystrons to produce the RF power for the main linac. On the one hand, this solution increases the cost of the main linac because the klystrons and modulators are more expensive than the drive-beam decelerator and also because a larger tunnel is needed to house the additional equipment. On the other hand, it avoids the substantial cost of the construction of the drive-beam complex and makes the linac more modular. One can therefore expect a competitive cost at low energies while the drive-beam solution leads to lower cost at high energies. The upgrade of the complex is cheaper with a drive-beam based design, since the additional cost to upgrade the drive-beam complex to feed a longer linac is relatively modest. However, an important advantage of the klystron-based design is that the main linac modules can easily be fully tested for performance when they are received. In contrast, the drive-beam option requires the construction of a substantial complex that can produce a \SI{100}{\ampere} drive beam before modules can be fully tested. The klystron-based option could therefore be implemented more rapidly than the drive-beam based solution.

\subsubsection{Design choice}
The klystron-powered design is based on a study~\cite{StagingBaseline} that uses the same optimisation tools as for the drive-beam based option.
The main linac model has been replaced with one that consists of a sequence of RF units, each powered by klystrons, see~\ref{sec:acc-technologies-klystronoption}, and the drive-beam complex has been removed.
A cost model for the klystrons and modulators is included.
Based on the conclusions of the study, a tentative accelerating structure and a parameter set have been chosen for this design. The optimum structure differs from the drive-beam based design. It is slightly shorter and has a smaller aperture. If one were to use the same accelerating structure as for the drive-beam based design, the expected cost would be about \SI{330}{MCHF} higher.

As can be seen in~\ref{t:scdkl1}, the beam emittance, energy spread and charge of the klystron-based design are very similar to the \SI{380}{\GeV} parameters, while the bunch is somewhat longer. The vertical emittance is also the same as for the drive-beam based design, while the horizontal emittance is smaller and proportional to the bunch charge. The number of bunches per train is significantly higher in order to produce the required luminosity.

The evolution of the vertical emittance along the collider is similar to the drive-beam based design,
while the horizontal emittance corresponds to the \SI{380}{\GeV} design. The horizontal and vertical emittances
remain below \SI{500}{\nm} and \SI{5}{\nm} at extraction from the damping ring, below \SI{600}{\nm} and \SI{10}{\nm} at injection
into the main linac and below \SI{630}{\nm} and \SI{20}{\nm} at the end of the main linac. At the interaction  point they will be
below \SI{660}{\nm} and \SI{30}{\nm}, respectively.

\begin{table}[!htb]
\caption{Key beam parameters of the klystron-based alternative at the collision point.}
\label{t:scdkl1}
\begin{center}
\begin{tabular}{l c l c}
\toprule
Parameter                  &   Symbol         &   Unit &    Stage 1  \\
\midrule
Centre-of-mass energy               & $\sqrt{s}$              &\si{\GeV}                                     & 380   \\
Repetition frequency                & $f_{\text{rep}}$        &\si{\Hz}                                     & 50  \\
Number of bunches per train         & $n_{b}$                 &                                              & 485  \\
Bunch separation                    & $\Delta\,t$             &\si{\ns}                                      & 0.5     \\
\midrule
Total luminosity                    & $\mathcal{L}$           &\SI{e34}{\per\centi\meter\squared\per\second} & 1.5 \\
Luminosity above \SI{99}{\percent} of $\sqrt{s}$ & $\mathcal{L}_{0.01}$    &\SI{e34}{\per\centi\meter\squared\per\second} & 0.9 \\
Total integrated luminosity per year& $\mathcal{L}_{\rm int}$ &\si{\per\fb}                                  & 180   \\
\midrule
Number of particles per bunch       & $N$                     &\num{e9}                                      & 3.87 \\
Bunch length                        & $\sigma_z$              &\si{\um}                                      & 60    \\
IP beam size                        & $\sigma_x/\sigma_y$     &\si{\nm}                                      & 119/2.9 \\
Normalised emittance (end of linac) & $\epsilon_x/\epsilon_y$ &\si{\nm}                                      & $\le 630/\le 20$ \\
Final RMS energy spread & & \si{\percent} & 0.35 \\
\midrule
Crossing angle (at IP)              &                         &\si{\mrad}                                    & 16.5   \\
\bottomrule
\end{tabular}
\end{center}
\end{table}

\subsubsection{Design implications}
For the klystron-based alternative, the main linac has been redesigned to evaluate the cost and verify the beam dynamics. No design optimisation has been performed for the other systems. It is expected that these have only minor impact on cost and system performance, as detailed below.

The RF design of the klystron-based injector has to differ from the drive-beam based case in order to accommodate the longer bunch train. However, the total charge per pulse is the same in both cases. Hence, to first order the same amount of installed RF is required. However, the choice of accelerating structure and pulse compressor in the injection system would be slightly different to obtain optimum efficiency.

The single bunch parameters at the entrance of the linac are very similar to the drive-beam based design. The main difference is the lower bunch charge, which helps the emittance preservation, and the smaller horizontal emittance, which needs to be achieved. Both parameters are very similar to the \SI{380}{\GeV} design and can thus be achieved with the corresponding design. The bunch length at the start of the main linac and afterwards is larger in the klystron-based design, which requires less compression in the RTML and eases the system requirements.

An optimised layout of the main linac has been developed and beam dynamics studies have been performed. They confirmed the expected results that the performance is the same as for the drive-beam case.

The beam delivery system design is the same for the baseline option and the klystron-based alternative,
since the beta-functions at the collision point are the same.

\subsection{Extension to higher energy stages}
\label{sect:HE_Intro}
The CLIC \SI{380}{\GeV} energy stage can be efficiently upgraded to higher energies, like the proposed \SI{1.5}{\TeV} and \SI{3}{\TeV} stages. This flexibility has been an integral part of the design choices for the first energy stage. The highest energy stage corresponds to the design described in the CLIC CDR~\cite{cdrvol1}, with minor modifications due to the first energy stages, as described below.
The only important difference to the CDR design is a new final focus system that has an increased distance between the last quadrupole of the BDS and the interaction point. This allows the magnet to be installed in the tunnel and outside of the detector.

\begin{figure}[h]
\begin{center}
\includegraphics[width=\textwidth]{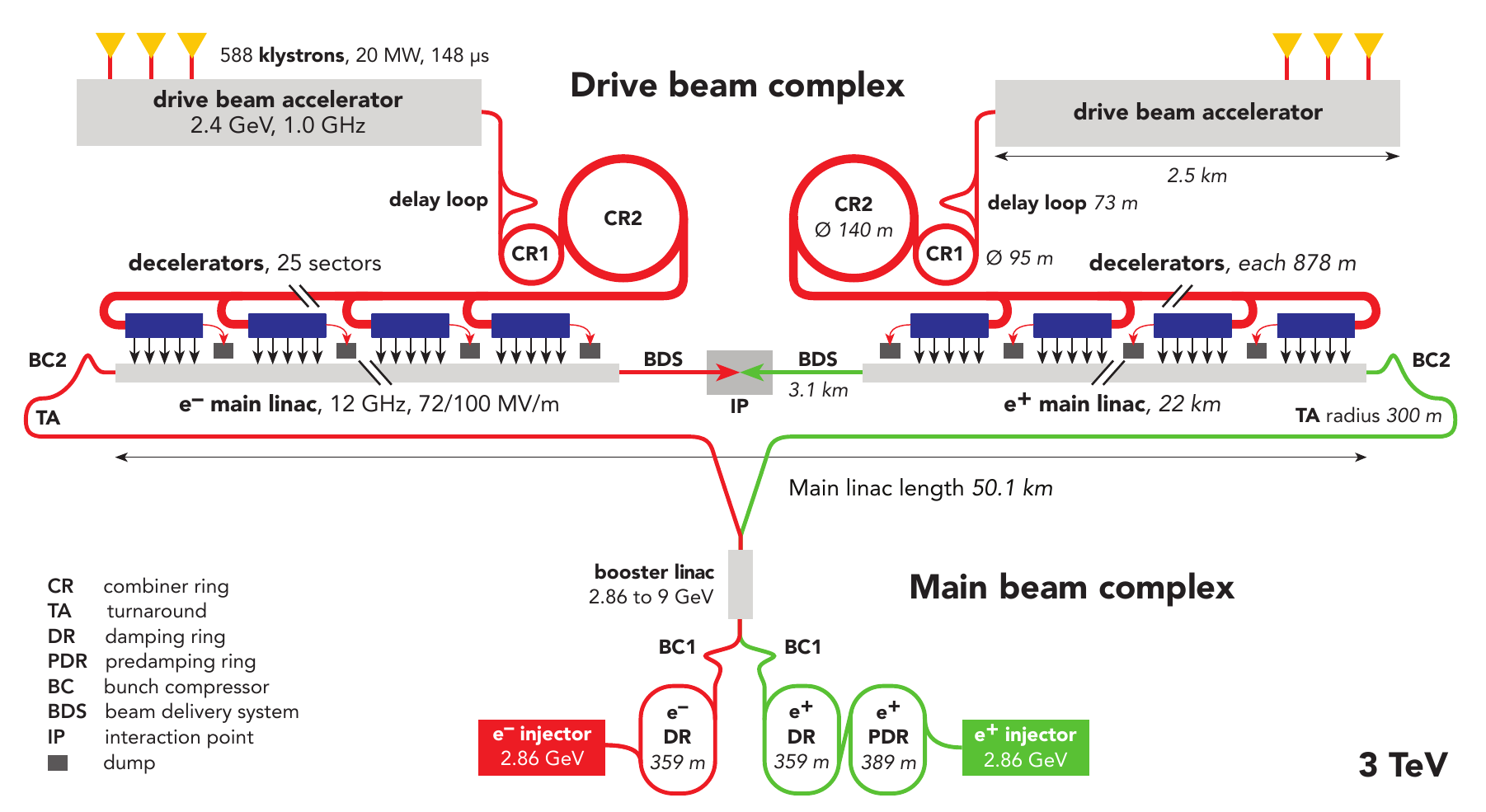}
\caption{Schematic layout of the CLIC complex at \SI{3}{\TeV}. \imcl}
\label{f:scdup0}
\end{center}
\end{figure}

\subsubsection{Baseline design upgrade}
The key parameters for the different energy stages of CLIC are given in~\ref{t:scdup1} and the schematic layout for the \SI{3}{\TeV} stage is
shown in~\ref{f:scdup0}.
The baseline concept of the staging implementation is illustrated in~\ref{f:scdup1}.
In the first stage, the linac consists of modules that contain accelerating structures that are optimised for this energy.
At higher energies these modules are reused and new modules are added to the linac.
First, the linac tunnel is extended and a new main-beam turn-around is constructed at its new end. The technical installations in the old turn-around and the subsequent bunch compressor are then moved to this new location.
Similarly, the existing main linac installation is moved to the beginning of the new tunnel.
Finally, the new modules that are optimised for the new energy are added to the main linac. Their accelerating structures have smaller apertures
and can reach a higher gradient of \SI{100}{\mega\volt/\meter}; the increased wakefield effect is mitigated by the reduced bunch charge and length.
The beam delivery system has to be modified by installing magnets that are suited for the higher energy and it will be extended in length. The beam extraction line also has to be modified to accept the larger beam energy but the dump remains untouched.
Alternative scenarios exist. In particular one could replace the existing modules with new, higher-gradient ones; however, this would increase the cost of the upgrade. In the following only the baseline is being discussed.

\begin{figure}[h]
\begin{center}
\includegraphics[width=\textwidth]{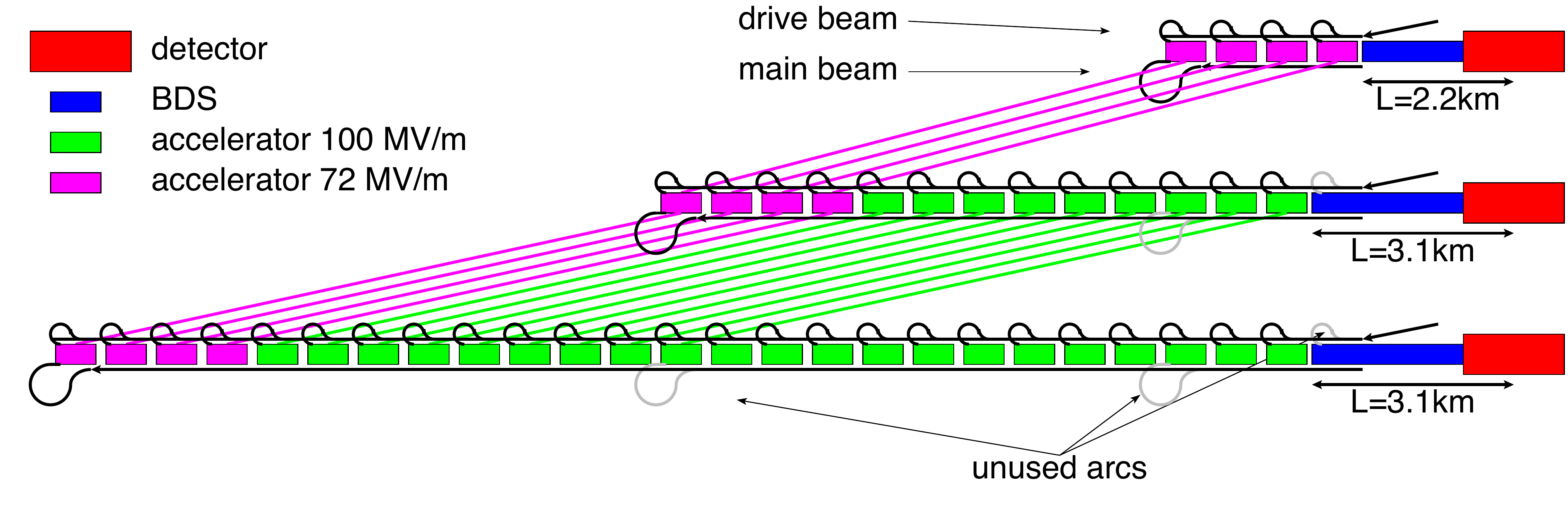}
\caption{The concept of the CLIC energy staging for the baseline design. \imcl}
\label{f:scdup1}
\end{center}
\end{figure}

The design of the first stage considers the baseline upgrade scenario from the beginning. For the luminosity target
at \SI{380}{\GeV}, the resulting cost increase of the first stage is \SI{50}{MCHF} compared to the fully optimised first energy stage (without the constraints imposed by a future energy upgrade beyond \SI{380}{\GeV}).
To minimise the integrated cost of all stages, the upgrades reuse the main-beam injectors and the drive-beam complex with limited modifications, 
and reuse all main linac modules.

In order to minimise modifications to the drive-beam complex, the drive-beam current is the same at all energy stages. The existing drive-beam RF units
can therefore continue to be used without modification. In addition, the RF pulse length of the first stage is chosen to be the same as
in the subsequent energy stages. This is important since the lengths of the delay loop and the combiner rings, as well
as the spacings of the turn-around loops in the main linac, are directly proportional to the RF pulse length.
Hence, the constant RF pulse length allows the reuse of the whole drive-beam combination complex.
For the upgrade from \SI{380}{\GeV} to \SI{1.5}{\TeV}, only minor modifications are required for the drive-beam production complex.
The drive-beam accelerator pulse length is increased in order to feed all of the new decelerators, and also its beam energy is increased by \SI{20}{\percent}. The energy increase is achieved by adding more drive-beam modules.
The pulse length increase is achieved by increasing the stored energy in the modulators to produce longer pulses.
The klystron parameters in the first energy stage have been chosen to be compatible with the operation using longer
pulses and higher average power. The remainder of the drive-beam complex remains unchanged, except that all magnets after the drive-beam linac need to operate at a \SI{20}{\percent} larger field, which is also foreseen in the magnet design. 
The upgrade from \SI{1.5}{\TeV} to \SI{3}{\TeV} requires
the construction of a second drive-beam generation complex.

The impact of the upgrades on the main-beam complex has also been minimised by design.
The bunches of the main-beam pulses have the same spacing at all energy stages, while at higher energies the number of bunches per train and
their charge is smaller. Therefore the main linac modules of the first stage can accelerate the trains of the second and third stage without modification. 
Since the drive-beam current does not change, also the powering of the modules is the same at all energies.
The upgrade to \SI{1.5}{\TeV} requires an additional 9 decelerator stages per side and the \SI{3}{\TeV} needs another 12.

Still some modifications are required in the main-beam complex. The injectors need to produce fewer bunches with a smaller charge
than before, but a smaller horizontal emittance and bunch length
is required at the start of the main linac. The smaller beam current requires less RF, so the klystrons can be operated at lower power and the emittance growth due to collective effects
will be reduced. The smaller horizontal emittance is mainly achieved by some adjustment of the damping rings.
The reduction of the collective effects that result from the lower bunch charge will allow to reach the new value with the same risk as in the first energy stage.

The preservation of the beam quality in the main linac is slightly more challenging at the higher energies. 
However, the specifications for the performance of alignment and stabilisation systems for the \SI{380}{\GeV} stage are based on the requirements for the \SI{3}{\TeV} stage. 
They are therefore sufficient for the high energy stages and no upgrades of these systems are required.

The collimation system is longer at \SI{1.5}{\TeV} and \SI{3}{\TeV} to ensure the collimator survival at the higher beam energies. Similarly the final focus system is slightly longer to limit the
amount of synchrotron radiation and emittance degradation in the indispensable bending of the beams. The systems have to be re-built using higher field magnets. However, the integration into the
existing tunnel is possible by design. The extraction line that guides the beams from the detector to the beam dump will also need to be equipped with new magnets.

\subsubsection{Upgrade from the klystron-based option}
\label{sect:HE_K_Upgrade}
The upgrade from a klystron-based first stage to higher energies is also possible by reusing the klystron-driven accelerating structures and the klystrons and by adding new drive-beam powered structures. In the klystron-based first energy stage, the single bunch parameters are the same as for the high energy stages, only the bunch charge will be slightly reduced at higher energies. Shorter bunch trains need to be accelerated at higher energies, which does not add any difficulty.

\begin{figure}[t]
\begin{center}
\includegraphics[width=\textwidth]{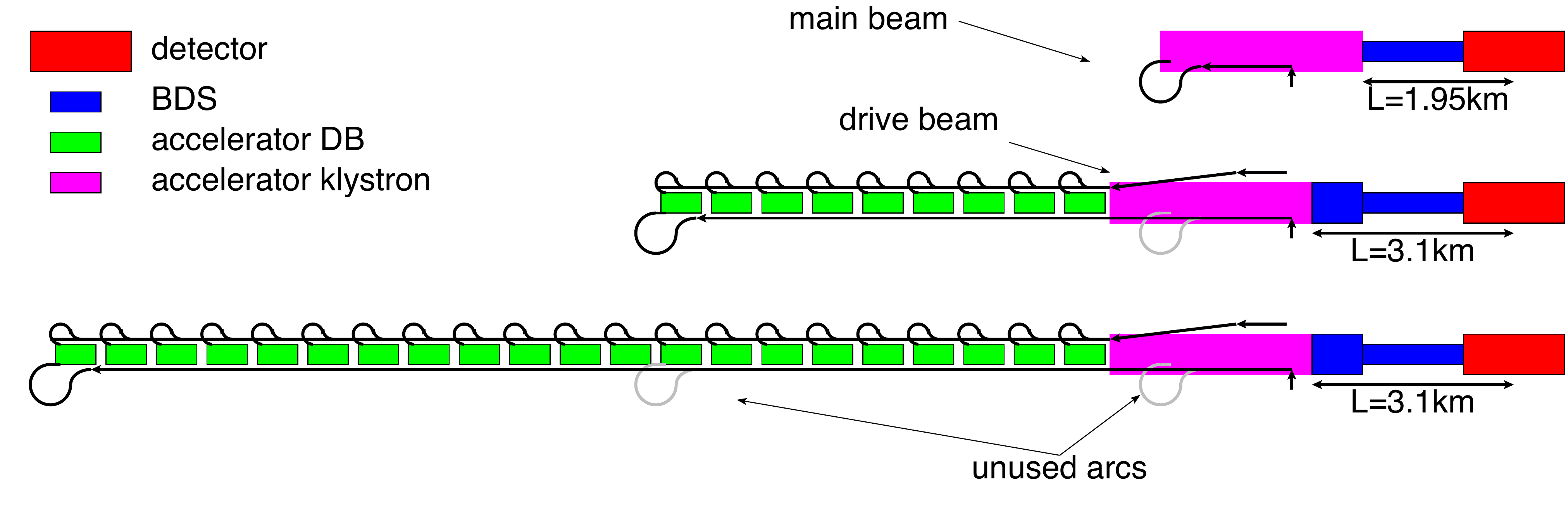}
\caption{The concept of the CLIC energy staging with a klystron-based first energy stage. \imcl}
\label{f:scdup2}
\end{center}
\end{figure}

An important difference with respect to the drive-beam powered first energy stage is the placement of modules. In order to provide the space for klystrons and modulators, the klystron-powered main linac tunnel has to be larger in radius than the tunnel housing the beam-driven acceleration. Therefore it appears best to extend the main linac for \SI{1200}{\meter} with a large tunnel and then continue with a smaller tunnel. All drive-beam powered modules are then placed in the smaller tunnel. The klystron-powered structures remain in the large tunnel. They need to be moved longitudinally slightly in order to adjust the lattice for the high energy, which requires longer quadrupoles with a wider spacing. The last \SI{1200}{\meter} of the linac is moved to the beginning of the large tunnel to provide the space for the high energy beam delivery system, see~\ref{f:scdup2}.

The impact of the energy upgrade on the main-beam injectors and damping rings is quite small. The bunch charge at \SI{3}{\TeV} is smaller than at \SI{380}{\GeV}; the difference is at the \SI{4}{\percent}-level, significantly smaller than for the upgrade of the drive-beam based machine. At higher energy, the number of bunches per beam pulse is also smaller, which is straightforward to accommodate.
The beam delivery system for klystron- and drive-beam based design are the same; hence the upgrade path is also the same.

\subsection{Accelerator technologies}
\label{sec:acc-technologies}
The CLIC accelerator is based on a similar set of technologies as already in use in other accelerators. Beam dynamic considerations dictate most of the requirements for these technologies and CLIC is expected to perform with very tight tolerances on most beam parameters. In this section, 
some of the most challenging technologies are mentioned, for which significant progress has been made since the publication of the CLIC CDR. Details on each technology including references can be found in the Project Implementation Plan~\cite{ESU18PiP}.  Most of these systems and concepts have been proven to work through the fabrication of prototypes and laboratory measurements. Some expert systems still need to be optimised for fabrication and/or routine operation. For some others the next challenge is to reduce the cost or the power consumption or to proceed towards large scale industrialisation.

\subsubsection{Main linac accelerating structures}
The main linac accelerating structures have to accelerate a train of bunches with a gradient of \SI{72}{\mega\volt/\meter} and a
breakdown rate of less than \SI{3e-7}{\per\meter}.
Damping features suppress higher-order modes, so-called transverse multi-bunch wakefields,
to avoid beam emittance growth; this enables high beam current and consequently
high RF-to-beam efficiency.
Finally, the accelerating structures must be built with micron precision tolerances and be equipped with special beam position monitors,
so-called wake monitors, in order to limit single-bunch emittance growth.

The overall optimisation of CLIC has been carried out following the insights from in-depth studies of the different aspects of the accelerating structure behaviour.
This optimisation has allowed the main parameters of the accelerating structure
to be determined, the detailed design to be made, and prototypes to be constructed and validated in both high-power and beam-based tests. 

CLIC accelerating structures are travelling wave with a tapered inner aperture diameter ranging
from \SI{8.2}{\mm} down to \SI{5.2}{\mm}, and are approximately \SI{25}{\cm} in length.
They are made from copper and operate at \SI{12}{\GHz}.
They are assembled from micron-precision disks that are bonded together.
Higher-order-mode suppression is provided by a combination of heavy damping, which is accomplished through four short terminated waveguides
connected  to each cell, and detuning accomplished through the iris aperture tapering. Photographs of the basic component disk, an assembled test prototype accelerating structure and a drawing of a full double structure assembly are shown in~\ref{fig:RFstructure-photos,fig:CLIC-structure-drawing}.

\begin{figure}[t]
\centering
\begin{subfigure}{.450\textwidth}
  \centering
  \includegraphics[width=\linewidth]{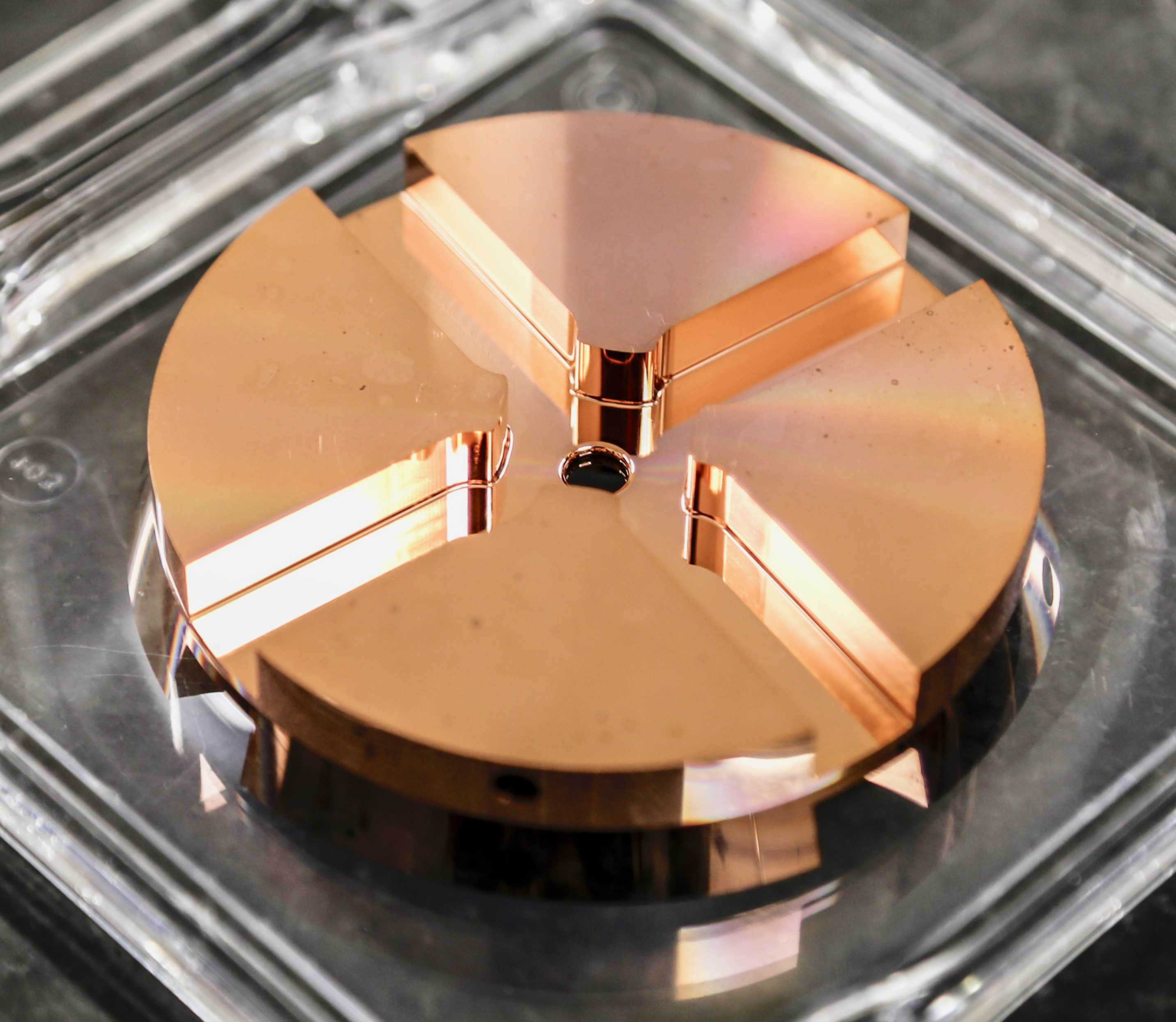}
   \caption{}
  \label{fig:RF-disk-photo}
\end{subfigure}
\hspace{.050\linewidth}
\begin{subfigure}{.450\textwidth}
  \centering
  \includegraphics[width=\linewidth]{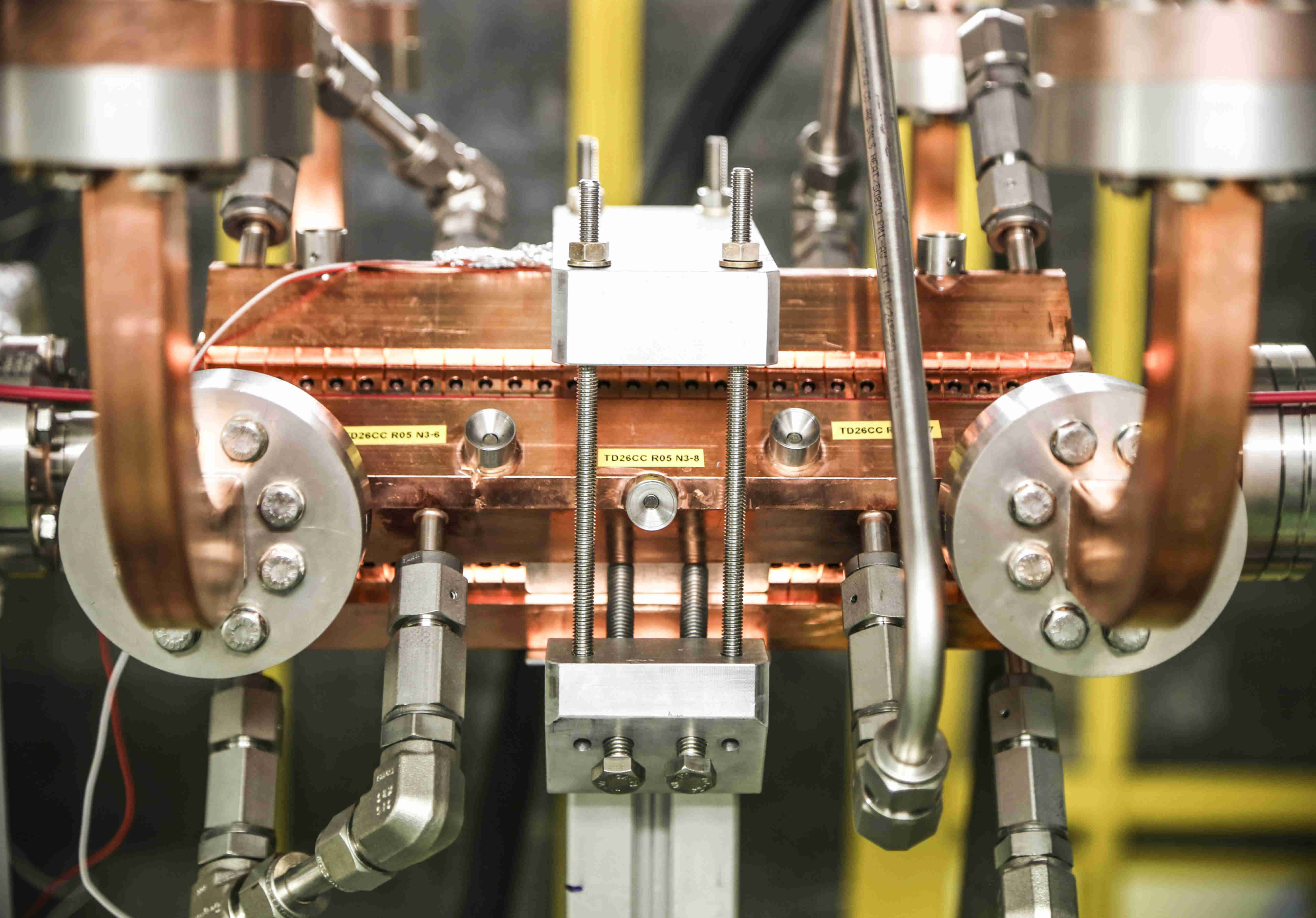}
   \caption{}
  \label{fig:CLIC-structure-photo}
\end{subfigure}
\caption{(a) The micron-precision disk which is the basic assembly block of the CLIC accelerating structures. Higher-order mode damping is provided by the four waveguides. 
(b) A prototype CLIC-G accelerating structure installed for high-gradient test. \imcl
}
\label{fig:RFstructure-photos}
\end{figure}

The different performance aspects have been validated in a series of dedicated tests. The most resource intensive has been high-gradient testing. The objective of these tests is to understand and determine high-field limits and to operate prototype accelerating structures for extended periods. These tests have been carried out in dedicated test stands, both at CERN and at KEK, which use klystron RF power sources in a configuration similar to the klystron-based version of CLIC. The tests involve conditioning the structures; that is, increasing the field level gradually to the nominal level, then operating them for extended periods at low breakdown rate. Over a dozen prototype \SI{3}{\TeV} accelerating structures, the so-called CLIC-G design, have been tested and a summary is shown in~\ref{fig:RFstructure-test-results}. The \SI{380}{\GeV} initial energy stage of CLIC requires a lower loaded accelerating gradient,
\SI{72}{\mega\volt/\meter}, than the \SI{3}{\TeV} stage. However, the iris aperture must be larger. The structures optimised for \SI{380}{\GeV} incorporate improvements understood from the high-gradient testing carried out up until now. They are currently in advanced stages of fabrication and will be tested soon.

In addition to these tests, an experiment to determine the effect of the heavy beam loading has been carried out using the CTF3 drive-beam injector beam. This experiment confirmed expectations of the effect and validated the design choices. Finally, the higher-order-mode suppression has been directly validated with beam in the FACET facility at SLAC~\cite{PhysRevAccelBeams.19.011001}.

\begin{figure}[t]
\begin{center}
  \centering
  \includegraphics[width=0.65\textwidth]{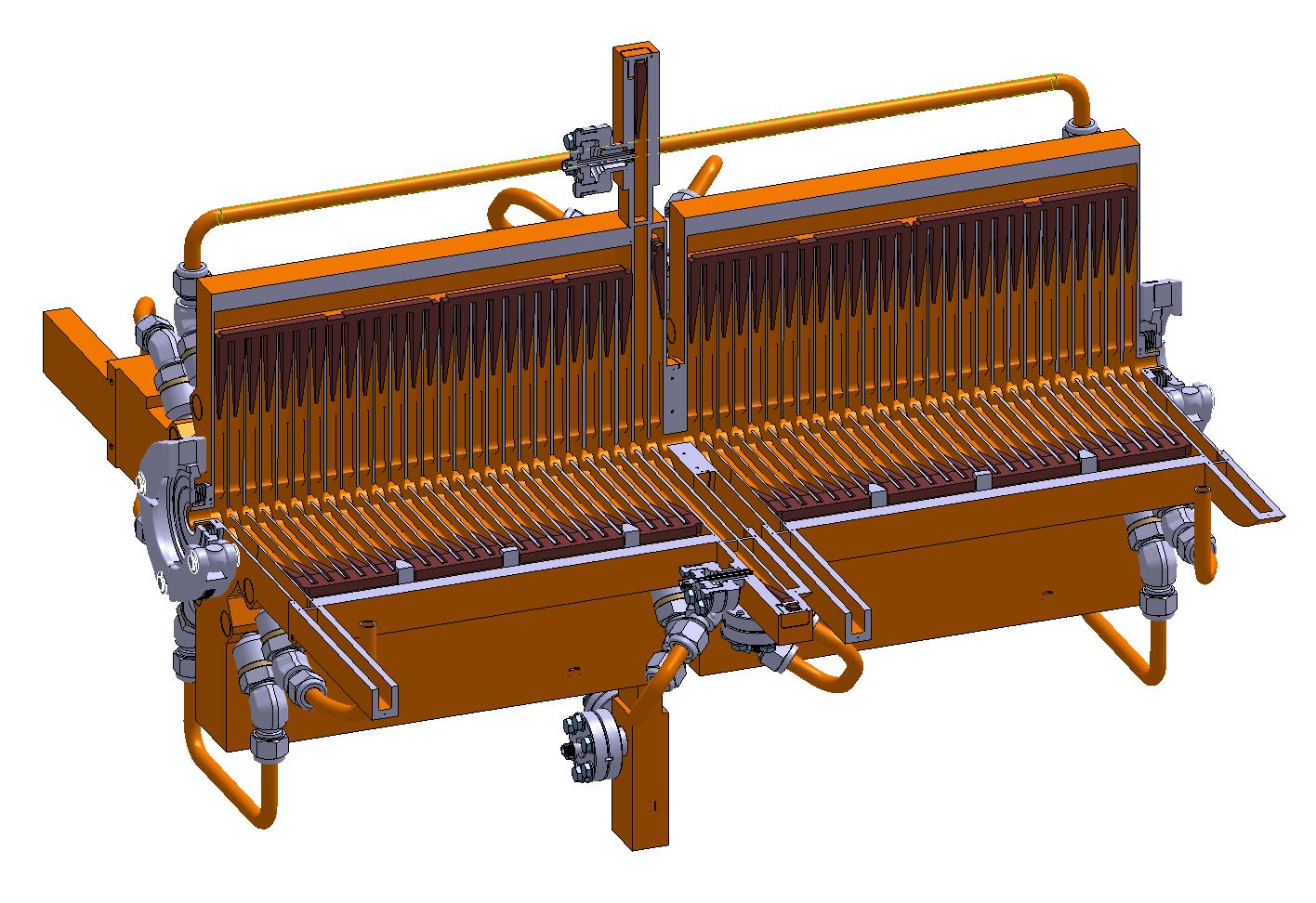}
 \caption{Assembly drawing of the double-structure acceleration unit. \imcl
}
\label{fig:CLIC-structure-drawing}
\end{center}
\end{figure}

\begin{figure}[htb]
\begin{center}
  \centering
  \includegraphics[width=0.75\textwidth]{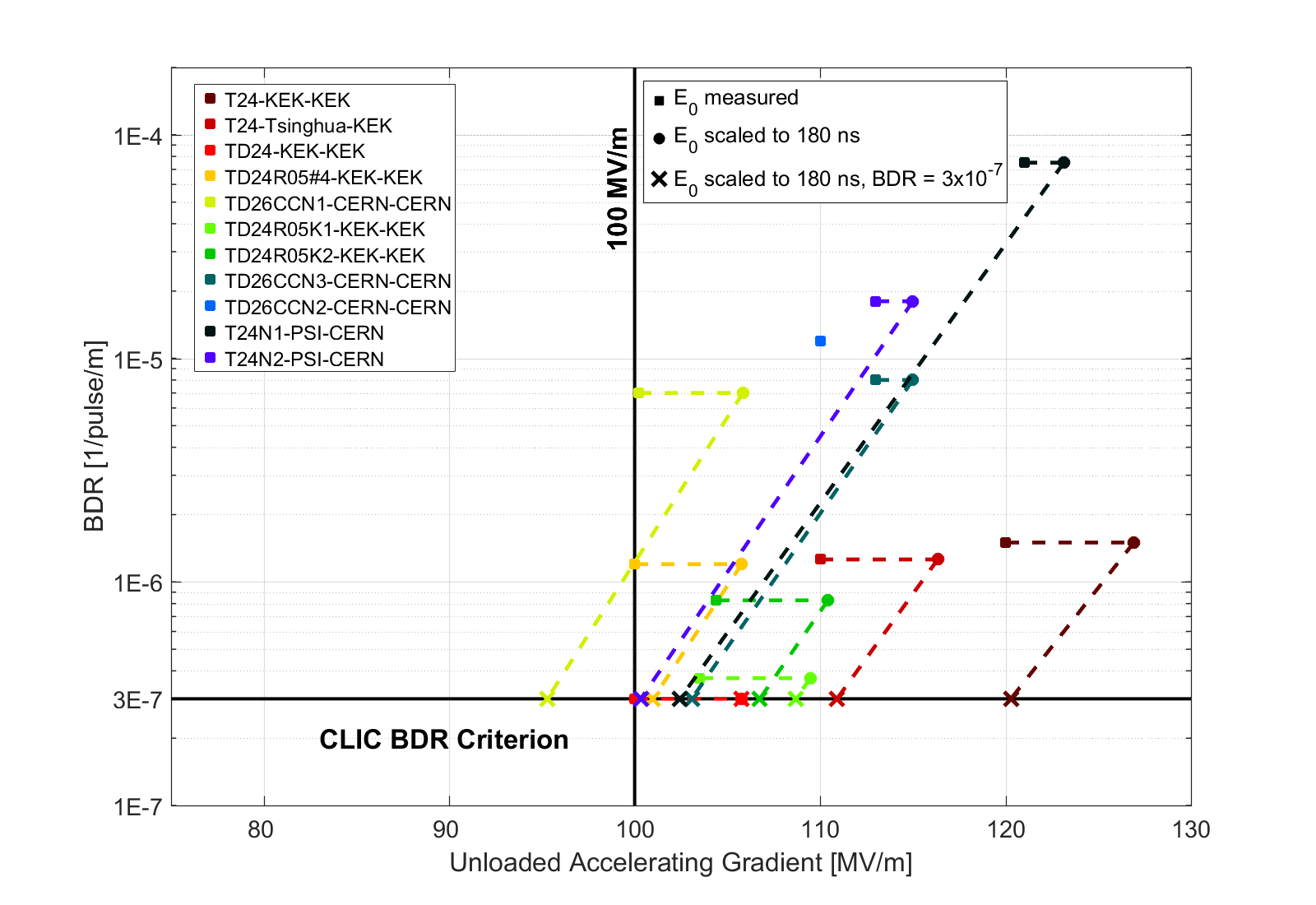}
 \caption{A summary of achieved performances of \SI{3}{\TeV} acceleration structures in tests. The vertical axis represents the breakdown rate per metre (BDR). The final operating conditions of the tests are indicated by squares. Known scaling is used to determine the performance for the nominal CLIC pulse duration (dashed lines connecting squares to circles) and subsequently for the CLIC-specified breakdown rate of \SI{3e-7}{\per\meter} (dashed lines connecting circles to crosses). \imcl
}
\label{fig:RFstructure-test-results}
\end{center}
\end{figure}

The accelerating structures represent an important contribution to the overall cost of CLIC and consequently costing and cost reduction are under active study. The micron tolerances as well as complexity related to the waveguide couplers and damping waveguide manifolds are the main cost drivers. Although the main focus of the prototypes described above has been high-gradient testing, important insights have been made on precision assembly and cost. More precise and lower cost alternatives based on these insights are now under design and fabrication. 

\subsubsection{RF power generation and distribution}
\label{sec:acc-technologies-RFpower}
Increasing the efficiency of the currently available klystrons is essential for CLIC, both for the two-beam and klystron-based CLIC options. For the drive-beam generation complex, two high-efficiency klystron prototypes in L-band technology have been developed in collaboration with industry, with the goal to obtain an efficiency above \SI{70}{\percent}. The first prototype,
using a 6-beam Multi-Beam Klystron (MBK), 
reached \SI{21}{\mega\watt} output power during the factory tests. 
Its efficiency of \SI{71.5}{\percent} remains remarkable high for a wide range of output power.
A second prototype built by another firm, 
based on a 10-beam MBK, also reached the required peak power and an efficiency of \SI{73}{\percent}.
However, it does not yet fulfil the requirements concerning stability and average power. 
Extensive testing of both prototypes continues~\cite{Marija:1981920}.

Modulator requirements for the drive beam were found to be  in an unexplored range, where specifications of fast pulse modulators (fast voltage rise and fall times to minimise power losses) and long pulse modulators (long voltage flat-top) have to be merged. The design effort for a suitable klystron-modulator topology has taken the high power electrical distribution over a $\sim$\SI{2}{\km} long drive beam into account. The solution found in this global optimisation imposes a modulator topology with a medium voltage DC stage and a voltage step-up pulse transformer~\cite{Marija:1981920}. Series and parallel redundancies have been studied and small scale prototypes have been designed and built. A full scale modulator prototype based on parallel redundancy topology has been designed and delivered to CERN from ETH Z\"{u}rich. First tests on an electrical dummy load demonstrated the feasibility of the voltage pulse dynamics up to \SI{180}{\kilo\volt}. This modulator represents the new state of the art in fast pulsed modulators with flat-top and medium voltage input.

For a klystron powered machine at \SI{380}{\GeV}, each X-band klystron will provide a peak RF power of \SI{50}{\MW} with a pulse width of \SI{1.6}{\us}
and a pulse repetition rate of \SI{50}{\Hz} at a frequency of \SI{11.9942}{\GHz}. These parameters are achievable using technology already available from industry~\cite{Sprehn:IPAC10}, as demonstrated in the operation of the X-band test facilities at CERN~\cite{Catalan-Lasheras:1742951}. As in the case of the L-band klystrons, and in collaboration with industry, a study is ongoing to improve the existing design of the klystron to achieve an efficiency of \SI{70}{\percent} while maintaining the required peak power~\cite{Baikov:2015}. Additionally, a study to replace the normal-conducting solenoid in the klystron with a superconducting solenoid is ongoing in collaboration with KEK.

The PETS are passive microwave devices that interact with the drive beam to generate RF power for two accelerating structures.
The power is collected and extracted at the downstream end, where a remotely controlled mechanism allows adjustment of the RF power
that flows into the accelerating structures. This flexibility allows sparking structures to be effectively switched off and also allows 
the structures in the main linac to be conditioned in parallel, each pair at their individual performance level. The PETS also contain damping waveguides
that are equipped with loads and avoid beam instabilities.
A total of sixteen PETS have been manufactured and tested in the two beam line of CTF3.

The RF power source for the main linac is connected to the accelerating structure through a network of waveguides that must transport RF power in excess of \SI{100}{\MW} with as little attenuation as possible. It controls the power produced in the PETS for the two-beam option or shapes the pulses from the klystron-modulator unit. The network distributes power among multiple accelerating structures, provides diagnostics and allows independent movement of the accelerating structure and power source. 

Fully equipped two-beam modules have been tested in a dedicated facilities with and without beam at CERN (see~\ref{fig:twobeam}).
Results obtained concerning alignment, vibrations, thermal stresses etc. have been fed back into the design of the next generation of modules.

\begin{figure}[t]
\begin{center}
  \centering
  \includegraphics[width=0.65\textwidth]{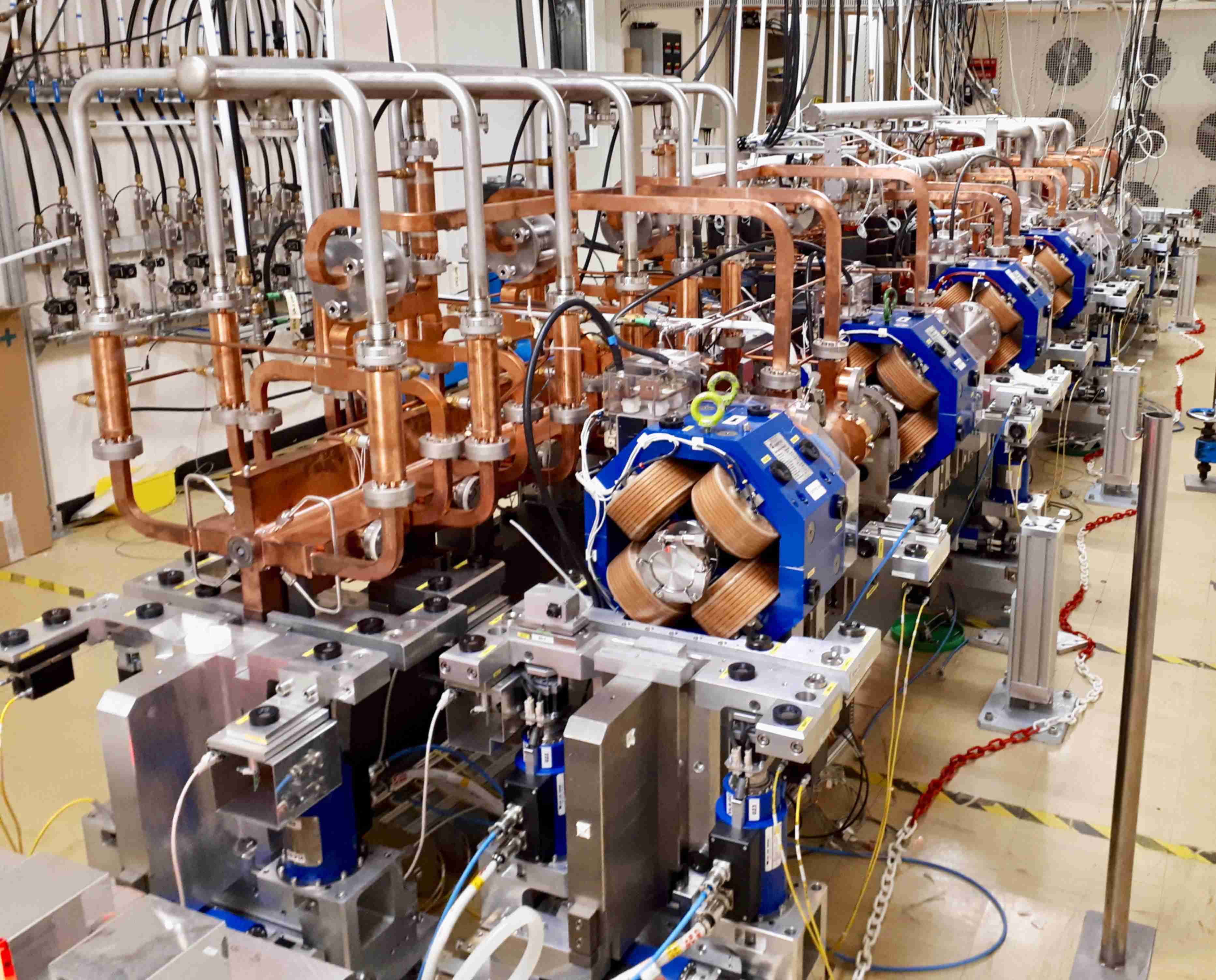}
 \caption{Two-beam module string used for alignment, thermomechanical stability and vacuum tests. The drive beam can be seen on the right, the main beam on the left. \imcl
}
\label{fig:twobeam}
\end{center}
\end{figure}

All of the necessary elements for the wave guide system (over-moded low loss transmission lines, mode converters, hybrids, pulse compressors, active phase shifters and power splitters, bends, direction couplers and loads) have been designed, fabricated and operated to full specifications, 
 in the two-beam test stand and the X-band test facility at CERN. 
\ref{fig:boc} shows some examples of recently produced components needed in the waveguide systems.
Although the waveguide network is not as technically challenging as the power source and accelerating structures, it represents an important cost element. Continuous efforts are being made to simplify the fabrication and assembly and reduce the cost~\cite{CatalanLasheras:2646747}.

\begin{figure}[htb]
\begin{center}
\begin{subfigure}{.33\textwidth}
  \centering
  \includegraphics[height=4.5cm]{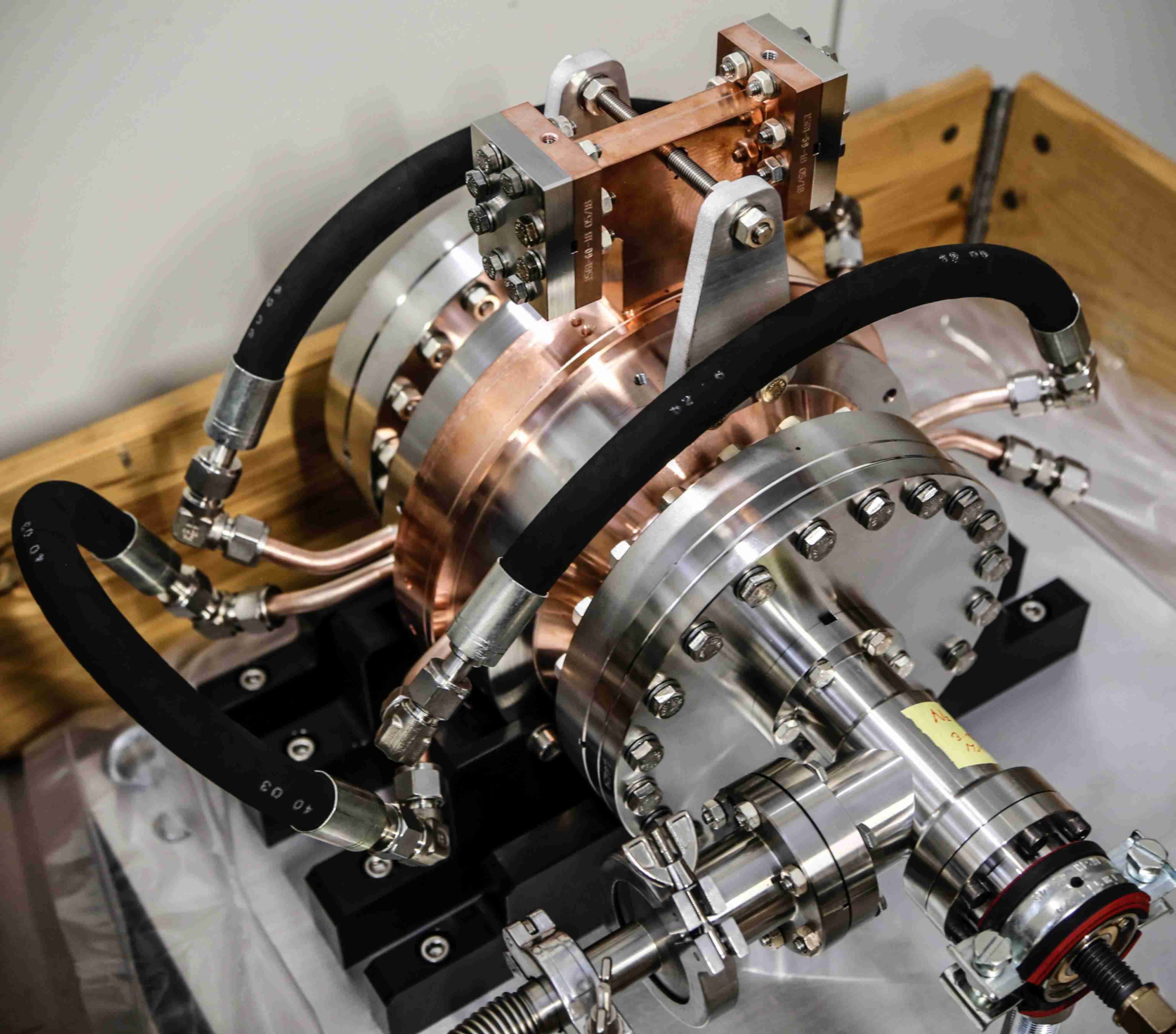}
  \caption{}
  \label{fig:boc1}
\end{subfigure}
\begin{subfigure}{.40\textwidth}
  \centering
  \includegraphics[height=4.5cm]{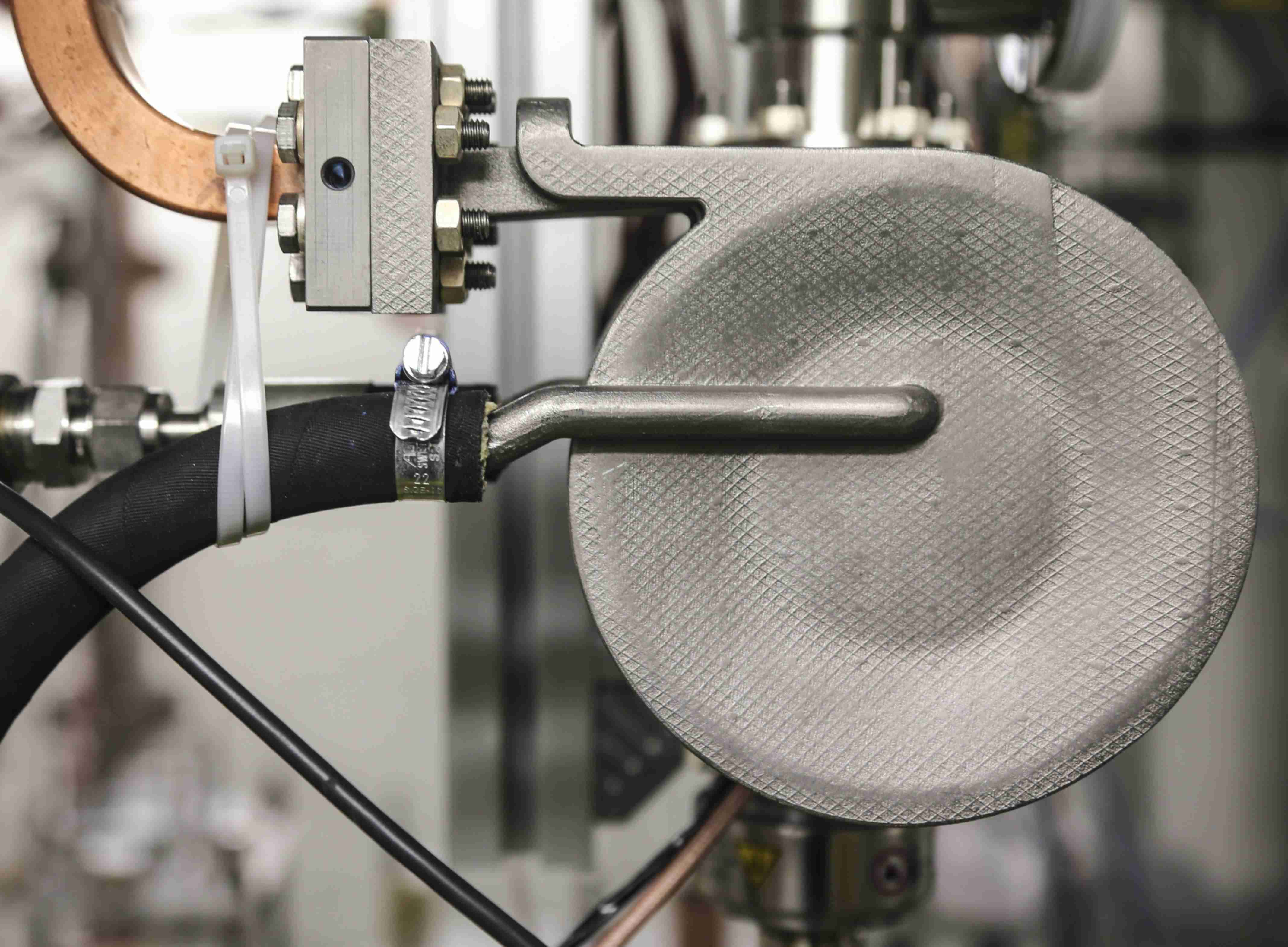}
  \caption{}
  \label{fig:boc2}
\end{subfigure}
\begin{subfigure}{.21\textwidth}
  \centering
  \includegraphics[height=4.5cm]{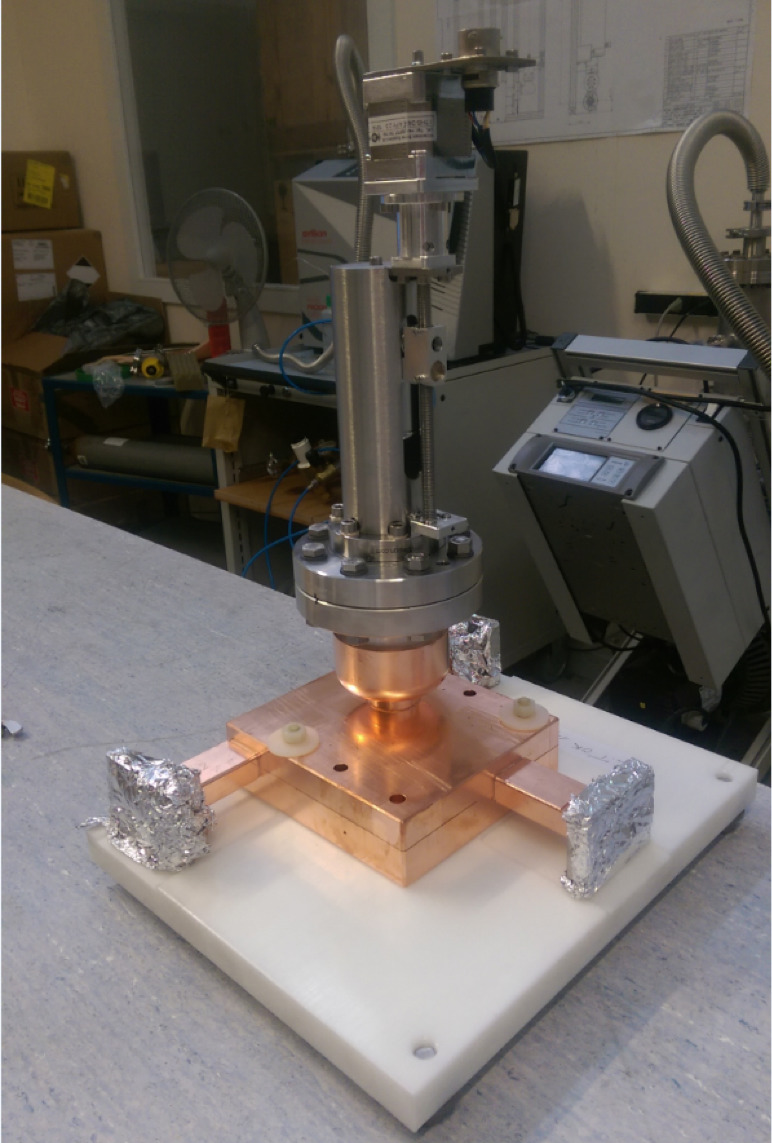}
  \caption{}
  \label{fig:boc3}
\end{subfigure}
 \caption{Components used in the waveguide system: (a) Barrel Open Cavity (BOC) pulse compressor (designed and manufactured by PSI), (b) compact 3D printed load and (c) variable power splitter. \imcl
}
\label{fig:boc}
\end{center}
\end{figure}

\subsubsection{Alignment and stabilisation}
\label{sec:acc-technologies-stab}
In order to preserve the luminosity, the total error budget allocated to the absolute positioning of the major accelerator components is \SIrange{10}{20}{\micro\meter}. For comparison, \SIrange{100}{500}{\micro\meter} are sufficient for LHC and HL-LHC.

The first ingredient of the CLIC alignment system is the Metrological Reference Network (MRN). Simulations have been carried out for the CLIC MRN~\cite{MainaudDurand:2289681}, considering stretched reference wires with a length of \SI{200}{\meter} and an accuracy of alignment sensors of \SI{5}{\micro\meter}. Simulations showed that the standard deviation of the position of each component with respect to a straight line was included in a cylinder with a radius smaller than \SI{7}{\micro\meter}. This was confirmed experimentally in a \SI{140}{\meter} long test facility. In order to achieve this accuracy, the sensors and active elements themselves were re-engineered in some cases. The performance of the capacitive Wire Positioning Systems and Hydrostatic Levelling Sensors was measured in the laboratory. Asymmetric cam movers with sub-micron displacement resolution have also been developed.

For the fiducialisation
and alignment of each element on the common 2-beam module support, a new strategy has been proposed, based on results obtained in the PACMAN project~\cite{Pacman2014,Caiazza:2280545} and on the development of an adjustment platform with five degrees of freedom~\cite{MainaudDurand:2018oas}. This strategy is based on individual determination of the axes of each component using metrological methods and a stretched wire. The absolute position of the wire can be measured to very high precision using a Coordinate Measuring Machine or a portable Frequency Scanning Interferometry system.

However, in order to maintain all the benefits of this very accurate alignment along the accelerator, absolute displacements of system elements caused by ground motion or vibrations during operation need to be avoided. As a first approximation, the integrated RMS displacement above a frequency of \SI{1}{\Hz} must stay below \SI{1.5}{\nm} in the vertical direction and below \SI{5}{\nm} in the horizontal direction for all main-beam quadrupoles (MBQ). For the final focus magnets, the integrated displacement above \SI{4}{\Hz} shall remain below \SI{0.14}{\nm} in the vertical plane. Besides an adapted civil engineering and a very careful design of the supporting systems of all the elements in the accelerator, an active vibration stabilisation system is required for all MBQ magnets along the main linac. Active stabilisation is based on a stiff support and piezo actuators that can reposition the magnet during the \SI{20}{\ms} between pulses  with high accuracy.  Five prototypes have been built with increasing complexity, mass and degrees of freedom. The fourth prototype reached the requirements for the main linac for a higher vibration background and for a nominal magnetic field and water-cooling. The last prototype (\ref{fig:mbq}) is a complete, fully integrated stabilisation system with an MBQ. Tests in the laboratory are in preparation. Most equipment used is commercially available, while the in-house developed components are technologically well within reach. Tests of a full system in a radiation environment are still outstanding.

\begin{figure}[h]
\begin{center}
  \centering
  \includegraphics[width=0.45\textwidth]{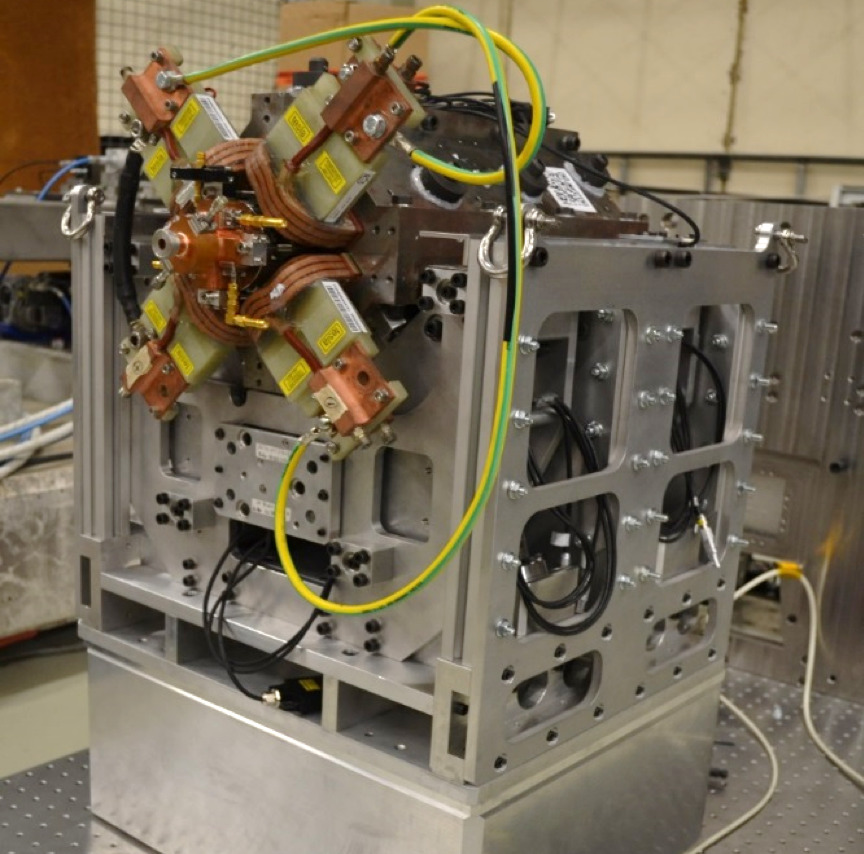}
 \caption{Main-beam quadrupole active stabilisation prototype. \imcl
}
\label{fig:mbq}
\end{center}
\end{figure}

Recent beam dynamics studies indicate that the shape of the transmissibility function is more important for the luminosity than the obtained integrated RMS displacement. 
This implies that a single combined control system, 
simultaneously taking measurements of ground motion and technical noise into account, is needed for the beam and for the stabilisation of the hardware. 
This understanding has triggered the development of adapted ground motion sensors for the stabilisation.

Indeed, commercial sensors including inertial sensors, geophones or broadband accelerometers, have two main limitations: they are not radiation hard and need to be re-designed to be integrated in a vibration control system.  CERN is collaborating with LAPP Annecy on the development of new sensors, based on new or combined methods, such as transducers, optical encoders, one-pass or multi-pass interferometers~\cite{Novotny:2017kww,Collette:2012} to measure the internal mass motion. At the same time, ULB Brussels is studying the replacement of the classical spring mass by an internal beam~\cite{Hellegouarch:2016,Balik:2301234}.

\subsubsection{Beam instrumentation}
In order to preserve low emittance beams over long distances, dispersion free-steering needs to be applied along the CLIC main linac. It relies on the use of cavity Beam Position Monitors (BPM) capable of achieving a few tens of nanometre precision in space, combined with a time resolution better than \SI{50}{\ns}. A low-Q cavity BPM for the CLIC main beam was constructed and tested at  CTF3. Its ability to measure the beam position for a 200 ns long train of bunches with a time accuracy better than \SI{20}{\ns} was demonstrated~\cite{Cullinan:2135975}.

CLIC relies also on colliding electron and positron bunches as short as \SI{150}{\femto\second}. The bunch length needs to be measured and controlled accurately with time resolution better than \SI{20}{\femto\second}. An R\&D programme was launched in 2009 to design and test non-invasive bunch length monitors using laser pulses and bi-refringent Electro-Optical (EO) crystals~\cite{Dabrowski:1058816,Berden:2007zz}. Based on such a technology, a new scheme, called Spectral Up-conversion, has been developed~\cite{Jamison:2010}. It directly measures the Fourier spectrum of the bunch using an optical spectrum imaging system, as the beam fields are printed onto a laser beam and up-converted from the far-IR-mid-IR spectrum to the optical region. The technique uses a long-pulse 
laser probe, transported through an optical fibre. This makes the system simpler and cheaper than the ultra-fast amplified systems of other EO schemes.

A breakthrough in beam size monitoring was achieved in \num{2011}, with the experimental measurement in ATF2 at KEK of the point-spread function of optical transition radiation, that allows for sub-micron resolution measurements using a simple, cheap and compact optical imaging system~\cite{PhysRevLett.107.174801}. In addition, Cherenkov diffraction radiation from long dielectrics has recently been tested at Cornell Electron Storage Ring (CESR) and ATF2 as an alternative to diffraction radiation from a small slit. It provides a very promising technique for non-invasive beam size measurements.

A high-performance and cost-efficient Beam Loss Monitor (BLM) system, based on optical fibre measuring Cherenkov light induced by lost charged particles, has been developed to monitor losses in the drive-beam decelerator sections~\cite{Kastriotou:2207310}. In particular, the study has addressed several key features of the BLM system such as the position resolution of the optical fibre detection system when using long electron pulses (i.e. \SI{200}{\ns})~\cite{Busto:2015}, and the crosstalk between losses from the main beam and the drive beam~\cite{Kastriotou:2015}.

\subsubsection{Vacuum system}
The original baseline for the vacuum system for the main linac, with long vacuum chambers providing pumping to several modules, was demonstrated in the laboratory module from the point of view of pumping performance. 
However, transverse forces from the vacuum system on the main-beam and drive-beam structure generated displacements which are not compatible with the CLIC requirements. 
Therefore, the current architecture of the vacuum system is  based on a combination of Non-Evaporable Getters (NEG) cartridge pumps combined with sputter ion pumps (\SI{100}{\liter\per\second} and \SI{5}{\liter\per\second}, respectively)
and NEG cartridge pumps (\SI{100}{\liter\per\second}). 
A set of Pirani and Penning gauges are installed on each beam line and in each module to complete the system~\cite{Garion:1407933}.

Technology originally proposed for the CLIC drive beam, e.g. a deformable RF bridge~\cite{Kastriotou:2012}, is now being implemented for LHC and HL-LHC. On the other hand, new vacuum technologies currently under development are considered for application at CLIC. Examples are the NEG-coated electroformed copper chambers~\cite{Amadora:2018}, permanent radiation-hard bake out systems, and Shape Memory Alloy connectors (see~\ref{fig:vac})~\cite{Niccoli:2017kzj,Niccoli:2017lom}.

\begin{figure}[h]
\begin{center}
\begin{subfigure}{.52\textwidth}
  \centering
  \includegraphics[height=4.6cm]{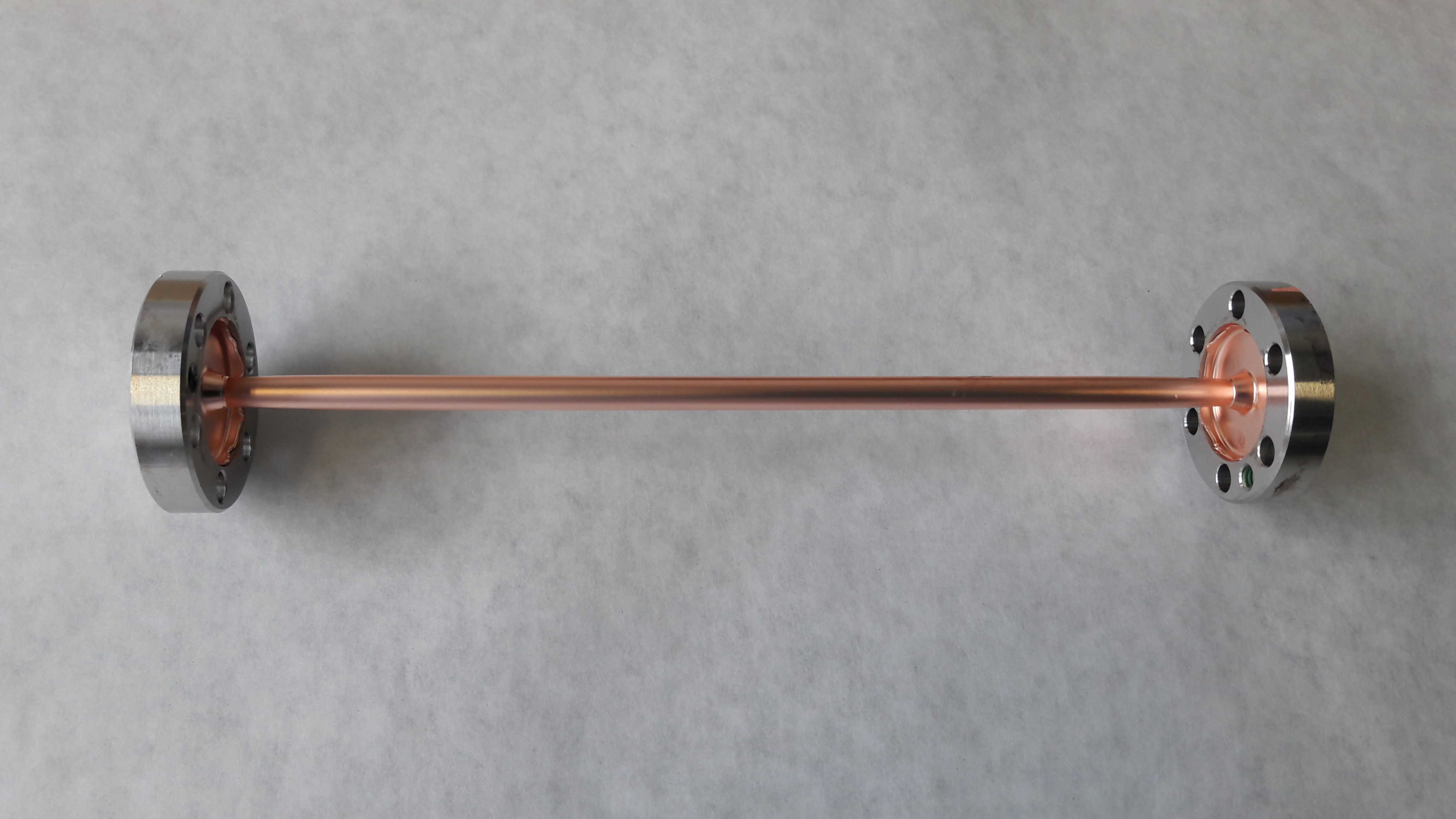}
  \caption{}
  \label{fig:vac1}
\end{subfigure}
\begin{subfigure}{.16\textwidth}
  \centering
  \includegraphics[height=4.6cm]{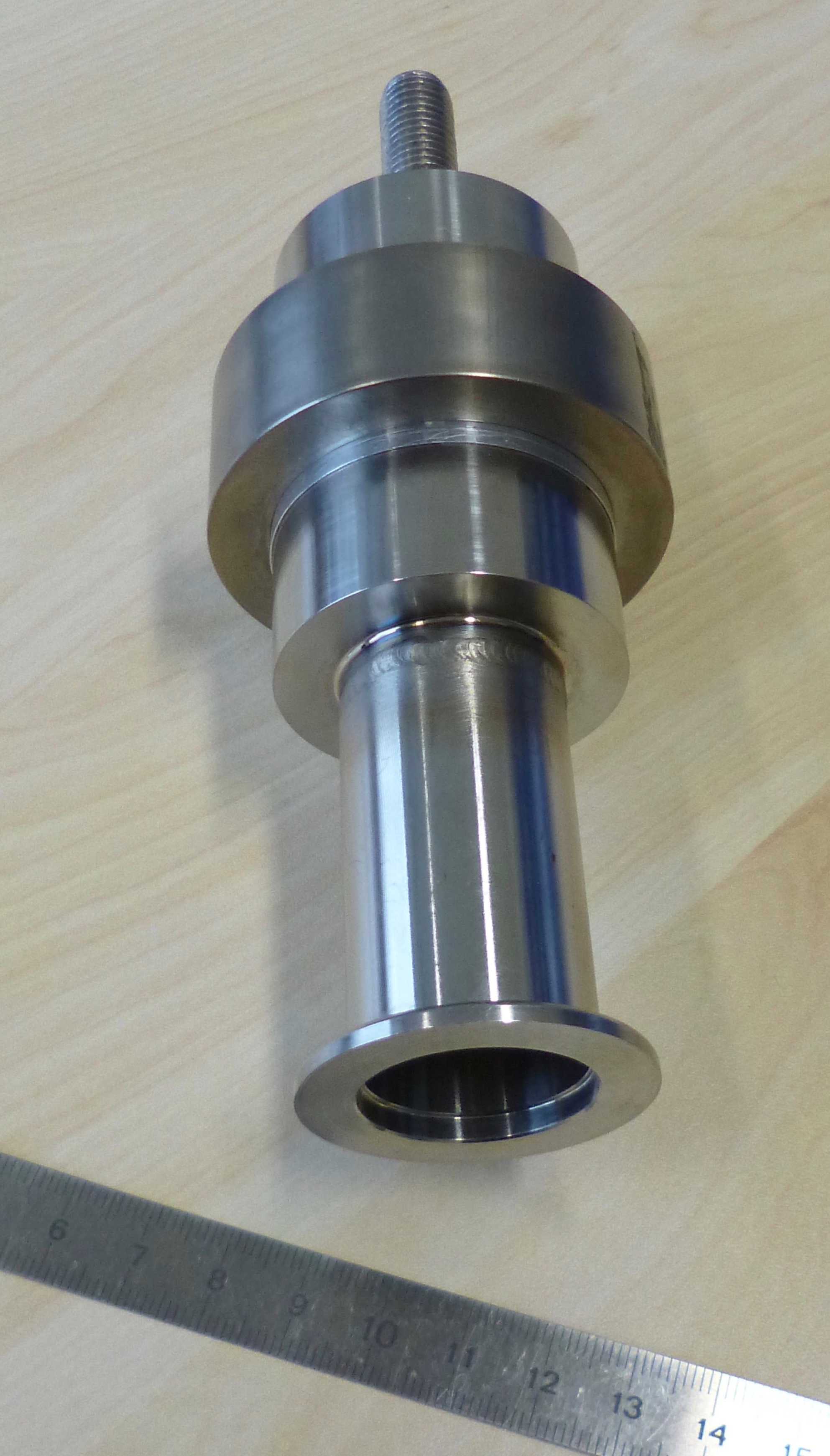}
  \caption{}
  \label{fig:vac2}
\end{subfigure}
\begin{subfigure}{.29\textwidth}
  \centering
  \includegraphics[height=4.6cm]{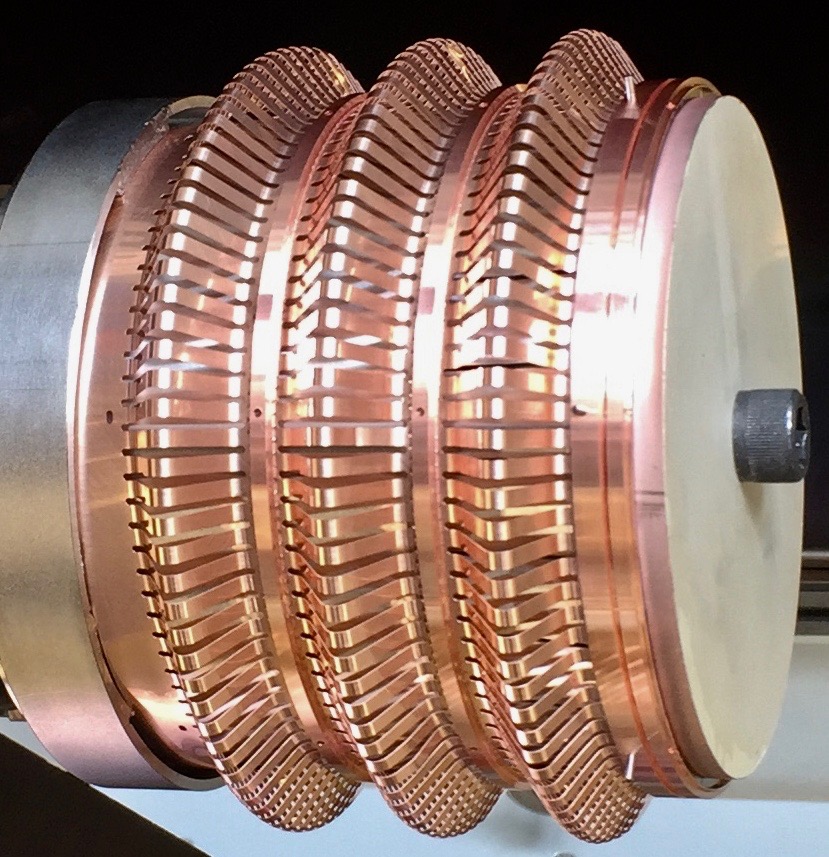}
  \caption{}
  \label{fig:vac3}
\end{subfigure}
 \caption{Examples of CLIC vacuum system components using new technologies: 
(a) electroformed copper chamber integrating stainless steel flanges, 
(b) ultrahigh vacuum coaxial Shape Memory Alloy connector 
and (c) deformable RF bridge. \imcl
}
\label{fig:vac}
\end{center}
\end{figure}

\subsubsection{Magnets}
Most of the magnets required for CLIC are normal-conducting electromagnets, well within the state of the art. However, their number and variety are well beyond current accelerator projects. For this reason, a significant effort has been invested in optimising the fabrication, assembly, and installation procedures. A total of 15 prototype electro-magnets have been manufactured and tested to verify the design choices, with system tests at CTF3 and in the laboratory. Given the large number of magnets in the CLIC complex, it is important to minimise costs and power consumption. Tuneable Permanent Magnets (PM) have been designed and manufactured for the quadrupoles in the main decelerator, in collaboration with the Daresbury laboratory~\cite{Shepherd:1461571,}. The design has been optimised for cost and industrialisation. The feasibility of the concept is now proven but studies on the radiation effects on the PM material are still needed before re-evaluating the baseline~\cite{Shepherd:2642418,Martinez:8264683}. Prototypes of the final quadrupole and sextupole, QD0 and SD0, have been manufactured using a hybrid technology (permanent magnets and electro-magnets) to increase the field with a reduced imprint. \ref{fig:mag1} shows examples of recently built prototypes.

\begin{figure}[t]
\begin{center}
\begin{subfigure}{.4\textwidth}
  \centering
  \includegraphics[height=6.3cm]{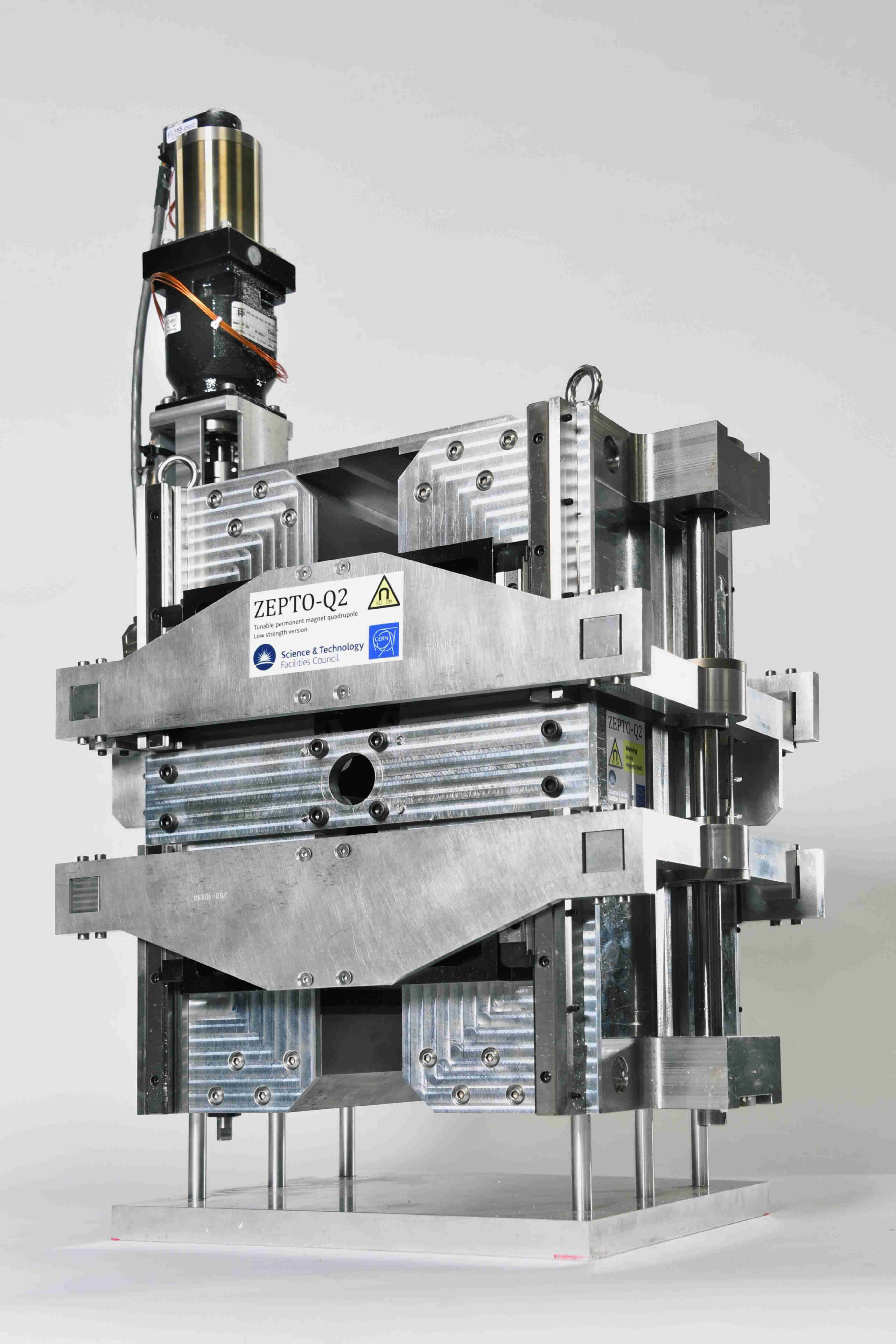}
  \caption{}
  \label{fig:mag11}
\end{subfigure}
\begin{subfigure}{.4\textwidth}
  \centering
  \includegraphics[height=6.3cm]{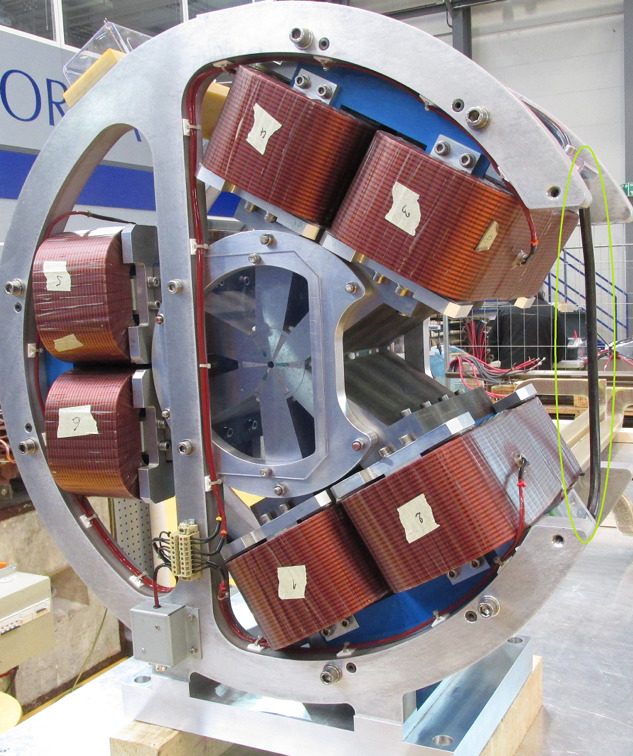}
  \caption{}
  \label{fig:mag12}
\end{subfigure}
 \caption{Prototypes of magnets for CLIC: (a) the tuneable permanent magnet quadrupole for the drive beam and (b) the hybrid SD0 final sextupole. \imcl
}
\label{fig:mag1}
\end{center}
\end{figure}

A special magnet, designed and currently being manufactured by CIEMAT, is the so-called longitudinal variable field magnet. This type of magnet will allow for a reduction in the total circumference of the damping ring of \SI{13}{\percent}, while preserving performance. The concept is being applied to the upgrade of light sources such as ESRF. However, the CLIC prototype is more challenging, as it is a tuneable permanent magnet combining dipole and quadrupole components, with a very high field of \SI{2.3}{\tesla} at its centre.

\begin{figure}[h]
\begin{center}
  \centering
  \includegraphics[height=6cm]{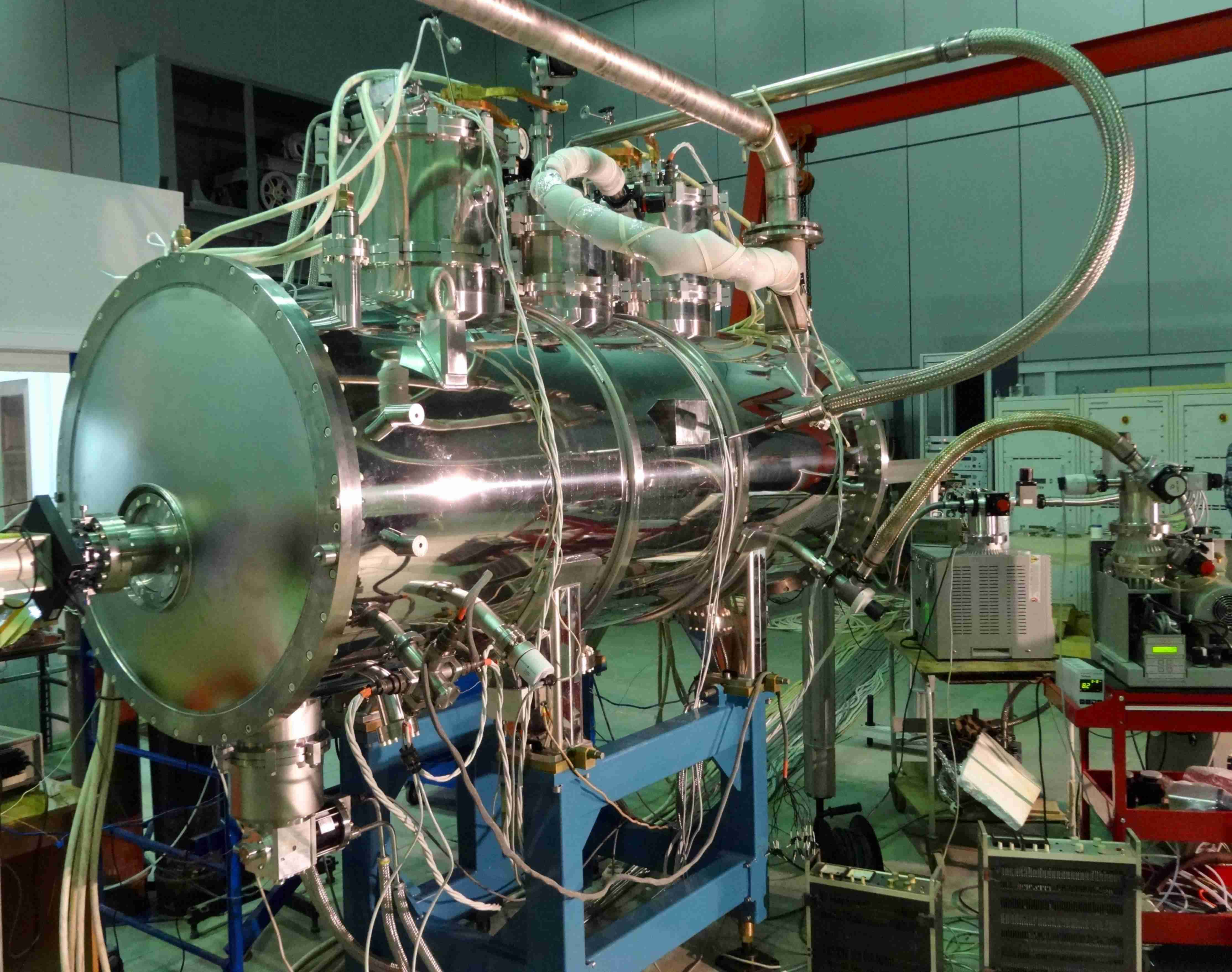}
 \caption{Superconducting wiggler being tested in BINP. \imcl }
\label{fig:mag2}
\end{center}
\end{figure}

The damping rings will contain a number of wigglers in each straight section to increase radiation damping and reduce the Intrabeam Scattering (IBS)
effect, thereby reaching an emittance which is at least an order of magnitude lower compared to planned or existing rings.  This is achievable by using superconducting wigglers. A Nb-Ti prototype (see~\ref{fig:mag2}) was manufactured by Budker Institute of Nuclear Physics in collaboration with CERN and the Karlsruhe Institute of Technology, where it is currently installed~\cite{Bernhard:2207403}. The prototype magnet was used to validate the technical design of the wiggler, in particular the conduction-cooling concept applied in its cryostat design. As part of the study, the expected heat load (several tens of Watt) due to synchrotron radiation from a future up-stream wiggler is simulated by heating the vacuum pipe with an electrical heater. A short model using Nb$_3$Sn, which will be able to reach a higher field and further reduce the damping ring circumference was also designed and manufactured. Further improvements on this prototype are under way.

Concerning pulsed magnets, the most challenging requirements come from the damping rings and the very high field uniformity and time stability required to extract the electron beam without deteriorating the final luminosity. The combined flat-top ripple and drop of the field pulse must be \SI{\pm2e-4}. In addition, the total allowable beam coupling impedance for each ring must be below \SI{1}{\ohm}. The damping ring extraction uses a strip-line kicker specifically designed for the CLIC characteristics. It is equipped with electrodes with a novel shape, called half-moon electrodes.  The electrode support, feedthroughs and manufacturing tolerances have been optimised to match the impedance during operation and to minimise the field inhomogeneity~\cite{Belver-Aguilar:2014xva,Belver-Aguilar:2017rnt}. A prototype of this kicker, shown in~\ref{fig:mag31}, has been manufactured in a collaboration between CERN, CIEMAT and IFIC in Spain~\cite{Pont:2018ulv}.

To power the strip-line kicker, an inductive adder (see~\ref{fig:mag32}) has been selected as a promising means of achieving the demanding specifications for the extraction kicker modulator of the damping ring. The inductive adder is a solid-state modulator, which can provide relatively short and precise pulses. The adder is assembled in layers each of which contributes linearly to the final voltage. Detailed research and development has been carried out on this device, which has the potential to be used also in other accelerators. Recent measurements on the prototype inductive adder
show that the flat-top stability achieved by applying modulation was 
\SI{\pm2.2}{\volt} over \SI{900}{\nano\second} at \SI{10.2}{\kilo\volt} output voltage. This pulse meets the stability
specifications for the damping ring extraction kicker~\cite{Holma:2018duk}.

\begin{figure}[h]
\begin{center}
\begin{subfigure}{.6\textwidth}
  \centering
  \includegraphics[height=5cm]{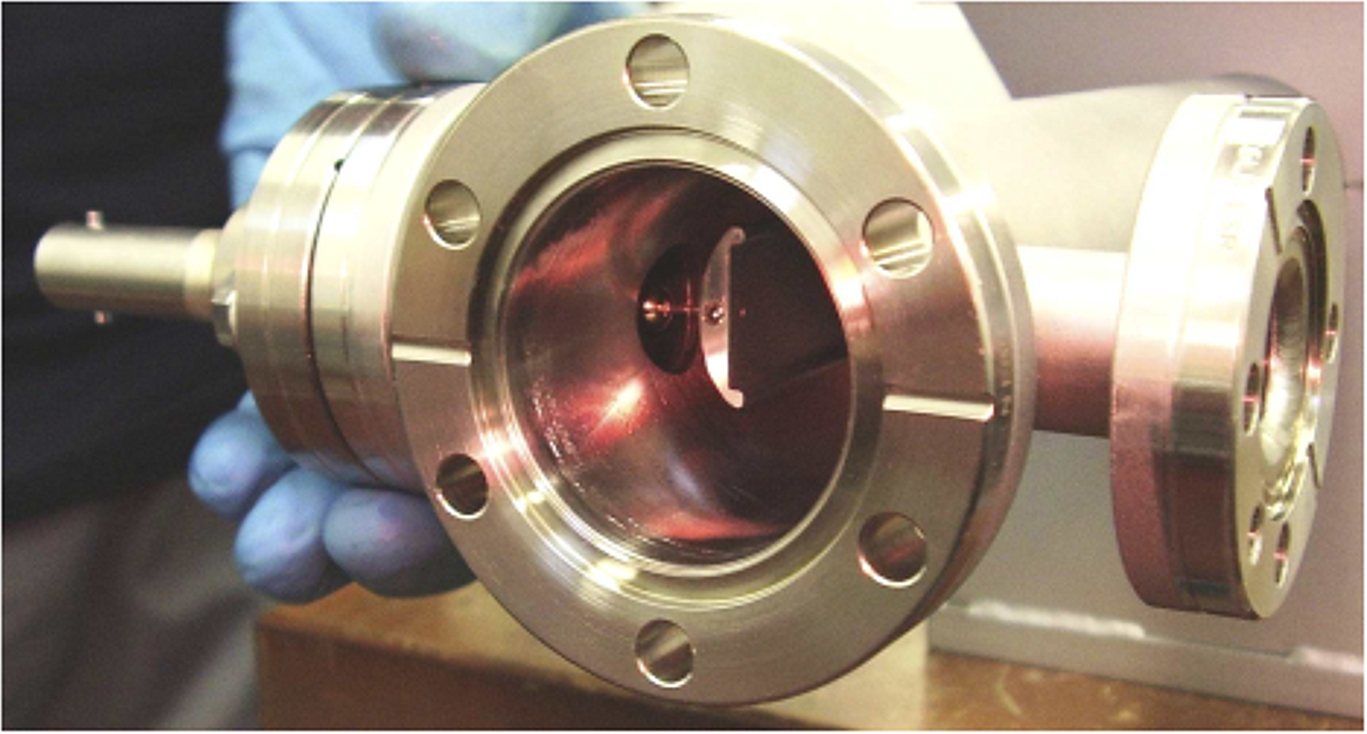}
  \caption{}
  \label{fig:mag31}
\end{subfigure}
\begin{subfigure}{.38\textwidth}
  \centering
  \includegraphics[height=5cm]{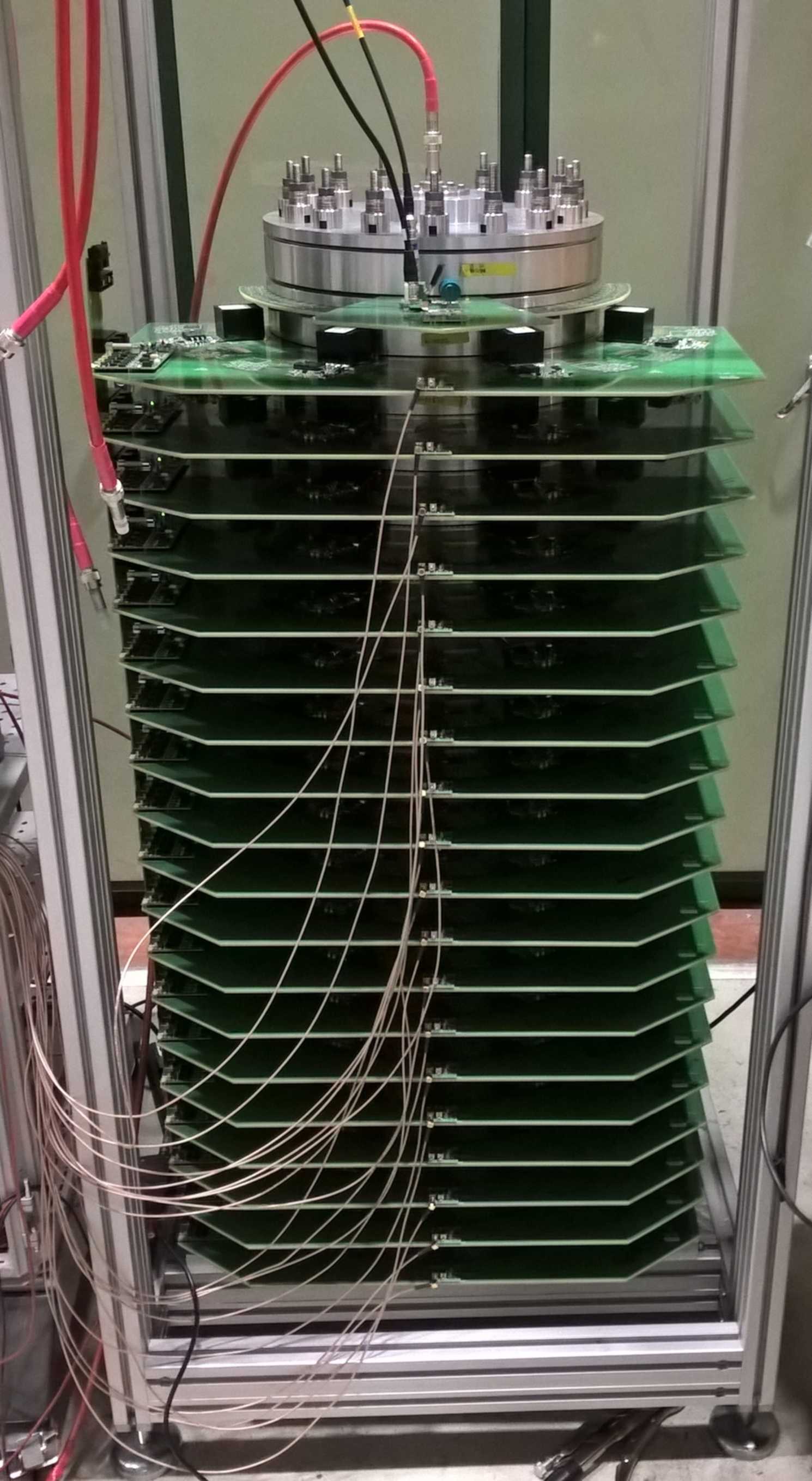}
  \caption{}
  \label{fig:mag32}
\end{subfigure}
 \caption{(a) Prototype strip-line kicker with optimised half-moon electrodes. (b) 20-layers inductive adder. \imcl
}
\label{fig:mag3}
\end{center}
\end{figure}

In order to complete its characterisation, the prototype strip-line kicker has been installed in the ALBA synchrotron to be tested with beam.
A first measurement indicates that the field homogeneity is within the desired range (\SI{\pm1e-4}) although the measurement error is still too large to quote definitive results.
In order to confirm the stability of the full system, the inductive adder will be also sent to ALBA and tested together with the strip-line kicker.

\subsubsection{Klystron-based main linac RF unit and module design}
\label{sec:acc-technologies-klystronoption}
Each main linac consists of a sequence of \num{1456} identical RF modules that are interleaved with
quadrupole modules to form the FODO lattice.
The RF module supports four pairs of accelerating structures, each with an
active length of \SI{0.46}{\meter} and a gradient of \SI{75}{\mega\volt/\meter}.

The RF power system per module consists of a two-pack solid-state modulator equipped with two \SI{53}{\MW} klystrons.
Two pulse compressor systems, which are equipped with linearising cavities, compress the \SI{2.0006}{\us}-long RF pulses of the klystrons
to \SI{334}{\ns}. The pulse is then distributed into the accelerating structures.
In~\ref{fig:k-unit} a klystron-based module is shown on the left, equipped with linearisation
and pulse compression cavities. On the right a top view of one RF unit in the Klystron and Main Linac tunnels.
High efficiency klystrons are considered in this scheme~\cite{Constable:2017syz}, operating with an efficiency in excess of \SI{60}{\percent}; the pulse compression device adopts Barrel Open Cavities providing a compression factor of \num{3.5} and delivering \SI{170}{\MW} RF power at their output to feed each of the four accelerating structures with \SI{40.6}{\MW}.

\begin{figure}[t]
\begin{center}
\begin{subfigure}{.49\textwidth}
  \centering
  \includegraphics[width=\linewidth]{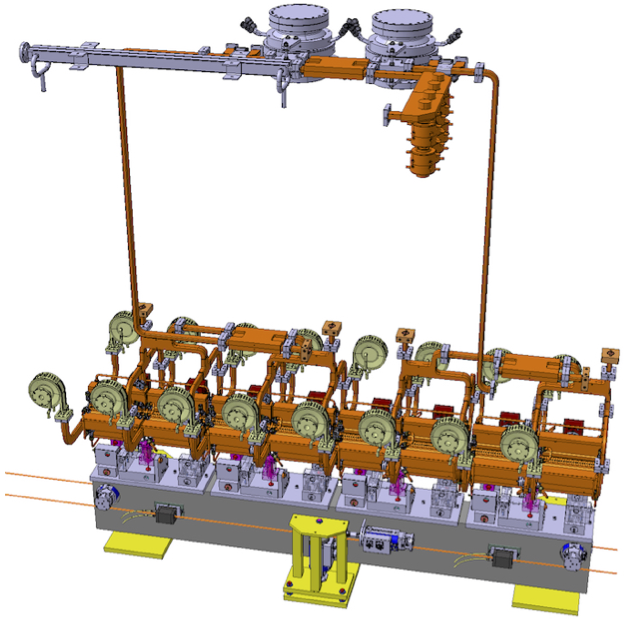}
   \caption{}
  \label{fig:k-unit1}
\end{subfigure}
\begin{subfigure}{.49\textwidth}
  \centering
  \includegraphics[width=\linewidth]{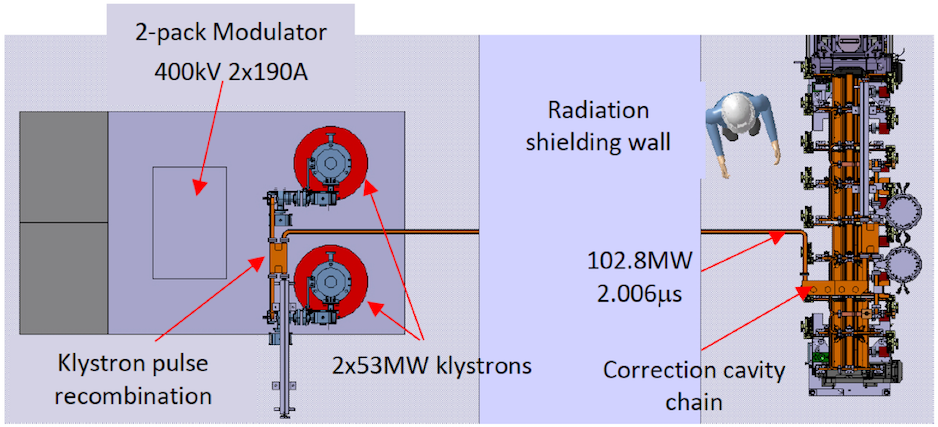}
   \caption{}
  \label{fig:k-unit2}
\end{subfigure}
\caption{(a) The klystron-based module with pulse compression and linearisation system.
(b) Top view of one RF unit in the Klystron and Main Linac tunnels. \imcl}
\label{fig:k-unit}
\end{center}
\end{figure}

\newpage
\newpage
\section{CLIC detector design, technologies and performance}
\label{sec:detector}
The CLIC detector layout and its technology choices are driven by the CLIC physics programme described in~\ref{sec:physics}, and by the experimental conditions at CLIC
described in~\ref{sec:expcond}.
The resulting detector concept is described in~\ref{sec:CLICdet}.
Details on the technology choices under investigation for the CLIC detector concept are given in~\ref{sec:technologies}
and the detector concept performance is detailed in~\ref{sec:CLICdetPerformance}.

\subsection{Experimental conditions at CLIC}
\label{sec:expcond}
The experimental conditions at CLIC are given by
the CLIC beam structure,
the presence of beam-induced backgrounds,
the low-rate environment in the $\Pep\Pem$ collisions, and
the beam particle energy spectrum at collision.

\paragraph{Beam structure}
Linear colliders operate in bunch trains.
For example, at \SI{380}{\GeV}, CLIC has a train repetition rate of \SI{50}{\Hz} with \num{352} bunches per train, each separated by \SI{0.5}{\ns}, resulting in a train duration of \SI{176}{\ns}.
One hard physics event is expected on average per bunch train. 
Beam-induced background events (described in detail below) giving significant energy deposits in the detector, can take place in several bunch crossings per train. 
The rates of physics and beam-induced background events, combined with the bunch separation, 
drive the timing requirements of the sub-detectors.
Full detector simulation studies were performed with a CLIC detector concept optimised to achieve low occupancies.
It was found that a hit time resolution of \SI{\sim5}{\ns} is needed in the vertex and tracking detectors, and \SI{1}{\ns} in the central calorimeters in order to sufficiently distinguish between energy deposits from hard physics events and those from beam-induced backgrounds.

The \SI{50}{\Hz} repetition rate and short bunch-train structure result in a low duty cycle, below \SI{0.001}{\percent} for all CLIC energy stages.
This offers the possibility of power pulsing the detector's front-end electronics,
which leads to a reduced power consumption and consequent reduction in 
the cooling infrastructure of the sub-detectors. 
This for instance results in a lower material budget for the vertex and tracking systems, which 
is of particular importance for the physics performance of the detector.  
Triggerless readout of the CLIC detector is foreseen at the end of each bunch train.

\paragraph{Beam-induced backgrounds}
In order to achieve high luminosities at CLIC, extremely small beam sizes and high bunch charges are required.
As an example, the transverse beam size at \SI{380}{\GeV} is about \SI{150 x 3}{\nm} in the horizontal and vertical direction, the bunch length is \SI{70}{\um}, and one bunch contains \num{5.2e9} particles, as shown in~\ref{t:scdup1}.
This leads to very high electromagnetic fields in the collision region that cause beam--beam interactions, leading to beam-induced background.
While most of these particles are produced at very small angles, some enter the detector region.
The two processes that produce significant fluxes of particles relevant for the CLIC detector design are incoherent $\Pep\Pem$ pairs and $\PGg\PGg\to$~hadron events~\cite{Chen:1992ax}.
The energy and polar angle distributions of the particles produced in these processes are shown in~\ref{fig:background} for a centre-of-mass energy of \SI{3}{\TeV}.
The detector occupancies caused by these background processes have an impact on the detector design choices, such as the diameter of the central beam pipe (thus the radius of the innermost
vertex detector layer) and on the sub-detector granularity.

The design effort for the CLIC detector has so far mainly focused on the \SI{3}{\TeV} case. The number of  $\PGg\PGg\to$~hadron events produced is reduced by a factor of almost 20 at \SI{380}{\GeV}, the incoherent $\Pep\Pem$ pairs are reduced by a factor of 5~\cite{CLICdet_performance}. Therefore, a detector layout with a smaller vacuum pipe and lower inner radius of the vertex detector barrel is being prepared for the first stage of CLIC operation. 

\begin{figure}[h]
\begin{center}
\begin{subfigure}{.49\textwidth}
  \centering
  \includegraphics[width=\textwidth]{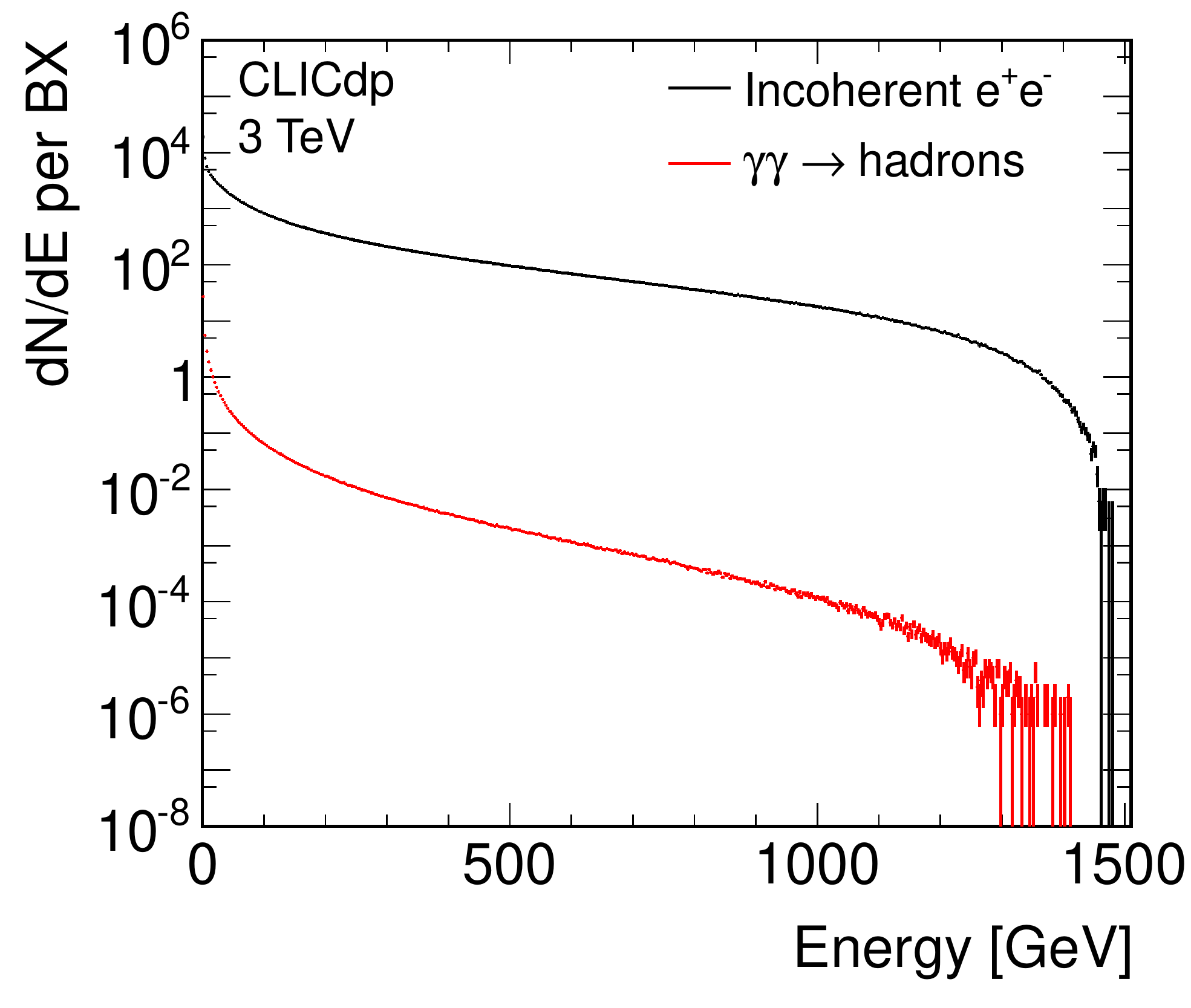}
  \caption{}
  \label{fig:background_energy}
\end{subfigure}
\begin{subfigure}{.49\textwidth}
  \centering
  \includegraphics[width=\textwidth]{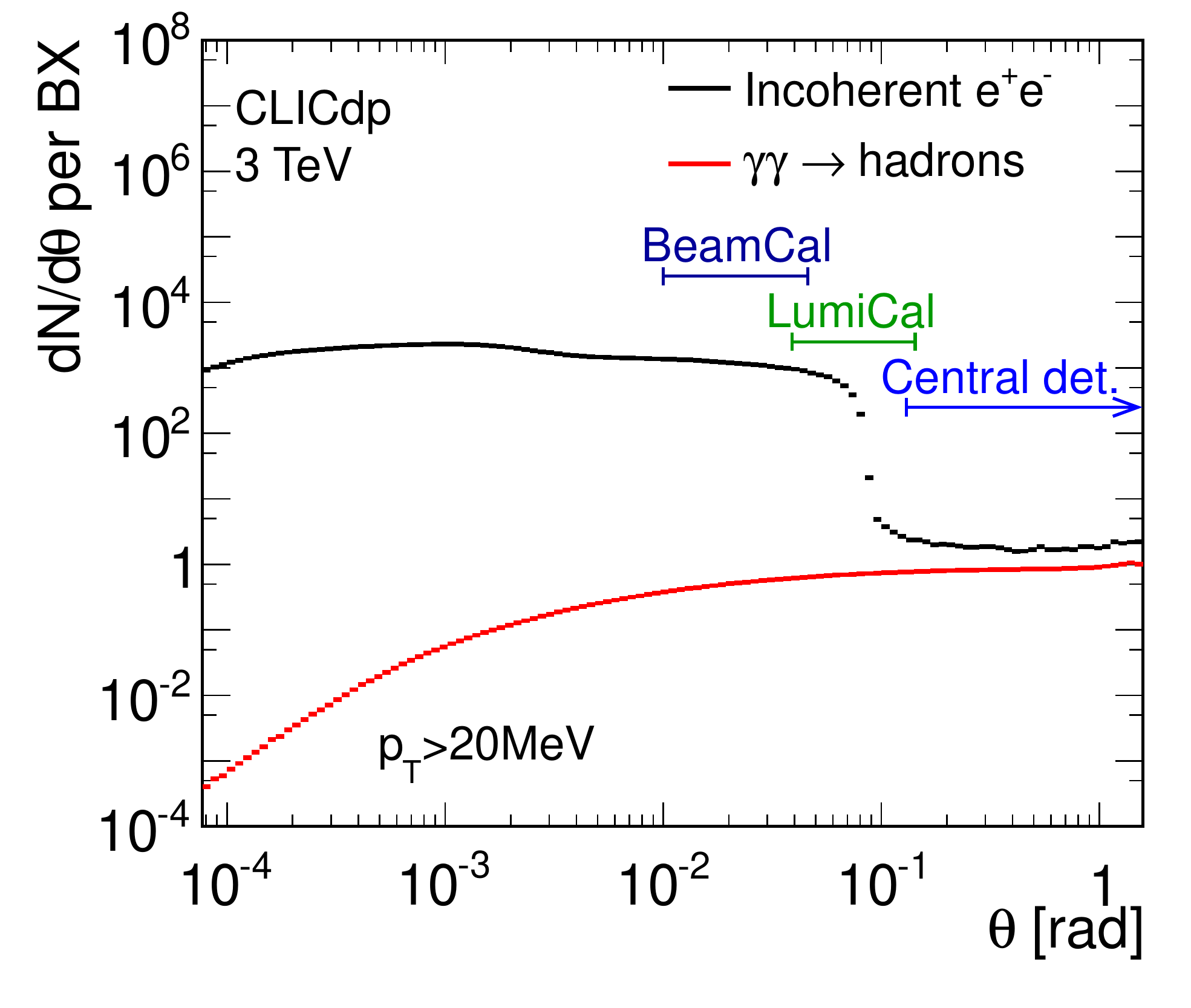}
  \caption{}
  \label{fig:background_theta}
\end{subfigure}
 \caption{(a) Energy distribution and (b) polar angle distribution per bunch crossing (BX) of beam-induced backgrounds.
  Both figures are for CLIC at \SI{3}{\TeV}.
  Generated particle distributions for $p_{\text{T}}>20\,$MeV are shown, including a \SI{2}{\GeV} c.m.\ threshold for \gghad.\imdp}
\label{fig:background}
\end{center}
\end{figure}

\paragraph{Clean environment in $\text{e}^\text{+}\text{e}^\text{-}$ collisions}
In hadron collisions such as at the LHC, large QCD backgrounds drive the design of the collider detectors.
The design and technology choices emphasise radiation hardness of many sub-detectors as well as complex trigger schemes.
In spite of the beam-induced backgrounds described above, $\Pep\Pem$ collisions provide a much cleaner environment than hadron collisions, such that radiation damage considerations are relevant only for the design of the very forward calorimeters.
Also, at linear $\Pep\Pem$ colliders, there is no need for triggers. 
This is possible due to the clean events, the low duty cycle, and the relatively low events rates described above.

An upper limit of the data volume per train and the data rate written to tape was estimated for the full CLIC detector including zero suppression and address encoding. 
The data volume per bunch train ranges from \SI{75}{\mega\byte} at \SI{380}{\GeV} to \SI{115}{\mega\byte} at \SI{3}{\TeV}. 
With a bunch-train repetition rate of \SI{50}{\hertz}, this results into data rates ranging from \SI{4}{\giga\byte/\second} at \SI{380}{\GeV} to \SI{6}{\giga\byte/\second} at \SI{3}{\TeV}~\cite{ESU18RnD}.
These numbers are mainly driven by the beam-induced backgrounds.

\paragraph{Collision energy and energy spread}
Due to beamstrahlung at the interaction point, a fraction of the incoming $\Pepm$ energy can be lost before the collision takes place leading to a reduction of the $\Pep\Pem$ collision energy.
Due to the beamstrahlung photons, $\Pepm\PGg$ and $\PGg\PGg$ collisions can also take place.
The resulting luminosity spectra for $\Pep\Pem$, $\Pepm\PGg$ and $\PGg\PGg$ collisions at two different CLIC centre-of-mass energies are illustrated in~\ref{fig:lumispectra}.
The effects caused by beamstrahlung increase with centre-of-mass energy. Initial State Radiation (ISR) creates an additional energy loss, which reduces further the collision energy.

\begin{figure}[h]
\begin{center}
\begin{subfigure}{.49\textwidth}
  \centering
  \includegraphics[width=\linewidth]{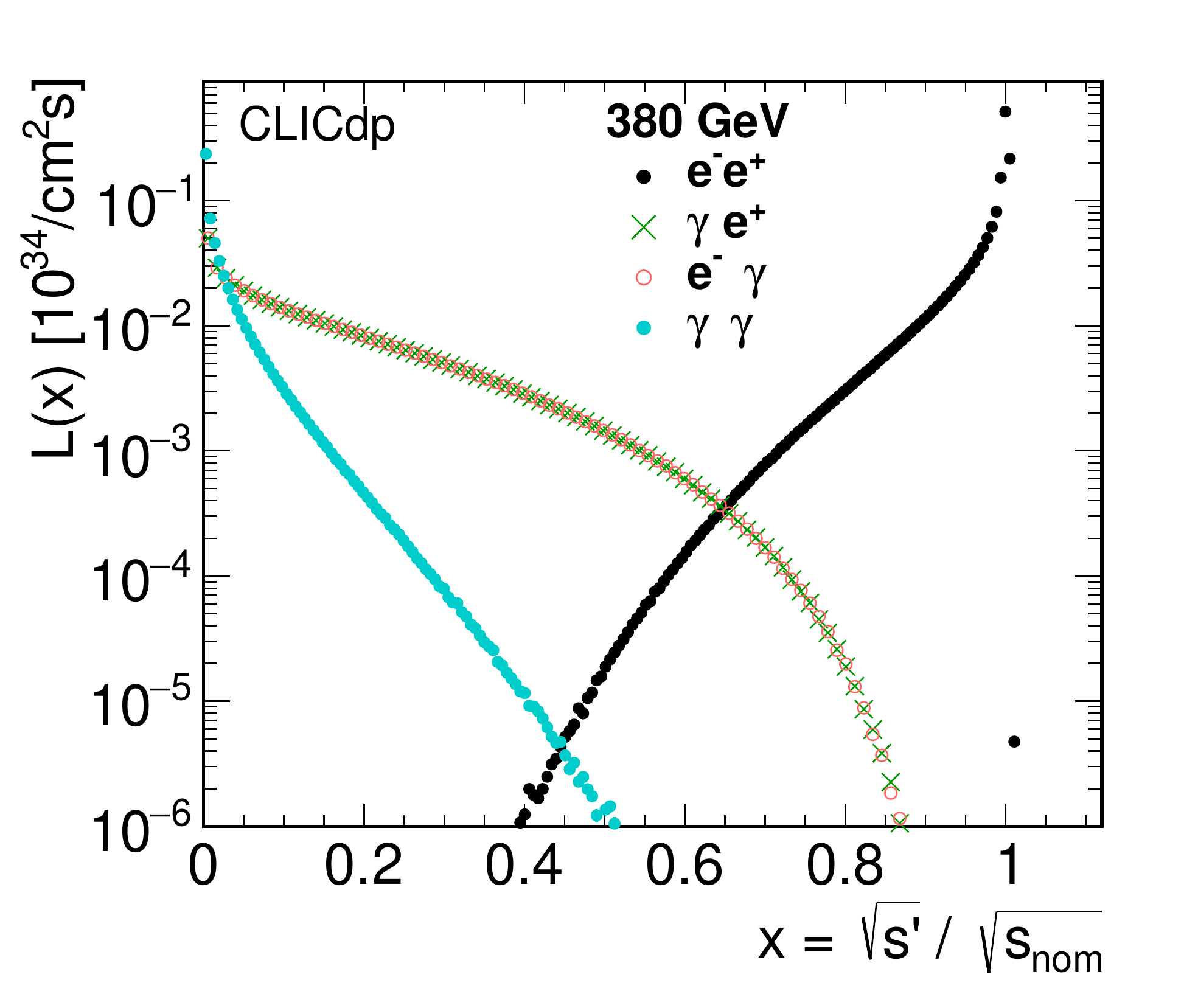}
   \caption{}
  \label{fig:luminosity_380gev}
\end{subfigure}
\begin{subfigure}{.49\textwidth}
  \centering
  \includegraphics[width=\linewidth]{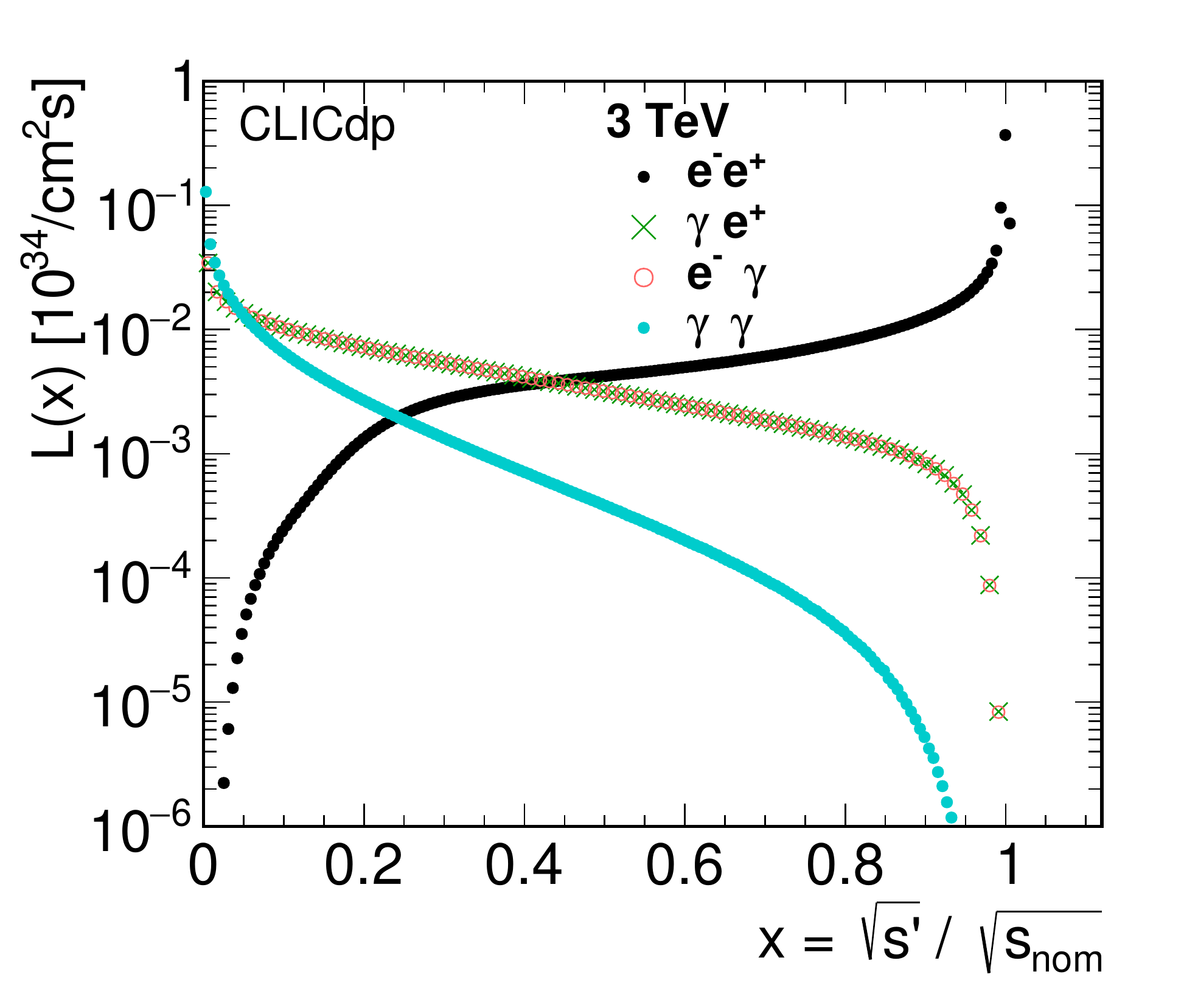}
   \caption{}
  \label{fig:luminosity_3tev}
\end{subfigure}
\caption{{\label{fig:lumispectra}}
Luminosity distributions for different types of collisions, (a) at $\sqrt{s_\text{nom}}=\SI{380}{\GeV}$ and (b) at $\sqrt{s_\text{nom}}=\SI{3}{\TeV}$~\cite{CLICdet_performance}.}
\end{center}
\end{figure}

\paragraph{Suppression of beam-induced backgrounds}
Particles from beam-induced backgrounds entering the central detector region have relatively low transverse momenta.
By applying \pT\ cuts on the reconstructed objects during physics analyses, the impact of beam-induced backgrounds can be reduced.
Beam-induced background can be further suppressed by making use of the hit time resolution of the different sub-detectors.
When combining hit timing information into cluster timing of the reconstructed particles, even tighter timing cuts can be applied.
Combined \pT\ and timing cuts optimised for the different detector regions are described in~\cite{cdrvol2, Brondolin:2641311}.
The tightness of the cuts, called \emph{Loose}, \emph{Selected}, and \emph{Tight}, can be adjusted to the event type under study and the centre-of-mass energy. For centre-of-mass energies of 380\,GeV and below, a special set of \emph{low energy Loose} cuts has been introduced.

For example, when considering $\epem\to\PQt\PAQt$ events at $\rootsprime=\SI{380}{\GeV}$, the average reconstructed
energy is \SI{370}{\GeV}. This is reduced slightly by the \emph{Loose} or \emph{Tight} timing cuts to \SI{366}{\GeV} or
\SI{357}{\GeV}, respectively, corresponding to \SI{98.9}{\percent} and \SI{96.5}{\percent} of the initial value. At the
same time, the average reconstructed energy of the \gghad{} background, which is initially \SI{45}{\GeV}, is reduced to
\SI{28}{\GeV} or \SI{62.2}{\percent} with the \emph{Loose} cuts and \SI{8}{\GeV} or \SI{17.8}{\percent} with the
\emph{Tight} cuts~\cite{Brondolin:2641311}.

\Cref{fig:tt_eventdisplay_380GeV,fig:tt_eventdisplay_3TeV} show the impact of such combined \pT{} and timing cuts for typical \mbox{$\epem\to\PQt\PAQt$} events at centre-of-mass energies of \SI{380}{\GeV} and \SI{3}{\TeV}.
Displayed are the reconstructed objects originating from the hard physics collision and from the \gghad{} interactions of 10 bunch crossings before and 20 bunch crossings after the hard event.

\begin{figure}[t]
\centering
\begin{subfigure}{.49\textwidth}
  \centering
  \includegraphics[width=\textwidth]{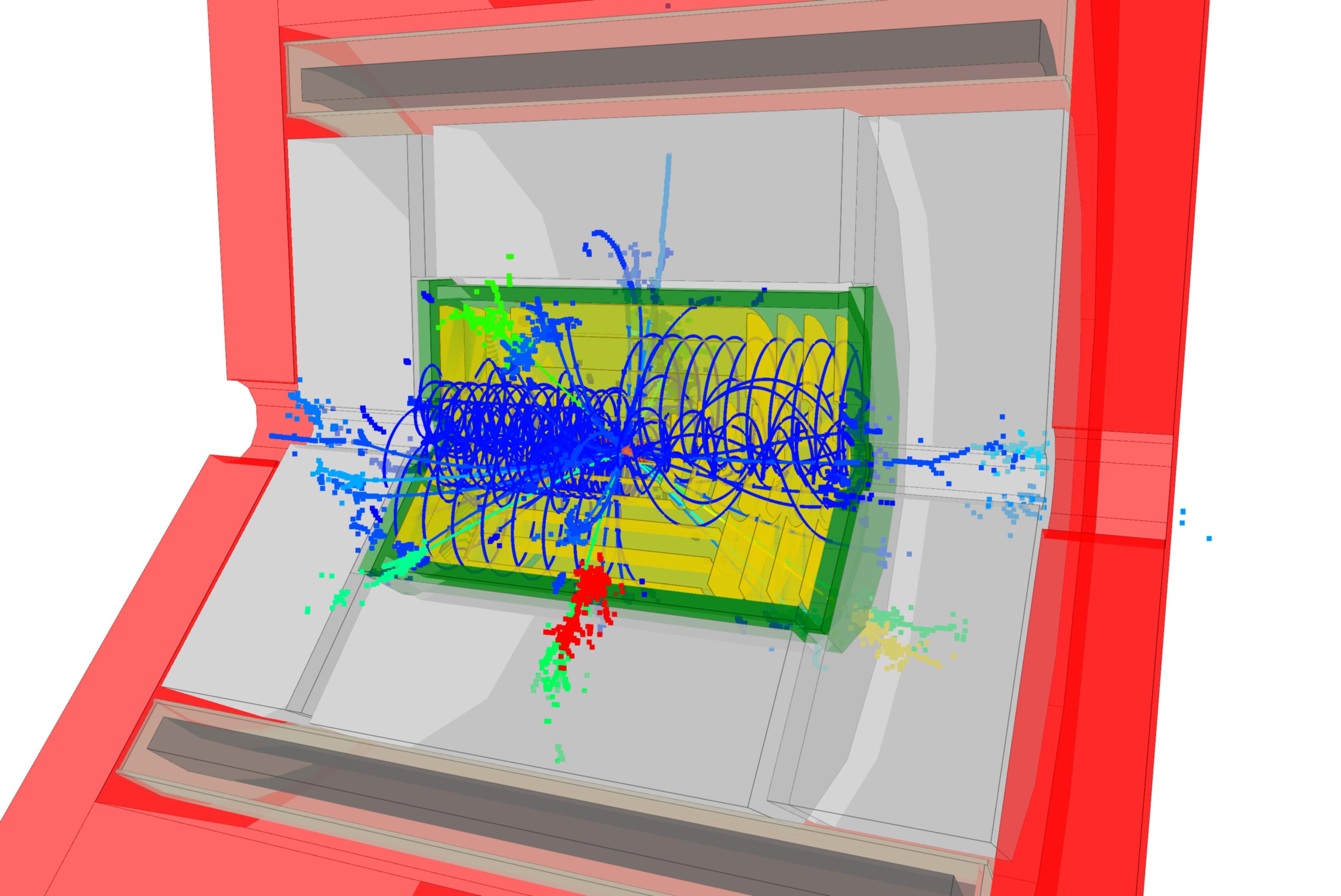}
   \caption{}
  \label{fig:tt_before}
\end{subfigure}
\begin{subfigure}{.49\textwidth}
  \centering
  \includegraphics[width=\textwidth]{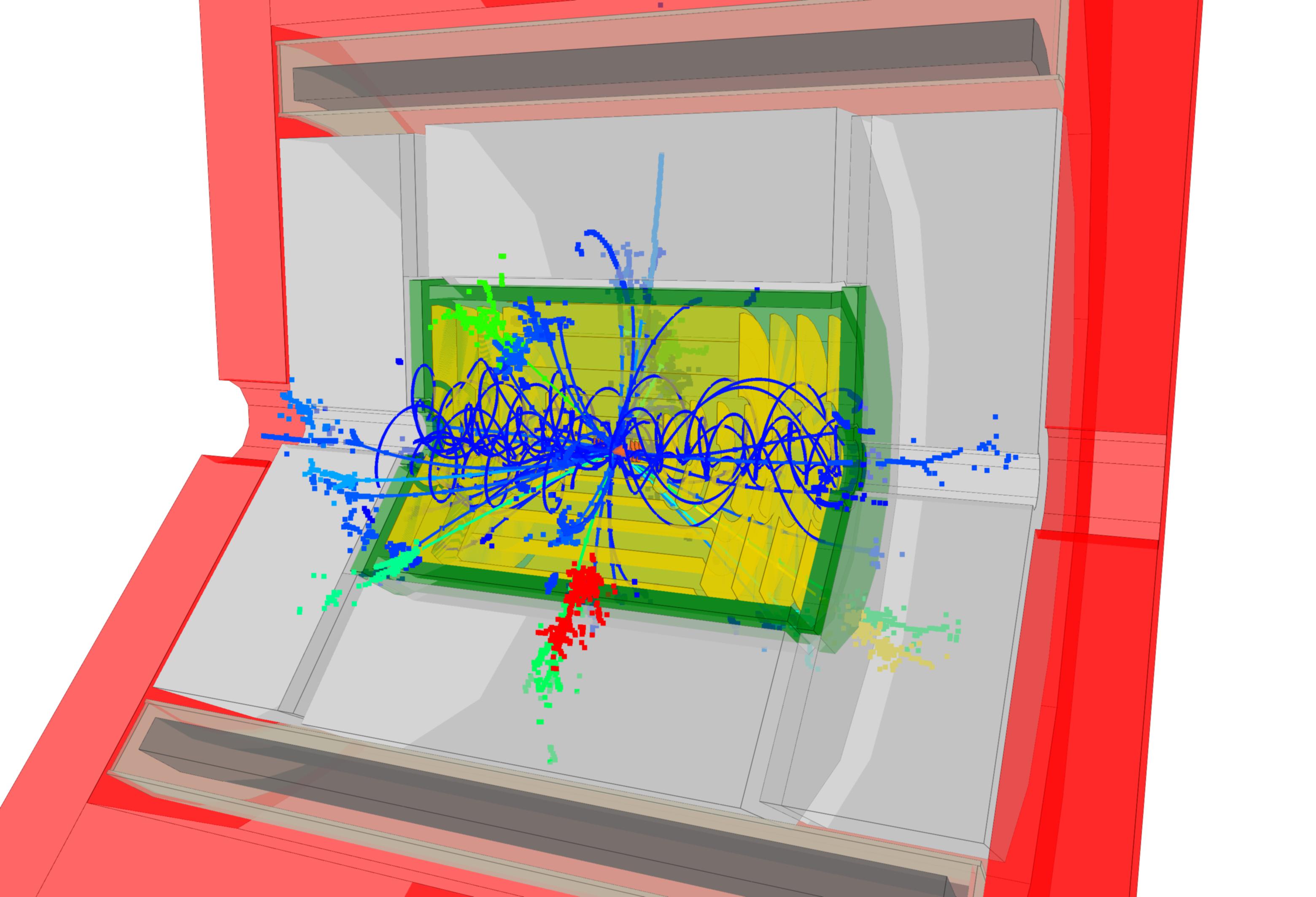}
   \caption{}
  \label{fig:tt_After}
\end{subfigure}
\caption{Event displays of $\epem\to\PQt\PAQt$ events at a centre-of-mass energy of \SI{380}{\GeV} (a) before, and (b)~after background suppression using \emph{Loose} \pT\ and timing cuts \label{fig:tt_eventdisplay_380GeV} \imdp}
\end{figure}

\begin{figure}[t]
\centering
\begin{subfigure}{.49\textwidth}
  \centering
  \includegraphics[width=\textwidth]{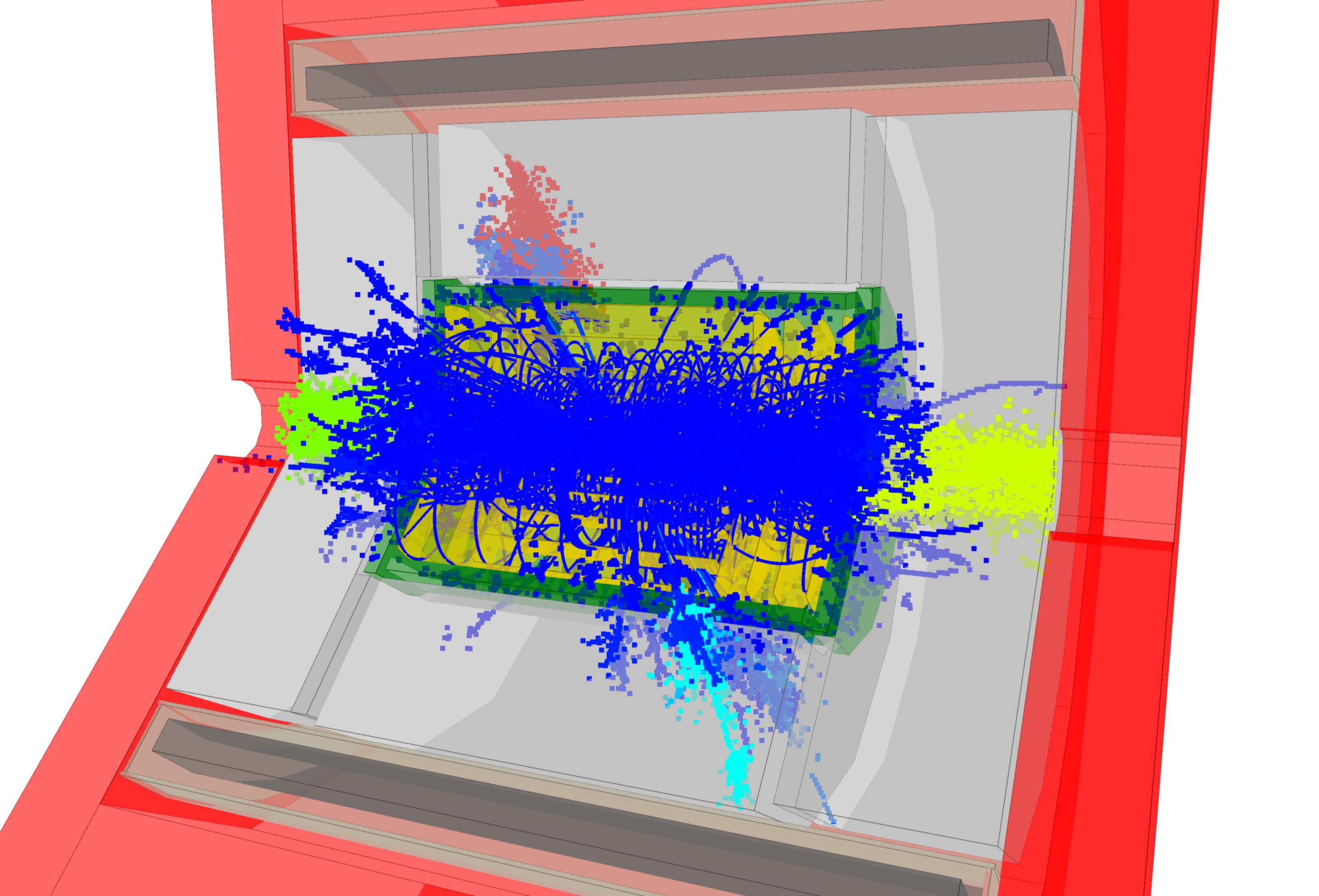}
   \caption{}
  \label{fig:tt3_before}
\end{subfigure}
\begin{subfigure}{.49\textwidth}
  \centering
  \includegraphics[width=\textwidth]{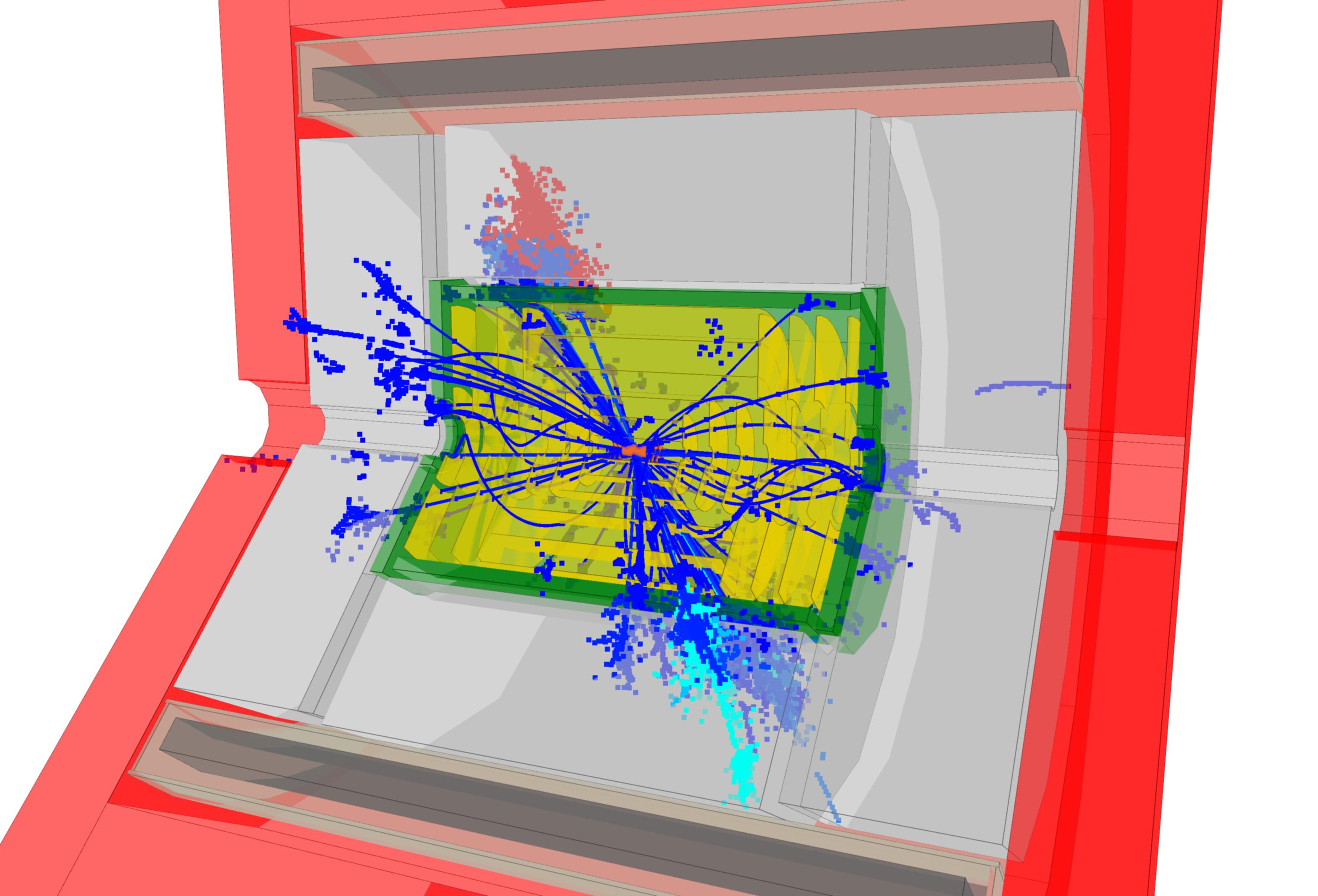}
   \caption{}
  \label{fig:tt3_After}
\end{subfigure}
\caption{Event displays of $\epem\to\PQt\PAQt$ events at a centre-of-mass energy of \SI{3}{\TeV} (a) before, and \newline (b) after background suppression using \emph{Tight} \pT\ and timing cuts.\label{fig:tt_eventdisplay_3TeV} \imdp}
\end{figure}

\subsection{Physics-driven detector requirements}
Besides being compatible with the CLIC operation conditions described above, the CLIC detector also needs to meet the physics performance targets. 
Motivated by precision physics measurements described in~\ref{sec:physics}, the targets used for the development of the CLIC detector are:
\begin{itemize}
\item excellent track-momentum resolution for high-momentum tracks in the barrel, at the level of $\sigma_{\pT}/\pT^2 \leq \SI{2e-5}{\per\GeV}$;
\item precise impact-parameter resolution, at the level of $\sigma_{d_0}^2 = (\SI{5}{\um})^2 + (\SI{15}{\um\GeV})^2/(p^2\sin^{3}\theta)$, to allow accurate reconstruction and enable flavour tagging with clean $\PQb$-, $\PQc$-, and light-quark jet separation;
\item jet-energy resolution for light-quark jets of $\sigma_E/E \leq \SI{3.5}{\percent}$ for jet energies in the range \SI{100}{\GeV} to \SI{1}{\TeV} ($\leq \SI{5}{\percent}$ at \SI{50}{\GeV});
\item detector coverage for electrons and photons to very low polar angles (\SI{\sim10}{\mrad}) to assist with background rejection.
\end{itemize}

\subsection{CLIC detector concept}
\label{sec:CLICdet}
The CLIC detector concept, referred to hereafter as CLICdet, is optimised for particle flow analysis.
It comprises a light-weight silicon-pixel vertex detector with a central barrel and forward petals in a spiral arrangement optimised for air cooling, a light-weight silicon tracker, and highly-granular electromagnetic (silicon-tungsten ECAL) and hadronic (scintillator-steel HCAL) calorimeters.
These detectors are surrounded by a superconducting solenoid providing a magnetic field of \SI{4}{\tesla}.
Beyond the solenoid, \mbox{CLICdet} comprises an iron yoke interleaved with detectors for muon identification.
The forward region of \mbox{CLICdet} close to the beam pipe is equipped with forward calorimeters, called LumiCal and BeamCal, optimised for the luminosity measurements and forward electron-tagging.
A more detailed description of \mbox{CLICdet} can be found in~\cite{CLICdet_note_2017}. CLICdet was optimised for operation at $\roots=\SI{3}{\TeV}$. As background rates at $\roots=\SI{380}{\GeV}$ are lower, some modifications to the inner detector layers are anticipated for the first energy stage~\cite{cdrvol2}.
A quarter-view of the cross section of the CLICdet concept is shown in~\ref{fig:quarter_view}. An enlarged view of the vertex detector is shown in~\ref{fig:vertex_3D_view}.
The forward region of the detector is presented in~\ref{fig:forward}.

An important change with respect to the CDR detector models~\cite{cdrvol2} is the location of the final focusing quadrupole QD0.
In order to enlarge the angular coverage of the HCAL endcap and thus to extend the physics reach of the CLIC detector, this quadrupole is located outside of the detector in the accelerator tunnel.
Nevertheless, for best luminosity performance the QD0 must be as close as possible to the interaction point.
The overall length of CLICdet has therefore been minimised by reducing the thickness of the iron yoke endcaps.
The missing iron is compensated by a set of end coils.
Both the position of the QD0 in the accelerator tunnel as well as the end coils are shown in~\ref{fig:QD0}.

\ref{fig:ExpArea} shows the experimental cavern around the interaction point as well as the service cavern, where the final assembly of the detector and maintenance work will take place.
Contrary to the push-pull scenario with two detectors, described in the CDR, operation with only one detector at CLIC is proposed here.

\begin{figure}[t]
\centering
\begin{subfigure}{.5225\textwidth}
  \centering
   \subcaptionbox{\label{fig:quarter_view}}{\includegraphics[width=\linewidth,trim={0 2.43cm 0 0},clip]{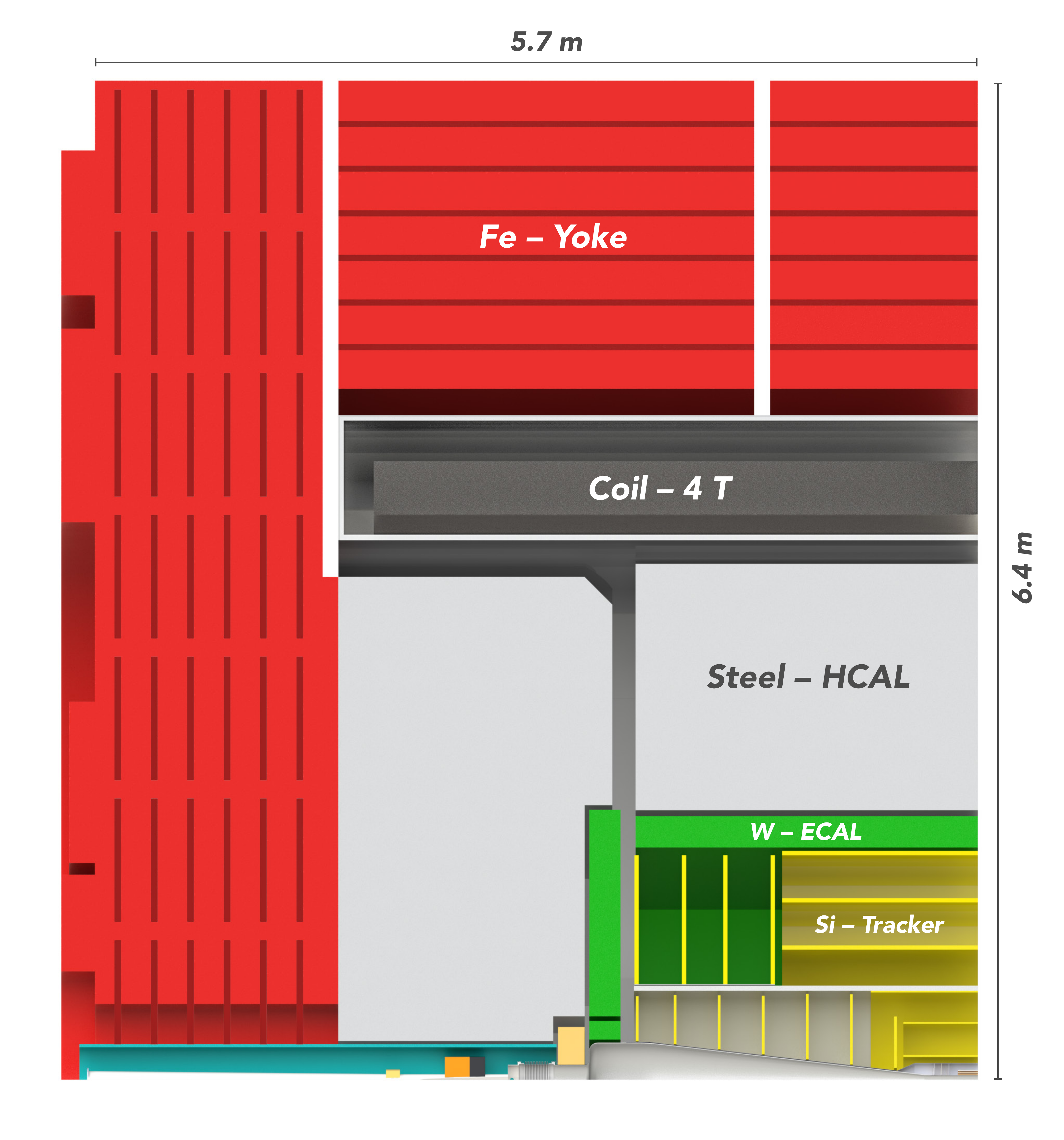}}
\end{subfigure}
\begin{subfigure}{.465\textwidth}
  \centering
   \subcaptionbox{\label{fig:vertex_3D_view}}{  \includegraphics[width=\linewidth]{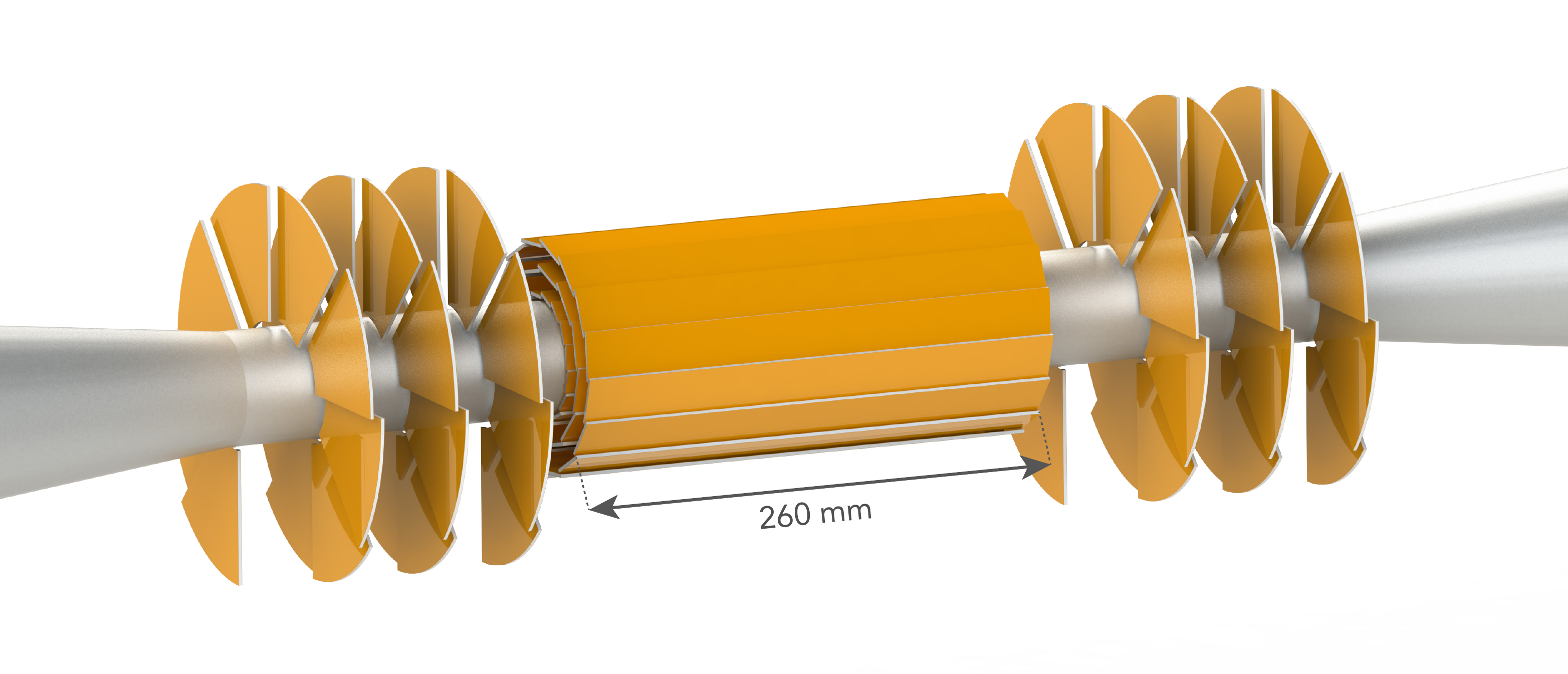}}
   \subcaptionbox{\label{fig:forward}}{  \includegraphics[width=\linewidth]{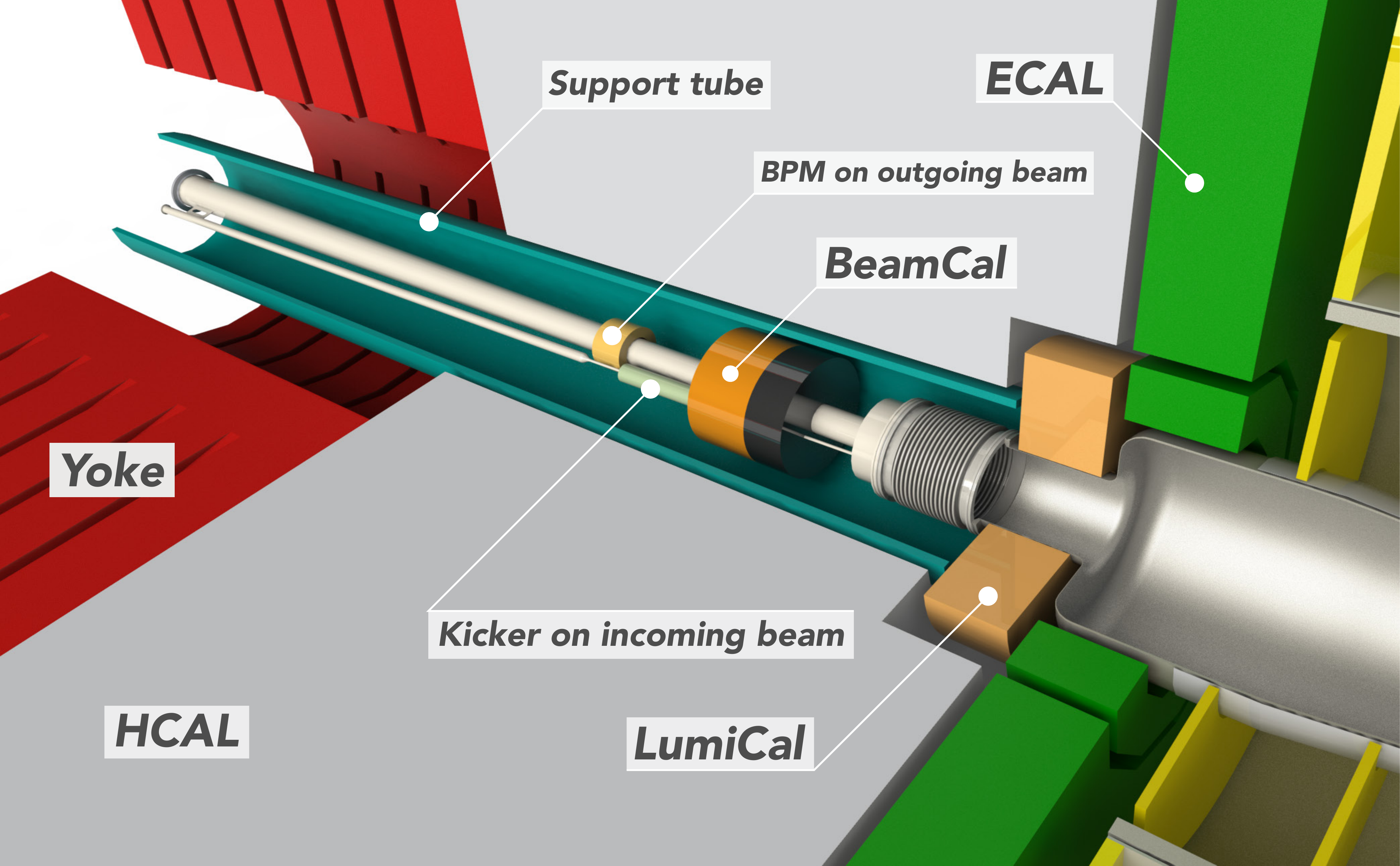}}
\end{subfigure}
\caption{(a) Longitudinal cross section showing a quadrant of CLICdet (side view).
The structures shown on the left of the image (i.e.\ outside of the yoke endcap) represent the end coils.\newline
(b) View of the vertex detector layout, with three double-layers in the barrel, and double-layer forward petals in a spiralling arrangement to facilitate air cooling. (c) Layout of the forward region of CLICdet. \imdp
}
\end{figure}

\begin{figure}[t]
\centering
\begin{subfigure}{.49\textwidth}
  \centering
  \includegraphics[width=\linewidth]{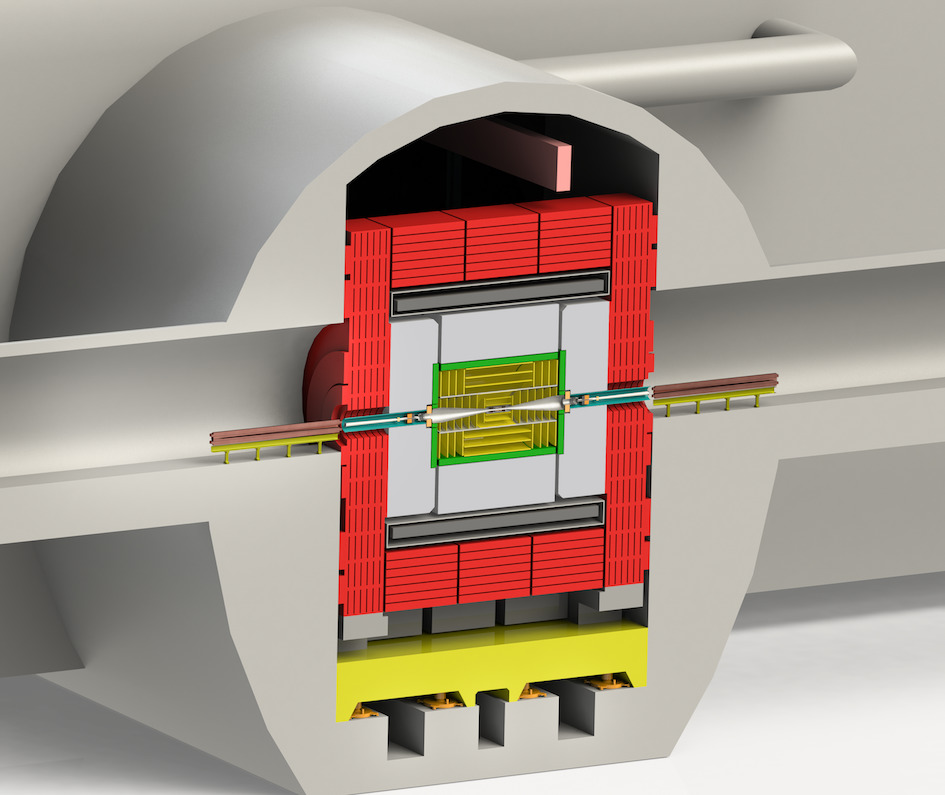}
  \caption{}
  \label{fig:QD0}
\end{subfigure}
\begin{subfigure}{.49\textwidth}
  \centering
  \includegraphics[width=\linewidth]{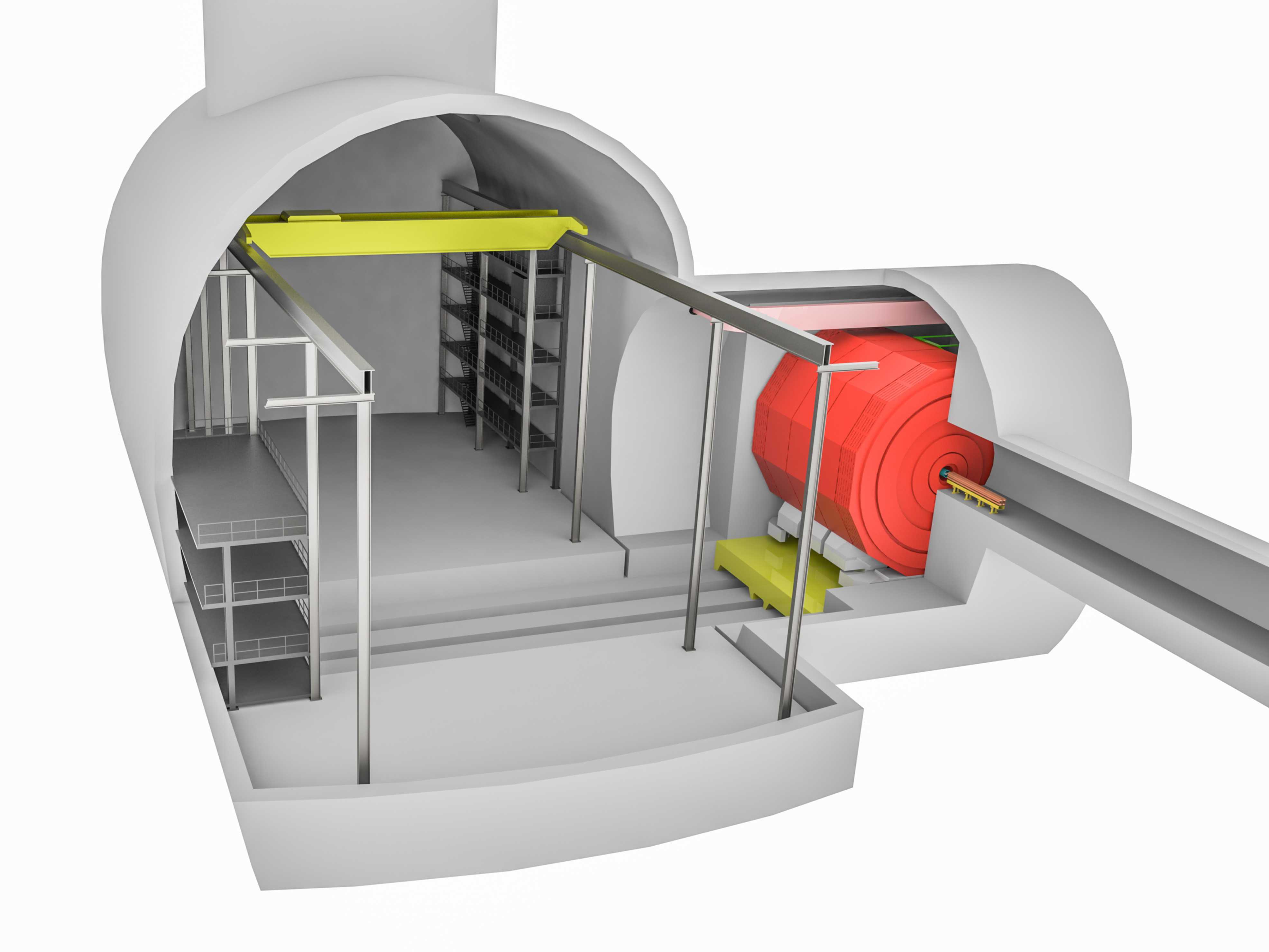}
  \caption{}
  \label{fig:ExpArea}
\end{subfigure}
\caption{
(a) The last focusing magnets QD0 are located just outside of CLICdet.\newline
(b) Experimental cavern and service cavern of CLICdet. \imdp
}
\end{figure}

\subsection{Detector technologies}
\label{sec:technologies}

A broad detector technology R\&D programme for CLIC has been ongoing for several years, as described in detail in~\cite{ESU18RnD}. In view of the time scales involved and the limited resources, the development targets those areas where CLIC requirements are the most challenging: the silicon vertex and tracking system, the high-granularity ECAL and HCAL calorimeter systems, as well as the compact very forward electromagnetic calorimeters LumiCal and BeamCal. The technology R\&D effort for the silicon vertex and tracking system is coordinated by the CLICdp collaboration, while the calorimeter developments are carried out within the CALICE and FCAL collaborations. For the muon identification system, CLIC requirements do not represent particular challenges. Therefore work was not yet initiated in this domain. The large superconducting detector solenoid, operating at 4\,T, has been the subject of design studies~\cite{cdrvol2} and first extrusion tests towards a reinforced conductor have been performed~\cite{Langeslag:2013lga,Shilon:2014oha}. With its 3.5\,m inner bore radius, technology challenges for the CLIC solenoid go beyond what was achieved for the CMS experiment. Development for the solenoid will therefore have to start as soon as decisions towards the realisation of CLIC are taken. Several general engineering studies, including detector assembly and access scenarios, were performed in the framework of the CDR~\cite{cdrvol2}. More recent engineering studies have focused mainly on the vertex and tracking system, in particular concerning low-mass supporting structures and air cooling~\cite{VillarejoBermudez:1982810,DuarteRamos:2138963}.

\subsubsection{Vertex and tracking technologies}
\label{sec:vertextrackingtech}
The CLIC physics objectives, combined with its experimental conditions, pose challenging technology requirements on the vertex and tracking system. For the vertex detector, the occupancy from beam-induced background particles requires the pixel size to be $\leq\,25\,\times 25\,\upmu\text{m}^2$. A hit position resolution of $3\,\upmu\text{m}$ is required, together with a hit time resolution of approximately 5\,ns. The material budget must be at most 0.2\%\,$X_0$ per layer, including sensors, electronics, supports and cabling. This limit in material budget implies a limit of $50\,\text{mW}/\text{cm}^2$ on power consumption including power pulsing, as this will allow for air cooling. While the individual requirements for the vertex detector can be met using state-of-the art technologies, the combination of all requirements is very challenging and requires new technological solutions. Compared with the vertex detector, the technology requirements on the CLIC tracking layers are more relaxed. Occupancies in the tracker impose strip length limits of 1\,mm to 10\,mm for an assumed strip pitch of $50\,\upmu\text{m}$~\cite{Nurnberg:2261066}. Anticipating future advances in integrated technologies, one can expect, however, that large pixels will be chosen for the tracker instead. A hit position resolution of $7\,\upmu\text{m}$ and a hit time resolution of $\sim$\,5\,ns are required, while the material budget is limited to 2\% $\xo$ per layer. Leak-less water cooling at sub-atmospheric pressure and at room temperature is currently foreseen for the tracker. The expected radiation exposure from non-ionising energy loss, dominated by \gghad{} events, leads to an equivalent neutron flux below $1\times 10^{11}\,\text{neq/(cm}^2\,\text{yr})$\,in the inner vertex layers. The total ionising dose is dominated by background from incoherent pairs and is less than $1\,\text{kGy/yr}$\,in the inner vertex layers.

In order to match the challenging vertex detector requirements, a comprehensive R\&D programme is ongoing. It involves simulations, ASIC and sensor designs, the construction and readout of small detector assemblies, laboratory tests and beam tests. It addresses  several state-of-the-art technology approaches and includes systematic performance mapping of parameters (e.g.\ sensor thickness) in order to fully understand options and dependencies. Detector simulation tools have been extended and refined accordingly in order to guide subsequent R\&D steps~\cite{Allpix2}. Both hybrid assemblies and depleted monolithic sensors are assessed. In addition, the feasibility of power pulsing, air cooling and ultra-thin support structures have been assessed for the vertex detector. While the R\&D programme focused initially on the vertex detector with small pixels of $25\,\times 25\,\upmu\text{m}^2$, some of the technologies are now also under consideration for the CLIC tracker. 

For the different development steps undertaken, suitable test assemblies were designed, built and tested. These are listed in~\ref{tab:SiModules}. Timepix and Timepix3 ASICs with $55\,\times 55\,\upmu\text{m}^2$ pixel pitch have been used in initial studies of hybrid assemblies in order to assess the effect of sensor thickness on the charge-collection efficiency and the position resolution. Subsequently, the CLIC project pioneered a first hybrid pixel detector ASIC for particle physics in 65 nm CMOS process technology and with small $25\,\times 25\,\upmu\text{m}^2$ pixel size. As a result CLICpix and its CLICpix2 upgrade are used in hybrid detector assemblies. In these assemblies either a thin silicon planar sensor is bump-bonded directly to the ASIC (DC coupling), or a HV-CMOS (High Voltage) sensor with embedded amplification is glued to the ASIC (capacitive coupling).  In another approach, monolithic CMOS technologies are explored. Recent progress with depleted monolithic CMOS technologies make them promising candidates for large-scale systems with low mass, together with facilitated production and reduced cost. In these technologies, the depleted signal formation region and the electronic readout circuitry are embedded in the same monolithic device. In the HV-CMOS technology studied, the pixel circuitry is embedded in a deep n-well that covers most of the pixel area. The deep n-well acts as a signal collecting electrode, while shielding the CMOS readout circuitry from the high voltage applied to the silicon bulk. On the other hand, HR-CMOS (High Resistivity) sensors are designed with a small collection electrode on top of a depleted high-resistivity epitaxial layer to achieve a small sensor capacitance for a large signal-to-noise (S/N) ratio and a low analogue power consumption. Monolithic Silicon-On-Insulator (SOI) wafers implement a layer of $\text{SiO}_{2}$ insulator between a high-resistivity sensor wafer and a thin low-resistivity electronics wafer. \ref{fig:vertextracker1,fig:vertextracker2,fig:vertextracker3} show a collection of images illustrating the silicon vertex and tracker R\&D effort.

\begin{table}[h]
\footnotesize
  \centering
\begin{threeparttable}
  \caption{\label{tab:SiModules} Summary of sensor assemblies and technologies explored in the framework of the CLIC vertex and tracker detector R\&D.}
  \begin{tabular}{c  *5{c}} 
    \toprule
test & type & coupling & cell size & active  & references  \\
assembly           &      &          &           & sensor    &             \\
           &      &          &           & thickness &             \\
           &     &           & $(\upmu\text{m}^2)$     & $(\upmu\text{m})$        &      \\
    \midrule
Timepix(3) + Si sensor & hybrid planar & bump-bonded & $55\,\times\,55$ & $50-500$ & \cite{TimepixTestbeamNote,ThesisNilou,ActiveEdge2018,Timepix3Calibration2018}   \\
 CLICpix + Si sensor & hybrid planar & bump-bonded & $25\,\times\,25$ & $50-200$ & \cite{CLICdp-Conf-2017-010}  \\
 CLICpix2 + Si sensor & hybrid planar & bump-bonded & $25\,\times\,25$ & 130, 200 & \cite{CLICdp-Conf-2018-003}  \\
 CLICpix + CCPDv3 & hybrid HV-CMOS & capacitive & $25\,\times\,25$ & $\sim30$ & \cite{CCPDv3_hynds,Buckland:2633983,VicenteBarretoPinto:2267848,HV-CMOS-2018}  \\
 CLICpix2 + C3PD & hybrid HV-CMOS & capacitive & $25\,\times\,25$ & $\sim30-100$ & \cite{c3pd_standalone,Kremastiotis:2017lhi}  \\
 ALICE investigator & HR-CMOS & monolithic & $28\,\times\,28$ & $\sim15-20$ & \cite{ThesisMagdalenaMunker,Munker:2646292} \\
 ATLASpix simple & HV-CMOS & monolithic & $40\,\times\,130$ & $\sim30-100$ & \cite{PERIC2018} \\
 Cracow SOI & SOI & monolithic  & $30\,\times\,30$ &  300, 500 & \cite{Bugiel:2310056} \\
 CLIPS & SOI & monolithic  & $20\,\times\,20$ & $100-500$ & in production \\
 CLICTD & HR-CMOS & monolithic & $30\,\times\,(8\,\times\,37.5)$~\tnote{a} & $<40$ & design phase~\cite{CLICdp-Conf-2018-008} \\
    \bottomrule
 \end{tabular}
\begin{tablenotes}
\item[a] The CLICTD cells are segmented in 8 sub-pixels in the long direction, in order to maintain the benefits of the small collection electrode.
\end{tablenotes}
\end{threeparttable}
\end{table}

Given the progress with the CLIC vertex and tracker R\&D, together with overall expected advances in semiconductor technologies worldwide, one can expect good prospects for reaching the CLIC objectives in due time. At the current phase of the CLIC  vertex and tracker R\&D, a number of conclusions can already be drawn from the obtained results:

\begin{itemize}
\item Good S/N ratios have been achieved for the detection of signals from thin $(50\,\upmu\text{m})$ fully-depleted planar sensors, sufficient for full detection efficiency and for satisfying the CLIC time-stamping requirements~\cite{TimepixTestbeamNote,ThesisNilou,Timepix3Calibration2018}.
\item In general the position resolution depends strongly on the pixel size and on the depleted sensor thickness. For planar sensors, small sensor thicknesses, which are needed to reach the low-mass requirements, go together with small charge sharing, limiting the achievable position resolution~\cite{TimepixTestbeamNote,ThesisNilou}. Sensor design with enhanced charge sharing is underway~\cite{elad_tipp2017}. More advanced ASIC process technologies (e.g.\ 28\,nm instead of 65\,nm) offer prospects for smaller pixel sizes and better position resolution.
\item Good progress was made towards reducing detector mass through the use of active-edge sensor technologies~\cite{ThesisNilou,ActiveEdge2018} and advances in Through-Silicon Via (TSV) interconnect technologies~\cite{Tick:2011zz,JINST-TSV-Campbell}.
\item Fine-pitch bump-bonding processes for hybrid silicon detector assemblies with pixel sizes as small as $25\,\times 25\,\upmu\text{m}^2$ are not readily available commercially at single die level. This slows down the R\&D process for this type of detector~\cite{CLICdp-Conf-2017-010,CLICdp-Conf-2018-003}. Capacitively coupled hybrid detectors have been assembled and operated successfully~\cite{CCPDv3_hynds,Buckland:2633983,HV-CMOS-2018,VicenteBarretoPinto:2267848,c3pd_standalone,Kremastiotis:2017lhi}. They, however, pose other challenges, such as uniform pressure and planarity during the bonding step, as well as the integration and cooling of larger module surfaces.
\item Promising results were obtained from tests using integrated technologies (SOI, HV-CMOS, HR-CMOS)~\cite{ThesisMagdalenaMunker,Munker:2646292,Bugiel:2310056,PERIC2018}. CLIC-specific fully integrated designs are underway (CLICTD~\cite{CLICdp-Conf-2018-008}, CLIPS). Integrated technologies offer a potential for high-precision performance over large surfaces, with a reduced material budget and at a lower cost.
\item Feasibility of power pulsing was demonstrated at the level of module-size low-mass powering demonstrators in the laboratory, including tests in a magnetic field~\cite{Blanchot:2014haa,Blanchot:2062429}. Power pulsing was also implemented successfully in hybrid ASICs and HV-CMOS sensors for CLIC. Power consumption levels below $50\,\text{mW}/\text{cm}^2$ have been achieved.
\item Feasibility of air cooling was demonstrated in simulation studies and in a full-scale CLIC vertex detector mock-up with realistic heat loads~\cite{DuarteRamos:2138963}.
\end{itemize}

\begin{figure}[h]
\centering
\begin{subfigure}{.400\textwidth}
  \centering
  \includegraphics[width=\linewidth]{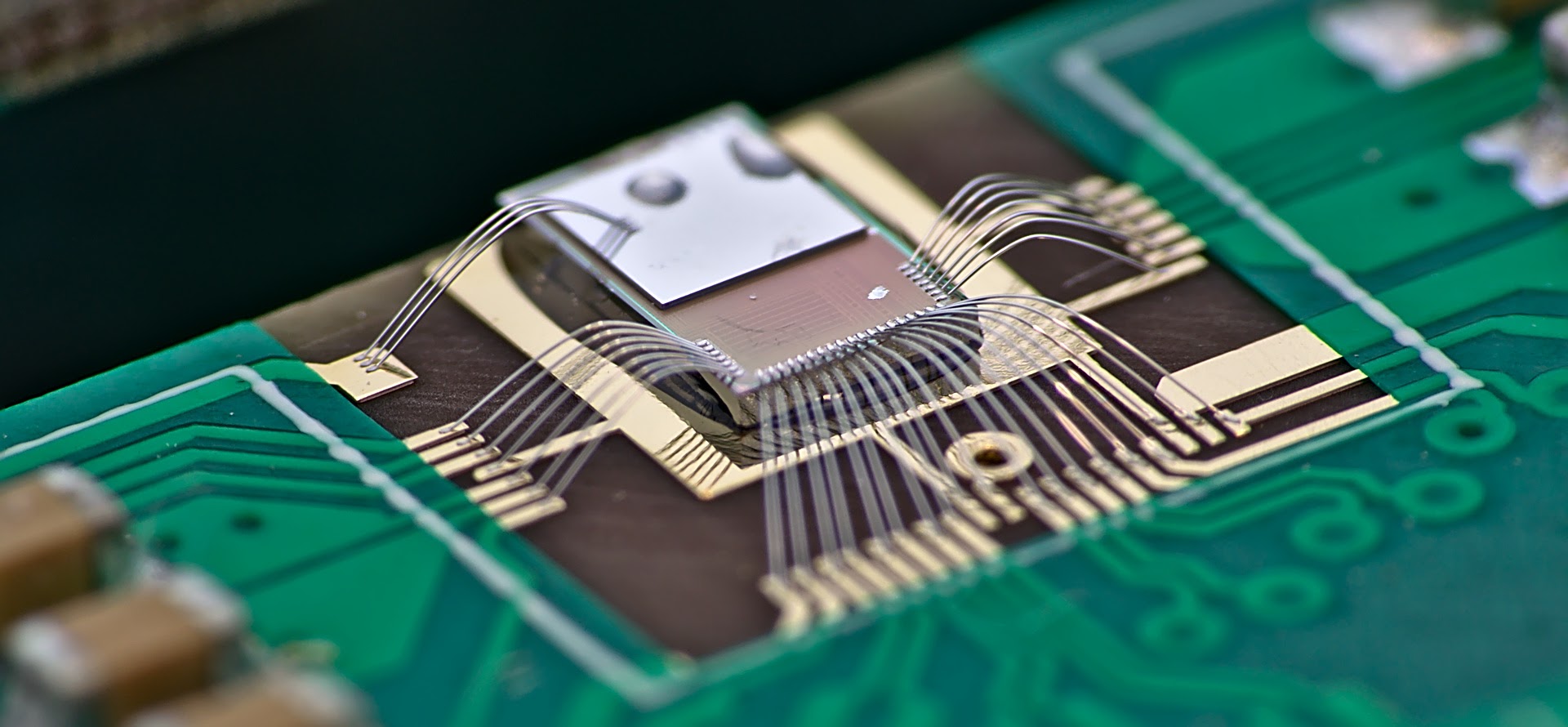}
   \caption{}
  \label{fig:CLICpix_planar}
\end{subfigure}
\begin{subfigure}{.590\textwidth}
  \centering
  \includegraphics[width=\linewidth]{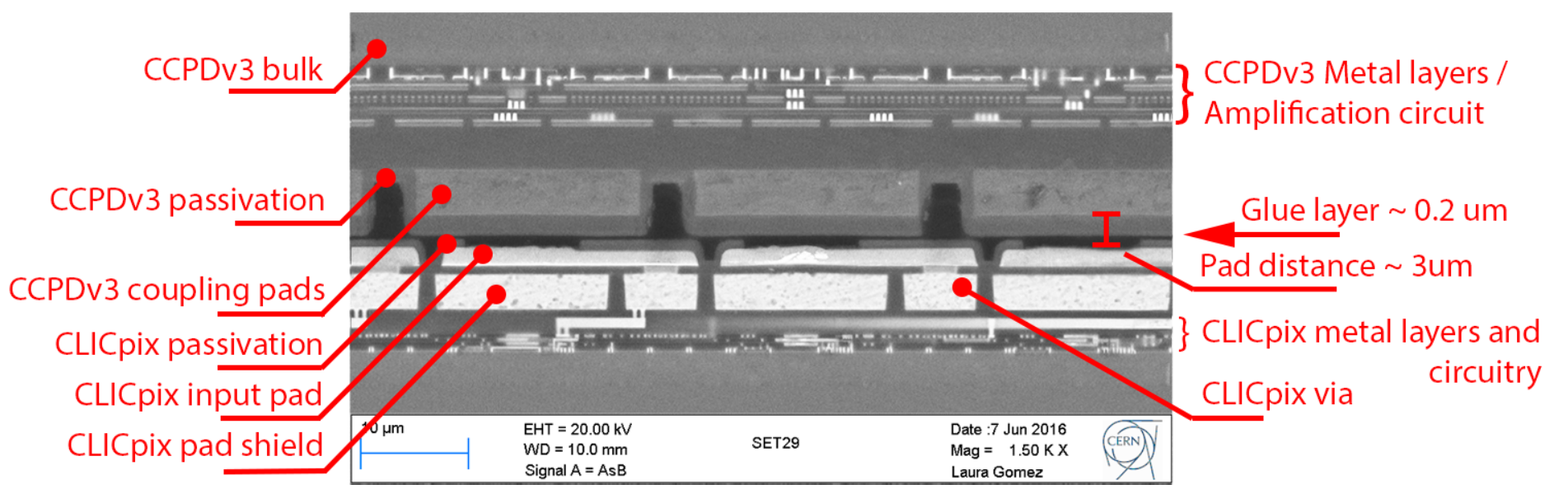}
   \caption{}
  \label{fig:ccpd_CLICpix}
\end{subfigure}
\caption{(a) Photograph of a $(50\,\upmu\text{m})$ thin planar sensor, shown on top, bump-bonded to a CLICpix ASIC, shown at the bottom. (b) Scanning electron microscope image of the cross section through a CCPDv3 and CLICpix capacitively-coupled assembly, showing the active HV-CMOS CCPDv3 sensor at the top, the CLICpix ASIC at the bottom, and the thin glue layer in the middle. \imdp
}
\label{fig:vertextracker1}
\end{figure}

\begin{figure}[h]
\centering
\begin{subfigure}{.220\textwidth}
  \centering
  \includegraphics[width=\linewidth]{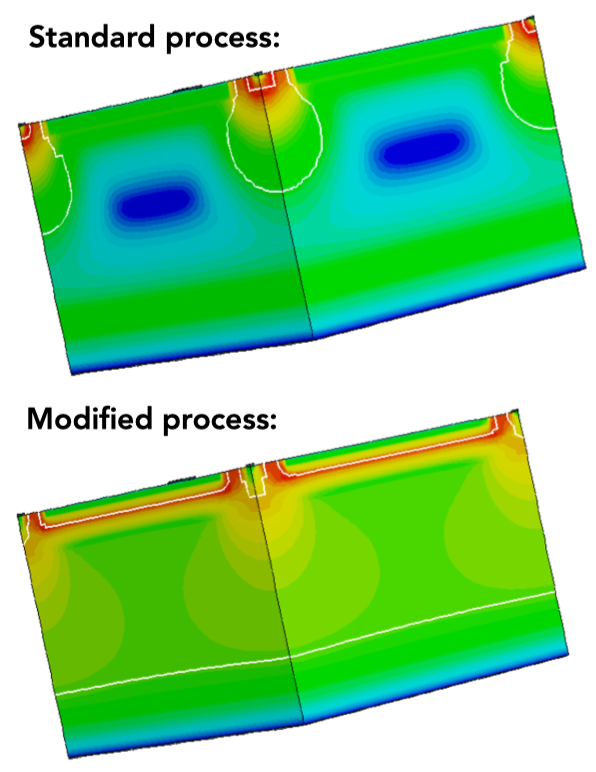}
   \caption{}
  \label{fig:TCAD_3D}
\end{subfigure}
\hspace{.030\linewidth}
\begin{subfigure}{.300\textwidth}
  \centering
  \includegraphics[width=\linewidth]{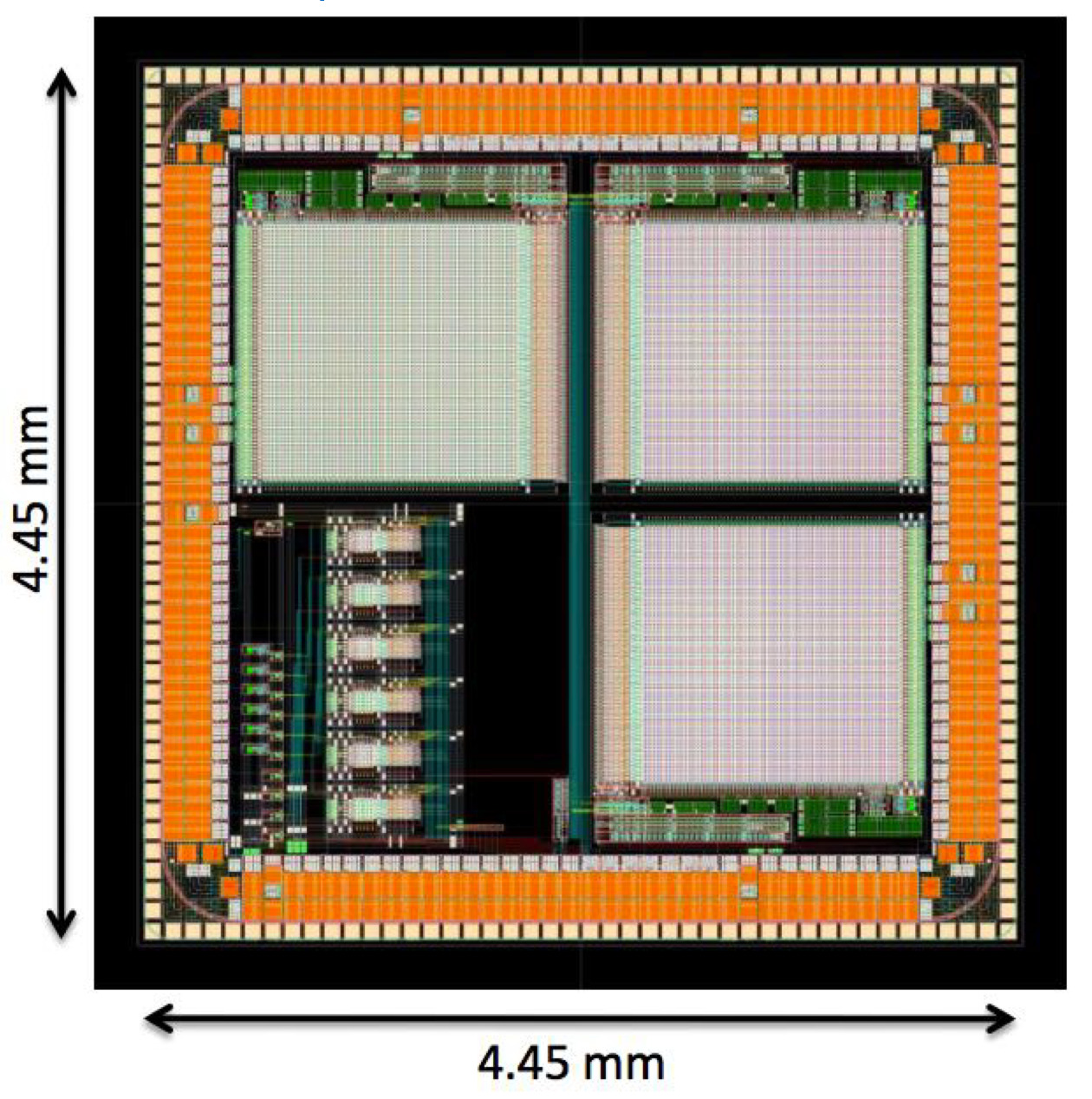}
   \caption{}
  \label{fig:SOI_CLIPS}
\end{subfigure}
\hspace{.030\linewidth}
\begin{subfigure}{.385\textwidth}
  \centering
  \includegraphics[width=\linewidth]{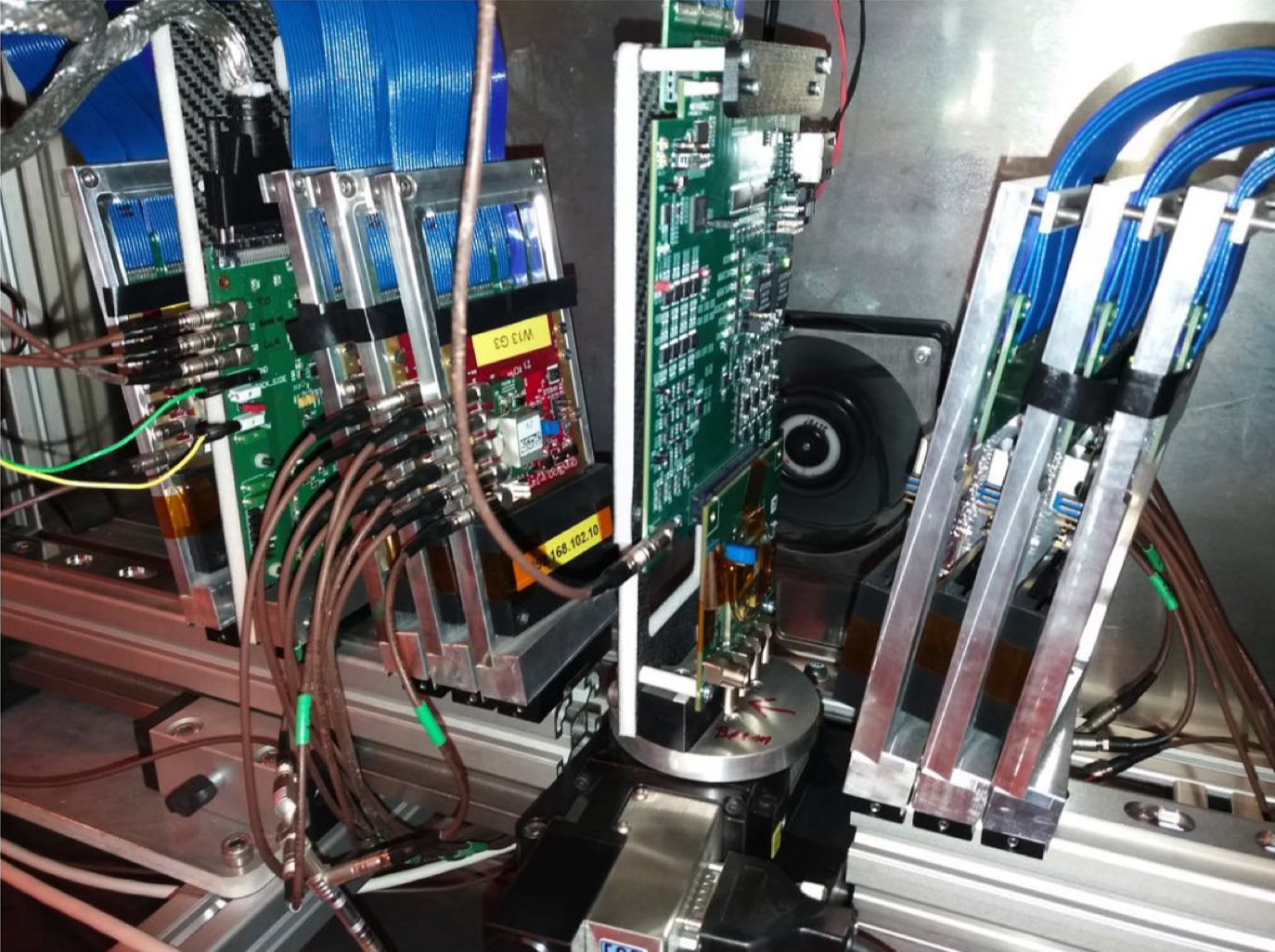}
   \caption{}
  \label{fig:Telescope}
\end{subfigure}
\caption{(a) Images of 3D TCAD (Technology Computer Aided Design) simulations of single-pixel electric field distributions for two HR-CMOS process variants.
(b) Sensor layout of the CLIPS fully monolithic sensor with $20\,\times 20\,\upmu\text{m}^2$ pixel sizes, designed in SOI technology.
(c) Photograph of the Timepix3 test beam telescope that was built for testing CLIC vertex and tracking R\&D assemblies. Seven Timepix3 reference layers are shown, together with two devices under test. \imdp
}
\label{fig:vertextracker2}
\end{figure}

\begin{figure}[h]
\centering
\begin{subfigure}{.380\textwidth}
  \centering
  \includegraphics[width=\linewidth]{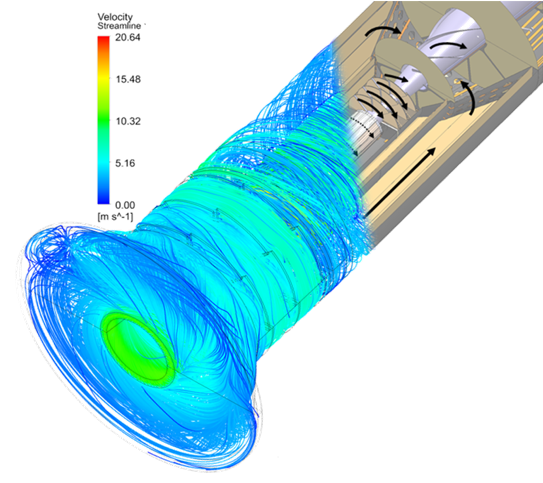}
   \caption{}
  \label{fig:cooling_illustration}
\end{subfigure}
\hspace{.050\linewidth}
\begin{subfigure}{.450\textwidth}
  \centering
  \includegraphics[width=\linewidth]{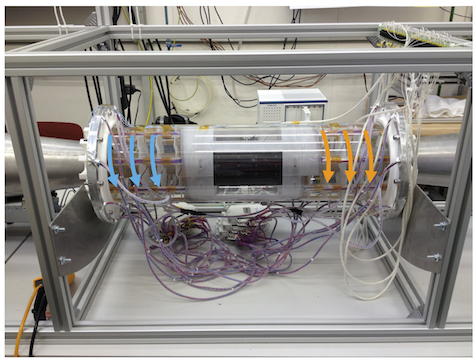}
   \caption{}
  \label{fig:vertex_mockup}
\end{subfigure}
\caption{(a) Composed image showing the vertex detector design, together with stream lines of cooling air as simulated by computational fluid dynamics.
(b) Photograph of a full-scale vertex detector mock-up, used to demonstrate the feasibility of air cooling. \imdp
}
\label{fig:vertextracker3}
\end{figure}

\subsubsection{Electromagnetic and hadronic calorimeters}

Requirements on the barrel and endcap calorimeter systems are driven by physics performance aims combined with the need to efficiently reject beam-induced background particles in the data. Highly-granular calorimetry together with performant particle flow analysis efficiently address both requirements. The CLICdet calorimeter system has been optimised accordingly, by combining simulation results with CALICE technology experiences on highly-granular calorimetry~\cite{ThesisStevenGreen,CLICdet_note_2017}. The electromagnetic calorimeter comprises a 40-layer sandwich of silicon pad sensors as active material interspersed with {1.9\,mm} tungsten plates, for a total depth of $22\,\xo$ and $1\,\lambdaint$. The silicon pads have a lateral size of $5\,\times 5\,\text{mm}^2$, and each active layer occupies only 3.15\,mm in depth, including space for readout and cabling. The hadronic calorimeter comprises a 60-layer sandwich of plastic scintillator active material interspersed with 19 mm thick steel plates, for a total depth of $7.5\,\lambdaint$. The scintillator tiles, with a thickness of  3\,mm and $3\,\times 3\,\text{cm}^2$ lateral size, are read out individually by silicon photomultipliers (SiPM). Each active layer covers 7.5 mm in depth. A time resolution of 1\,ns is required for individual calorimeter hits in ECAL and HCAL. A large dynamic range is required, covering large energy deposits from high-energy showers, as well as from single minimum-ionising particles.

Technologies for both the silicon-tungsten (SiW) ECAL and the scintillator-steel analogue HCAL (AHCAL) have been validated through the construction and tests of successive prototypes. Initial prototypes, the so-called physics prototypes~\cite{Ecal-phys-proto,AHcal-phys-proto}, served to implement the core technology features in devices. 
These prototypes are large enough to assess their response to showers from individual particles and to provide detailed shower data for validating the expected jet performance through event reconstruction with PandoraPFA particle flow analysis~\cite{thomson:pandora, Marshall:2013bda, Marshall2013153,Marshall:2015rfaPandoraSDK}. Also they served to gain deeper understanding of core technological aspects like mechanical assembly, embedded electronics, signal development, noise, power pulsing (SiW ECAL), calibration and systematic effects~\cite{CaliceOverview2016}. For the 30-layer SiW ECAL the results show an energy resolution for electrons at the level of $16.6\%\,\text{/}\,\sqrt{E}$ with a constant term of $1.1\%$~\cite{Adloff:2008aa}. Furthermore, a full separation of close-by particles down to a distance of 2.5\,cm was demonstrated~\cite{CaliceOverview2016}. A 38-layer AHCAL prototype with more than 7000 readout cells was built and tested in beams of different particles over a wide energy range. This was the first device to use SiPMs on a large scale. The imaging capabilities of the calorimeter allow exploiting shower substructures, such as using MIP tracklets for calibration purposes or using knowledge of local hit density for improving the energy resolution through software compensation~\cite{software-compensation-NIM}. Exposure to pions in the range 10\,-\,80\,GeV yields an energy resolution at the level of $57.6\%\,\text{/}\,\sqrt{E}$ with a constant term of $1.6\%$. This result is further improved to $44.3\%\,\text{/}\,\sqrt{E}$ with a constant term of $1.8\%$ through software compensation, which gives a lower weight to energy deposits in high-density regions.

The next generation of highly-granular calorimeter prototypes are so-called technological prototypes. Their design includes lessons learned from the first generation prototypes and, in addition, includes engineering constraints and scalability features for the construction, that will also be needed for the final detectors. One example is the SiPM-on-tile hadronic calorimeter prototype, comprising 38 detection layers for a total of nearly 22000 scintillator tiles (see~\ref{fig:AHCAL-photo})~\cite{Sefkow:2018rhp}. For this prototype, construction and quality assurance processes have been optimised and automatised. For example, the SiPMs are integrated in the readout boards, the scintillators are produced and wrapped via automated processes and the assembly makes use of automated pick-and-place devices. Improved electronics readout with auto-triggering capability, nanosecond-level timing capabilities and power pulsing is included. \ref{fig:Electron_event_AHCAL,fig:Hadron_event_AHCAL} show event displays of a 100\,\GeV electron and a 100\,\GeV hadron, respectively, recorded with the SiPM-on-tile calorimeter prototype at the CERN SPS test beam.

Likewise, recent prototypes of the SiW ECAL integrate more of the engineering and scalability aspects, which will be needed for the future silicon ECAL calorimeter system with tens of millions of channels. For the current prototypes, electro-mechanical challenges were overcome in order to fit the active layers into thin slots ($<4\,\text{mm}$) between the absorber layers. Robotic procedures are used for the assembly. Modules with $5\,\times\,5\,\text{mm}^2$ cell sizes and advanced electronics readout features are currently undergoing beam tests. Two sensor thicknesses are used, $320\,\upmu\text{m}$ and $650\,\upmu\text{m}$, in order to assess the optimal signal-to-noise ratio.

CALICE prototype results confirm the performances expected from simulations, thereby giving overall confidence in the performance predictions for the CLIC detector.

\begin{figure}[t]
\centering
\begin{subfigure}{.330\textwidth}
  \centering
  \includegraphics[width=\linewidth]{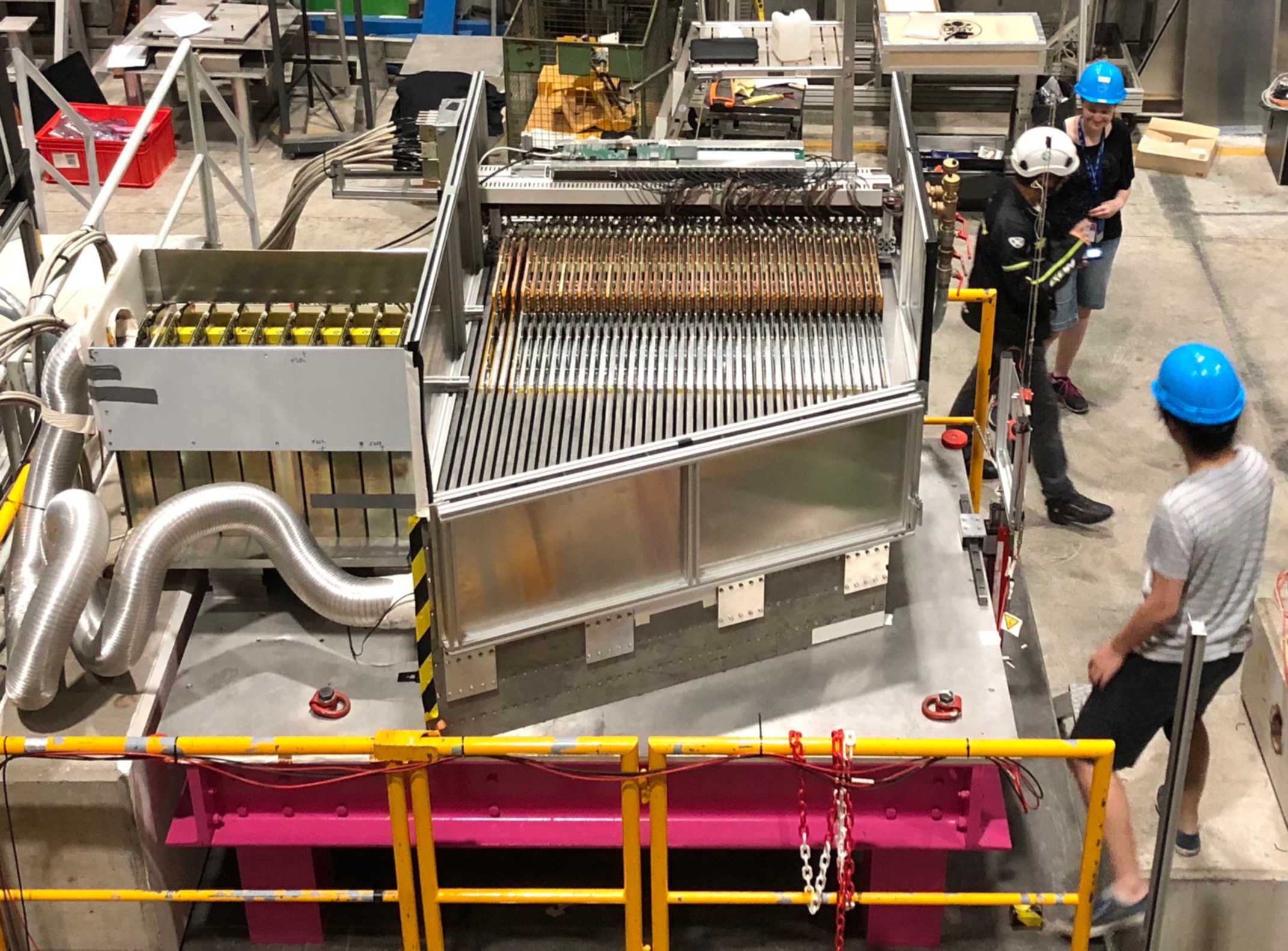}
   \caption{}
  \label{fig:AHCAL-photo}
\end{subfigure}
\hspace{.030\linewidth}
\begin{subfigure}{.285\textwidth}
  \centering
  \includegraphics[width=\linewidth]{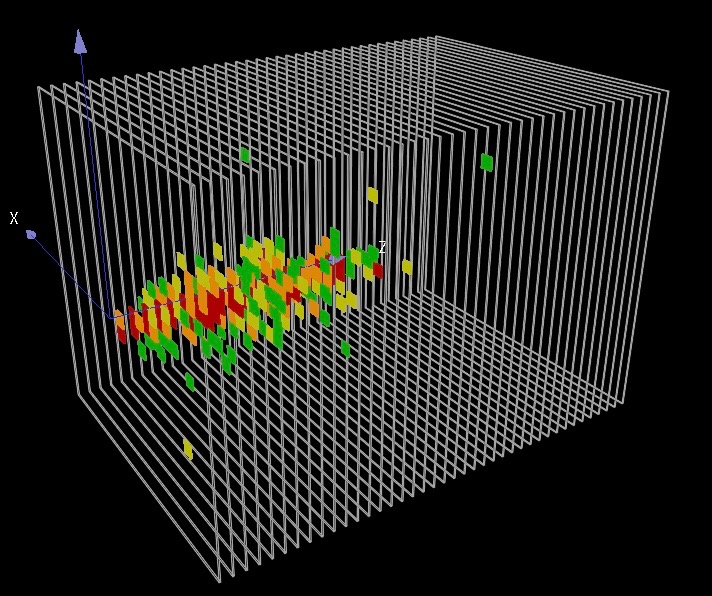}
   \caption{}
  \label{fig:Electron_event_AHCAL}
\end{subfigure}
\hspace{.030\linewidth}
\begin{subfigure}{.285\textwidth}
  \centering
  \includegraphics[width=\linewidth]{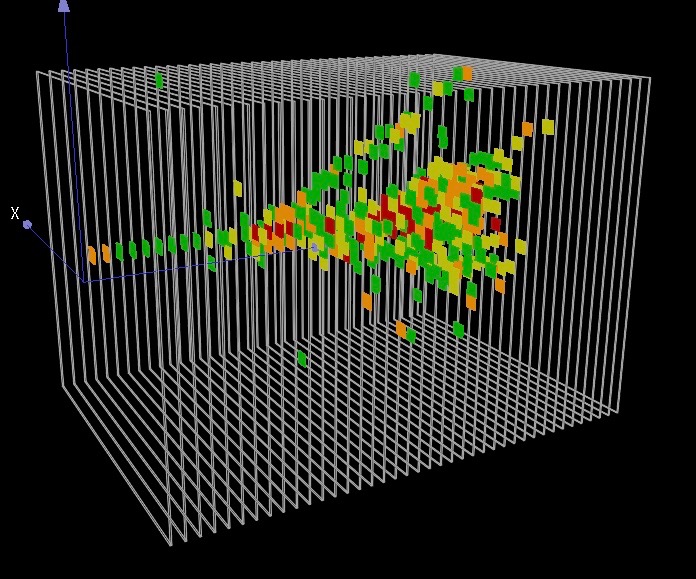}
   \caption{}
  \label{fig:Hadron_event_AHCAL}
\end{subfigure}
\caption{(a) Photograph of the CALICE SiPM-on-tile hadronic calorimeter technological prototype, comprising 38 detection layers for a total of nearly 22000 scintillator tiles.
(b) Event display of a 100\,GeV electron.
(c) Event display of a 100\,GeV hadron. \imgc{CALICE}
}
\label{fig:AHCAL2018}
\end{figure}

\subsubsection{Very forward calorimeters}
The very forward calorimeters, LumiCal and BeamCal, are compact fine-grained cylindrical electromagnetic calorimeters centred on the outgoing beams at 2.5\,m and 3.2\,m from the interaction point, see~\ref{fig:forward}. Their main functions include tagging of very forward-going electrons and photons, as well as in-situ measurement of the luminosity based on Bhabha scattering. Their geometrical acceptance range spans $\theta\,=\,10-46\,\text{mrad}$ for BeamCal and $\theta\,=\,39-134\,\text{mrad}$ for LumiCal. In view of large beam-induced backgrounds in this region, the very forward calorimeters are exposed to high radiation levels. This holds in particular for BeamCal for which radiation resistance for an ionising dose of up to $1\,\text{MGy}$ per year is required. Angular precision for the measurement of high-energy electromagnetic showers (up to 1.5\,TeV) call for a very small Moli\`{e}re radius of $\sim$1\,cm and a large dynamic range. Both devices are based on 40-layer sandwich designs comprising 300\,$\upmu$m thick semiconductor detectors (in <1\,mm gaps) interleaved with 3.5\,mm thick (1\,\xo) tungsten absorbers. The detector development for the very forward calorimeters is carried out in the framework of the FCAL collaboration. Detector development for BeamCal principally concentrates on radiation studies of sensor materials (GaAs, sapphire, SiC and silicon diode sensors). Recent LumiCal milestones are the construction and beam tests of compact 4-layer and 8-layer LumiCal prototypes with silicon sensors. The 4-layer prototype uses custom-designed FCAL electronics, albeit with a limited number of readout channels, whereas the 8-layer prototype is based on existing APV25 readout ASICs~\cite{FCALsensorplaneperformance,FCAL-multilayer-EPJC2018,APV_ieee,APV_nima}. For the latter, ultra-compact readout layers were achieved, covering only $650\,\upmu\text{m}$ in depth including connections to the readout electronics located outside the active area.  As a result, a small effective Moli\`{e}re radius of $8.1\,\pm\,0.3\,\text{mm}$ was measured~\cite{FCAL:ICHEP18} using electron beams of 1--5\,GeV. Current FCAL prototypes use silicon pad sizes of $\sim$0.3$\,\text{cm}^2$. In view of the high occupancies expected at CLIC, significantly reduced pad sizes will be an advantage. Ongoing R\&D on monolithic CMOS silicon sensors with small pixel sizes (see~\ref{sec:vertextrackingtech}) opens future perspectives towards pixelised (analogue or digital) LumiCal sensor layers, offering compactness as well as large dynamic range for the measurement of high-energy electromagnetic showers. While radiation tolerances of up to 1\,MGy have been achieved for BeamCal for several sensor technologies, further R\&D is needed to fully assess the performance of the various sensor materials and means to integrate them in BeamCal. The forward calorimeters also require very high alignment accuracy of $10\,\upmu\text{m}$ in transverse direction and $100\,\upmu\text{m}$ in longitudinal direction, for which initial concepts have been devised.

\subsection{Detector performance}
\label{sec:CLICdetPerformance}

In this section a summary of the CLICdet detector 
performance is presented for single particles, complex events and jets.
Individual particles are used to probe track reconstruction and particle identification, while the reconstruction of particles inside jets and the flavour tagging performances are tested in di-jet events. 
Jet energy and angular resolution as well as W--Z mass separation are studied in di-jet samples.

The CLICdet detector geometry is described with the DD4hep
software framework~\cite{frank15:ddg4} and simulated in Geant4~\cite{Agostinelli2003,
  Allison2006, Allison2016186}.
The reconstruction software is implemented in the linear collider
Marlin-framework~\cite{MarlinLCCD}. The reconstruction algorithms use geometry information provided by DD4hep~\cite{sailer17:ddrec}. The reconstruction starts with the
overlay of \gghad{} background events, corresponding to 30 bunch crossings around the physics
event~\cite{LCD:overlay}. Subsequently, the hit positions in the tracking detectors are smeared with
Gaussian distributions according to the expected resolutions.
Tracks are reconstructed using the ConformalTracking algorithm~\cite{Leogrande:2630512}.  
Particles are reconstructed and identified using the PandoraPFA particle flow algorithms~\cite{thomson:pandora, Marshall:2013bda, Marshall2013153,Marshall:2015rfaPandoraSDK}, combining information from tracks, calorimeter clusters, and hits in the muon system. The detector response to each type of Pandora particle flow object (PandoraPFO) -- charged hadrons, photons, neutral hadrons, electrons, and muons -- is calibrated separately with type-specific calibration constants.
After full particle flow reconstruction, particles from beam-induced backgrounds are suppressed through p$_T$-dependent timing cuts described above. 
Vertex reconstruction and heavy-flavour tagging is
performed by the LCFIPlus program~\cite{Suehara:2015ura}. Larger simulation and reconstruction samples were produced with the iLCDirac grid production tool~\cite{ilcdirac13}.
A more detailed description of the software tools, analysis methods and a complete set of performances are given in~\cite{CLICdet_performance}.
Selected examples are presented below.

\paragraph{Tracking performance}
\ref{fig:res} shows the transverse momentum resolution and the transverse impact parameter resolution, obtained with isolated muon tracks of different momenta and polar angles. The precise measurement of leptonic final states requires a transverse momentum resolution of the order of $2 \times 10^{-5}$~\GeV$^{-1}$ for high-energy particles~\cite{cdrvol2}, which is achieved  in the central detector region as shown in~\ref{fig:mom_res_p}. Similar performances are achieved with isolated electrons and pions.
Efficient flavour tagging, relying on the precise reconstruction of primary and secondary vertices, requires excellent impact parameter resolution. The targeted transverse impact parameter resolution, depicted as dashed lines in~\ref{fig:d0_res_theta}, is clearly reached for high-energy muons in the central detector region. 
This result is 
closely linked to the single point resolution in the vertex detector: varying this parameter from the nominal 3\,\micron to 5~\micron leads to a $d_{0}$ resolution degradation by 50\%.
The material budget  in the vertex detector impacts the 10\,\GeV tracks in the forward detector region, and the 1\,\GeV tracks at any angle (cf.~\ref{fig:d0_res_theta}).
In additional studies,
the longitudinal impact parameter resolution is found to be much smaller than the longitudinal bunch length (44\,\micron at 3\,\TeV collision energy) for high-energy muons at all polar angles and reaches a minimum of 1.5\,\micron for 100\,\GeV muons at 90\degrees.
Polar and azimuthal angular resolutions of, respectively, 0.05\,mrad and 0.025\,mrad are achieved for high-energy muons in the central detector region.

\begin{figure}[t]
\centering
\begin{subfigure}{.5\textwidth}
  \centering
  \includegraphics[width=\linewidth]{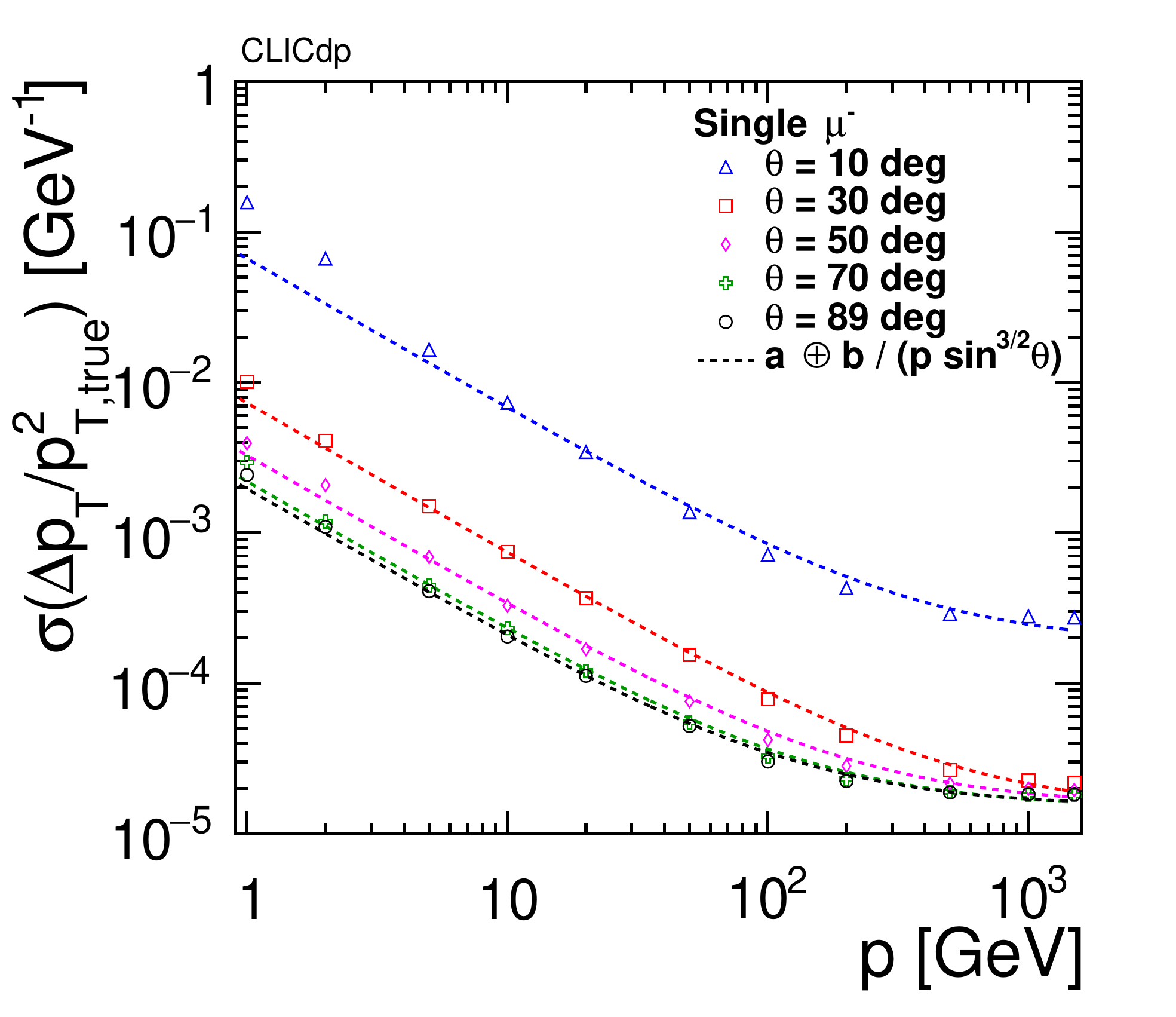}
  \caption{}
  \label{fig:mom_res_p}
\end{subfigure}%
\begin{subfigure}{.5\textwidth}
  \centering
  \includegraphics[width=\linewidth]{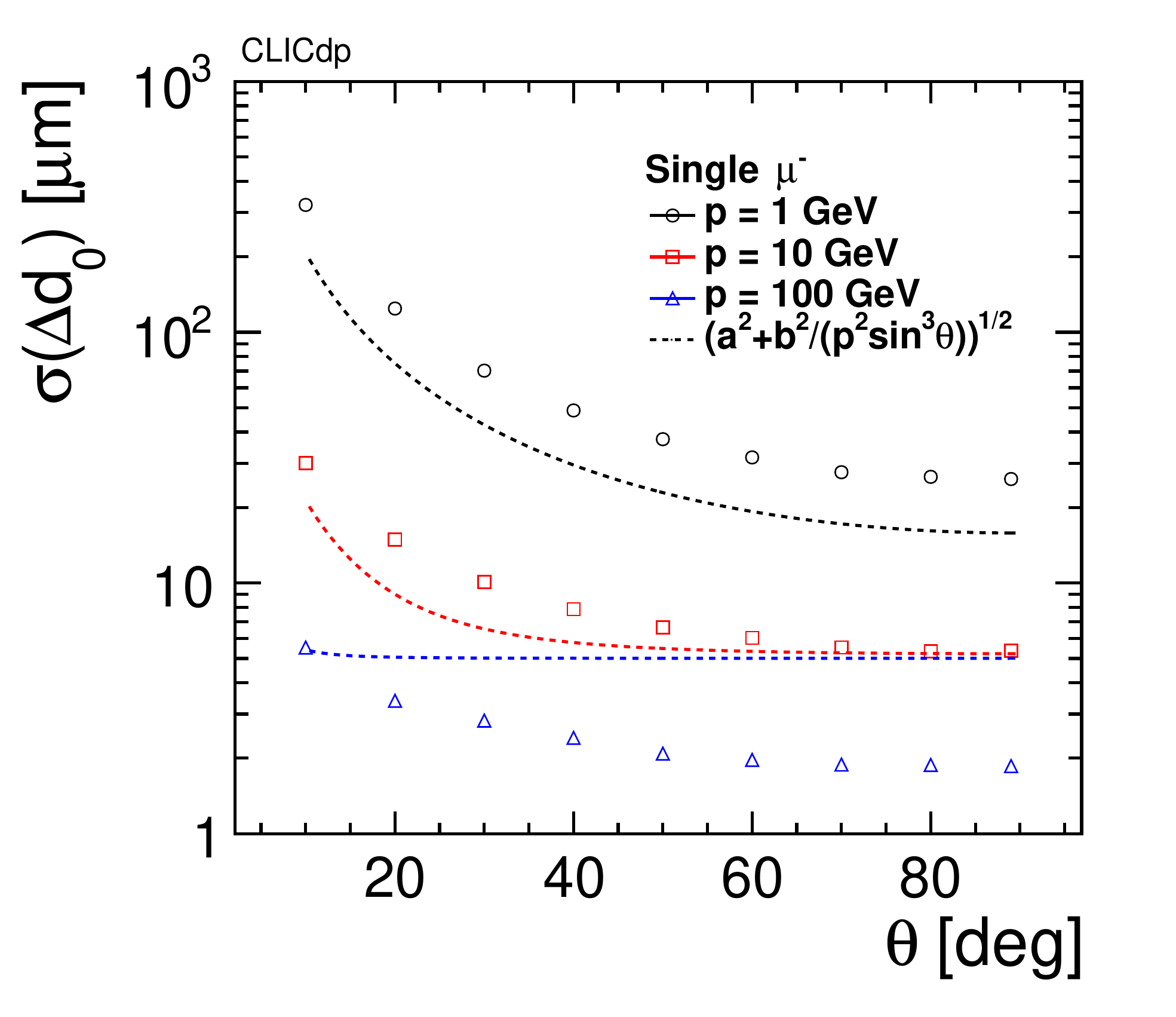}
  \caption{}
  \label{fig:d0_res_theta}
\end{subfigure}
\caption{(a) Transverse momentum resolution as a function of momentum for muons at polar angles $\theta$ = 10\degrees, 30\degrees, 50\degrees, 70\degrees, 89\degrees and (b) transverse impact parameter resolution as a function of polar angle for muons with momenta of 1, 10 and 100\,GeV~\cite{CLICdet_performance}. The lines in (a) represent the fit of each curve with the parameterisation as inserted in the figure. In figure (b), the lines show the detector performance goals with $a$ = 5~\micron and $b$ = 15~\micron GeV used in the parameterisation inserted~\cite{cdrvol2}.}
\label{fig:res}
\end{figure}

Tracking performance has been studied in detail for prompt and displaced (i.e. not originating from the interaction point) tracks~\cite{CLICdet_performance}.
A reconstruction efficiency of 100\% has been demonstrated for single particles (muons, pions, electrons),
with an efficiency loss of up to 1\% observed only in the very forward region ($\theta$ = 10\degrees). 
Similarly, tracks from displaced particles are found to be reconstructed with good efficiency.

The tracking efficiency and fake rate inside jets have been tested in \bb{} events at 3~\TeV centre-of-mass energy. The results are shown in~\ref{fig:3TeV_bb_pt} as a function of transverse momentum. Fully efficient tracking down to \pT $\simeq$ 1~\GeV is observed, then the efficiency decreases to 92\% at \pT = 100~\MeV. This value is reduced to 80\%, when beam-induced background particles from \gghad{} events as produced at the 3~\TeV CLIC stage are overlaid (cf.~\ref{fig:bbbar3TeV_eff_pt}). In these studies, only tracks are considered, whose associated Monte Carlo particle is separated from all other particles by at least 0.02 rad, in order to limit confusion in pattern recognition. However, the efficiency loss for less separated tracks is found to amount to only 2--3\%, independently on the beam-induced background.

The fake rate is shown in~\ref{fig:bbbar3TeV_fake_pt}. In events without backgrounds, it increases with transverse momentum from 0.3\% at 1~\GeV to roughly 1\% above 10~\GeV, where more straight tracks lead to increased confusion in pattern recognition. The effect of background is particularly large for \pT < 1~\GeV, and the fake rate reaches a maximum of 6\% at 100~MeV. The origin of this fake rate is found to be mostly due to tracks in the region 10\degrees < $\theta$ < 20\degrees, and due to two particles separated by less than 0.04 rad. The fake rate for displaced tracks amounts to a few percent
for the whole range of radii of production vertices probed, while the beam-induced background is found to have a small impact.
\newcommand{\zlike}{\ensuremath{\PZ/\HepParticle{\PGg}{}{\ast}}\xspace}
For additional investigations, using \zlike{} events of different masses decaying into light quarks and \ttbar{} events at 3~\TeV, performance very similar to that described above for \bb{} has been found.

\begin{figure}[t]
\centering
\begin{subfigure}{.5\textwidth}
  \centering
  \includegraphics[width=\linewidth]{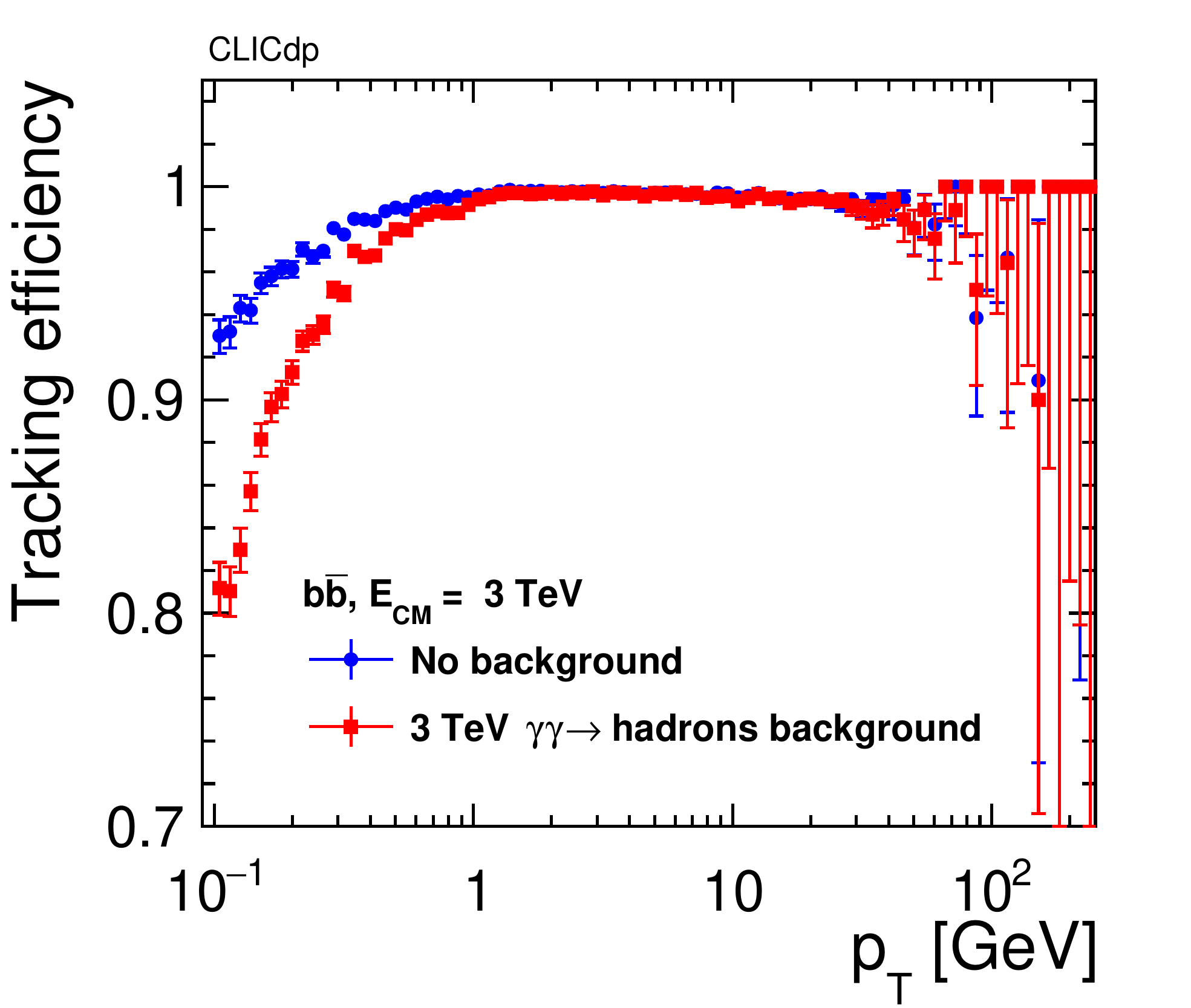}
  \caption{}
  \label{fig:bbbar3TeV_eff_pt}
\end{subfigure}%
\begin{subfigure}{.5\textwidth}
  \centering
  \includegraphics[width=\linewidth]{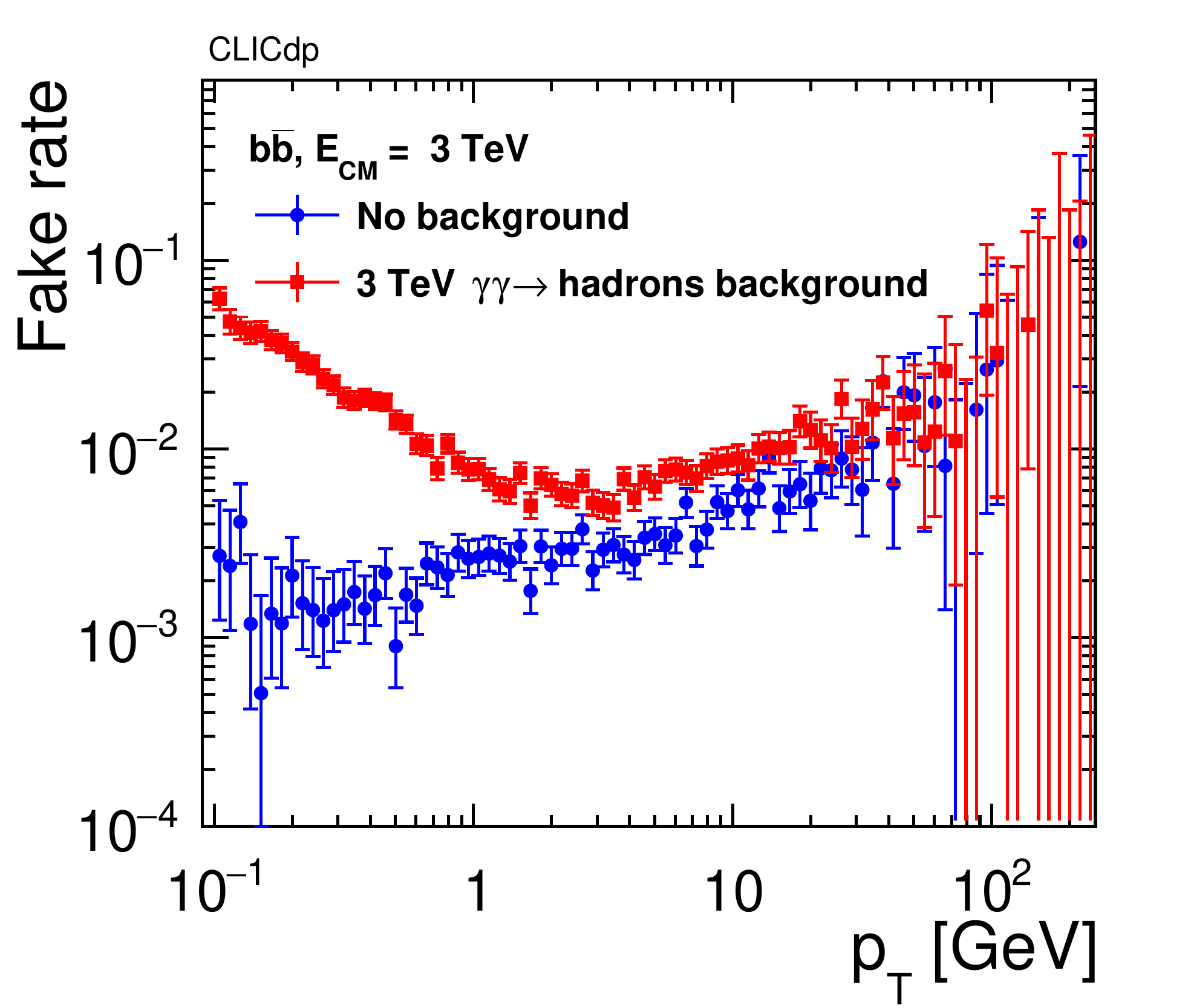}
   \caption{}
  \label{fig:bbbar3TeV_fake_pt}
\end{subfigure}
\caption{(a) Tracking efficiency and (b) fake rate as a function of \pT{} for \bb{} events at 3\,TeV, 
with and without \gghad{} background overlay~\cite{CLICdet_performance}. Note that the errors bars indicate statistical uncertainties, which are naturally larger at high \pT{} due to the particle spectrum in the sample used.}
\label{fig:3TeV_bb_pt}
\end{figure}

\paragraph{Calorimetry performance}
The PandoraPFA particle identification efficiency is studied over a wide range of energies in all regions of polar angle~\cite{CLICdet_performance}, and more than 90\% efficiency is found for all particle types, all energies, and for polar angles from $15^\circ-165^\circ$. In the case of muons,
the identification efficiency is larger than 99\% from 10\,GeV up to 1.5\,TeV.  The impact of beam-induced backgrounds from \gghadrons is investigated in \ttbar{} events at 3\,TeV. Considering the W bosons leptonically decaying into muons and electrons in these events, the muon identification is largely unaffected, while the efficiency for electrons is reduced by 3--5\%.

Di-jet samples from $\zlike \rightarrow \PQq\PAQq$ (with \PQq{}$=$\PQu, \PQd or \PQs quarks), simulated without initial state photon radiation and at several centre-of-mass energies, are used to study the performance of jet reconstruction. Software compensation~\cite{Tran:2017tgrSoftwareCompensation} is applied to hadron clusters in the HCAL, in order to improve their energy measurement. 
The jet energy resolution is studied by comparing the response of Monte Carlo (MC) truth particle-level jets (clustering stable particles excluding neutrinos) to those reconstructed at detector level (clustering PandoraPFOs), 
using the VLC algorithm~\cite{Boronat:2016tgdVLC} in exclusive mode to force the event into two jets. The $\gamma$ and $\beta$ of VLC parameters are fixed to 1.0 and the radius parameter is set to $R=0.7$.
The two reconstructed jets are required to be matched to each of the MC truth particle-level jets within an angle of $10^\circ$.

The resulting jet energy resolution is shown in~\ref{fig:JER_100_3000_VLC7} for several jet energies as function of the $|\cos\theta|$ of  the quark.
The performance goal~\cite{cdrvol2} of 3--4\% jet energy resolution at high energies is achieved in the barrel $(|\cos\theta|$<0.7) and endcap region with the exception of the most forward angles.
For low energy jets (50\,GeV),  the jet energy resolution is around 4.5--5.5\%. 
For very forward jets with $0.975<|\cos\theta|<0.985$, the jet can be partly outside of the tracker volume.
This leads to a large tail to lower reconstructed energies, and is reflected in jet energy resolution values which reach up to 20\%. 

In events where 3\,\TeV beam-induced backgrounds from \gghadrons are overlaid on the physics event, \emph{Tight} selection cuts are applied to the PandoraPFOs prior to jet clustering~\cite{cdrvol2}. \ref{fig:JER_100_3000_VLC7_3TeVBG} shows the jet energy resolution for di-jet events with 3\,\TeV \gghadrons background overlaid. 
A degradation of the jet energy resolution is observed for all jet energies, however
for high energy jets, this is limited to less than 0.5\% for most of the $|\cos\theta|$ range. 
Since hadrons from beam-induced backgrounds tend to be produced more in the forward direction,
 the impact is larger for jets with $|\cos\theta|>0.80$. 

\begin{figure}[t]
\centering
\begin{subfigure}{.5\textwidth}
  \centering
  \includegraphics[width=\linewidth]{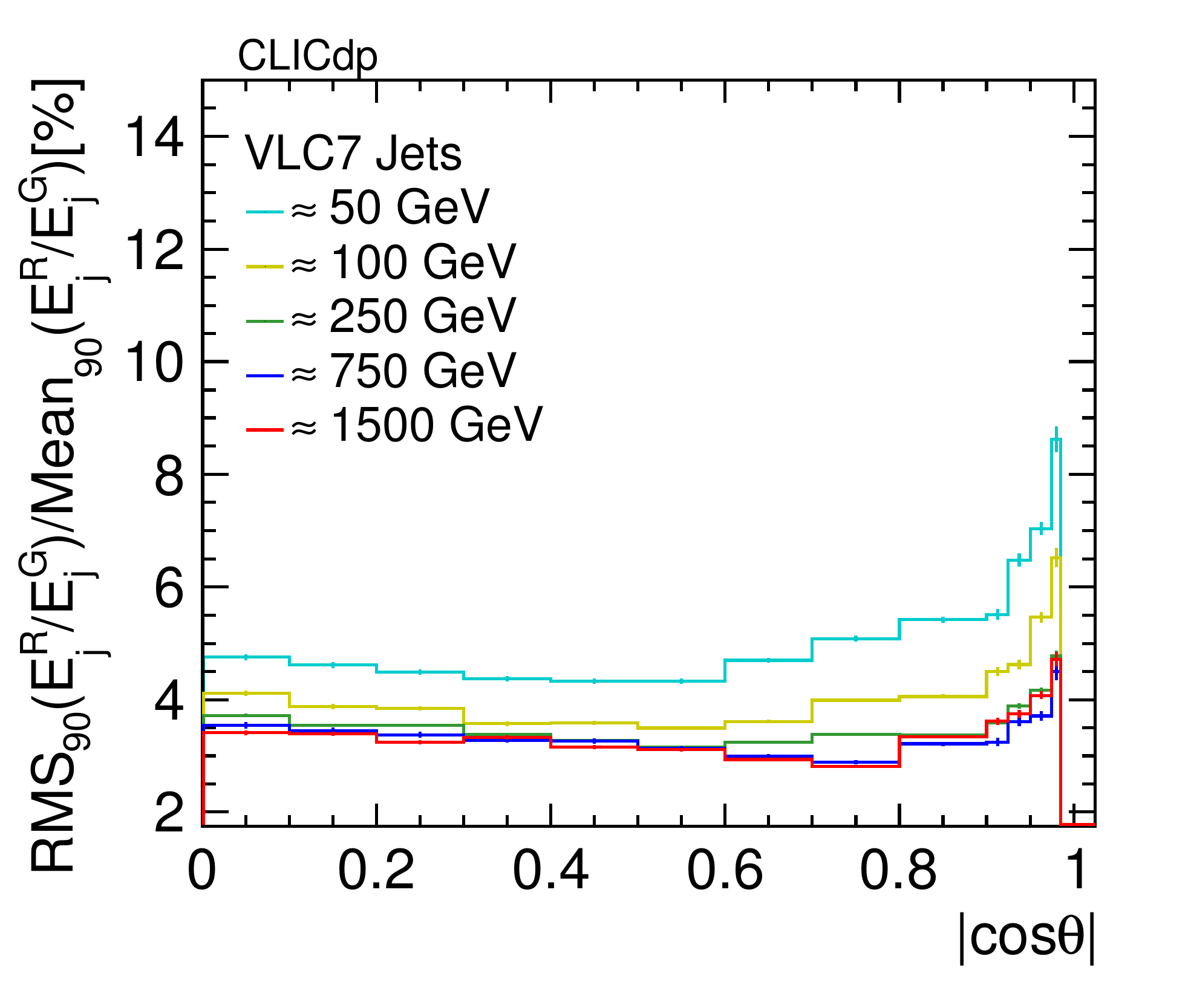}
  \caption{}
  \label{fig:JER_100_3000_VLC7}
\end{subfigure}%
\begin{subfigure}{.5\textwidth}
  \centering
  \includegraphics[width=\linewidth]{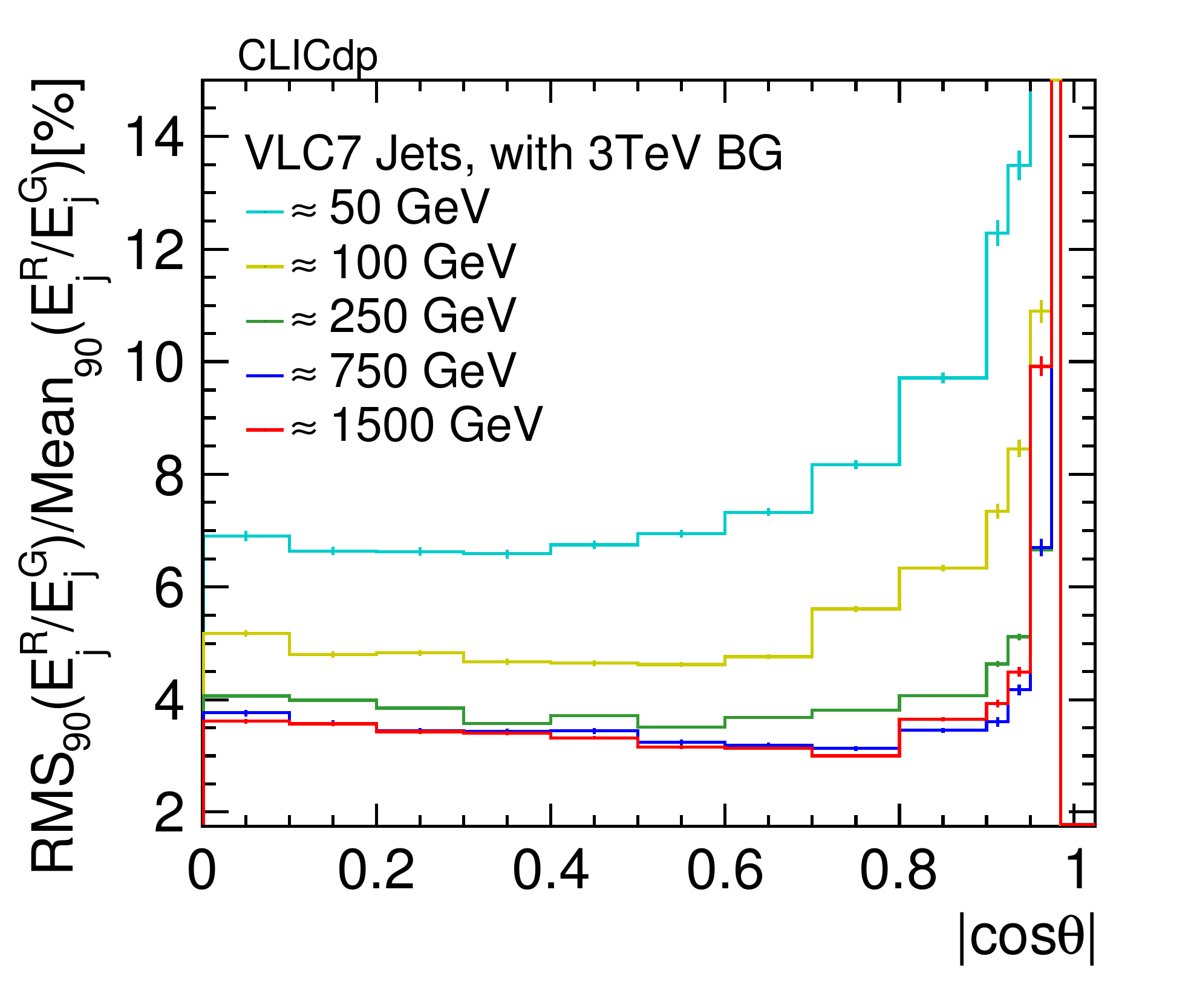}
   \caption{}
  \label{fig:JER_100_3000_VLC7_3TeVBG}
\end{subfigure}
\caption{(a) Jet energy resolution for various jet energies as function of $|\cos\theta|$ of the quark for events without and (b) with \gghad{} background conditions at 3 TeV. RMS$_{90}$ is used as a measure of the jet energy resolution~\cite{CLICdet_performance}.}
\end{figure}

\begin{figure}[t]
\centering
\begin{minipage}[l]{0.49\textwidth}
\includegraphics[width=1.0\textwidth]{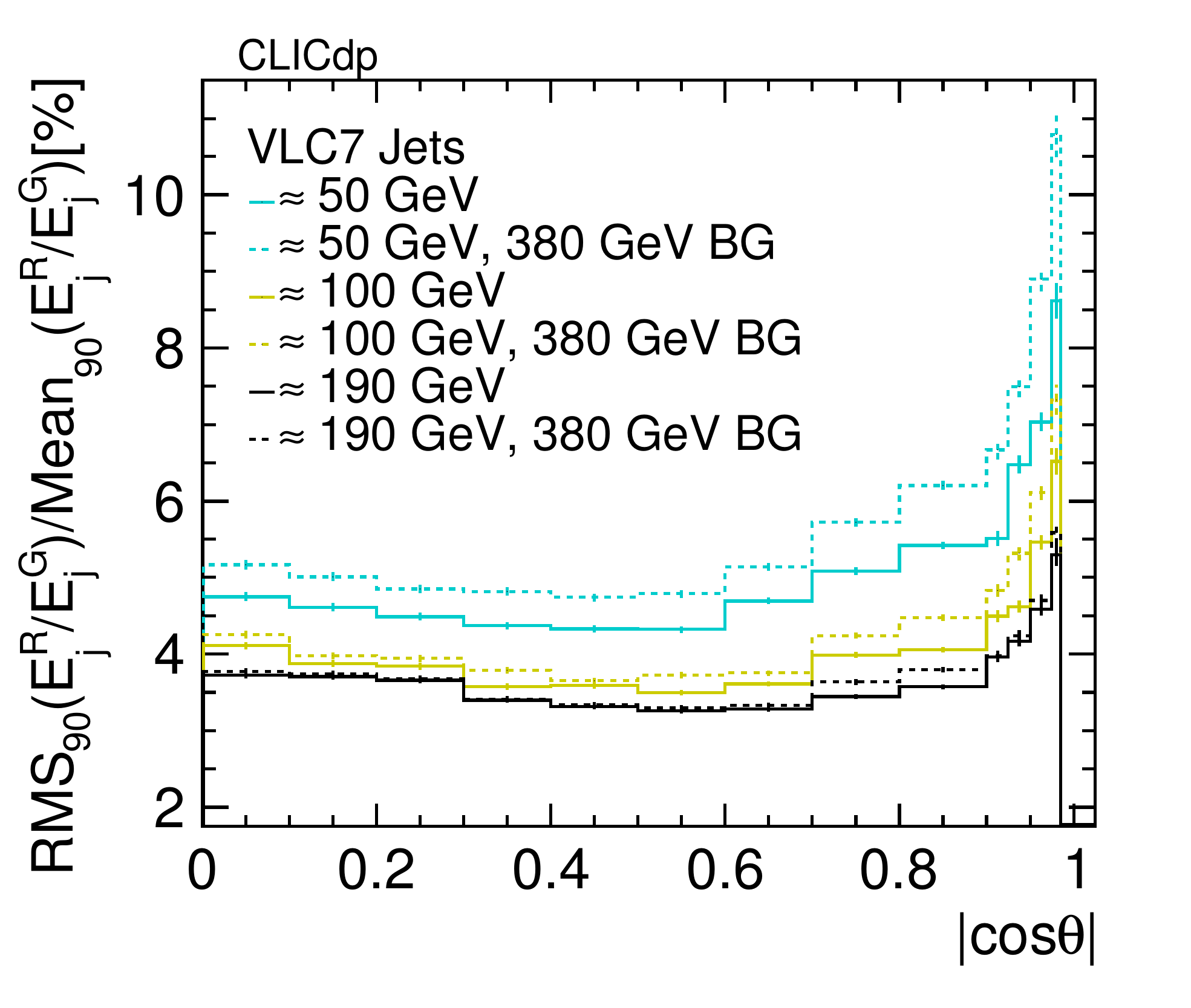}
\end{minipage}
\caption{Jet energy resolution for various jet energies as function of $|\cos\theta|$ of the quark with and without 380\,GeV \gghadrons background overlaid on the physics di-jet event. RMS90 is used as a measure of the jet energy resolution~\cite{CLICdet_performance}.}
\label{fig:jet_resp_jets_wO_380}
\end{figure}

 Beam-induced backgrounds at 380\,GeV CLIC are expected to be significantly smaller than at 3\,TeV. 
For this reason less strict \emph{low energy loose} selections~\cite{Brondolin:2641311} are used for 380\,GeV jet resolution studies.
As shown in~\ref{fig:jet_resp_jets_wO_380}, overlaying the 380\,\GeV beam-induced background levels has little impact on the jet energy resolution for most jet energies, except for very forward jets, where an increase of 0.5--1\% is observed. Even for 50 GeV jets, only a small degradation of the jet energy resolution to about 5\% is observed in the barrel, and to 6--9\% for endcap and forward jets. 
In further studies, jet angular resolutions have been found to be below $1^{\circ}$  in azimuth $\phi$,  and in polar angle $\theta$  values below $0.5^{\circ}$ have been found for jet energies above 100\,GeV.

The precise reconstruction of masses of resonances in hadronic channels over wide ranges of energies is a challenging task. Di-jet masses from hadronic decays of W and Z bosons are studied using simulated di-boson events, in which only one of the bosons decays into di-quarks, i.e.\  $\PZ\PZ\to\nu\bar{\nu}\PQq\PAQq$\ and $\mathrm{WW}\rightarrow l\nu \PQq\PQq$. The boson energies in this study vary from 125\,GeV, where both bosons are created almost at rest, up to 1\,TeV, where the bosons are heavily boosted. The event is clustered in two VLC jets with the same parameter setting and input selection as used in the jet energy resolution studies. In the WW events the charged lepton from the leptonically decaying W is removed prior to jet clustering. A cut is imposed on the polar angle of both MC truth jets $|\cos\theta|<0.9$ to ensure that jets are well contained within the detector acceptance. The di-jet mass distributions are fitted with a Gaussian, iteratively changing the limits of the fit range to 2~$\sigma$ around each side of the mean of the fit, until the fitted $\sigma$ stabilises within 5\%. As an example,~\ref{fig:WW_ZZ_500_DiJetMass} shows the di-jet mass distributions for W and Z bosons with $E=500\,\mathrm{GeV}$ with the Gaussian fits in events without and with the overlay of 3\,TeV beam-induced backgrounds from \gghadrons. The ideal Gaussian separation between the reconstructed W and Z di-jet masses is derived using the overlap fraction between both Gaussian curves, which is defined as the fraction of W (Z) bosons which are above (below) the intersection point of the di-jet mass distributions. Without backgrounds, a separation between 2.0 and 2.5 $\sigma$ can be achieved, corresponding to overlap fractions of 15-19\%. In the presence of beam-induced backgrounds as expected for CLIC at 3\,TeV, the separation achieved is between 1.7 and 2$\sigma$, with overlap fractions between 19 and 23\%. For low energy bosons the impact of 380\,GeV beam-induced background levels is evaluated, leading to a modest decrease of the separation from 2.1 to 2.0$\sigma$.

\begin{figure}[tbp]
\centering
\begin{subfigure}{.5\textwidth}
  \centering
  \includegraphics[width=\linewidth]{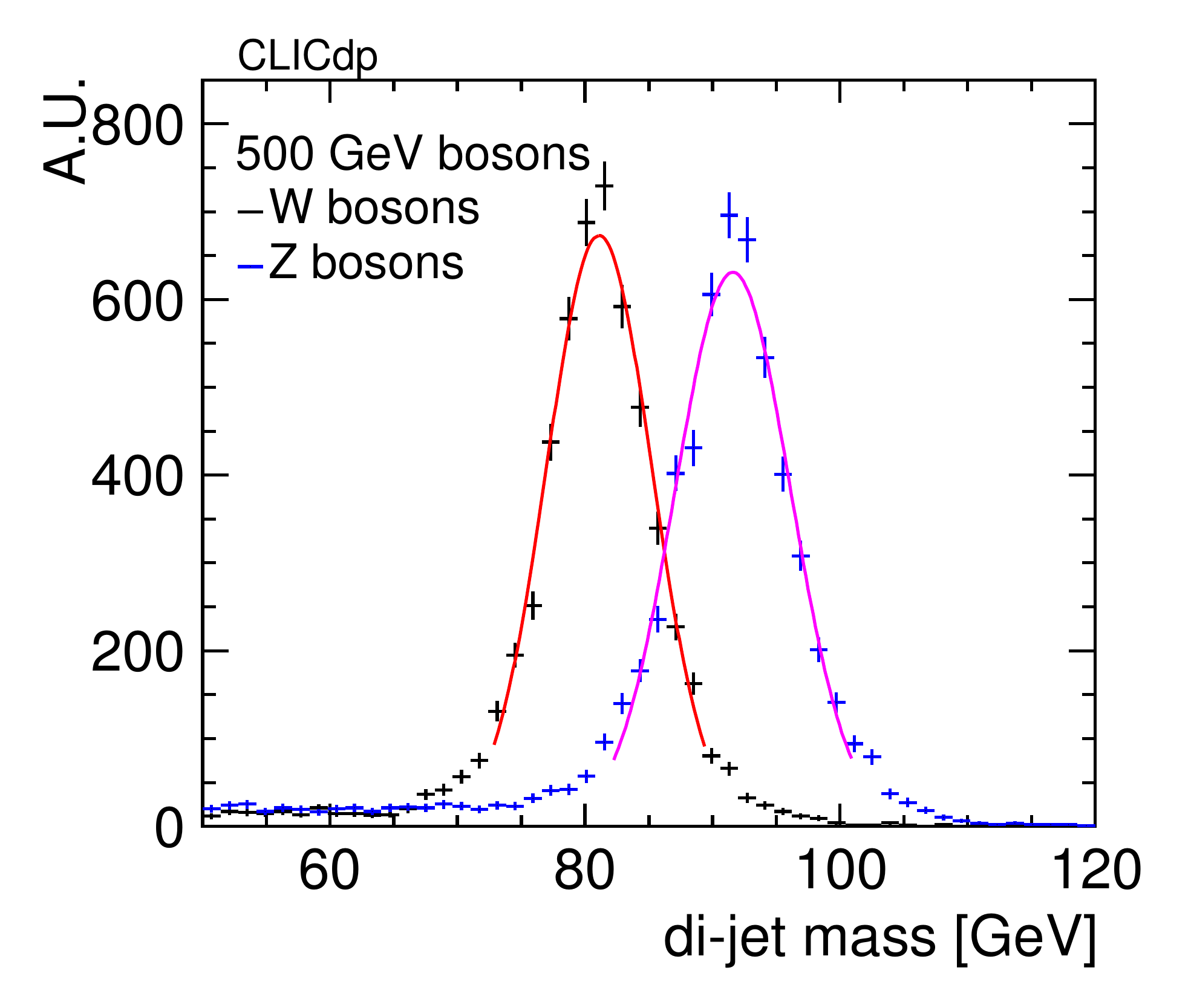}
  \caption{}
  \label{fig:WW_ZZ_500_VLC7}
\end{subfigure}%
\begin{subfigure}{.5\textwidth}
  \centering
  \includegraphics[width=\linewidth]{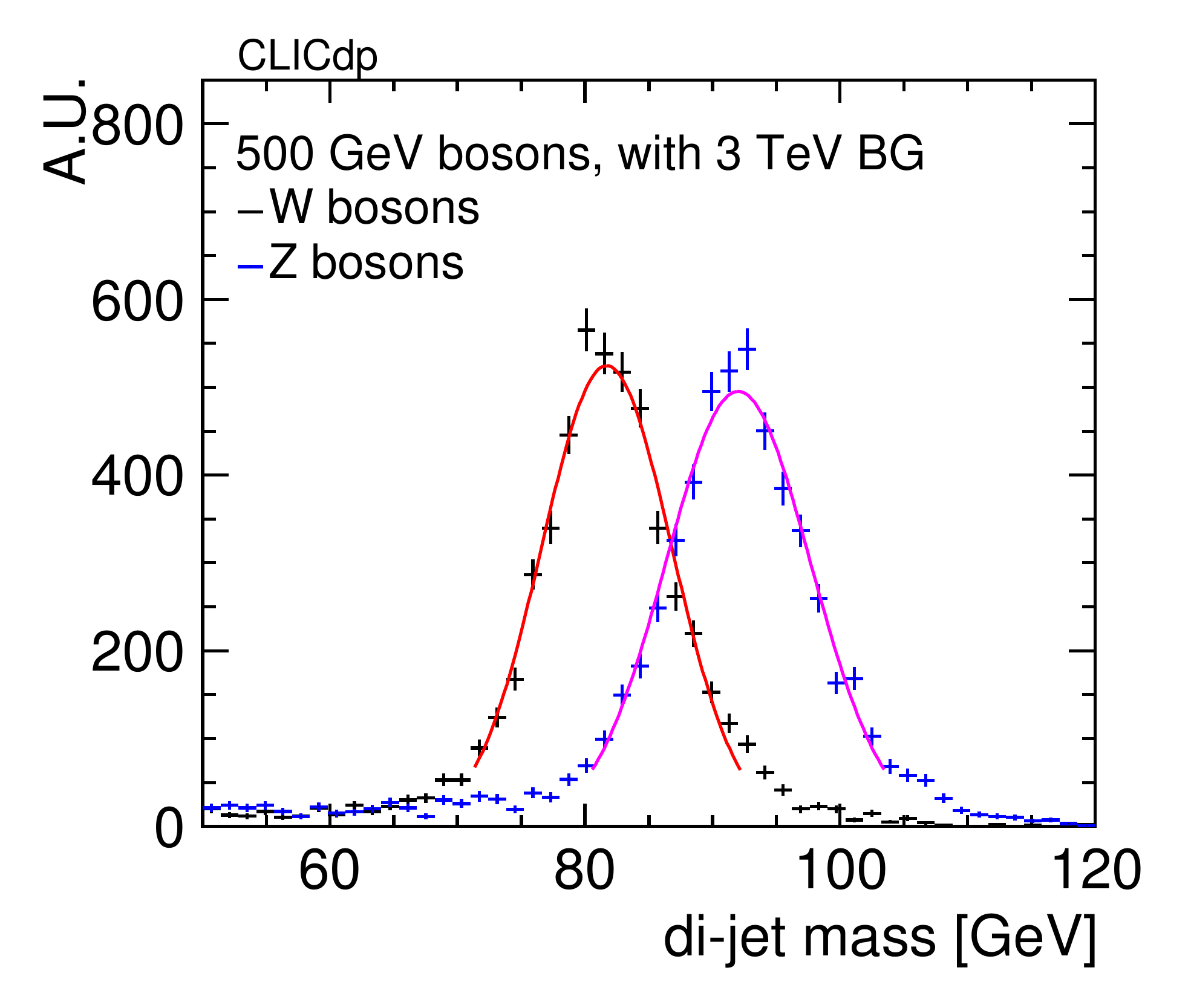}
   \caption{}
  \label{fig:WW_ZZ_500_VLC7_3TeVBG}
\end{subfigure}
\caption{Di-jet mass distributions of hadronically decaying W and Z with $E=500\,\mathrm{GeV}$ in $\mathrm{WW}\rightarrow l\nu qq$ and $\mathrm{ZZ}\rightarrow \nu\bar{\nu}q\bar{q}$ events, together with Gaussian fits of the di-jet mass for events,  (a) without beam-induced backgrounds and (b) with overlay of 3\,\TeV beam-induced backgrounds from \gghadrons~\cite{CLICdet_performance}.}
  \label{fig:WW_ZZ_500_DiJetMass}
\end{figure}

\paragraph{Flavour tagging performance}
To investigate detector performances in terms of flavour tagging, di-jet events at 500\,\GeV centre-of-mass energies (at the 3 TeV CLIC) with a mixture of polar angles between 20\degrees and 90\degrees have been simulated and reconstructed. Tracks and selected PandoraPFOs are used as input to the LCFIPlus vertex finder.
The beauty (charm) misidentification probability is assessed separately for charm (beauty) and light-flavour contamination. The effect of 3\,\TeV \gghad{} background is also evaluated. 
Without background overlaid, at 80\% beauty identification efficiency the misidentification amounts to 10\% as charm and 1.5\% as light-flavour jets.
If the \gghad{} background is included, the performance is slightly worse, with 13\% and 2\% identified as charm and light-flavour jets, respectively (\ref{fig:b_mis_eff}).
Similarly, at 80\% charm identification efficiency the misidentification is 25\% as beauty jets without, 30\% with background overlaid (\ref{fig:c_mis_eff}). In this case, the same misidentification rate holds for light-flavour jets.

In order to estimate the impact of track reconstruction on the flavour tagging, the same study has been performed using the true (Monte Carlo) pattern recognition~\cite{CLICdet_performance}. The results indicate that both beauty and charm tagging can be improved by optimising the pattern recognition.
In particular, in beauty tagging a reduction of misidentification of a b-quark as c- or light-quark by 20 to 30\% can be expected.

\begin{figure}[tbp]
\centering
\begin{subfigure}{.5\textwidth}
  \centering
  \includegraphics[width=\linewidth]{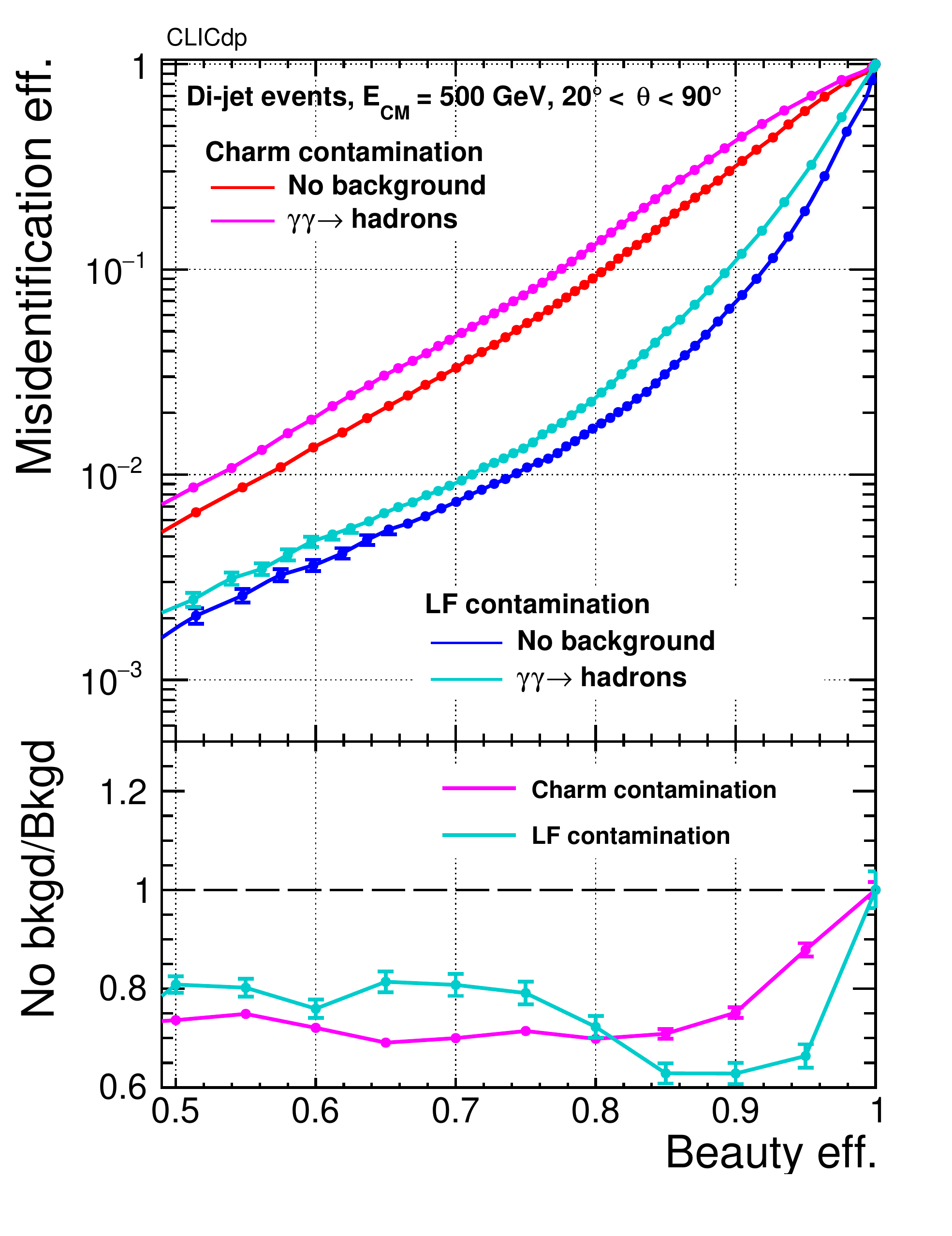}
  \caption{}
  \label{fig:b_mis_eff}
\end{subfigure}
\begin{subfigure}{.5\textwidth}
  \centering
  \includegraphics[width=\linewidth]{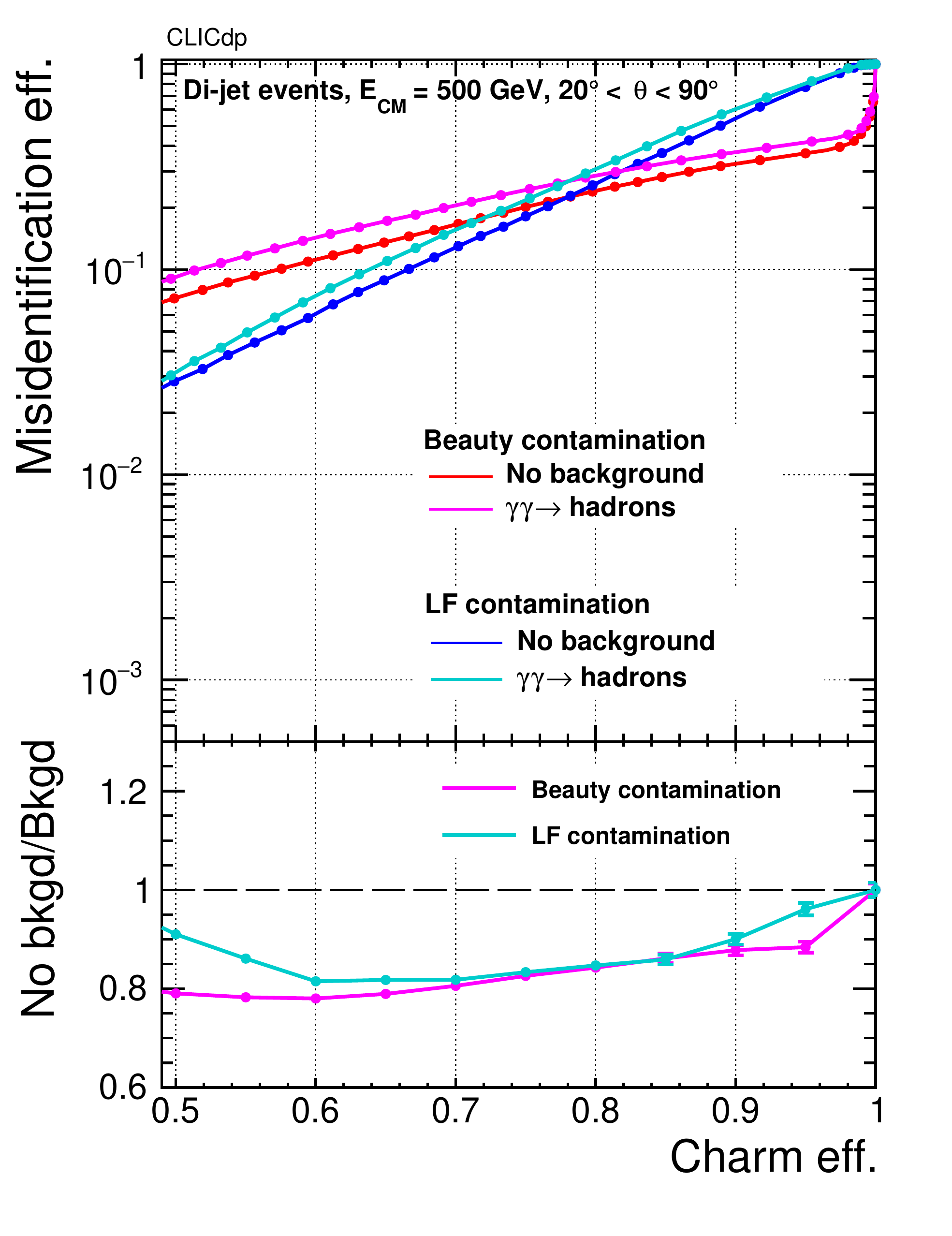}
   \caption{}
  \label{fig:c_mis_eff}
\end{subfigure}
\caption{Global performance of (a) beauty tagging and (b) charm tagging 
for jets in di-jet events at $\roots = 500\,\GeV$ with a mixture 
of polar angles between 20\degrees{} and 90\degrees{}. 
A comparison of performance with and without \gghad{} background is presented. 
On the y-axis, the misidentification probability and the ratio of the 
misidentification probabilities with and without 3\,\TeV \gghad{} background are given~\cite{CLICdet_performance}.}
\label{fig:beff_ceff_background}
\end{figure}

\cleardoublepage
\textcolor{white}{ }
\newpage
\section{CLIC project implementation}
\label{sec:project}
 The general concept of the CLIC accelerator staging and the parameters of the three centre-of-mass stages are  described in~\ref{sec:accelerator}. 
This section describes the implementation of the CLIC accelerator.
\ref{sect:IMP_Stages} summarises the civil engineering and infrastructure. 
\ref{sect:IMP_Sched} describes the schedule for the CLIC programme, from the start of construction to the end of operation at 3\,TeV, while~\ref{sect:IMP_Cost,sect:IMP_Power} cover the cost, and the power and energy consumption of the accelerator.

\subsection{The CLIC stages and construction}
\label{sect:IMP_Stages}

\begin{figure}[h!]
\centering
\includegraphics[width=0.6\textwidth]{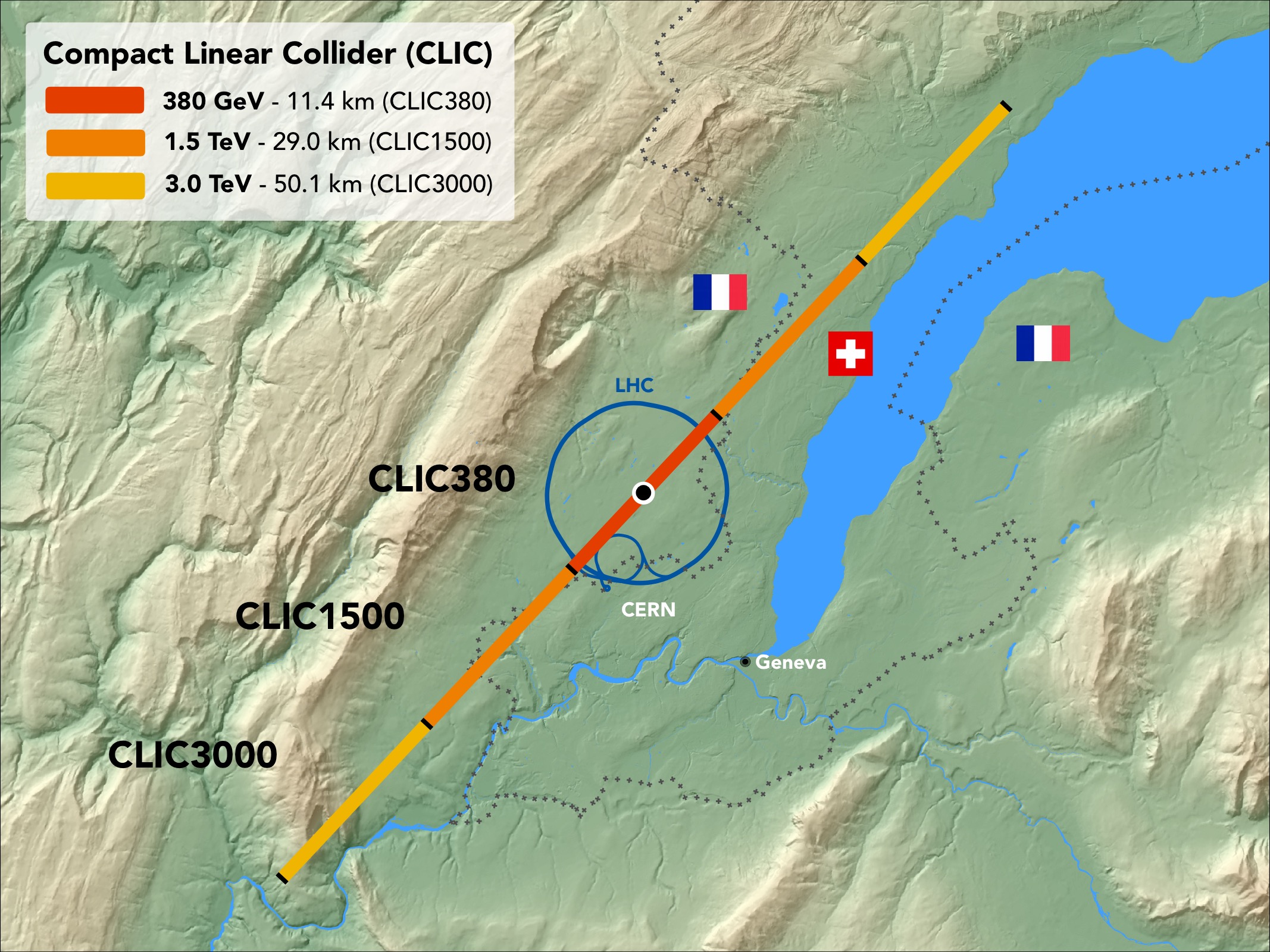} 
\caption{\label{fig_IMP_1} The CLIC main linac footprint near CERN, showing the three implementation stages. \imcl}
\end{figure}
 
The CLIC accelerator is foreseen to be built in three stages with centre-of-mass energies of 380\,GeV, 1.5\,TeV
and 3\,TeV as schematically shown in~\ref{fig_IMP_1}. 
\ref{t:scdup1} in~\ref{sec:accelerator} summarises the main accelerator parameters for the three stages.
The accelerator extension from 380\,GeV to higher energies is described in~\ref{sect:HE_Intro}. 
The installation and commissioning schedules are presented in~\ref{sect:IMP_Sched}. 
More details about the CLIC accelerator and the staged implementation can be found in~\cite{ESU18PiP}. 

Along with the optimisation of the accelerator complex for 380\,GeV, the civil engineering and infrastructure designs
have been revised, 
maintaining an optimal path for extending the facility 
to higher energies. These studies are summarised in the following.

\subsubsection{Civil engineering and infrastructure}

The civil engineering design has been optimised for the 380 GeV stage including: the tunnel 
length and layout, an optimised injection complex, and a siting optimisation for access shafts and their associated structures. 
For the klystron option, a larger tunnel diameter is needed and a detailed layout study was completed.

Previous experience from the construction of LEP and LHC has shown that the sedimentary rock 
in the Geneva basin, known as molasse, provides suitable conditions for tunnelling. Therefore, boundary 
conditions were established so as to avoid the limestone of the Jura mountain range and 
to avoid siting the tunnels below Lake Geneva, whilst maximising the portion of tunnel located in the molasse.
Based on the regional geological and surface data, and using a bespoke digital modelling Tunnel 
Optimisation Tool (TOT) developed specifically for CLIC, a 380\,GeV solution has been found that 
can be readily
upgraded to the higher energy stages at 1.5\,TeV and 3\,TeV.
\ref{fig_CEIS_3} shows the simplified geological profile of the CLIC accelerator stages. 
The 380\,GeV and 1.5\,TeV stages are located entirely in molasse rock. 
The solution shown is both optimised for 380\,GeV and provides a realistic upgrade 
possibility for the 1.5\,TeV and 3\,TeV stages.
\begin{figure}[htb!]
\centering
\includegraphics[width=\textwidth]{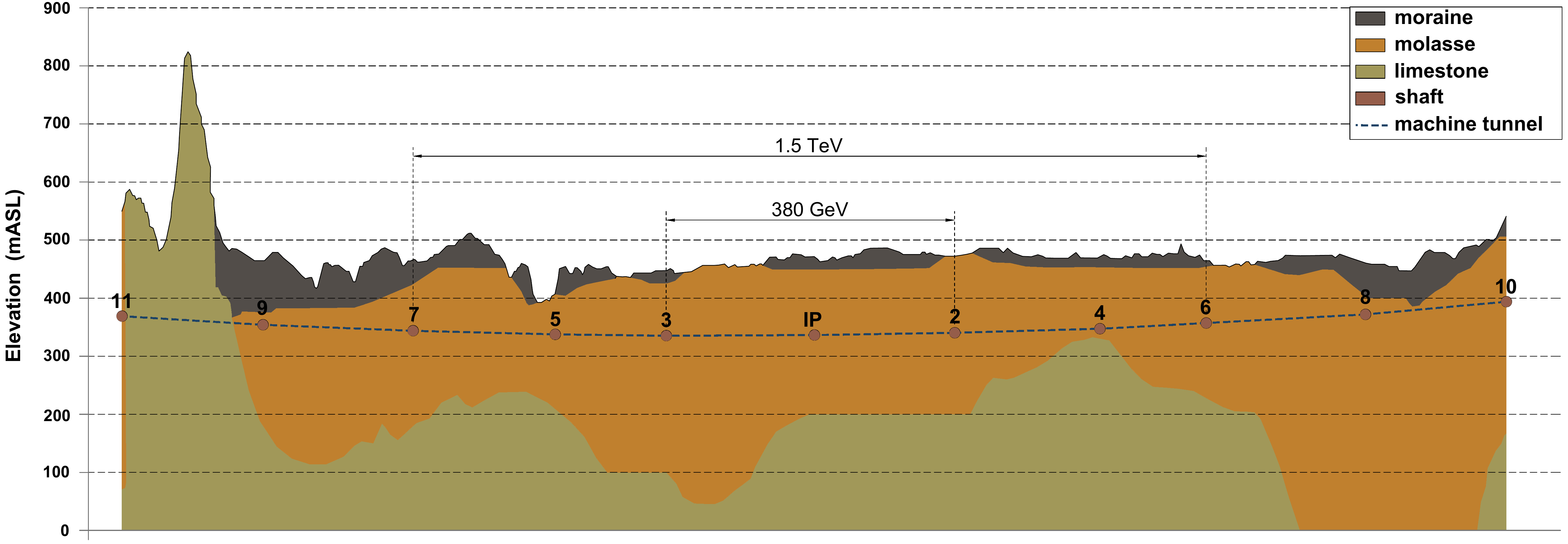}
\caption{\label{fig_CEIS_3} Geological profile of the CLIC three-stage main tunnel. \imcl}
\end{figure}

An initial boundary condition for the civil engineering layout was to 
concentrate the drive-beam and main-beam injectors and the interaction point on the CERN Pr\'{e}vessin site. As shown 
in~\ref{fig_CEIS_4} a solution was found in which the injection complex and the experimental area can be 
located entirely on CERN land. 
\begin{figure}[htb!]
\centering
\includegraphics[width=\textwidth]{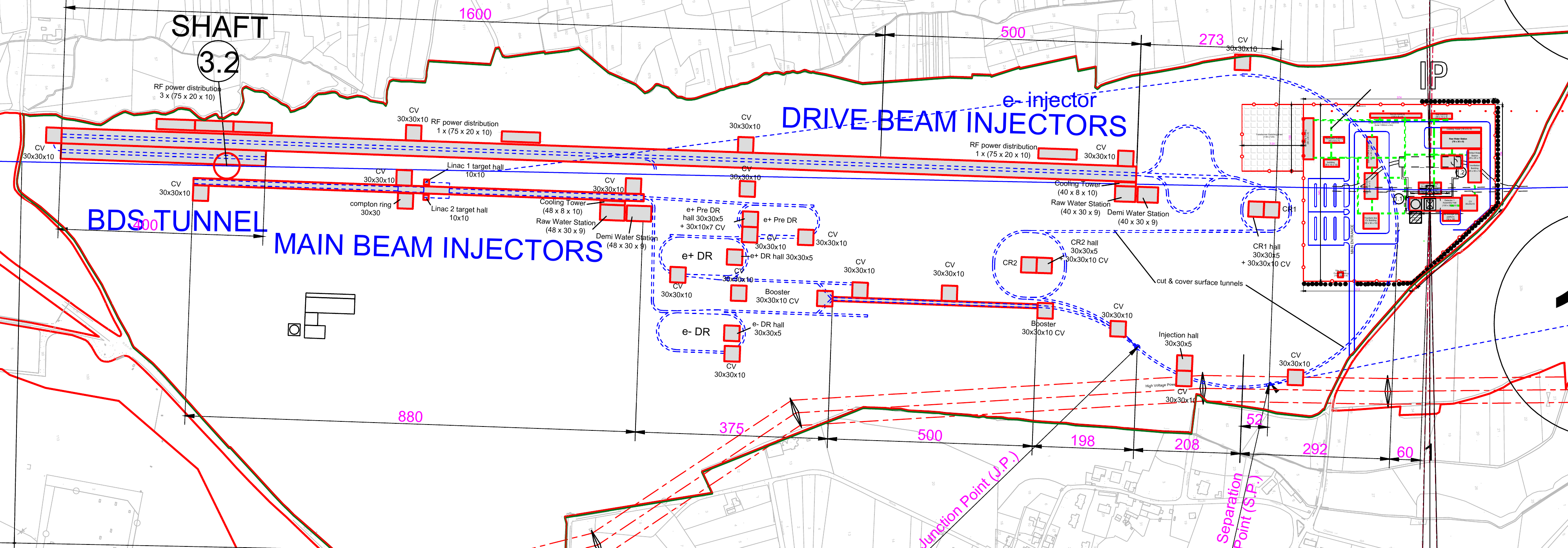}
\caption{\label{fig_CEIS_4} Schematic layout of the injectors, damping rings and drive beam complex centred on the CERN Pr\'{e}vessin site. \imcl}
\end{figure}

For the baseline design with drive beam a tunnel with a 5.6\,m internal diameter is required to house the 
two-beam modules and all the necessary services, as shown in~\ref{fig:CEIS_7a}.  
For the klystron design a 10\,m internal diameter tunnel is required (\ref{fig:CEIS_7b}) to house
both the accelerating modules and the klystron gallery separated by a 1.5\,m thick shielding wall.  
In order to minimise the impact of vibrations on the accelerating modules, the services compartment will be located below the 
klystron gallery. 
\begin{figure}[h!]
\centering
\begin{subfigure}{.49\textwidth}
\includegraphics[height=0.7\textwidth]{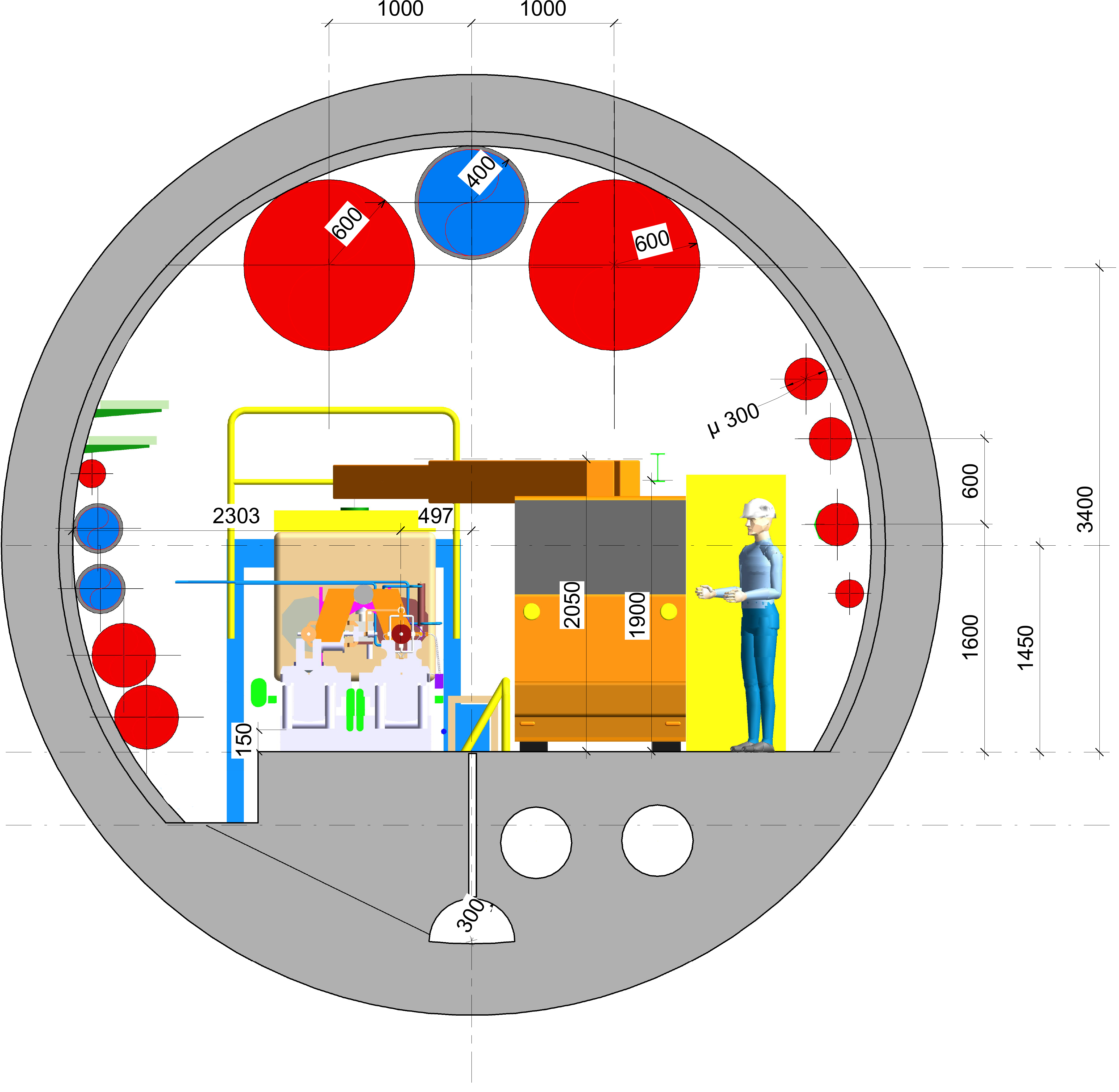}
   \caption{}
\label{fig:CEIS_7a}
\end{subfigure}
\begin{subfigure}{.49\textwidth}
\includegraphics[height=\textwidth]{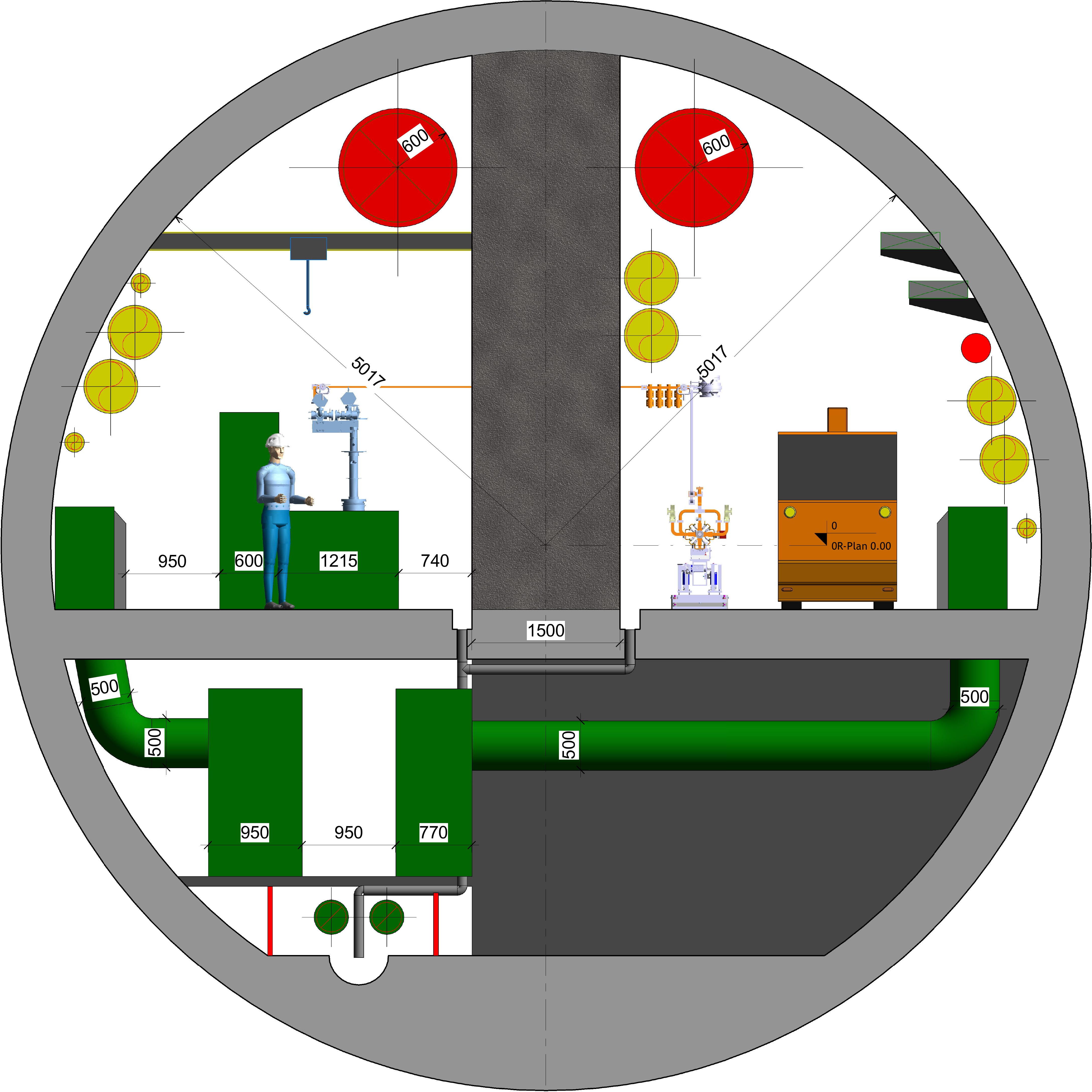}
   \caption{}
\label{fig:CEIS_7b}   
\end{subfigure}
\caption{(a) Main linac cross section for the drive-beam design and (b) the klystron-based option. The relative sizes are represented in the figure. \imcl}
\end{figure}

The detector and service caverns are connected to one another by an escape tunnel that leads to a safe zone in 
each of the caverns. The service cavern is accessible via a shaft with 12\,m internal diameter. 
\ref{fig:ExpArea} provides a view of the interaction region and the service cavern.

The klystron-based option allows significant civil engineering simplifications in the area of the injection complex 
since no drive-beam facility is needed for the first stage. The drive-beam turnarounds have also been
removed. On the other hand, the increase in the tunnel diameter and hardware complexity due to the 
klystron gallery increases the civil engineering and infrastructure challenges underground.

It is foreseen that all of the tunnels will be constructed using tunnel boring machines (TBMs).
For TBM excavation in a sector with good rock conditions, a single pass pre-cast lining is adopted. 
The beam delivery system (BDS) will remain the same for both the two-beam and the klystron 
designs. However, for reasons of tunnelling efficiency, the cross-section of the BDS tunnel for the klystron
design will have an internal diameter of 10\,m, thus allowing the same TBMs to be used for both the 
main linac and the BDS tunnel. 

The infrastructure needs for the accelerator have been updated, and further details have been added to the studies made 
for the CLIC CDR in 2012. Detailed information can be found in~\cite{ESU18PiP}, and a summary is given here: 

\begin{itemize}
\item The electrical network is composed of a transmission and a distribution level. 
The transmission level brings the power from the source of the European Grid to the CLIC sites and access points.
This network typically operates at high voltage levels of 400\,kV, 135\,kV and 63\,kV. 
The distribution level distributes the power from the transmission level to the end users at low and medium 
voltage levels comprised in the range of 400\,V to 36\,kV. Emergency power is also included. 
\item The cooling and ventilation systems have been studied according to the required heat load for accelerator operation. Their main architecture and technical implementations have been defined, covering both surface and underground facilities, as well as safety issues such as smoke extraction in the tunnels. The studies provide input to the civil engineering, installation planning, cost and power estimates, and schedules. 
\item The transport, logistics and installation activities cover many items (e.g modules, magnets, RF units, vacuum pipes, beam dumps, cooling and ventilation equipment, electrical cables, cable trays and racks) and were studied starting from the unloading of components upon arrival at the CERN site. 
The most demanding aspects of transport and handling concern the installation of the underground equipment in both the two-beam and the klystron designs.
\item Safety systems, access systems and radiation protection systems have been studied and are included in the schedules, cost and power estimates, covering all areas from injectors to beam-dumps. A hazard identification and mitigation analysis shows that fire protection is the dominant safety-related implementation issue.
\end{itemize}
The above studies, carried out by the CERN civil engineering and infrastructure groups, follow 
the standards used for other accelerator implementations and studies at CERN (e.g.\ HL-LHC, FCC). 
The standardisation applies to all items listed above, including their cost, power and schedule estimates.

\subsubsection{Annual and integrated luminosities} 

Estimates of the integrated luminosities are based on an annual operational scenario~\cite{Bordry:2018gri}. 
After completion of CLIC commissioning, it is estimated that 185 days per year will be used for operation, with 
an average accelerator availability of 75\%, thus yielding physics data taking during 1.2~$\times$~10$^7$ seconds annually. 
The remaining time is shared between maintenance periods, technical stops and extended 
shutdowns as discussed in~\ref{sect:IMP_Power}. The yearly luminosity and the cumulative integrated 
luminosity for the three stages of the CLIC programme are shown in~\ref{fig_IMP_5}.
A luminosity ramp-up of three years (10\%, 30\%, 60\%) is assumed for the first stage and two years (25\%, 75\%) for 
subsequent stages. 
Prior to data-taking at the first stage, commissioning of the individual systems and one full year of commissioning with beam are foreseen. 
These are part of the construction schedule.
The beam-polarisation scheme foreseen for the CLIC programme is described in~\ref{sec:polarisation}.

\begin{figure}[h!]
\centering
\begin{subfigure}{.49\textwidth}
\includegraphics[width=\textwidth]{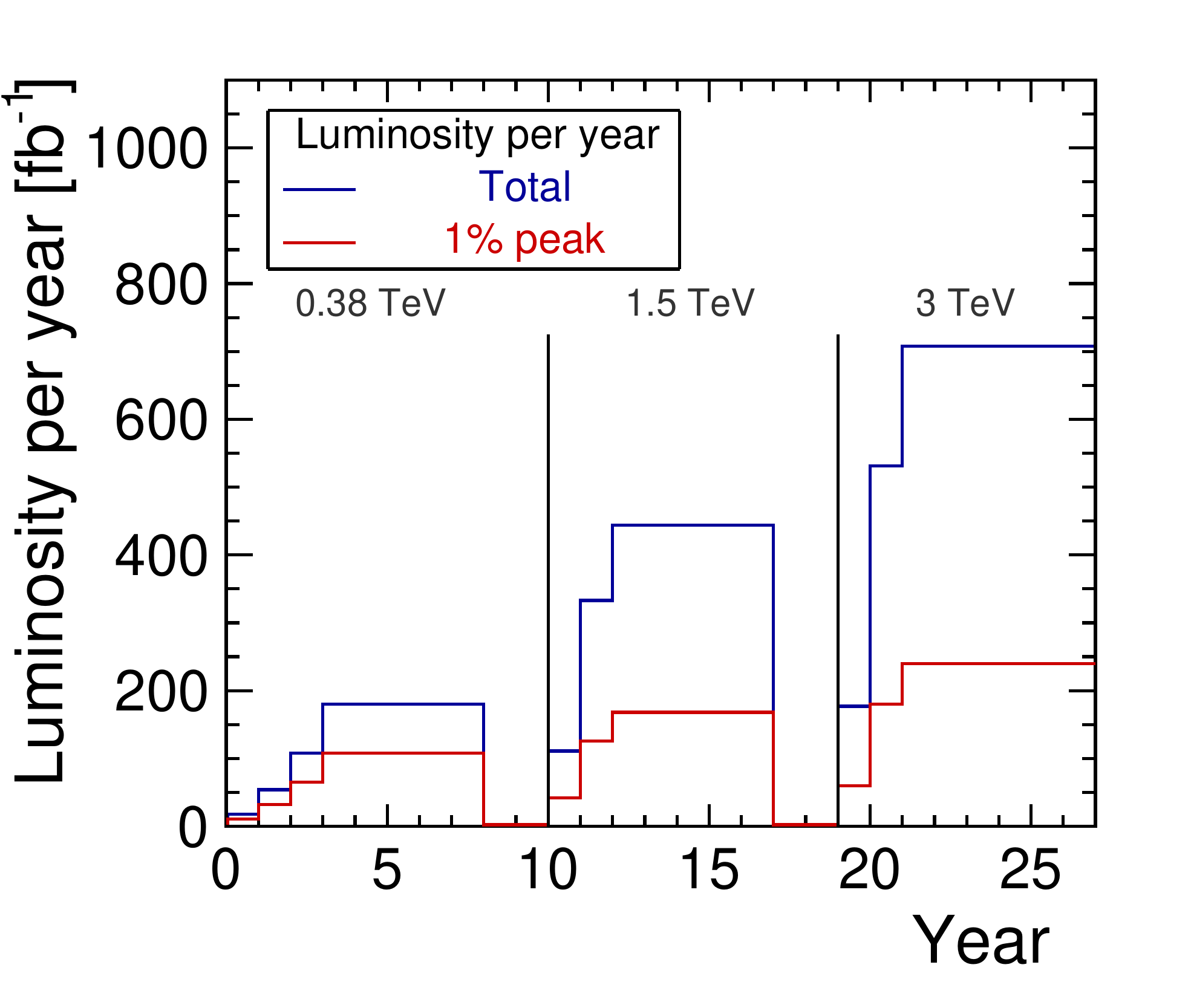}
   \caption{}
\end{subfigure}
\begin{subfigure}{.49\textwidth}
\includegraphics[width=\textwidth]{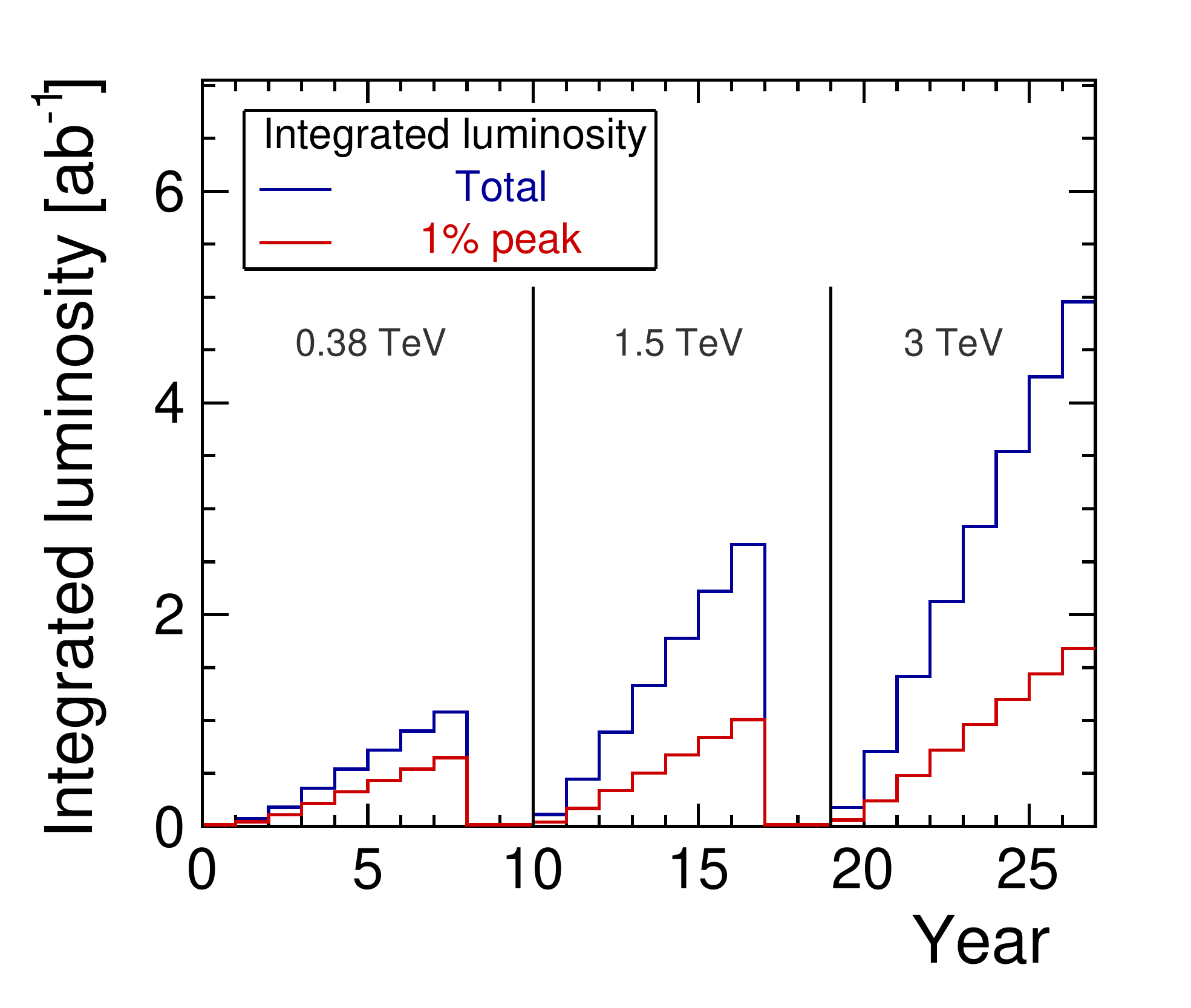}
   \caption{}
\end{subfigure}
\caption{\label{fig_IMP_5} (a) Luminosity and (b) integrated luminosity per year in the 
proposed staging scenario, for the total luminosity in blue and the luminosity at centre-of-mass energies above 99\% of the nominal centre-of-mass energy in red. Years are counted from the start of physics running. From~\cite{Roloff:2645352}.}
\end{figure}

\subsection{Construction and operation schedules}
\label{sect:IMP_Sched}
The construction schedules presented in this section are based on the same methodologies as those used for the CLIC CDR~\cite{cdrvol1}. 
Following input from equipment experts and the CERN civil engineering and infrastructure groups, 
small adjustments were made to the construction and installation rates used for the schedule estimates. 
Details about the various parameters used can be found in~\cite{ESU18PiP}. The installation is followed by hardware commissioning, final alignment and commissioning with beam.

\subsubsection{380\,GeV drive-beam schedule}

The schedule for the first stage of CLIC at 380\,GeV, based on the drive-beam design, is shown in~\ref{fig_IMP_6}. It comprises the following time-periods:

\begin{itemize}
\item  Slightly more than five years for the excavation and tunnel lining, the installation of the tunnel infrastructures, and the accelerator equipment transport and installation.
\item  Eight months for the system commissioning, followed by two months for final alignment.
\item  One year for the accelerator commissioning with beam.
\end{itemize}

In parallel, time and resources are allocated for the construction of the drive-beam surface building, 
the combiner rings, damping rings, main-beam building and experimental areas, and their corresponding system installation and commissioning, as shown in~\ref{fig_IMP_6}. 

\begin{figure}[h!]
\centering
\includegraphics[width=0.8\textwidth]{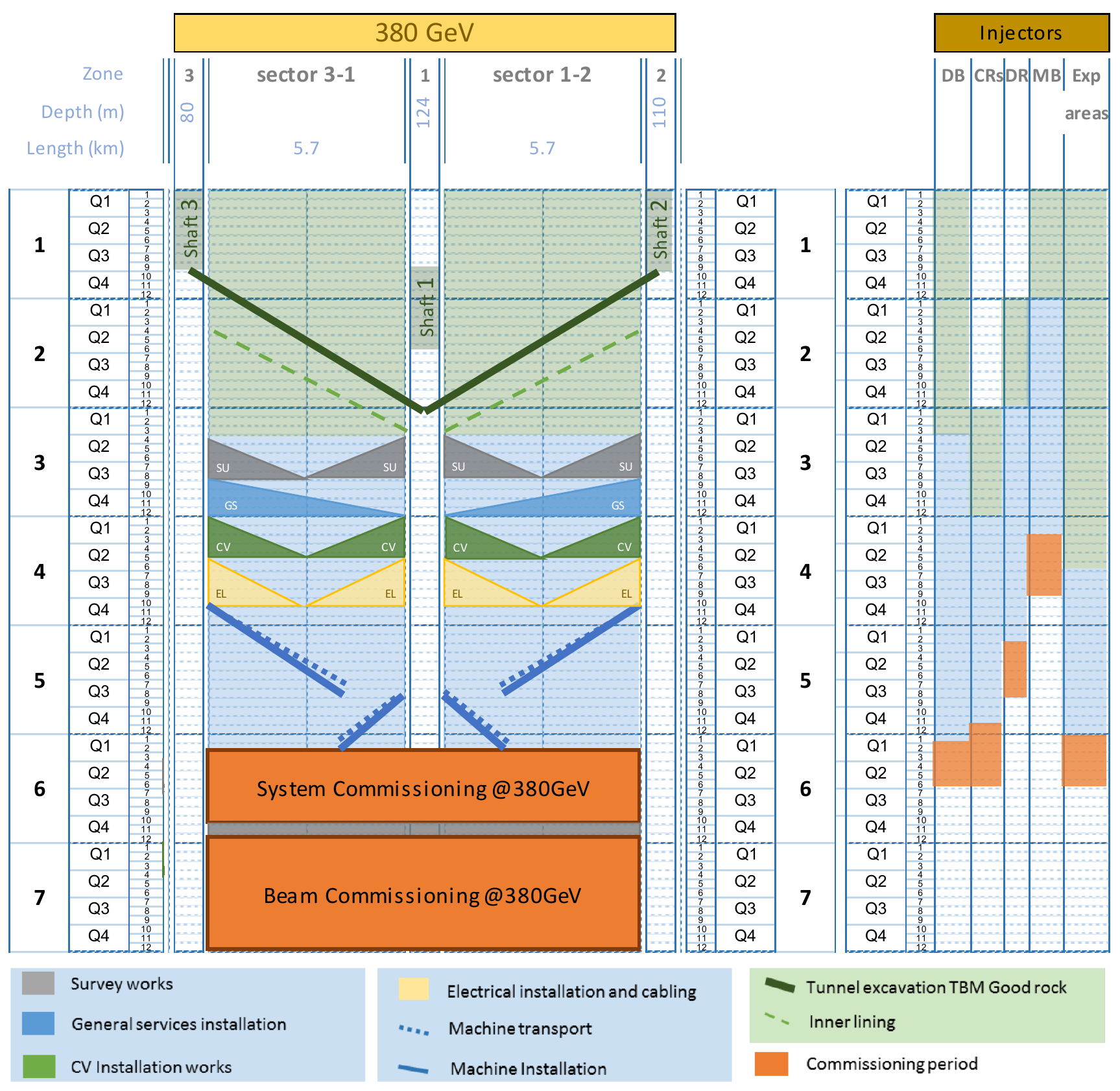}
\caption{\label{fig_IMP_6} Construction and commissioning schedule for the 380\,GeV drive-beam based CLIC facility. 
The vertical axis represents time in years. The abbreviations are introduced in~\ref{scd:clic_layout}. \imcl}
\end{figure}

\subsubsection{380\,GeV klystron-driven schedule}

In this scheme the RF power is provided by X-band klystrons and modulators, installed underground all along the main linac. 
The total time for installation is slightly different from the drive-beam case. 
The surface buildings and installations are reduced to those exclusively needed for the main beam and experimental area, reducing the surface construction activities correspondingly. 
On the other hand, the installation time in the main tunnel is longer, due to the RF units and the additional infrastructures required. 
Even though it is possible to work in parallel in the main linac tunnel and in the klystron gallery, the overall transport, 
installation and handling logistics are more time consuming. The time needed for construction, installation and commissioning is eight years, compared to seven years for the drive-beam option at the same CLIC energy of 380\,GeV.

\subsubsection{Schedules for the stages at higher energies and the complete project}

In both cases discussed above, the 380\,GeV collider is designed to be extended to higher energies. 
Most of the construction and installation work can be carried out in parallel with the data-taking at 380 GeV. 
However, it is estimated that a stop of two years in accelerator operation is needed between two energy stages. 
This time is needed to make the connection between the existing machine and its extensions, 
to reconfigure the modules used at the existing stage for their use at the next stage, 
to modify the beam-delivery system, to commission the new equipment and to commission the entire new accelerator complex with beam. 

As the construction and installation of the 1.5\,TeV and subsequent 3\,TeV equipment cover periods of 4.5 years, 
the decision about the next higher energy stage needs to be taken after $\mathrm{\sim}$4-5 years of data taking at the existing stage, 
based on physics results available at that time.  
The corresponding scenario is shown in~\ref{fig_IMP_9} for the drive-beam based scenario. 
A more detailed breakdown of the full project schedule can be found in~\cite{ESU18PiP}. 
The overall upgrade schedule is very similar for the case in which the first stage will be powered by klystrons.

In a schedule driven by technology and construction, the CLIC project would cover 34 years, counted from the start of construction. 
About 7 years are scheduled for initial construction and commissioning and a total of 27 years for data-taking at the three energy stages, 
which includes two 2-year intervals between the stages. 

\begin{figure}[h!]
\centering
\includegraphics[width=\textwidth]{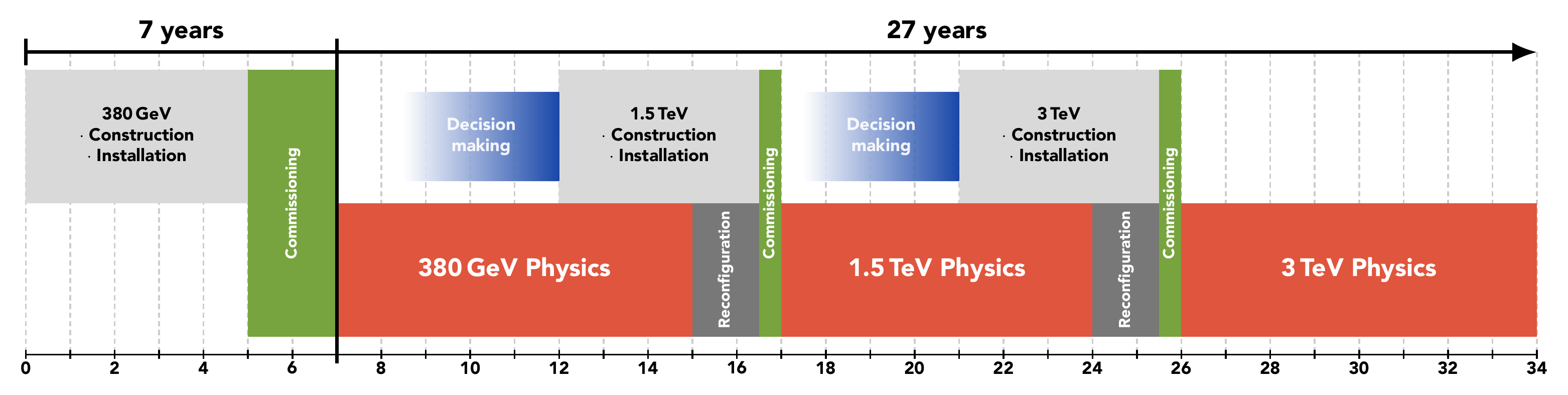}
\caption{\label{fig_IMP_9} Technology-driven CLIC schedule, showing the construction and commissioning period and the three stages for data taking.
The time needed for reconfiguration (connection, hardware commissioning) between the stages is also indicated. \imcl}
\end{figure}

\subsubsection{Concluding remarks on the schedule}

The schedule for construction and installation shows that the CLIC project can be implemented well in time for first collisions in 2035, provided it can be launched by 2026. 
The most critical CLIC technology-specific items driving the schedule are the main-beam module production and installation, as well as the RF units. 
The other schedule drivers, such as the tunnelling, the buildings and the infrastructures are more common, similar to other projects at CERN and elsewhere.

\subsection{Cost estimate}
\label{sect:IMP_Cost}

For the cost estimate of CLIC the methodology used is the same as for previous CLIC cost estimates and the estimates of other projects, such as the LHC experiments and the Reference Design Report and Technical Design Report of the International Linear Collider (ILC)~\cite{Phinney2007,Adolphsen:2013kya}. Previous CLIC cost estimates were reported in the CLIC CDR~\cite{cdrvol3} for two different implementation options at 500\,GeV. An initial cost estimate for the first stage at 380\,GeV was presented together with the introduction of the corresponding CLIC energy staging scenario in~\cite{StagingBaseline}. Since then, many CLIC optimisation studies have been undertaken with a particular focus on cost reduction, as reported in~\ref{sec:accelerator}. The resulting cost estimates, as well as the methodologies and assumptions used have been presented in November 2018 to a cost review panel composed of international experts. After recommendations on minor issues by the review panel, the estimates have been updated accordingly. As the cost of the accelerator is significantly larger than the cost of the experiment, this Section focuses on the accelerator when presenting the methodologies and the various aspects of the outcome. The resulting estimated cost of the 380\,GeV stage is presented, together with an estimate for upgrading to higher energies. The same tools and basic principles are applied for estimating the cost of the experiment, and the results are presented in~\ref{sec:detectorcost}.   

\subsubsection{Scope and method}

CLIC is assumed to be a CERN-hosted project, constructed and operated within a collaborative framework with participation and contributions from many international partners. Contributions from the partners are likely to take different forms (e.g.\ in kind, in cash, in personnel, from different countries, in different currencies or accounting systems). Therefore a "value and explicit labour" methodology is applied. The value of a component or system is defined as the lowest reasonable estimate of the price of goods and services procured from industry on the world market in adequate quality and quantity and satisfying the specifications. Value is expressed in a given currency at a given time. Explicit labour is defined as the personnel provided for project construction by the central laboratory and the collaborating institutes, expressed in Full Time Equivalent (FTE) years. It does not include personnel in the industrial manufacturing premises, as this is included in the value estimate of the corresponding manufactured components. The personnel in industrial service contracts that are part of the accelerator construction, outside CERN or at CERN, are also accounted for in the value estimate of the corresponding items.

For the value estimate, a bottom-up approach is used, following the work breakdown structure of the project, starting from unit costs and quantities for components, and then moving up to technical systems, subdomains and domains. This allows accounting for all aspects of the production process and the application of learning curves for large series. 
For some parts (e.g.\ standard systems), cost scaling from similar items is used, implying that 
detailed knowledge on the work breakdown is not required, but rather estimators characterising the component. 

The basic value estimate concerns the construction of the 380\,GeV CLIC stage on a site close to CERN, where the 380\,GeV stage of CLIC constitutes a project in itself.
As a consequence, large-series effects expected on unit costs -- learning curves and quantity rebates -- remain limited to the quantities required for the completion of the 380\,GeV stage. Estimates are provided both for the drive-beam based and the klystron-based options, together with the corresponding incremental value for upgrading to higher energies.

The value estimates given cover the project construction phase, from approval to start of commissioning with beam. 
They include all the domains of the CLIC complex from injectors to beam dumps, together with the corresponding civil engineering and infrastructures. Items such as specific tooling required for the production of the components, reception tests and pre-conditioning of the components, and commissioning (without beam) of the technical systems, are included.
On the other hand, items such as R\&D, prototyping and pre-industrialisation costs, acquisition of land and underground rights-of-way, computing, and general laboratory infrastructures and services (e.g.\ offices, administration, purchasing and human resources management) are excluded. Spare parts are accounted for in the operations budget. The value estimate of procured items excludes VAT, duties and similar charges, taking into account the fiscal exemptions granted to CERN as an Intergovernmental Organisation.

The uncertainty objective for the final outcome is $\mathrm{\pm}$25\%. To this aim, uncertainties on individual items are grouped in two categories. The first one, \textit{technical uncertainty}, relates to technological maturity and likelihood of evolution in design or configuration. The second category, \textit{commercial uncertainty}, relates to uncertainty in commercial procurement. Based on a statistical analysis of LHC procurements this uncertainty is estimated as $50\%/n$, where $n$ is the number of expected valid bids for each component~\cite{Lebrun2010}.

The CLIC value estimates are expressed in Swiss franc (CHF) of December 2018. Consequently, individual entries are escalated in time according to appropriate indices, as published by the Swiss federal office of statistics.
Furthermore, the following average exchange rates have been applied: 1 EUR=1.13 CHF, 1 CHF=1 USD, 1 CHF=114 JPY. More detailed information on the costing tool, on escalation and currency fluctuations, and on the individual cost uncertainty factors applied can be found in the CLIC project plan~\cite{ESU18PiP}.

\subsubsection{Value estimates and cost drivers}

The breakdown of the resulting cost estimate up to the sub-domain level is presented in~\ref{Tab:Cost} 
for the 380\,GeV stage of the accelerator complex, both for the baseline design with a drive beam and for the klystron-based option. 
\ref{fig_IMP_10} illustrates the sharing of cost between different parts of the accelerator complex. 
The injectors for the main-beam and drive-beam production are among the most expensive parts of the project, together with the main linac, and the civil engineering and services.

\begin{table}[ht]
\caption{Cost breakdown for the 380\,GeV stage of the CLIC accelerator, for the drive-beam baseline option and for the klystron option.}
\label{Tab:Cost}
\centering
\begin{tabular}{l l S[table-format=4.0] S[table-format=4.0]}
\toprule
\multirow{2}{*}{Domain} & \multirow{2}{*}{Sub-Domain} & \multicolumn{2}{c}{Cost [\si{MCHF}]} \\
 &  & {Drive-beam} & {Klystron} \\ \midrule
 \multirow{3}{*}{Main-Beam Production} & Injectors & 175 & 175 \\
 & Damping Rings & 309 & 309 \\
 & Beam Transport & 409 & 409 \\ \hline
\multirow{3}{*}{Drive-Beam Production} & Injectors & 584 &  {---} \\
 & Frequency Multiplication & 379 & {---}  \\
 & Beam Transport & 76 &  {---} \\ \hline
\multirow{2}{*}{Main Linac Modules}  & Main Linac Modules & 1329 & 895 \\
 & Post decelerators  & 37 &  {---}  \\ \hline
Main Linac RF  & Main Linac Xband RF & {---} & 2788 \\ \hline
\multirow{3}{*}{\makecell[l]{Beam Delivery and \\ Post Collision Lines}}   & Beam Delivery Systems & 52 & 52 \\
 & Final focus, Exp. Area & 22 & 22 \\
 & Post-collision lines/dumps & 47 & 47 \\ \hline
Civil Engineering & Civil Engineering & 1300 & 1479 \\ \hline
\multirow{4}{*}{Infrastructure and Services}  & Electrical distribution  & 243 & 243 \\
 & Survey and Alignment & 194 & 147 \\
 & Cooling and ventilation  & 443 & 410 \\
 & Transport / installation & 38 & 36 \\ \hline
\multirow{4}{*}{\makecell[l]{Machine Control, Protection \\ and Safety systems}} & Safety systems  & 72 & 114 \\
  & Machine Control Infrastructure & 146 & 131 \\
 & Machine Protection & 14 & 8 \\
 & Access Safety \& Control System & 23 & 23 \\ \midrule
\bfseries Total (rounded) & & \bfseries 5890 & \bfseries 7290 \\
\bottomrule
\end{tabular}
\end{table}

\begin{figure}[h!]
\centering
\begin{tikzpicture}[font=\sffamily]
\sansmath
\definecolor{col1}{RGB}{164,164,164}
\definecolor{col2}{RGB}{205,125,45}
\definecolor{col3}{RGB}{119,153,213}
\pgfplotstableread[col sep=&, header=true]{
scenario & {Main-Beam Production} & {Drive-Beam Production} & {Main Linac Modules } & {Main Linac RF } & {Beam Delivery, Post-Collision Lines  } & {Civil Engineering} & {Infrastructure and Services} & {Machine Control, Protection and Safety systems}
{380 GeV Drive-beam} & 893.475542 & 1039.186 & 1365.382083 & 0 & 121.128 & 1300.454403 & 916.767 & 255.409
{380 GeV Klystrons} & 893.475542 & 0 & 895.030311 & 2787.538141 & 121.128 & 1478.812503 & 836.027 & 275.923
}\mydata
\begin{axis}[
ybar stacked,
ymin = 0,
bar width=35pt,
width= 10cm,
legend style={cells={align=left}, draw=none, at={(1.06,0.60)},anchor=west,/tikz/every even column/.append style={column sep=0.5cm}},
legend cell align={left},
legend image code/.code={\draw[draw=none] (0cm,-0.1cm) rectangle (0.3cm,0.1cm);},
ylabel={MCHF},
symbolic x coords={{380 GeV Drive-beam},{380 GeV Klystrons}},
xtick=data,
enlarge x limits=0.7,
enlarge y limits={value=.15,upper},
y tick label style={/pgf/number format/.cd,%
scaled y ticks = false,
set thousands separator={},
fixed},
every node near coord/.style={black,/pgf/number format/.cd,fixed zerofill,precision=0,set thousands separator={}}
]
\addplot+[fill=blue!60,draw=none] table[x=scenario,y index=1]    from \mydata;
\addlegendentry{Main-Beam Production};
\addplot+[fill=cyan!60,draw=none] table[x=scenario,y index=2]    from \mydata;
\addlegendentry{Drive-Beam Production};
\addplot+[fill=yellow!60,draw=none] table[x=scenario,y index=3]    from \mydata;
\addlegendentry{Main Linac Modules};
\addplot+[fill=orange!60,draw=none] table[x=scenario,y index=4]    from \mydata;
\addlegendentry{Main Linac RF};
\addplot+[fill=red!60,draw=none] table[x=scenario,y index=5]    from \mydata;
\addlegendentry{Beam Delivery, Post Collision Lines};
\addplot+[fill=blue!60!cyan!60,draw=none] table[x=scenario,y index=6]    from \mydata;
\addlegendentry{Civil Engineering};
\addplot+[fill=cyan!60!yellow!60,draw=none] table[x=scenario,y index=7]    from \mydata;
\addlegendentry{Infrastructure and Services};
\addplot+[,fill=red!60!cyan!60,draw=none] table[x=scenario,y index=8]    from \mydata;
\addlegendentry{Machine Control, Protection \\ and Safety systems};

\node[pin={5890}] at (0,500) {};
\node[pin={7290}] at (100,640) {};
\end{axis}
\end{tikzpicture}
\caption{\label{fig_IMP_10} Cost breakdown for the 380\,GeV stage of the CLIC accelerator, for the drive-beam baseline option and for the klystron option. \imcl}
\end{figure}
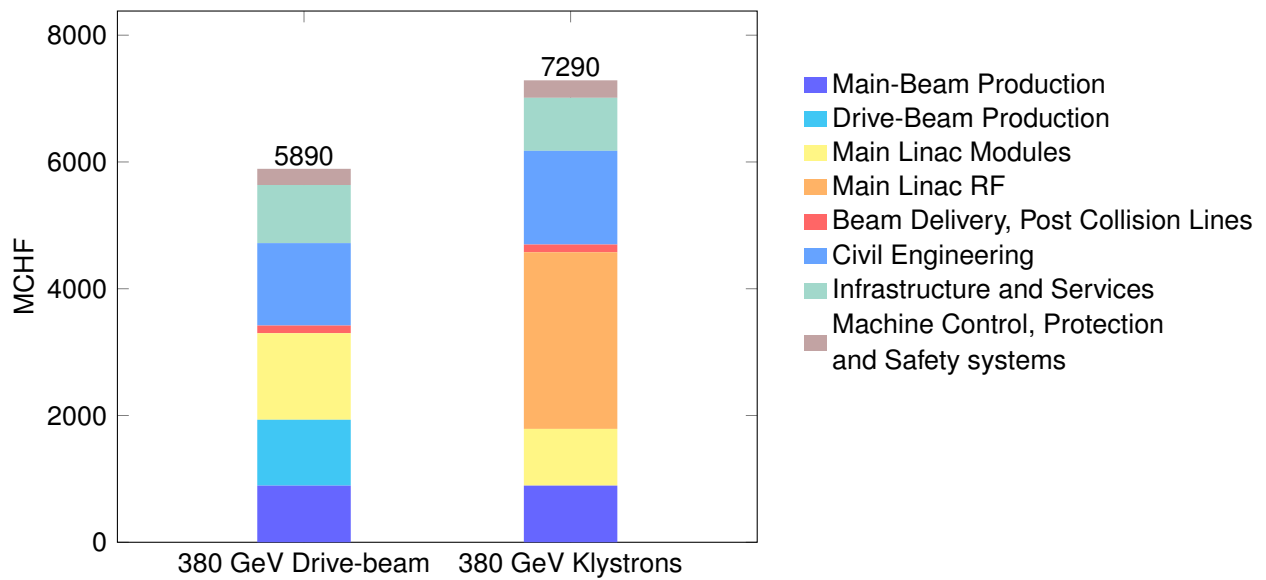

Combining the estimated technical uncertainties yields a total (1$\sigma$) error of 1270\,MCHF for the drive-beam based facility, and 1540\,MCHF when using klystrons.
In addition, the commercial uncertainties, defined above, need to be included. They amount to 740\,MCHF and 940\,MCHF for the drive-beam and klystron-based options, respectively.
The total uncertainty is obtained by adding technical and commercial uncertainties in quadrature. Finally, for the estimated error band around the cost estimate, the resulting total uncertainty is used on the positive side, while only the technical uncertainty is used on the negative side~\cite{cdrvol3}. The cost estimate for the first stage of CLIC including a 1$\sigma$ overall uncertainty is therefore:
\begin{center}
\begin{tabular}{lcc}
CLIC \SI{380}{\GeV} drive-beam based &:& $5890^{+1470}_{-1270}\,\si{MCHF}$\quad ;\\ \\
CLIC \SI{380}{\GeV} klystron based &:& $7290^{+1800}_{-1540}\,\si{MCHF}$\quad.
\end{tabular}
\end{center}

The difference between the drive-beam and klystron-based estimates is mainly due to the current cost estimates for the X-band klystrons and corresponding modulators. 
The increased diameter of the main linac tunnel, required to host the RF gallery in the klystron-based option, also contributes to the cost-difference. 
By reducing the X-band RF costs by 50\% in the klystron option, the overall cost of the two options becomes similar. 
To achieve such a reduction would require a dedicated development programme together with industry for X-band klystrons and associated modulators.
There is still room for possible gains through optimising the accelerating structure parameters, klystron design and luminosity performance. The cost of the klystron-based option is more affected by the luminosity specification than the drive-beam option.

The cost composition and values of the 1.5\,TeV and 3\,TeV stages have also been estimated. The energy upgrade to 1.5\,TeV has a cost estimate of $\sim \text{5.1}\,\text{billion CHF}$, including the upgrade of the drive-beam RF power needed for the 1.5\,TeV stage. In the case of expanding from a klystron-based initial stage this energy upgrade will be 25\% more expensive.
A further energy upgrade to 3\,TeV has a cost estimate of $\sim \text{7.3}\,\text{billion CHF}$, including the construction of a second drive-beam complex.

The CLIC technical cost drivers have been identified, together with potential cost mitigation alternatives. These will be addressed in the next phase of the CLIC project as discussed in~\ref{sec:objectives}.
In general, further cost reduction studies will require close collaboration with industry. Beyond technical developments, optimal purchase models need to be defined, optimising the allocation of risks and production responsibilities between industry, CERN and collaboration partners in each case. In particular, the module production and RF units have a potential for cost reduction. For a klystron-based implementation, the cost reductions of the RF system are of crucial importance.

\subsubsection{Labour estimates}

A first estimate of the explicit labour needed for construction of the CLIC accelerator complex was obtained~\cite{cdrvol3} 
by assuming a fixed ratio between personnel and material expenditure for projects of similar nature and size.
Scaling with respect to the LHC - a CERN-hosted collider project of similar size to CLIC - provides a good estimator. 
Data from the LHC indicate that some 7000\,FTE-years were needed for construction, for a material cost of 3690\,MCHF (December 2010), corresponding to about $1.9\,\text{FTE-year}/\text{MCHF}$. 
About 40\% of this labour was scientific and engineering personnel, and the remaining 60\% worked on technical and project execution tasks.

In terms of complexity, the different CLIC sub-systems resemble the LHC case. 
Therefore, following the LHC approach outlined above, construction of the 380\,GeV stage of the CLIC accelerator complex would require 11500\,FTE-years of explicit labour. 
It is worth noting that this preliminary result is rather similar to the $1.8\,\text{FTE-year}/\text{MCHF}$ derived for the ILC~\cite{Adolphsen:2013kya}. Although the RF technology differs between ILC and CLIC, the main elements of the accelerator complex are similar in the two projects. 

\subsubsection{Value estimate and cost drivers of the CLIC detector}
\label{sec:detectorcost}
The methodology used for estimating the cost of the CLIC detector~\cite{CLICdet_note_2017} is similar to the one used for the accelerator complex, and is based on the detector work breakdown structure~\cite{clicdet_cost}. Some differences in the approach, given by the specificities of the detector, are detailed in the CDR~\cite{cdrvol3}.
A breakdown of the value estimate for the CLIC detector is given in~\ref{Tab:Det_Cost}. The main cost driver is the cost of the silicon sensors
for the 40-layer Electromagnetic Calorimeter (ECAL). For example, a 25\% reduction in the cost of silicon per unit of surface would reduce the overall detector cost by more than 10\%. Alternative designs for ECAL are feasible, but will reduce the detector performance (e.g. worse energy resolution for photons~\cite{CLICdet_note_2017}).

\begin{table}[ht]
\caption{Cost estimate of the CLIC detector~\cite{clicdet_cost}.}
\label{Tab:Det_Cost}
\centering
\begin{adjustbox}{width=0.85\linewidth,center}
\pgfplotstableread[header=true]{
name z
Vertex 13
{Silicon Tracker} 43
{Electromagnetic Calorimeter} 180
{Hadronic Calorimeter} 39
{Muon System} 16
{Coil and Yoke} 95
{Other} 11
{ } {nan}
{Total } 397
}\data

\pgfplotstableset{create on use/error/.style={
    create col/expr={\thisrow{z}
    }
  }
}

\pgfplotsset{select coords between index/.style 2 args={
    x filter/.code={
        \ifnum\coordindex<#1\def\pgfmathresult{}\fi
        \ifnum\coordindex>#2\def\pgfmathresult{}\fi
    }
}}

\newcommand{\errplot}{%
\begin{tikzpicture}[trim axis left,trim axis right]
\begin{axis}[y=-\baselineskip,
xbar,
xmax=50,
width             = 7.5cm,
axis y line=none,
ytick             = \empty,
xtick={0,10,...,50},
xticklabels = {0,10\%,20\%,30\%,40\%,50\%},
axis x line*      = bottom,
]
\addplot+[xbar,fill=orange,draw=black,select coords between index={0}{7}] table [x expr ={\thisrowno{1}/397*100},y expr=\coordindex]{\data};
\end{axis}
\end{tikzpicture}
}

\pgfplotstablegetrowsof{\data}
\let\numberofrows=\pgfplotsretval

\pgfplotstabletypeset[columns={name,error,z},
  col sep = comma,
  every head row/.style = {before row=\toprule, after row=\midrule},
  every last row/.style = {after row=[0ex]\bottomrule, before row=[1ex]\midrule},
  columns/name/.style = {string type, column name=System},
      every row 7 column 2/.style={
        postproc cell content/.style={
          @cell content=\textcolor{white}{##1}
        }
      },
  columns/error/.style = {
    column name = {Cost fraction},
    assign cell content/.code = {
    \ifnum\pgfplotstablerow=0
    \pgfkeyssetvalue{/pgfplots/table/@cell content}
    {\multirow{\numberofrows}{6cm}{\errplot}}%
    \else
    \pgfkeyssetvalue{/pgfplots/table/@cell content}{}%
    \fi
    }
  },
  columns/z/.style    = {column name = {Cost [\si{MCHF}]},   column type={S[table-format=2.1]}, string type},
]{\data}

\end{adjustbox}
\end{table}

\subsubsection{Operation costs}

A preliminary estimate of the CLIC accelerator operation cost, with focus on the most relevant elements, is presented here. 
The material cost for operation is approximated by taking the cost for spare parts as a percentage of the hardware cost of the maintainable components.
These annual replacement costs are estimated at the level of:
\begin{itemize}
\item 1\% for accelerator hardware parts (e.g.\ modules).
\item 3\% for the RF systems, taking the limited lifetime of these parts into account. 
\item 5\% for cooling, ventilation, electronics and electrical infrastructures etc. (includes contract labour and consumables)
\end{itemize}
These replacement/operation costs represent 116 MCHF per year.

An important ingredient of the operation cost is the CLIC power consumption and the corresponding energy cost, which is discussed in~\ref{sect:IMP_Power} below. 
This is difficult to evaluate in CHF units, as energy prices are likely to evolve. The expected energy consumption of the 380\,GeV CLIC accelerator, operating at nominal luminosity, corresponds to 2/3 of CERN's current total energy consumption. 

Concerning personnel needed for the operation of CLIC, one can assume efforts that are similar to large accelerator facilities operating today. Much experience was gained with operating Free Electron Laser linacs and light-sources with similar technologies.
As CLIC is a normal-conducting accelerator operated at room temperature, one can assume that the complexity of the infrastructure, and therefore the maintenance efforts, compare favourably with other facilities.
The maintenance programme for equipment in the klystron galleries is demanding, but is not expected to impact strongly on the overall personnel required for operation.
The ILC project has made a detailed estimate of the personnel needed to operate ILC, yielding 640\,FTE. This number includes scientific/engineering (40\%), technical/junior level scientific staff (40\%) and administrate staff (20\%) for the operation phase~\cite{Adolphsen:2013kya,Evans:2017rvt}. The difference between a 250\,GeV and a 500\,GeV ILC implementation was estimated to be 25\%. In the framework of CERN, these numbers would distribute across scientific/engineering/technical staff, technical service contracts, fellows and administrative staff. 
The level of CLIC operational support required is expected to be similar to the ILC estimates.  

Given the considerations listed above, one can conclude that operating CLIC is well within the resources deployed for operation at CERN today. 
Operating CLIC concurrently with other programmes at CERN is also technically possible. This includes LHC, as both
accelerator complexes are independent. Building CLIC is not destructive with respect to the existing CERN accelerator complex. Electrical grid connections are also independent. The most significant limitation will therefore be the resources, in particular personnel and overall energy consumption.

\subsection{Power and energy consumption}
\label{sect:IMP_Power}

The nominal power consumption at the 380\,GeV stage has been estimated based on the 
detailed CLIC work breakdown structure.
This yields for the drive-beam option a total of 168\,MW for all accelerator systems and services,
taking into account network losses for transformation and distribution on site. The breakdown 
per domain in the CLIC complex (including experimental area and detector) 
and per technical system is shown 
in the left part of~\ref{fig_IMP_11}. 
Most of the power is used in the drive-beam and main-beam injector complexes, comparatively little in the main linacs. 
Among the technical systems, the RF represents the major consumer.
For the klystron-based version the total power consumption is very similar at 164\,MW as shown in the right part of~\ref{fig_IMP_11}.

These numbers are significantly reduced compared to earlier estimates due to optimisation of the injectors for 380\,GeV,
introducing optimised accelerating structures for this energy stage, significantly improving the RF efficiency, and consistently using
the expected operational values instead of the full equipment capacity in the estimates. 
For the 1.5 and 3.0\,TeV stages these improvements have not been studied in detail and the power estimates from the CDR are used~\cite{cdrvol3}. 

\begin{figure}[h!]
\centering
\begin{adjustbox}{width=\linewidth}
\begin{tikzpicture}[font=\sffamily,lines/.style={draw=none},scale=1,align=left]
\sansmath
\pie [
text = legend,
radius = 4.5,
sum=auto,
every only number node/.style={text=white},
style={lines},
pos={0,0},
    color={
    blue!60, 
    cyan!60, 
    yellow!60, 
    orange!60, 
    red!60, 
    blue!60!cyan!60, 
    cyan!60!yellow!60, 
    red!60!cyan!60, 
    red!60!blue!60, 
    orange!60!cyan!60 
    },
] {
1/  Main-beam injectors,
1 /  Main-beam damping rings,
1 /  Main-beam booster and transport,
1 /  Drive-beam injectors,
1 /  Drive-beam frequency multiplication and transport,
1 /  Two-beam acceleration,
1 /  Main linacs (klystron),
1 /  Interaction region,
1 /  Infrastructure and services,
1 /  Controls and operations
}

\pie [rotate = 90,
radius = 4.5,
sum=auto,
every only number node/.style={text=white},
style={lines},
    color={
    blue!60, 
    cyan!60, 
    yellow!60, 
    orange!60, 
    red!60, 
    blue!60!cyan!60, 
    red!60!cyan!60, 
    red!60!blue!60, 
    orange!60!cyan!60 
    }
] {
6/  ,
53/ ,
9/ ,
45/ ,
14/ ,
3/ ,
4/ ,
33/ ,
1/ 
}
\pie [rotate = 90,
radius = 4.5,
sum=auto,
every only number node/.style={text=white},
style={lines},
pos={18,0},
    color={
    blue!60, 
    cyan!60, 
    yellow!60, 
    cyan!60!yellow!60, 
    red!60!cyan!60, 
    red!60!blue!60, 
    orange!60!cyan!60 
    }
] {
6 /,
41 /,
8 / ,
75 /,
4 / ,
29 / ,
1 / 
}
\node at (18,5) {\Large Klystron-based option: 164\,MW};
\node at (0,5) {\Large Drive-beam option: 168\,MW};
\end{tikzpicture}
\end{adjustbox}
\caption{\label{fig_IMP_11} Breakdown of power consumption between different domains of the CLIC accelerator in \si{\MW} at a centre-of-mass energy of \SI{380}{\GeV}, for the drive-beam option on the left and for the klystron option on the right. The contributions add up to a total of \SI{164}{\MW} and \SI{168}{\MW} in the two cases. \imcl}
\end{figure}
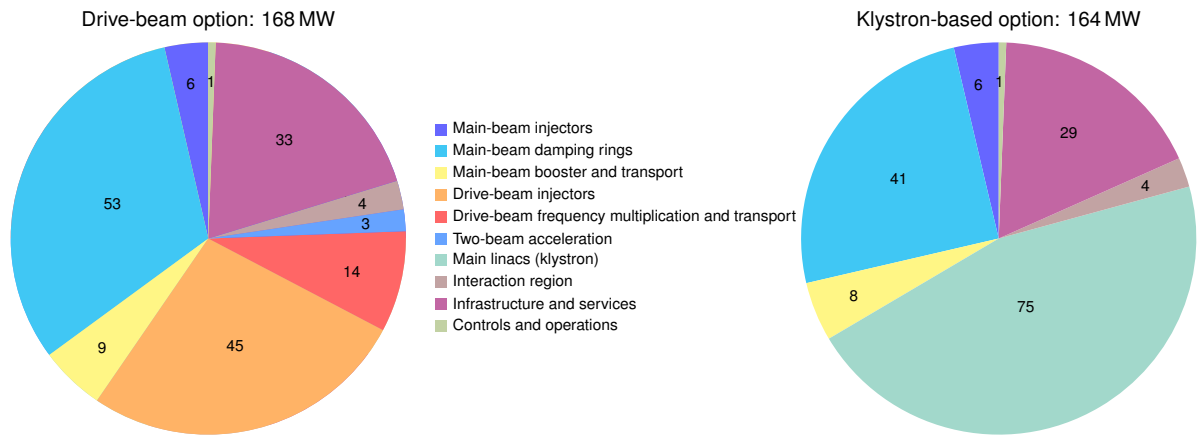

\begin{table}[ht]
\caption{Estimated power consumption of CLIC at the three centre-of-mass energy stages and for different operation modes. The \SI{380}{\GeV} numbers are for the drive-beam option and have been updated as described in \ref{sect:IMP_Power}, whereas the estimates for the higher energy stages are from~\cite{cdrvol3}.}
\label{Tab:Power}
\centering
\begin{tabular}{S[table-format=4.0]S[table-format=3.0]S[table-format=2.0]S[table-format=2.0]}
\toprule
{Collision energy [\si{\GeV}]}  & {Running [\si{\MW}]} &  {Standby [\si{\MW}]} & {Off [\si{\MW}]} \\
\midrule
380     &  168 & 25   & 9 \\
1500    &  364  & 38  & 13 \\
3000    &  589  & 46  & 17 \\
\bottomrule
\end{tabular}
\end{table}

\ref{Tab:Power} shows the nominal power consumption in three different operation modes of CLIC, including the "running" mode at the different energy stages, 
as well as the residual values for two operational modes corresponding to short ("standby") and long ("off") beam interruptions. 
Intermediate power consumption modes exist, for example when a part of the complex is being tested, or during 
transitional states as waiting for beam with RF on. 
The contribution of these transitional states to the annual energy consumption is dealt with by averaging between "running" and "standby" 
for certain periods, 
as described below.

\subsubsection{Energy consumption}
\begin{figure}[t]
\centering
\begin{adjustbox}{width=0.6\linewidth}
\begin{tikzpicture}[font=\sffamily,lines/.style={draw=none},]
\sansmath
\pie [rotate = 90,
scale font=false,
radius = 4,
text = legend,
sum=auto,
every only number node/.style={text=white},
style={lines},
    color={
    red!60,
    orange!60,
    yellow!60,
    red!60!blue!60,
    red!60!cyan!60,
    cyan!60!yellow!60
    },
] {
120 / Annual shutdown,
30 / Commissioning,
10 / Technical stops,
20 / Machine development,
46 / Fault induced stops,
139 / Data taking
}
\end{tikzpicture}
\end{adjustbox}
\caption{\label{fig_IMP_12} Operation schedule in a "normal" year (days/year). \imcl}
\end{figure}
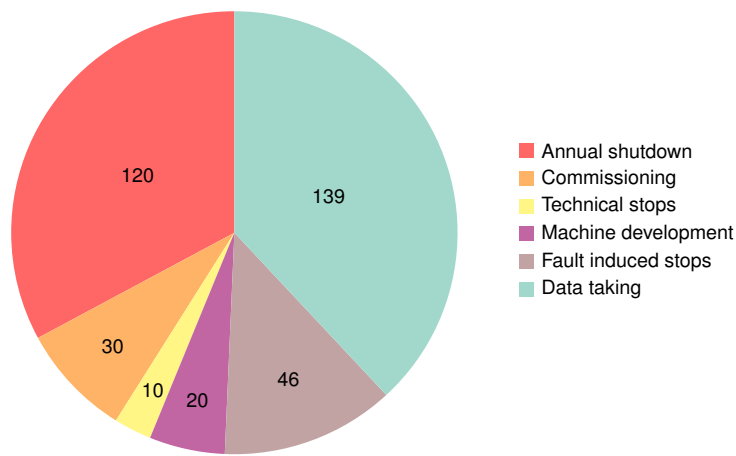

Estimating the yearly energy consumption from the power numbers requires an 
operational scenario, which is detailed in~\cite{Bordry:2018gri} and depicted in~\ref{fig_IMP_12}. In any "normal" year, i.e.\ once CLIC has been fully commissioned, 
the scenario assumes 120 days of annual shutdown, 30 days for beam-commissioning, and 30 days of scheduled maintenance, 
including machine development and technical stops (typically 1 day per week, or 2 days every second week). This leaves 185 days of operation for physics, for which 75\% availability is assumed,
 i.e.\ 46 days of fault-induced stops. This results in 139 days, or 1.2~$\times$~10$^7$ seconds, per year for physics data taking. 

In terms of energy consumption the accelerator is assumed to be "off" during 120 days and "running" during 139 days. 
The power consumption during the remaining time, covering commissioning, technical stops, machine development and fault-induced stops 
is taken into account by estimating a 50/50 split between "running" and "standby".
In addition, one has to take reduced operation into account in the first years at each energy stage to allow systematic tuning up of all parts of the accelerator complex. A luminosity ramp-up of three years (10\%, 30\%, 60\%) in the first stage and two years (25\%, 75\%) in subsequent CLIC stages is considered.
For the energy consumption estimate we change the corresponding reduction in "running" time to a 50/50 mixture of the two states mentioned above, resulting in a corresponding energy consumption ramp-up.

The evolution of the resulting electrical energy consumption over the years is illustrated in~\ref{fig_IMP_13}. For comparison, CERN's current energy consumption is  approximately 1.2\,TWh per year, of which the accelerator complex uses around 90\%.

\begin{figure}[h!]
\centering
\includegraphics[width=0.5\textwidth]{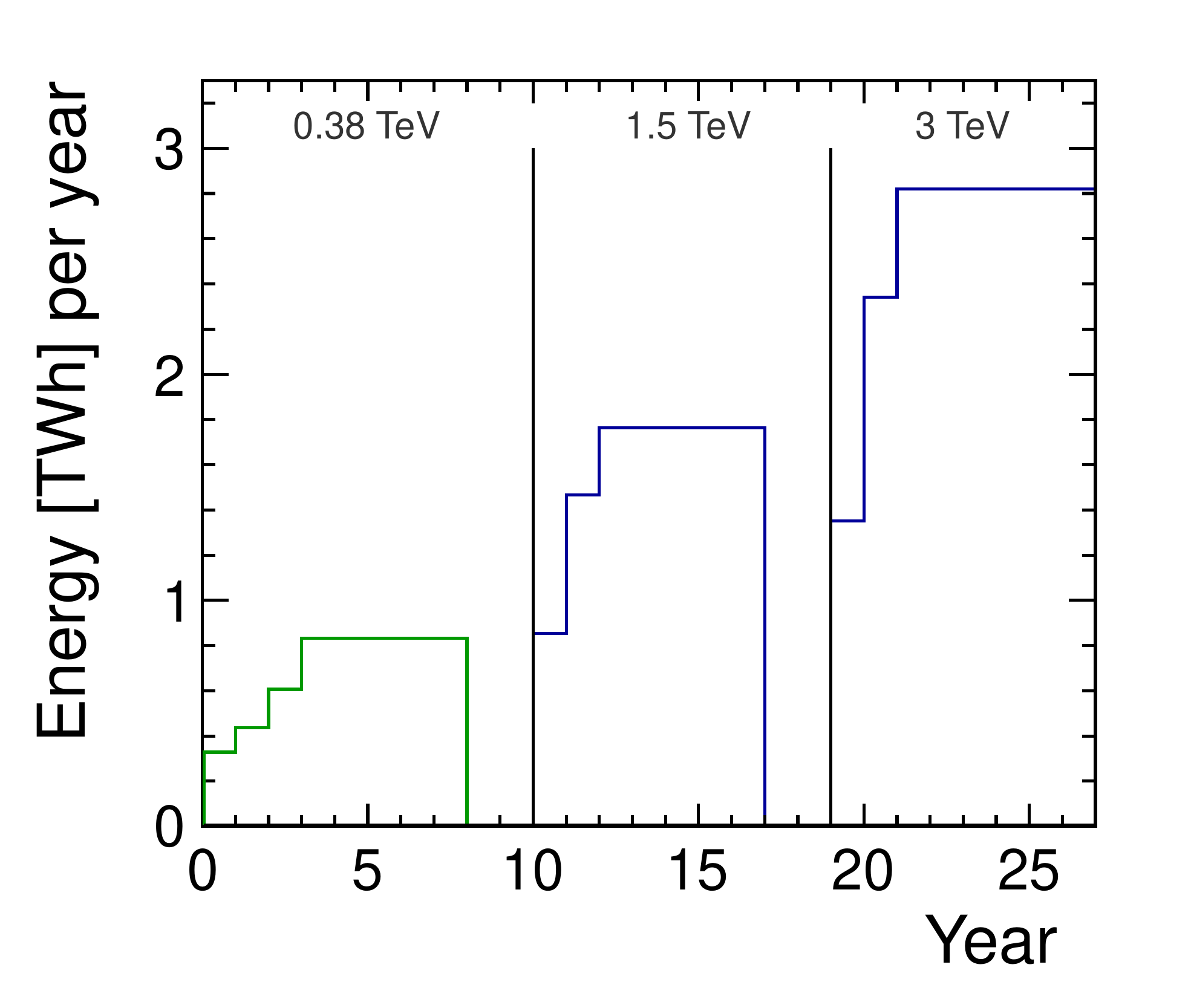}
\caption{\label{fig_IMP_13} Estimated yearly energy consumption of CLIC. The initial stage estimate is revised in detail (green), while numbers for the higher energy stages are from~\cite{cdrvol3} (blue). \imcl}
\end{figure}

\subsubsection{Power reduction studies and future prospects}
Since the CDR~\cite{cdrvol1} in 2012 the CLIC collaboration has systematically explored power reduction and technical system optimisation across the complex. As a result the power estimate is reduced by around 35\% for the initial stage. The main contributors to the reduced estimate are:
\begin{itemize}
\item The accelerating structures were optimised for 380\,GeV and corresponding luminosity, impacting among others on RF power needs and the machine length. The optimisation was done for cost but it was also shown that cost and power are strongly correlated.
\item The injector systems and drive-beam facility were optimised to the 380 GeV parameters taking into account R\&D on various technical systems, for example reducing the number of drive-beam klystrons to around 60\% of earlier designs.
\item High efficiency klystron studies have reached a maturity such that 70\% efficiency can be taken as the baseline.
\item Permanent magnets can partly replace electromagnets.
\item Nominal settings of RF systems, magnets and cooling have consistently been used, analysing the power consumption when running at full luminosity. This replaces earlier estimates which, in some cases, were based on maximum equipment capacity.
\end{itemize}

In summary, the estimate of the power consumption can be considered to be detailed and complete for the initial 380 GeV stage. The estimates for the higher energy stages have not been scrutinised in order to include the saving measures listed above. Also for the initial stage further work can lead to additional savings.
This concerns in particular the damping ring RF power, where further studies are needed before a revised baseline can be introduced. 
The total power consumption of the damping rings (53\,MW) is dominated by the RF system (45\,MW). 
In the present design, the power efficiency of the RF system is rather low due to high peak power requirements for compensation of transient beam loading effects. 
Work is ongoing to improve the design and reduce the peak RF power requirements by introducing an optimum modulation of both phase and amplitude of the input RF signal. 
This may result in a significant (up to a factor of 2) reduction of the damping ring RF system power consumption, potentially reducing the overall damping ring power consumption to around 30\,MW.

\newpage
\section{Future opportunities}
\label{sec:opportunities}

A key advantage of a linear collider is the extendibility in energy. CLIC can provide electron-positron collisions with centre-of-mass energies of up to 3\,TeV. The usage of novel approaches for an upgrade of CLIC might allow to reach even higher centre-of-mass energies. In the following section, the physics motivation and possible accelerator technologies for such an upgrade are introduced.

\subsection{Physics motivation}

An increase of the centre-of-mass energy beyond 3\,TeV would enhance the physics potential even further beyond the capabilities of the baseline CLIC programme described in \autoref{sec:physics}. The aim of such a collider would be direct and indirect searches for phenomena beyond the Standard Model. The discussion that follows focuses on the motivation for a 10\,\TeV electron-positron collider.

The centre-of-mass energy dependence in the range up to 30\,TeV for many important Standard Model processes in electron-positron collisions is shown in \autoref{fig:sm_cross_sections_30tev}. Above the kinematic threshold, the cross sections for Higgsstrahlung and two-fermion production (e.g.\ $\epem \to \PQt\PAQt$) scale as $1/s$. A similar energy dependence is visible for $\PW$-boson pair production. This is a first indication that the desired integrated luminosities at 10\,TeV would exceed those for the baseline CLIC energy stages.

\begin{figure}[h]
\centering
\begin{minipage}[l]{0.59\textwidth}
\includegraphics[width=1.0\textwidth]{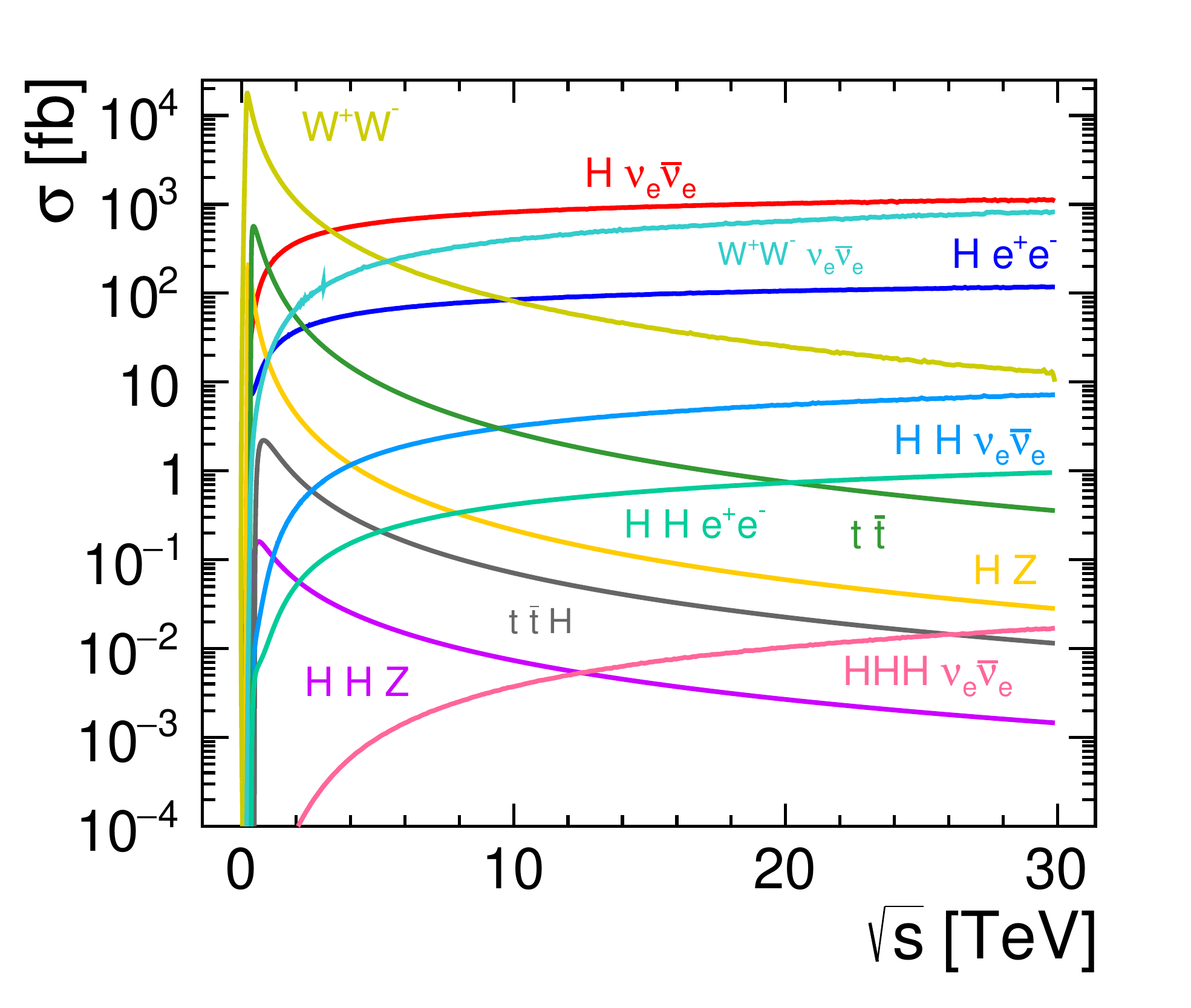}
\end{minipage}
\caption{Cross section as a function of centre-of-mass energy for the main Standard Model processes at a very high-energy $\epem$ collider. The values shown correspond to unpolarised beams and include the effect of Initial State Radiation (ISR). \imdp}
\label{fig:sm_cross_sections_30tev}
\end{figure}

On the other hand, the rate of events with final states produced in $\PW\PW$ or $\PZ\PZ$ boson fusion rises approximately as $\log(s)$. For example, the cross section of the dominant contribution to double-Higgs production, $\epem \to \PH\PH\PGne\PAGne$, is about a factor 4 larger at 10\,TeV compared to 3\,TeV. Although the dependence of the cross section on the Higgs self-coupling decreases somewhat with energy, a significant improvement of the knowledge of the Higgs self-coupling is expected for an integrated luminosity of a few ab$^{-1}$ at 10\,TeV. Even higher centre-of-mass energies of a few tens of TeV would also give access to triple Higgs production.

The indirect sensitivity to New Physics of Higgs and $\PWp\PWm$ production is illustrated using Standard Model effective field theory (see also \autoref{sec:physics_bsm}). In \autoref{fig:eft_10tev} the sensitivities of the three baseline energy stages of CLIC are compared to 4\,ab$^{-1}$ collected at a 10\,TeV $\epem$ collider. The sensitivies to the scales of four dimension-6 operator coefficients, defined as $\Lambda/\sqrt{c}$, are shown. The results are based on the fit described in~\cite{Ellis:2017kfi}, with the linear dependence on the coefficients now computed more accurately. The projections used as input are largely obtained from benchmark analyses based on full detector simulations~\cite{ClicHiggsPaper}. The projections for 3\,TeV are extrapolated to 10\,TeV assuming that the shape of the beamstrahlung spectrum is the same for both energies. Generally, new physics scales well beyond the centre-of-mass energy of the collider can be probed. The 10\,TeV stage enhances the reach for some operators by almost a factor 2 compared with 3\,TeV. In particular, the measurement of the Higgsstrahlung cross section at the highest possible energy is important for the reach on $\bar{c}_W-\bar{c}_B$, $\bar{c}_{HW}$ and $\bar{c}_{HB}$. The reach on $\bar{c}_{3W}$ shown here decreases at higher energy due to helicity suppression of the linear interference term, but will also grow with energy at the quadratic level or if the interference is recovered by suitable differential measurements.

A very high-energy $\epem$ collider also provides unique opportunities for direct searches for new states. In \autoref{fig:higgsino_10tev} the number of generic Higgsino (doublet of massive Dirac fermions with hypercharge 1/2) pair production events is shown as a function of the Higgsino mass for different assumptions on the integrated luminosity. Due to the absence of QCD backgrounds, $\epem$ collisions are especially suitable for the discovery of electroweak states. The number of events produced is independent of the Higgsino mass except very close to the kinematic threshold. Hence a discovery would be possible for masses of almost up to 5\,TeV, which exceeds the capabilities of a hadron collider even with a centre-of-mass energy of the order of 100\,TeV. A percent-level measurement of the Higgsino pair production cross section would typically be possible with an integrated luminosity of a few ab$^{-1}$.

\begin{figure}[h]
\centering
\begin{subfigure}{0.52\textwidth}
  \centering
  \includegraphics[width=\textwidth]{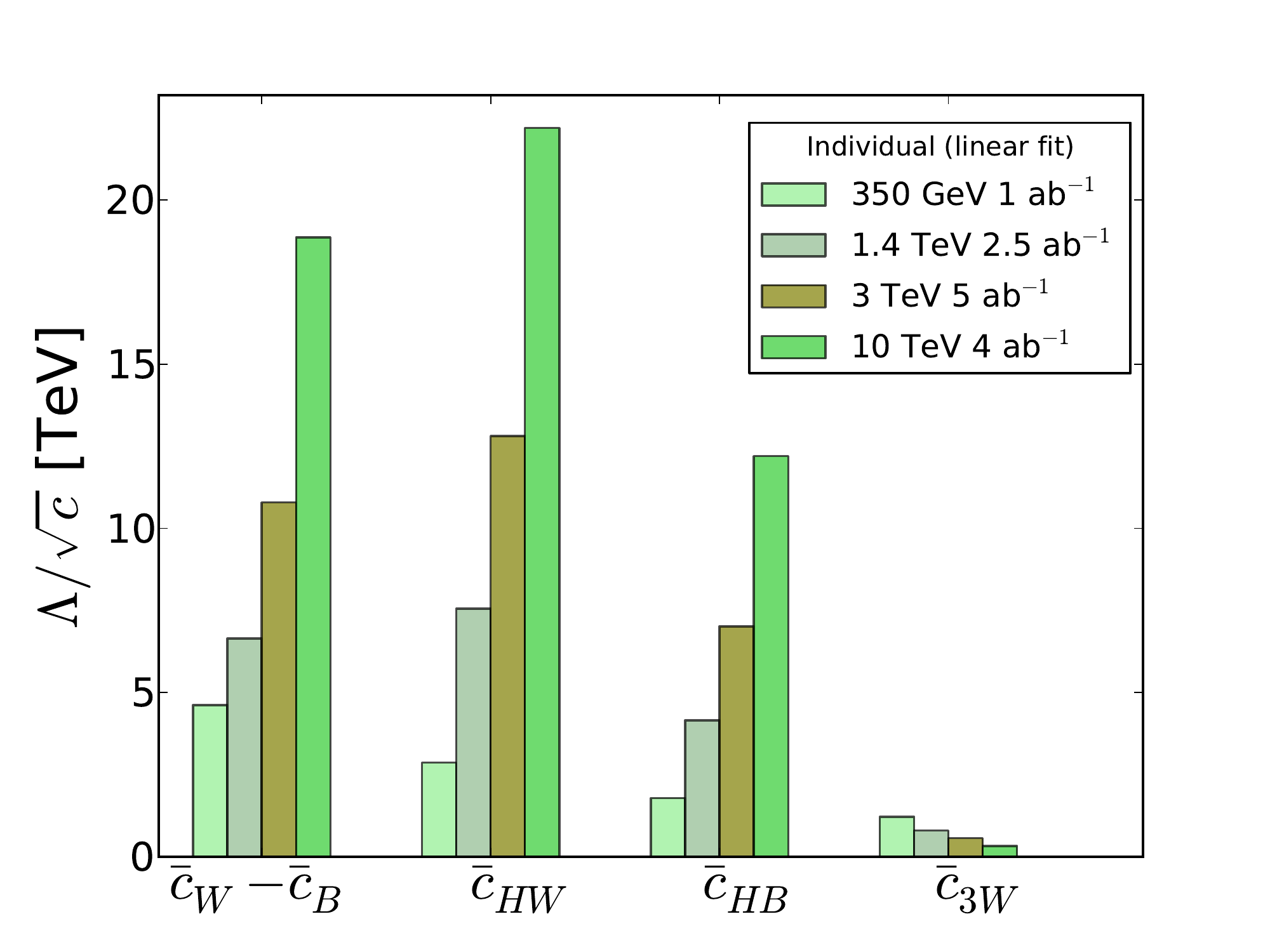}
  \caption{}
  \label{fig:eft_10tev}
\end{subfigure}
\begin{subfigure}{0.45\textwidth}
  \centering
  \includegraphics[width=\textwidth]{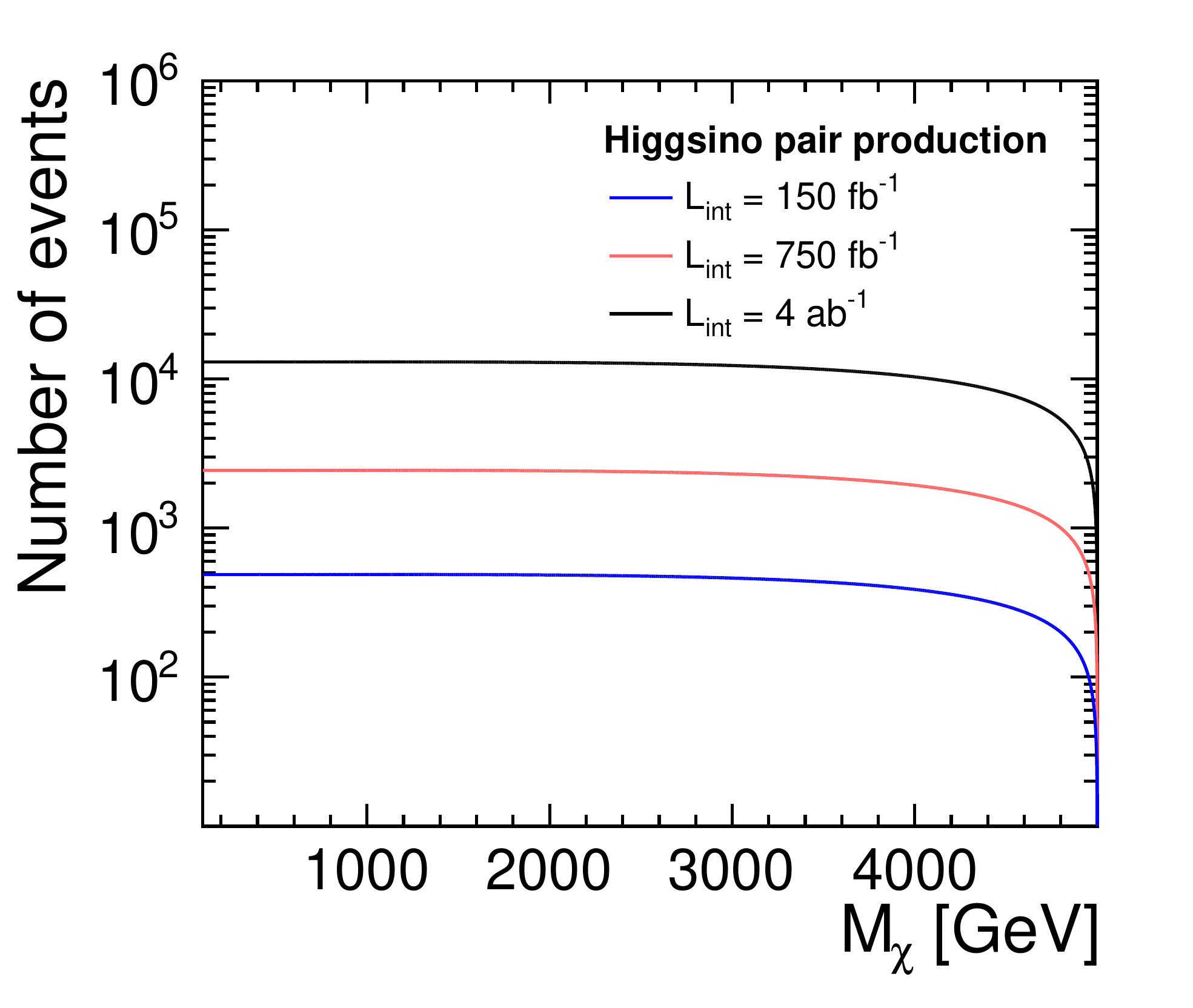}
  \caption{}
  \label{fig:higgsino_10tev}
\end{subfigure}
\caption{Examples for the New Physics potential of a 10\,TeV $\epem$ collider at \roots = 10\,TeV. (a) Sensitivities of Higgs boson and $\PW^{+}\PW^{-}$ production at 350\,GeV, 1.4\,TeV, 3\,TeV and 10\,TeV to the scales of various dimension-6 operator coefficients (based on~\cite{Ellis:2017kfi}). (b) Number of Higgsino pair production events as a function of the Higgsino mass for three different assumptions on the integrated luminosity at \roots = 10\,TeV \imdp.}
\label{fig:bsm_sensitivity_10tev}
\end{figure}

In conclusion, the sensitivity of an $\epem$ collider for new phenomena increases strongly with its centre-of-mass energy. A 10\,TeV collider collecting an integrated luminosity of a few ab$^{-1}$ would provide unique physics capabilities. This programme would be complementary to the baseline energy stages of CLIC.

Although the results shown in this section assume unpolarised beams, electron beam polarisation would enhance the capabilities of a 10\,TeV collider in a similar way as for the baseline CLIC energy stages. Left-handed electron beam polarisation would significantly enhance the cross sections for double Higgs and Higgsino production. Many indirect searches profit from precision measurements of polarisation asymmetries.

\subsection{Opportunities for extension based on future technologies}
CLIC technology is mature and can provide collision energies in the range of up to 3\,TeV, at affordable cost at a site close to CERN.
In the future, novel technologies may make it possible to extend the CLIC energy range, for example to 10\,TeV.
In particular, novel acceleration technologies could replace parts of CLIC technology in the main linacs in the future,
even if they are currently not yet mature enough for collider applications.
The two main relevant technologies are dielectric accelerating structures and acceleration using plasma.
Both can either use a laser to produce the accelerating field or an electron beam.
Also the use of protons is being investigated to produce high fields in a plasma~\cite{c:awake}, however no concept
exists at this time for an electron-positron collider based on this technology.

These technologies must have high power efficiency and maintain excellent beam quality in order
to achieve a luminosity at 10\,TeV that is similar to the CLIC performance at 3\,TeV. The corresponding studies are only beginning and important work is required before conclusions could be drawn 
on the feasibility of reaching the luminosity goals.
However, the CLIC design work aims to ensure that the CLIC collider is consistent with upgrades using such technologies. Therefore a dedicated CLIC working group was
established to ensure this goal. As the technologies are not mature, no detailed design can be made, and only a general compatibility can be ensured.

\subsubsection{General concept}
The proposed concepts for novel technology colliders consist of two linacs pointing at each other, similar to CLIC
except for the technology used in the main accelerator.

The CLIC tunnels, hosting the main linac and BDS of each beam,
are laser straight and cross at an angle of 20\,mrad~\cite{Schulte2001}.
The crossing angle is optimal for collisions at 3\,TeV, and is
likely to be a good choice for higher energy collisions as well.
The configuration avoids bent beam trajectories except what is required for the functionality of the beam delivery system.
This important feature maintains the beam quality.
Therefore all of the tunnels and corresponding infrastructures can continue to be used. This includes the
main linacs, the beam delivery system, the detector hall and the post collision line.

As an example, if after the operation of the 1.5\,TeV CLIC stage, the copper structures in the main linacs were replaced
by dielectric or plasma-based acceleration with an effective gradient of 1\,GV/m, one would be able to reach 10\,TeV. The beam delivery system would certainly also need to be upgraded for this energy.

The single bunch parameters used by the novel concepts are similar to those for CLIC, with the exception of the
laser-driven dielectric acceleration. Hence, one can expect that the CLIC injectors, which provide 9\,GeV low emittance
electron and positron beams, can be re-used. However, some modifications might be required
to adapt to a time structure and bunch length optimised for the future technologies. 
The injectors are an important part of the overall cost at 380\,GeV.

\subsubsection{Dielectric accelerating structures}
The dielectric structures are in principle hollow cylinders with a dielectric coating
and are studied in different laboratories for RF frequencies ranging from X-band
to THz~\cite{Cros2017}.

A collaboration led by Argonne National Laboratory explores beam-driven X-band acceleration, i.e. a frequency very similar to CLIC.
It might be possible that the gradient exceeds that of the current CLIC technology and the structures are cheaper to fabricate.
The concept is similar to the CLIC scheme and uses a drive beam
to generate the RF power. Argonne and the CLIC collaboration are assessing the ultimate performance of
this technology and the options to use it for CLIC energy upgrades.

One can imagine the following scenarios of using this technology in CLIC.
If higher gradients can be reached at sufficient RF pulse lengths, 
CLIC components could be replaced in an existing main linac with higher gradient versions to increase the energy.
Even if the technology does not reach higher performance but is cheaper, one can use it to reduce the
cost of the energy upgrades.
In these scenarios a large part or even all of the CLIC complex can be reused, including the drive beam, and the difference with the use of the current CLIC technology is very small.
The luminosities and time structure are expected to be similar to the CLIC values.

If studies show that the RF pulse length would have an optimum significantly different from the current CLIC value, some effort is required to
adapt the injectors and the drive-beam complex accordingly, but one can still expect to reuse the largest part of the collider.

Direct laser acceleration in silicon chips has been proposed -- the ``system on a chip'' type technology~\cite{England2014}.
The suggested beam parameters differ strongly from those of CLIC, and an assessment of a potential upgrade path is therefore challenging
at this point. It might be that mainly the infrastructure could be reused.

\subsubsection{Plasma-based acceleration}
Beam-driven plasma wakefield accelerators (PWFA) use an electron drive beam to deplete a region of a plasma from electrons
in order to generate strong electric fields that accelerate one bunch of the main beam. They have demonstrated
very high gradients of more than 50\,GV/m over almost a metre~\cite{Blumenfeld2007} and good efficiency for transferring power from the drive beam to the main beam
of more than 30\%~\cite{Litos2014}. A reasonable goal for the effective gradient, i.e. including the filling factor, is 1\,GV/m~\cite{Cros2017}. A tentative
proposed scheme can be found in~\cite{Adli2013}.

To produce sufficient luminosity, the total beam current in plasma-based colliders must be of the same order of magnitude as in the CLIC case.
The same holds for the bunch charge. Differently from CLIC, only single bunches are accelerated within each pulse. Therefore
the time between subsequent collisions in the interaction point is much larger than the 0.5\,ns in CLIC and rather of the order of $100\,\upmu\text{s}$.
This might require some modifications of the main-beam generation complex but it is likely that a large part could be reused.

One can consider using the CLIC drive-beam complex to generate the high-power drive beams for the plasma acceleration.
This might require additional return arcs, the rearrangement of the linac components, and potentially the replacement of the accelerating structures with
modified versions. However, the costly RF power system can be reused fully. An example can be found in~\cite{Schulte2017}.

Laser-driven plasma wakefield accelerators~\cite{Leemans2014} work in a similar fashion, except that
a laser beam replaces the drive beam to generate the fields. So one can still anticipate that a similar fraction of the CLIC complex can be maintained
except for the drive-beam complex.

While dielectric-based acceleration can be applied to electrons and positrons equally well, this is not necessarily the case for plasma-based
acceleration.  When the electron or laser drive beam passes the plasma, the electrons are expelled. The remaining positive ions
focus an electron main beam but they defocus a positron main beam. Possible solutions to this problem are being studied, such as hollow plasmas
or the acceleration of the positrons at a longitudinal location with a high density of electrons close to the centre.
However, the validity of these solutions remains to be demonstrated and the electron and positron linacs might have important differences.
An alternative approach that is being studied is to use electron beams in both linacs and convert them to
very high energy photons with a laser. Depending on the exact parameter choices, the colliding photon beams would typically carry up to about 80\% of the
original beam energy and reach luminosities that are comparable to the positron-electron options. The energy spectrum would be significantly broader. The CLIC interaction region design is compatible with
housing a $\gamma$--$\gamma$ collider~\cite{Telnov2005}.
Another option that was considered is to use a conventional linac for the positrons and plasma acceleration for the electrons. In this case the positron energy might be lower than the electron energy, which is less interesting from a physics point of view. The gain in centre-of-mass energy would be compromised and the large event boost would pose additional demands on the detector.

\subsubsection{Luminosity enabling technologies}
High luminosity is key for high energy colliders and requires excellent beam quality.
Studies of the beam quality are in their very early stages for the novel technologies, with the exception of the
beam-driven dielectric acceleration. 
However, tolerances on alignment and stability of the beam and the components will be very tight in both the transverse and longitudinal planes, considerably exceeding the requirements for CLIC~\cite{Schulte2016}. 
Whether these can be achieved is one of the main feasibility issues for collider studies based on novel technologies.

For CLIC, significant effort and resources have been put into developing technologies that address the beam quality.
These include the demonstration of a system that can stabilise the CLIC final quadrupoles to the sub-nanometre scale~\cite{Janssens:2015ekh,c:stab2},
the demonstration of phase feedback on the 50\,fs level~\cite{c:phasetest}, the development
of beam based alignment techniques~\cite{Latina2014,Pfingstner2014}, 
and metrology and static alignment techniques~\cite{Pacman2014}.
The development of these new methods and precision tools in the context of CLIC are important steps towards the future use of novel acceleration technologies.
In addition, the experience gained by the operation of CLIC will be the foundation for addressing the same issues at the
even more challenging level required for novel acceleration technologies.

\cleardoublepage
\textcolor{white}{ }
\newpage

\section{CLIC objectives for the period 2020--2025}
\label{sec:objectives}

\subsection{Accelerator complex}
\label{sect:NextPhase}

The project implementation for CLIC foresees an initial five-year preparation phase prior to a  construction start envisaged by 2026. 
The overall schedule towards first beams by 2035 is shown in~\ref{fig_IMP_14}. 
This leaves a 2-year margin in addition to the construction and commissioning period estimated in the technology-driven schedule shown in~\ref{fig_IMP_9}.
\begin{figure}[h!]
\centering
\includegraphics[width=\textwidth]{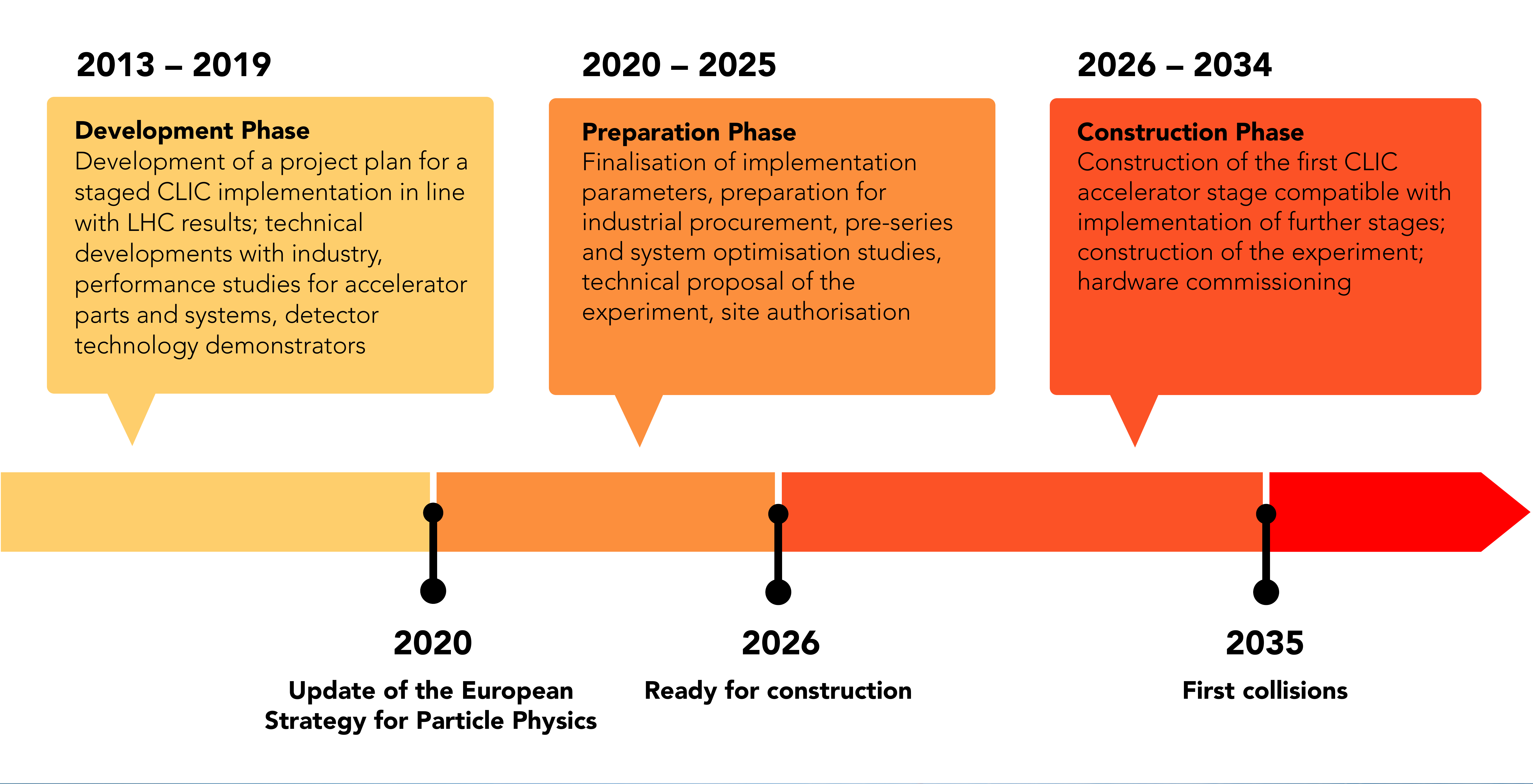}
\caption{Schematic view of the CLIC implementation schedule, with first collisions in 2035. \imcl}
\label{fig_IMP_14}
\end{figure}

In order to analyse the priorities for the preparation phase, the following project risks and mitigations have been considered:
\begin{itemize}
\item  Performance: The dominant performance risk is related to the luminosity.  Luminosity performance is based on technical performance and
reliability as well as design robustness and system redundancy. Risk mitigation implies further studies at design and technical level, including on variation of parameters such as temperatures, mechanical instabilities and vibrations, magnetic fields, etc. Most importantly, performance validations in normal-conducting Free Electron Laser (FEL) Linacs and other compact linac systems will provide powerful demonstrations and new benchmarks for reliability, technical parameters, simulation and modelling tools on the timescale of 2020--2025.  
\item  Technical systems: 
The main technical risks are related to RF sources, the X-band components, and overall system integration for the main linac. 
Reliable, efficient and cost-effective klystrons, modulators and X-band structures are components which are crucial for the machine. 
Additional thermo-mechanical engineering studies of the main linac tunnel, integrating all components, are important in order to further improve the understanding
of the mechanical and thermal stability needed for CLIC. 
In addition, further system tests (beyond what has been achieved with CTF3) of the high-power drive beam would be most desirable.
\item  Implementation: 
Principal risks are associated with the industrial production of large numbers of  modules and the civil engineering. 
Work during the preparation phase includes qualifying companies for industrial production and optimising the work distribution and 
component integration. The module installation and conditioning procedures need to be refined and further verified. 
Cost control is crucial and is an integral part of these studies. This requires work on optimising the 
risk sharing models between industry, CERN and collaborative partners for the most critical and costly components.
Detailed site-specific design work related to civil engineering and infrastructure needs to be performed.
\end{itemize}

\subsubsection{Accelerator programme overview} 
To address these issues the forthcoming preparation phase will comprise further design, technical and industrial developments,
with a focus on cost, power and risk reduction, in preparation for the Technical Design Report. System verifications in FEL linacs and low emittance rings will be increasingly important. The governance structure and the international collaboration agreements for the construction phase will be prepared during this time. 

Civil engineering and infrastructure preparation will become increasingly detailed during the preparation phase.
An environmental impact study and corresponding public enquiry will be needed as a prerequisite to authorisations for construction. 
Experience from the LEP and LHC projects indicates that approximately two years will be needed for such formal steps, as required by the procedures in the CERN host states. 

The key elements of the CLIC accelerator activities during the period 2020--2025 are summarised in~\ref{tab:CLIC2025}.

\begin{table}[h]
\begin{center}
\caption{Main CLIC accelerator objectives and activities in the next phase.}
\label{tab:CLIC2025}
\small
\begin{tabular}{c c}
\toprule
\textbf{Activities} & \textbf{Purpose} \\
\midrule
\multicolumn{2}{c}{\textbf{Design and parameters}}\tabularnewline  
\begin{minipage}[t]{0.46\columnwidth}
Beam dynamics studies, parameter optimisation, cost, power, system verifications in linacs and low emittance rings
\end{minipage} &
\begin{minipage}[t]{0.46\columnwidth}
Luminosity performance and reduction of risk, cost and power
\end{minipage} \tabularnewline
\\
\multicolumn{2}{c}{\textbf{Main linac modules}}\tabularnewline 
\begin{minipage}[t]{0.46\columnwidth}
Construction of 10 prototype modules in qualified industries, two-beam and klystron versions, optimised design of the modules with their supporting infrastructure in the main linac tunnel
\end{minipage} & %
\begin{minipage}[t]{0.46\columnwidth}
Final technical design, qualification of industrial partners, production models, performance verification 
\end{minipage} \tabularnewline
\\
\multicolumn{2}{c}{\textbf{Accelerating structures}}\tabularnewline
\begin{minipage}[t]{0.46\columnwidth}
Production of $\sim50$ accelerating structures, including structures for the modules above
\end{minipage} & %
\begin{minipage}[t]{0.46\columnwidth}
Industrialisation, manufacturing and cost optimisation, conditioning studies in test-stands  
\end{minipage} \tabularnewline
\\
\multicolumn{2}{c}{\textbf{Operating X-band test-stands, high efficiency RF studies}}\tabularnewline  
\begin{minipage}[t]{0.46\columnwidth}
Operation of X-band RF test-stands at CERN and in collaborating institutes for structure and component optimisation, further development of cost-optimised high efficiency klystrons 
\end{minipage} & %
\begin{minipage}[t]{0.46\columnwidth}
Building experience and capacity for X-band components and structure testing, validation and optimisation of these components, cost reduction and increased industrial availability of high efficiency RF units 
\end{minipage} \tabularnewline
\\
\multicolumn{2}{c}{\textbf{Other technical components}}\tabularnewline
\begin{minipage}[t]{0.46\columnwidth}
Magnets, instrumentation, alignment, stability, vacuum   
\end{minipage} &
\begin{minipage}[t]{0.46\columnwidth}
Luminosity performance, costs and power, industrialisation 
\end{minipage} \tabularnewline
\\
\multicolumn{2}{c}{\textbf{Drive-beam studies}}\tabularnewline  
\begin{minipage}[t]{0.46\columnwidth}
Drive-beam front-end optimisation and system tests to $\sim20\,\text{MeV}$
\end{minipage} &
\begin{minipage}[t]{0.46\columnwidth}
Verification of the most critical parts of the drive-beam concept,  further development of industrial capabilities for L-band RF systems
\end{minipage} \tabularnewline
\\
\multicolumn{2}{c}{\textbf{Civil Engineering, siting, infrastructure}}\tabularnewline  
\begin{minipage}[t]{0.46\columnwidth}
Detailed site specific technical designs, site preparation, environmental impact study and corresponding procedures in preparation for construction
\end{minipage} &
\begin{minipage}[t]{0.46\columnwidth}
Preparation for civil engineering works, obtaining all needed permits, preparation of technical documentation, tenders and commercial documents
\end{minipage} \tabularnewline
\\
\bottomrule
\end{tabular}
\end{center}
\end{table}

\subsubsection{Programme implementation, technology demonstrators and collaboration}
The design studies and technical work for CLIC is broadly shared among the CLIC collaboration partners. The CLIC (accelerator) collaboration currently comprises 53 institutes from 31 countries~\cite{clic-study}.

The potential for collaborative projects is increasing with the current expansion in the field of Free Electron Laser (FEL) linacs and next-generation light sources. In particular, the increasing use of X-band technology, either as the main RF technology or for parts of the accelerators (deflectors, linearisers), is of high relevance for the next phase of CLIC.
Construction, upgrades and operation of FEL linacs and conventional light sources, several of which are located at laboratories of CLIC collaboration partners, provide many opportunities for common design and component developments, and for acquiring crucial system test experience. Furthermore, the fact that there are significant resources invested in such accelerators world-wide, provides excellent opportunities for building up industrial capabilities and networks.

X-band RF systems and structure manufacturing used to be exclusively available in the US and Japan. However, today there are fourteen
institutes capable of developing and testing X-band structures. All are working together on optimising the technology.
The increasing number of qualified companies for accelerating structure manufacturing, together with the growing industrial availability
of RF systems, make it easier for new groups to engage in these technologies.
As a consequence, several smaller accelerators using X-band technology are in a proposal or technical preparation phase. In this context it is important to mention the SPARC 1\,GeV X-band linac at INFN~\cite{Diomede:IPAC18}, a possible upgrade of CLARA at Daresbury~\cite{CLARA-upgrade}
and the CompactLight~\cite{CompactLight} FEL study. The CompactLight design study is co-financed by the European Commission. It involves 24 partners preparing technical designs for compact FELs based on X-band linacs at energies ranging from 6\,GeV down
to small room-size systems for X-ray production through Inverse Compton Scattering (e.g.\ SmartLight~\cite{SmartLight}).
Furthermore, the implementation of a 3.5\,GeV X-band linac (eSPS) has been proposed~\cite{Akesson:2640784} at CERN. It aims at injecting electrons into the SPS for further acceleration, and suggests implementing the linac during the period 2019--2024. 

The growing use of CLIC technology allows for implementing
several of the CLIC project activities described in~\ref{tab:CLIC2025} in the form of collaborative projects
together with the projects and technology partners mentioned above. Nevertheless, the principal ingredient to a successful preparation phase for CLIC,
and the ability to team up with such partner projects, is an increase in resources for CLIC at CERN.

\subsection{Detector and physics}
\label{sect:NextPhase_dp}
Until now the CLIC detector and physics studies have essentially covered two phases. Work towards the CDR (2009--2012) focused on understanding the CLIC physics potential, mostly still without input from LHC data, and the experimental conditions at CLIC. Two ILC detector concepts were adapted to CLIC  conditions. Physics studies using these detectors confirmed that CLIC delivers high-precision measurements, despite the luminosity spectrum and the presence of significant beam-induced background. These studies also led to understanding the detailed detector performance requirements, as described in~\ref{sec:detector}. During the period 2013--2018 the focus of the studies changed. In addition to ongoing Linear Collider detector technology R\&D, well adapted to CLIC in the domain of calorimetry, CLIC-specific detector R\&D is being performed in the areas of silicon vertex and tracker R\&D, power pulsing and low-mass cooling, as reported in~\ref{sec:technologies}. The R\&D currently aims principally for validated ''technology demonstrators", rather than full prototypes. Detector simulation studies have led to a single optimised CLIC detector concept together with improved software tools for simulation and event reconstruction (see~\ref{sec:CLICdet,sec:CLICdetPerformance}). Physics studies have followed the evolution of the physics landscape, resulting in a focus on Higgs studies, top-quark studies and the assessment of the CLIC potential for new physics in a context where LHC has not yet identified any clear BSM signals and where the scope for theoretical interpretations is still very large (see~\ref{sec:physics}).

If CLIC prepares for construction to start by 2026, the extent of the detector activities will need to increase significantly. While, owing to limited resources, technology developments currently focus only on the most challenging detectors, in the next phase all aspects of the experiment need to be addressed. Investments and priorities will be driven by time considerations, according to estimated lead times for the R\&D, prototyping, and industrialisation, and taking detector construction schedules into account. For example, general infrastructures, large supporting structures, magnet yoke and magnet coils have to be installed early, followed by the calorimeters, the muon detectors and finally the inner tracking system. Therefore, the vertex and tracking detectors, the readout electronics, the data transmission and computing facilities can still profit from future technology advances, while the design and technologies for the superconducting solenoid coil have to be frozen earlier.

The main areas of activity for the period 2020--2025 will therefore comprise:

\begin{itemize}
\item    Detector engineering, detector integration and technical coordination (including assembly and maintenance scenarios, services, general infrastructures, safety aspects, industrialisation, schedules, costing, etc.);
\item    Superconducting solenoid design (including demonstrators of the various technology aspects and subsequent industrialisation);
\item    Electromagnetic, hadronic and forward calorimeters (design to CLIC specifications, corresponding demonstrators and full prototypes, industrialisation aspects);
\item    Muon detectors (design to CLIC specifications and corresponding demonstrators);
\item    Vertex and tracking detectors (technology development, demonstrators, full modules, full design).
\end{itemize}
These sub-detector projects will include the corresponding on-detector and off-detector electronics developments and data-transmission studies.  Physics studies will continue in parallel, while the development and deployment of software tools will remain in pace with the needs of the project phase. 

In line with the level of objectives, the CLICdp collaboration (currently 30 institutes from 18 countries~\cite{clic-study}) will grow significantly and its structure will evolve accordingly, incorporating the necessary legal and organisational frameworks for agreements on formal commitments and sharing of deliverables.

\clearpage
\section{Summary}
\label{sec:summary}
In this document, the research and development on the Compact Linear Collider is summarised, with emphasis on recent studies and R\&D for the CLIC accelerator complex, improvements to the CLIC detector concept, and developments in the domain of the CLIC physics potential.
CLIC is foreseen to be built and operated in stages. This report provides details of an updated staging scenario,
which is optimised for physics performance, and contains assumptions about commissioning and running time per year
which were recently harmonised with those of other future CERN  projects. 
For the first stage with a centre-of-mass energy of $\roots=380\,\GeV$ an integrated luminosity of 1\,ab$^{-1}$ is foreseen. 
This is followed by operation at 1.5\,TeV with 2.5\,ab$^{-1}$, and by a third stage at 3\,TeV with 5\,ab$^{-1}$  integrated luminosity.
This CLIC physics programme spans over 25--30 years.
The updated baseline also specifies $\pm 80$\% electron polarisation, with the sharing between the two longitudinal polarisation states optimised for the best physics reach at each energy stage.

The construction and operation of CLIC is described, with the two-beam acceleration scheme as baseline scenario.
Normal-conducting high-gradient 12\,GHz accelerating structures are powered via a high-current drive beam. 
The accelerating structures will be operated in the range of 70 to 100\,MV/m, resulting in a total accelerator length of 11\,km for the 380\,GeV stage and 50\,km for 3\,TeV.
For the first energy stage, an alternative scenario is presented, with X-band klystrons powering the main-beam accelerating structures.
Details of an implementation of CLIC near CERN are described, generally with emphasis on the 380\,\GeV stage. These include results on civil engineering studies, construction and upgrade schedules, electrical networks, cooling and ventilation, transport and safety aspects.  

Beam experiments and hardware tests described in this report demonstrate that the CLIC performance goals can be met.
For instance, accelerating gradients of up to 145\,MV/m are reached with the two-beam concept at CTF3,
and breakdown rates of the accelerating structures well below the limit of $3 \times 10^7\text{m}^{-1}$ are stably achieved at X-band test platforms.
High luminosities can be achieved by using nanometre beam sizes. This requires low-emittance beams as well as novel alignment and stabilisation techniques.
There is substantial progress in all of these domains: performances as needed for the CLIC damping rings are achieved by modern synchrotron light sources;
special alignment procedures for the main linac are now available; sub-nanometre stabilisation of the final focus quadrupoles is demonstrated.
In general, beam physics studies, technical developments and system tests for CLIC resulted in significant progress in recent years. Reductions in cost and energy consumption have been among the main objectives of these developments, resulting in a better energy efficiency of the 380\,GeV stage, with power around 170\,MW, together with a lower estimated cost, now around \mbox{\num{6} billion \si{CHF}}.

The CLIC detector layout and the technology choices for the different sub-detectors are described in this report.
The detector characteristics are driven by the CLIC physics programme and by the experimental conditions at CLIC.
CLIC detector simulation studies have led to a new, optimised CLIC detector concept CLICdet, using an improved software suite for event simulation and reconstruction.
CLICdet is optimised for particle flow with a light-weight vertex and tracking system, highly-granular calorimeter systems, 
a 4\,T solenoid and a return yoke equipped with detectors for muon identification.
CLICdet also has very forward calorimeters for luminosity measurements and forward electron tagging.
Due to the beam structure of CLIC with a very low duty cycle below 0.001\%, triggerless readout can be applied, and 
it is possible to operate the sub-detectors with power pulsing and with cooling concepts optimised for very small material budgets.

Detector R\&D activities have validated technology demonstrators for vertex and tracking detectors as well as for the foreseen calorimeter concepts.
The calorimeter R\&D for CLIC is pursued within the CALICE and FCAL collaborations.
Synergies in detector R\&D with other projects are exploited, such as with the HL-LHC detector upgrades.
The validation of the detector technology includes laboratory measurements and test beam experiments.
In addition to tests targeting detector performance parameters, such as the energy or position resolution, the powering and cooling concepts were validated.
Concepts of ultra-light mechanical support structures and assembly procedures have been established and studied.

The physics potential of CLIC is assessed through full detector simulation studies of benchmark physics processes, using the CLIC detector including beam-induced backgrounds. 
In parallel, dedicated phenomenological studies using parameterised detector performance are being pursued. Results from both full and fast simulation studies are reported.
It is shown that the initial stage of CLIC at $380\,\GeV$ gives access to precision measurements of the Standard Model Higgs boson and the top quark. 
To this end the first stage also foresees  a top-quark pair-production threshold scan around $350\,\GeV$.
The second stage at $1.5\,\TeV$ opens more Higgs channels including $\ttbar\PH$, double-Higgs production, and rare decays, and allows direct sensitivity to many BSM models.  
The third stage at $3\,\TeV$ gives the best sensitivity to new physics and double-Higgs production through Higgs self-coupling.

CLIC accelerator technology has reached a mature state and is increasingly being put to use in accelerator projects around the globe.
A detector design concept exists, and technology demonstrators for the sub-detectors have been built.
A work-plan for the preparation phase towards building CLIC is outlined in this report.
The CLIC accelerator and detector can be ready for a construction start around 2026.
First collisions at the 380\,GeV energy stage would then take place towards 2035.
CLIC provides excellent sensitivity to Beyond-Standard-Model physics effects, through direct searches and via a broad set of precision Standard Model physics measurements that reach well beyond the projections for HL-LHC.
In summary, CLIC represents a compelling opportunity for the post-LHC era.

\section*{Acknowledgements}
This work benefited from services provided by the ILC Virtual Organisation, supported by the national resource providers of the EGI Federation. 
This research was done using resources provided by the Open Science Grid, which is supported by the National Science Foundation and the U.S. Department of Energy's Office of Science.
This work was supported by
the European Union's Horizon 2020 Research and Innovation programme under Grant Agreement No.\,654168 (AIDA-2020);
the European Union's Horizon 2020 Research and Innovation programme under Grant Agreement No.\,777431 (CompactLight);
the European Union's Horizon 2020 Marie Sklodowska-Curie Research and Innovation Staff Exchange programme under Grant Agreement No.\,645479 (E-JADE); 
the National Commission for Scientific and Technological Research (CONICYT), Chile;
the DFG cluster of excellence ``Origin and Structure of the Universe'', Germany;
the Federal Ministry of Education and Research (BMBF), Germany under Grant Agreement No.\,05H18VKRD1; 
the Israel Science Foundation (ISF);
the I-CORE Program, Israel;
the Israel Academy of Sciences;
the Programma per Giovani Ricercatori ``Rita Levi Montalcini'' of the Ministero dell'Istruzione, dell'Universit\`a e della Ricerca (MIUR), Italy; 
the Research Council of Norway;
the National Science Centre, Poland, HARMONIA project under contract UMO-2015/18/M/ST2/00518 and OPUS project under contract UMO-2017/25/B/ST2/00496;
the Polish Ministry of Science and Higher Education under contract No.\,3501/H2020/2016/2 and 3812/H2020/2017/2; 
the Ministry of Education, Science and Technological Development of the Republic of Serbia under contract No.\,OI171012;
the Spanish Ministry of Economy, Industry and Competitiveness under projects MINEICO/FEDER-UE, FPA2015-65652-C4-3-R, FPA2015-71292-C2-1-P and FPA2015-71956-REDT; 
the Generalitat Valenciana under grant PROMETEO/2018/060, Spain; 
the IFIC, IFCA, IFT and CIEMAT grants under the Centro de Excelencia Severo Ochoa and Maria de Maeztu programs, SEV-2014-0398, MDM-2017-0765, SEV-2016-059, MDM-2015-0509, Spain;
the Swedish Research Council;
the Swiss National Science Foundation FLARE and FORCE grants 147463, 141146, 135012, 131428, 125272 and 126838;
the Scientific and Technological Research Council of Turkey (TUBITAK) under grant number 118F333; 
the UK Science and Technology Facilities Council (STFC), United Kingdom;
and the U.S. Department of Energy, Office of Science under contract DE-AC02-06CH11357.

\newpage

\printbibliography[title=References]

\end{document}